\documentclass[10pt,onehalfspacing]{myphdthesis}
\usepackage{nomencl} 
\usepackage{rotating}
\makenomenclature
\usepackage{mathrsfs}
\usepackage{mathtools}
\usepackage[multiple]{footmisc}
\newcommand{\reduceVspace}{\vspace*{-10mm}}
\newcommand{\increaseVspace}{\vspace*{10mm}}
\usepackage{hhline} 
\title{Separating Simultaneous Seismic Sources using Robust Inversion of Radon and Migration Operators}
\author{Amr Ahmed Mahmoud Ibrahim}
\degree{Doctor of Philosophy}
\specialization{Geophysics}
\copyrightyear{2015}
\copyrightseason{Fall}
\supervisor{Mauricio D.~Sacchi, Physics}
\committeeone{Jeffrey Gu, Physics}
\committeetwo{Erik Rosolowsky, Physics}
\committeethree{Richard Marchand, Physics}
\committeefour{Ivan Mizera, Mathematics}
\external{Eric Verschuur, Faculty of Applied Sciences, Delft University of Technology}
\hypersetup{
  pdfauthor = {Amr Ahmed Mahmoud Ibrahim},
  pdftitle = {Separating simultaneous seismic sources using robust Radon and migration operators, Ph. D. Thesis},
  pdfsubject = {Physics with a specialization in Geophysics},
  pdfkeywords = {Radon Transform, thesis, robust, inversion, source separation}
}
\begin{document}
\frontmatter
\pagenumbering{gobble}
\maketitle
\pagenumbering{roman}
\setcounter{page}{2}
\makesignature
\begin{abstract}

The advent of high density 3D wide azimuth survey configurations has greatly increased the cost of seismic acquisition. Simultaneous source acquisition presents an opportunity to decrease costs by reducing the survey time. Source time delays are typically long enough for seismic reflection energy to decay to negligible levels before firing another source. Simultaneous source acquisition abandons this minimum time restriction and allows interference between seismic sources to compress the survey time. Seismic data processing methods must address the interference introduced by simultaneous overlapping sources. 

Simultaneous source data are characterized by high amplitude interference artefacts that may be stronger than the primary signal. These large amplitudes are due to the time delay between sources and the rapid decay of seismic energy with arrival time. Therefore, source interference will appear as outliers in de-noising algorithms that make use of a Radon transform. This will reduce the accuracy of Radon transform de-noising especially for weak signals.  Formulating the Radon transform as an inverse problem with an $\ell_1$ misfit makes it robust to outliers caused by source interference. This provides the ability to attenuate strong source interference while preserving weak underlying signal.

In order to improve coherent signal focusing, an apex shifted hyperbolic Radon transform (ASHRT) is used to remove source interferences. ASHRT transform basis functions are tailored to match the travel time hyperbolas of reflections in common receiver gathers. However, the ASHRT transform has a high computational cost due to the extension of the model dimensions by scanning for apex locations. By reformulating the ASHRT operator using a Stolt migration/demigration kernel that exploits the Fast Fourier Transform (FFT), the computational efficiency of the operator is drastically improved. 

Moreover, the computational efficiency of the Stolt-based ASHRT operator allows us to extend the model dimension to fit seismic diffractions with the same accuracy as seismic reflections. The Asymptote and Apex Shifted Hyperbolic Radon Transform (AASHRT) can better focus diffracted energy by extending the basis functions to account for the asymptote time shift associated with diffractions.  This transform is used to interpolate synthetic data that contain significant amount of seismic diffractions. The results of the interpolation tests show that the AASHRT transform is a powerful tool for interpolating seismic diffractions.

\end{abstract}

\begin{preface}

A version of the work in chapter 4 of this thesis has been published in the journal paper Ibrahim, A. and M. D. Sacchi. ``Simultaneous source separation using a robust Radon transform.'' Geophysics 79 (2014): V1--- V11.  In this publication, I was responsible for designing the processing algorithms, generating the data examples and writing the manuscript.  Dr. Sacchi was the supervisory author and was involved in the concept formulation and manuscript editing. 

A version of the work in chapter 5 of this thesis has been published in the journal paper Ibrahim, A. and M. D. Sacchi. ``Fast simultaneous seismic source separation using Stolt migration and demigration operators." Geophysics 80 (2015): WD27---WD36. In this publication, I was responsible for designing the processing algorithms, generating the data examples and writing the manuscripts. Dr. Sacchi was the supervisory author and was involved in the concept formulation and manuscripts editing.  

Parts of the research work presented in chapter 6 of this thesis are part of research collaboration with Dr. Paolo Terenghi at Petroleum Geo-Services (PGS), UK. A version of our work has been published in the peer-reviewed conference proceeding Ibrahim, Amr, Mauricio D. Sacchi, and Paolo Terenghi. ``Wavefield reconstruction using a Stolt-based asymptote and apex shifted hyperbolic Radon transform." 85th Annual International Meeting, SEG, Expanded Abstracts (2015): 3836---3841. In this publication, I was responsible for generating the data examples and writing the manuscript. Dr. Terenghi and Dr. Sacchi were the supervisory authors and were involved in the concept formulation and manuscripts editing. The synthetic data examples in this work were generated by Dr. Terenghi and processed by me.  Applications of the proposed Radon transform is included in the patent application Hegge, Robertus Franciscus, Paolo Terenghi, Andreas Klaedtke, and Amr Ibrahim. ``Prediction and substraction of multiple diffractions." Provisional Patent application no. 62/195,459 (2015).  

\end{preface}

\begin{dedication}
\emph{To my parents and my teachers}
\end{dedication}

\begin{acknowledgements}
First and foremost, I would like to thank my supervisor Dr. Mauricio D. Sacchi for his great supervision, guidance and patience during my PhD years. He was always encouraging and supportive during all the stages of my research.  I am honored to be one of his students. 
I would like to thank all the members of my PhD committee for taking the time from their busy schedules to review my thesis. Special thanks to Dr. Jeff Gu for his course on global seismology which was my first real geophysics course. My term paper for his course was my first interaction with Radon transforms and its applications. 
I would like to acknowledge Dr. Daniel Trad at for his helpful discussion and comments about my research. His work on Radon transform and Stolt migration was the inspiration for much of the work in this thesis. He was always ready to provide me with advice and guidance when I ask for it. 
My special thanks to Dr. Paolo Terenghi at PGS UK office for his inspiring discussions and ideas about the applications of my work to seismic interpolation. Our collaboration on extending Radon transform to handle seismic diffractions added substantially to my thesis. I am also grateful for his support during my internship at Petroleum Geo-Services (PGS), UK. I am also very thankful for Dr. Andreas Klaedtke the section manager at PGS UK for his constructive comments about my research. 
I would also like to thank WesternGeco company for making the Mississippi Canyon data available to our research group.  
I would like to thank all my current and former colleagues at the Signal Analysis and Imaging Group for their help and useful discussions. Special thanks to Nadia Kreimer for helping me with using Seismic Unix generate seismic images and cubes. I would like to thank Aaron Stanton who was always ready to answer all my questions about coding and seismic processing. I would like to thank Emmanuel Bongajum for his help during the start my PhD and taking the time to explain many new concepts to me. 
At last but not least, I would like to thank my mother, father, siblings and my dear Heba for support and their patience during my long studies in Canada.   
 
\end{acknowledgements}

\maketoc 
\makelot 
\makelof 
\makelos 
\mainmatter
\chapter[Introduction]
{
	Introduction
}

Four geophysical methods are commonly used for oil and gas exploration, gravity, aero-magnetic, electromagnetic and seismic methods. Gravity and aero-magnetic methods provide low resolution subsurface images that cover large area with comparatively low cost. Therefore, they are mainly used as initial reconnaissance methods to locate areas of interest for oil and exploration. Electromagnetic methods which were widely used in the mining industry, have been used recently in oil and gas exploration \citep{Ashcroft2011}. Electromagnetic methods can be used to detect the presence of hydrocarbons in the Earth subsurface. However, subsurface images estimated by electromagnetic methods have limited vertical resolution compared to images estimated by seismic methods. Therefore, the majority of oil and gas exploration activities use seismic exploration methods to estimate subsurface images.  
 
Seismic exploration methods can be classified according to the mode of seismic wave transmission into seismic refraction and seismic reflection methods. Seismic refraction methods are usually used to explore shallow depths and require large spatial separation between source and receivers. Therefore, seismic refraction methods are not widely used in oil and gas exploration \citep{Gadallah2005}. On the other hand, seismic reflection methods are the backbone of current oil and gas exploration activities. Exploration using seismic reflection methods consists mainly of two steps, acquiring seismic reflection data and processing this data to estimate the subsurface image. The acquisition step begins by firing an artificial seismic source (or sources) near the Earth surface to generate seismic waves that can travel deep into the subsurface. Seismic waves will travel through the subsurface with part of its energy getting reflected at the boundaries between different rock layers. These reflections are caused by the changes in rock properties between the subsurface layers. Changes in wave velocity and rock density across the different rock layers will change the acoustic impedance. These sudden changes in acoustic impedance will cause seismic reflections at the boundaries between the different subsurface layers. 
The amplitude of reflected seismic waves is recorded as a function of the wave arrival time at the surface by an array of sensors (receivers). These recorded seismic traces that belong to one seismic source is known as a seismic shot gather. The seismic shot gather experiment is repeated by spatially displacing the source and receivers according to a predefined acquisition geometry \citep{Ikelle2005}.   

In seismic data processing, the recorded seismic data passes through a sequence of processing steps. These steps are designed to remove undesired signals (such as noise and multiples), remove the source and receiver signatures (deconvolution) and regularize the sampling grid (interpolation). Finally, an imaging algorithm uses the seismic data to estimate a subsurface image, a process commonly known as seismic migration \citep{Berkhout1982,Biondi2006}. All the shot gathers can be combined (stacked) prior to imaging in post-stack migration algorithms. Alternatively, shot gathers data can be used separately to estimate partial images of the subsurface in pre-stack migration algorithms. In either case, seismic migration algorithms use seismic reflections recorded at the surface to trace back the reflected waves to its subsurface reflection points. The boundaries between subsurface layers are imaged by mapping all reflections recorded at the surface back to their subsurface reflections points. Additional information about the rock properties of the subsurface layers can be estimated from the amplitude variation with offset (distance between receiver and source) /angle (incidence angle of wave) \nomenclature{AVO/AVA}{Amplitude Versus Offset/Angle}(AVO/AVA).  Subsurface images and AVO/AVA information are used to delineate geological structures and locate possible oil and gas accumulations. The accuracy of this information will increase success rate for drilling new oil and gas reserves and reduce the overall production cost.  Additionally, seismic methods can be used to monitor and guide oil production in enhance oil recovery techniques such as steam-assisted gravity drainage (SAGD) \nomenclature{SAGD}{Steam-Assisted Gravity Drainage} \citep{Eastwood1994,Byerley2009}. In this case, the seismic imaging process is repeated over time (time lapse/ 4D seismic) to monitor the changes in reservoir rock properties due to enhanced oil recovery. This will provide the information needed to control and optimize the enhance oil recovery process.  

The accuracy and the economic efficiency of seismic methods are important for the success of both exploration and production. Seismic imaging accuracy is directly related to the amount of recorded seismic reflections that are produced by the subsurface target (seismic illumination).  Seismic acquisitions are usually designed to maximize the amount of the information recorded about the targeted subsurface layers. This usually requires increasing the sampling density and/or the range of receivers and sources. Traditionally, seismic surveys increase the range and the sampling density of seismic receivers on the expense of seismic sources. This imbalance between the receiver and the source sides of seismic acquisitions is especially prominent in the towed streamer marine acquisition \citep{De-Kok2002}. This imbalance is the direct result of the relatively lower cost of deploying more seismic receivers to the cost of deploying more seismic sources. Furthermore, increasing the number of sources increases the total survey time and thereby increases its cost. The cost of seismic acquisition can be a significant factor that affects the design and implementation of seismic exploration surveys. Seismic acquisition cost can be high enough to impact whether or not developing a hydrocarbon reservoir is economical \citep{Beasley2008LeadingEdge}.

In recent years, seismic acquisition costs are becoming a more significant factor in the design of seismic surveys.  The industry is shifting to explore more difficult to image subsurface areas. Exploring these areas requires expensive 3D wide azimuth surveys to increase the seismic illumination of the subsurface and improve imaging accuracy. Wide azimuth surveys increase seismic illumination by increasing the density and range of both the receivers and sources grids. The cost of seismic acquisition is related to many factors and it is a complicated topic to cover in detail. However, there is a general agreement that seismic acquisition cost is proportional to the total survey time \citep{Beasley1998}. Therefore, designing efficient seismic acquisitions that reduce survey time is an intense topic of research in seismic exploration. Improving acquisition efficiency to reduce survey time is the most often cited reason for developing simultaneous seismic sources \citep{Beasley1998, Berkhout2008, Hampson2008, Fromyr2008}. In order to understand simultaneous seismic sources and the motives behind its development a brief introduction to seismic acquisitions is needed.

\section{Seismic acquisition} 

\begin{figure}[htbp]
\centering
\includegraphics{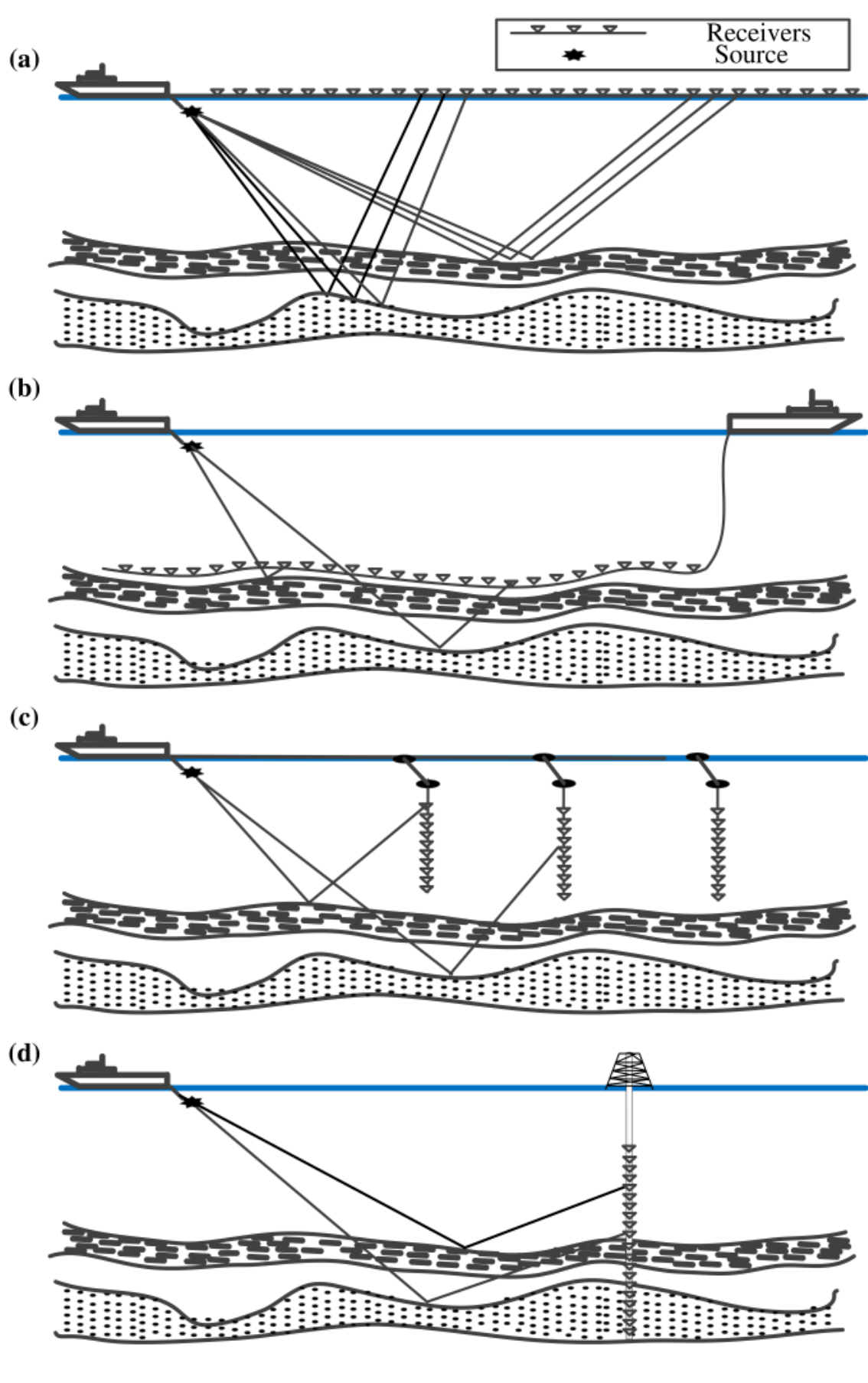}
\caption{The distribution of sources and receivers in marine acquisition. (a) Towed streamer. 
(b) Ocean bottom seismic/Ocean bottom cable (OBS/OBC). (c) Vertical cable. (d) Walkaway VSP.}
\label{ch1_marine_acq_confgs}
\end{figure}

Seismic acquisitions are usually sorted with respect to the acquisition environment into marine (off-shore) and land (onshore) acquisitions. In marine seismic acquisitions, the most commonly used configurations are towed streamer, ocean bottom sensors/cable (OBS/OBC)\nomenclature{OBS/OBC}{Ocean Bottom Seismic/Cable}, vertical cable, walkaway vertical seismic profile (VSP) \nomenclature{VSP}{Vertical Seismic Profile} \citep{Ikelle2005}. In towed a streamer configuration, seismic receivers are contained in cables which are called streamers and are floated horizontally near the sea surface.   In the OBS/OBC configuration, the receivers are distributed along the sea floor either as separate sensors (OBS) or contained in cables (OBC). In both vertical cable and walkaway VSP, seismic receivers are distributed vertically below the seismic source. In vertical cable configuration, the receivers are located in the water column.  On the other hand, the receivers in a walkaway VSP are located in the borehole deep into the subsurface. Figure \ref{ch1_marine_acq_confgs} displays the schematics of these four different marine acquisition configurations. In land seismic acquisitions, the most commonly used configurations are surface seismic, vertical cable and vertical seismic profile (VSP).  In surface seismic acquisition, seismic sources and receivers are placed along the Earth surface or buried at shallow depth. In both vertical cable and VSP, the sources are at the surface and the receivers are placed vertically in a borehole.  However, the borehole in the VSP acquisition is as deep as the targeted oil reservoir \citep{Ikelle2005}. 

The geometry of the source and receiver grids plays a major role in determining both the acquisition cost and the imaging accuracy.  In the simple 2D acquisition geometry, seismic sources and receivers are deployed and displaced along a straight line which reduces the cost considerably. In 2D acquisition, the seismic wavefield is sampled over two dimensions, the arrival time (which represents the depth) and one spatial dimension along the surface. This simplifies data processing by ignoring the 3D nature of the reflected seismic wavefield. However, 2D acquisition geometry reduces the subsurface imaging accuracy especially for complicated subsurface structures. Since the easy-to-find oil and gas reserves are exhausted, the industry is forced to explore the more difficult to image subsurface targets. These targets require the accurate subsurface imaging provided by 3D seismic acquisition geometries. Therefore, 2D seismic acquisitions are replaced by the more advanced 3D acquisition geometries \citep{Ikelle2005,Long2010}.

\begin{figure}[htbp]
\centering
\includegraphics{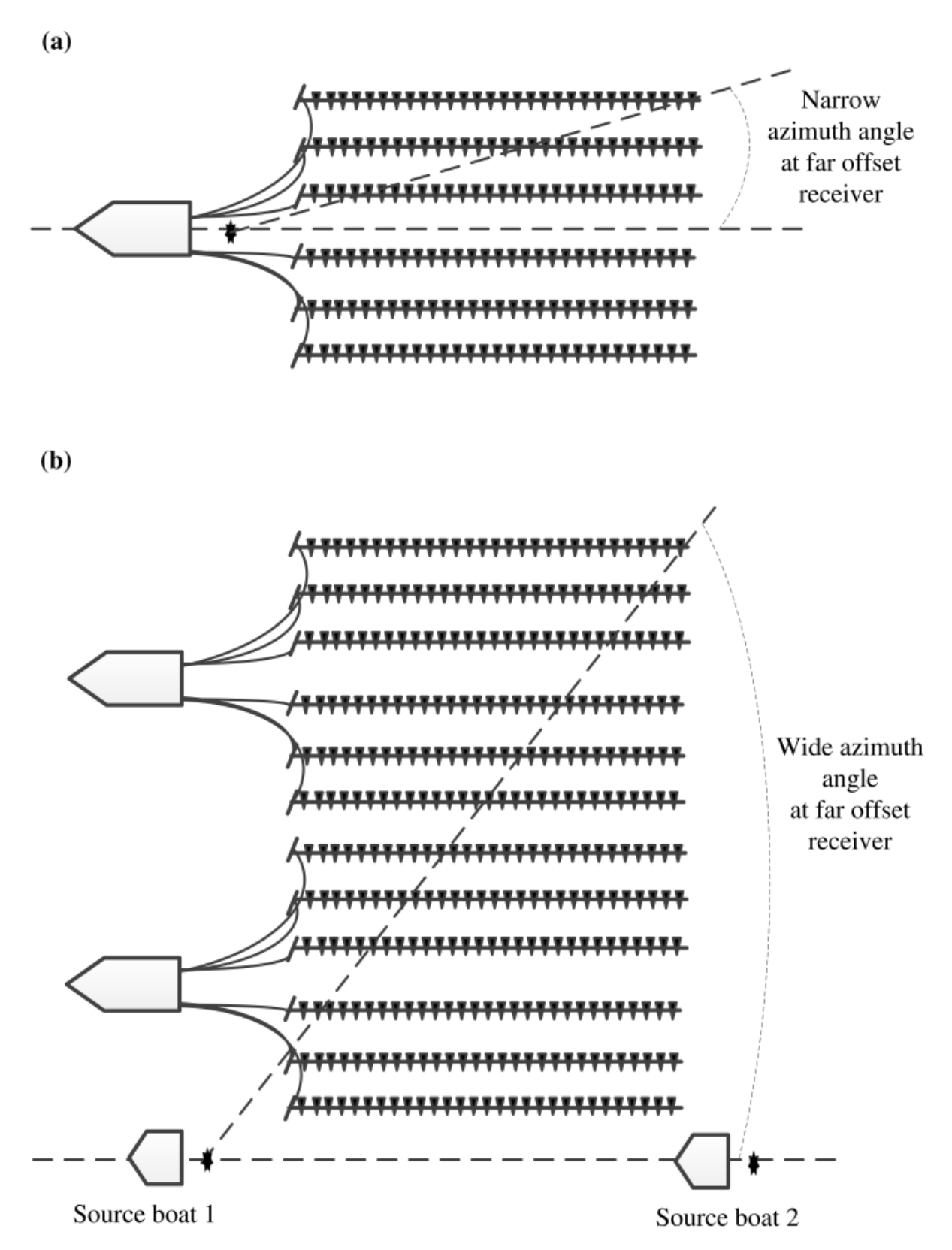}
\increaseVspace
\caption{ (a) Conventional narrow azimuth (NAZ) streamer survey. 
(b) Wide azimuth (WAZ) streamer survey using two boats.}
\label{ch1_azimuth}
\end{figure}

Another factor that plays a major role in determining both seismic acquisition cost and the imaging accuracy is the environment of the target area. Challenging exploration areas such as deep water, sub-salt structures, areas affected by basalts or areas affected by thick carbonates require better subsurface seismic illumination than what is available from the standard narrow azimuth (NAZ)\nomenclature{NAZ}{Narrow Azimuth survey} 3D surveys \citep{Long2010}. The receiver azimuth angle is the angle between source-receiver direction and survey direction (see Figure \ref{ch1_azimuth}). Azimuth angle represents the direction of seismic illumination of the subsurface target. Azimuth angles distribution is important for seismic survey design since accurate imaging requires uniform subsurface illumination. In land surface surveys, seismic sources and receivers are physically decoupled. Therefore, it is theoretically possible to achieve a perfect azimuth (PAZ) survey \nomenclature{PAZ}{Perfect Azimuth survey}\citep{Long2010}. However, the high cost of land acquisition results in sparse sampling of both the sources and the receivers grids. 

This situation is different for marine streamer surveys which are the most commonly used survey type in seismic acquisitions. Marine streamer surveys acquire seismic data in overlapping shot gathers in which the sources and receivers cannot be moved independently from each other. Additionally, the number of towed streamers is limited and the streamers have narrow spatial spread along the crossline direction (the direction perpendicular to the streamer).  Moreover, restrictions on the minimum boat speed and minimum time interval between successive sources make dense source sampling expensive. All these reasons generally results in towed streamer surveys having acceptable receiver density while the source density is lower than desirable. Therefore, with the exception of the near offset receivers, the range of receiver azimuth angles in standard towed streamer survey is limited (see Figure \ref{ch1_azimuth}a). This indicates that the subsurface target is only illuminated from one particular direction which can lead to unacceptably poor imaging. The seismic illumination problem is more severe for targets with significant wave scattering like salt bodies \citep{Fromyr2008}. Moreover, steeply dipping subsurface structures tend to increase seismic illumination non-uniformity due to the backward geometry of the wave trajectory in the subsurface \citep{Beasley1998patent}. Therefore, the demand for advanced azimuth acquisitions has been increasing in recent years which increased seismic acquisition cost significantly. 
  
There are several types of advance azimuth acquisition surveys such as wide azimuth survey (WAZ)\nomenclature{WAZ}{Wide Azimuth survey} \citep{Michell2006}, rich azimuth (RAZ)\nomenclature{RAZ}{Rich Azimuth survey} \citep{Howard2006}, multi azimuth survey \nomenclature{MAZ}{Multi Azimuth survey} \citep{Keggin2006} or full azimuth survey using coil shooting \citep{Ross2008}. The design of advanced azimuth surveys is outside the scope of this thesis. However, the common aim of all these advanced azimuth acquisitions is to increase the azimuth coverage by increasing sampling range and density of receives and/or sources \citep{Long2010}. Figure \ref{ch1_azimuth} shows an example of conventional NAZ survey using one boat and a wide azimuth survey using two streamer boats and two source boats. Increasing the number of sources/streamers will increase azimuth coverage and consequently improve the subsurface illumination but it will also increase the cost considerably. The high cost of advanced azimuth surveys results mainly from employing more boats to cover the same area. Moreover, increasing the total number of seismic sources will increase the total survey time and increase the operational costs. The minimum time constraint on firing sequential seismic sources is the main bottleneck for reducing the total survey time. In conventional acquisition, seismic sources are fired sequentially with long time delay between successive sources. The time delay has to be long enough so that the energy of reflections will decay to negligible levels before firing another source. Simultaneous seismic sources acquisition abandons this minimum time restriction and allows interference between seismic sources. 

Simultaneous seismic sources are especially beneficial for reducing the cost of advanced azimuth surveys that use multiple source boats. Additionally, simultaneous seismic sources can be used to increase the total number of source to improve illumination without significant cost increase. The main drawback of simultaneous seismic sources is source interferences that can negatively impact the imaging accuracy. There are two main approaches for processing simultaneous seismic sources data. In the first approach, each source data is separated and then a conventional processing sequence is used to image the subsurface. In the second approach, the imaging algorithms are modified to handle source interferences and image simultaneous seismic sources data directly without source separation.  Research in recent years showed promising results for simultaneous source separation and imaging. Therefore, simultaneous seismic sources are becoming an accepted technique by the oil and gas industry.

\section{Motivations}

Despite the development of several migration algorithms that can image simultaneous source data directly without source separation, access to pre-stack single source data is still necessary for AVO/AVA studies \citep{Peng2013,Chengbo2014,Davies2015}. Also, the industry still prefers to evaluate the quality of separated source data prior to imaging. Current source separation methods can be sorted into two main categories: denoising-based methods \citep{Akerberg2008,Huo2009,Maraschini2012,Ibrahim2014GEO} and inversion-based methods \citep{Wason2011,Cheng2015}. Denoising-based methods treat source interferences as noise and remove them by differentiating coherent primary signals from incoherent source interferences. Several denoising methods use inversion and impose the sparsity of seismic data in an appropriate transform domain to suppress source interferences. On the other hand, the inversion-based separation methods formulate the source separation problem as an inversion problem. As expected, the source separation inversion problem is an ill-posed problem and an inversion constraint is needed. Again, data sparsity in an appropriate transform domain can serve as an inversion constraint to estimate the solution. 

The sparse inversion constraint is widely used in both denoising-based and inversion-based source separation methods. However, the inversion cost function for denoising-based separation method is different from the inversion-based separation method. There are two main differences between the cost functions of the two source separation categories. The first difference is related to the size of the inversion problem. Inversion-based separation methods estimate all single sources from their combined simultaneous source data in one inversion problem. On the other hand, the denoising-based separation methods estimate only a portion (a seismic gather) of the single source data in one inversion problem. Also, the inversion-based separation methods use the combined operators of blending and the sparse transform while denoising-based methods use the sparse transform operator only. The size of the problem and the operator computational cost is important since iterative sparse solvers require large number of repetitive forward/adjoint operator computations. Therefore, denoising-based separation methods are more computationally efficient than the inversion-based separation methods. Additionally, denoising-based separation methods divide the separation problem into smaller inversion problems which is beneficial for parallel computation and quality control. 

The second difference between the two categories is in the treatment of the noise term in the data. In the inversion-based methods, the noise that appears in the misfit between estimated and observed data is due to small fitting errors, random noise and signals not accounted for by the transform basis functions. In this case, the standard least-squares ($\ell_2$) norm misfit is efficient since it can handle small errors that follow a Gaussian probability distribution.  On the other hand, the noise that appear in the misfit between the estimated and the observed data in the denoising-based separation is composed from source interferences. In this case, the misfit cannot be assumed to be small or follow a Gaussian probability distribution. Therefore, one cannot use the $\ell_2$ norm misfit. Furthermore, source interferences will appear as incoherent noise with large amplitudes in common receiver gathers due to the random time delays between sources. In this case, source interferences will appear as outliers for the denoising algorithms that use common receiver gathers. These outliers can degrade the solution accuracy especially for the low amplitude primary signal mixed with large amplitude source interferences. \cite{Claerbout1973} pioneering work studied the problem of outliers that arise in seismic processing due to erratic seismic noise. They suggested using an $\ell_1$ norm misfit which is more robust to outliers instead of the conventional $\ell_2$ norm misfit. \cite{Lynn1987} provided an early study for the unintentional source interferences by different seismic crews in the Gulf of Mexico and north sea. They noted the distorting effects of high amplitude interferences and suggested using $\ell_1$ norm misfit for AVO/AVA studies of data contaminated by source interferences. Several researchers noted the deficiency of denoising-based source separation when strong interferences mix with weak signals \citep{Akerberg2008,Beasley2008SEG,Kim2009}. \cite{vanBorselen2012} noted that even Radon transforms that exploit the power of stacking can have difficulty distinguishing weak coherent signal from strong incoherent interferences. For these reasons, incorporating robust inversion into denoising-based source separation can improve accuracy and preserve weak signals \citep{Ibrahim2013SEG,Ibrahim2014GEO}.  

Another important factor that impacts source separation is the choice of a suitable transform for sparse denoising. The main justification for using the sparsity constraint is that seismic data can be represented by few (sparse) coefficients in the domain of a suitable transform. Sparse inversion have proved to be a powerful tool in seismic data processing especially for denoising \citep{Rodriguez2012, Yu2015} and interpolating \citep{Sacchi1998,Zwartjes2007a}. However, the choice of a suitable transform to sparsely represent the observed data is critical for the success of sparse inversion applications. The transform ability to focus seismic data into a sparse transform model is directly related to the similarity between the transform basis functions and the data main components. For this reason, it is desirable to tailor the transform basis functions to resemble the main components of seismic data as close as possible. The design of an efficient mathematical transformation that can sparsely represent the data is an important and extensive research topic. In general, these transforms can be divided into two main categories, model-driven transforms and data-driven transforms \citep{Zhu2015}. Model-driven transforms are based on a formulated mathematical model of the data which leads to pre-defined basis functions (dictionary). For example, the mathematical model for Fourier transform assumes that the data is composed from sinusoidal basis functions.  Several model-driven transforms have been used for sparse inversion of seismic data such as Fourier \citep{Sacchi1998}, Radon \citep{Thorson1985,Sacchi1995}, wavelet \citep{Sinha2005}, curvelet \citep{Herrmann2008}, seislet \citep{Fomel2010} and contourlet \citep{Shan2009}. On the other hand, data-driven transforms learns the transform basis functions from the data and do not have pre-defined transform basis functions. For example, principal component analysis (PCA)  \nomenclature{PCA}{Principal Component Analysis} transforms the data into new coordinate system \citep{Jolliffe2002}. The new coordinate system is chosen such that the eigenvalues of the data are aligned with the coordinate axes in decreasing order of magnitude. Data-driven transforms learn its basis functions from the data which tailor the transform basis functions for sparse representation.  Therefore, data-driven transforms can be more adaptable to changes in the structure of observed seismic data than the model-based transforms. Recently, data-driven transforms have been used in seismic data processing for denoising \citep{Chen2015,Beckouche2014,Zhu2015}, interpolation \citep{Yu2015} and source separation \citep{Cheng2015}. However, data-driven transforms have a high computational cost and include explicit storage of the transform basis functions learned from the training data sets used in dictionary learning. Therefore, it is desirable to design computationally efficient model-based transforms with pre-defined basis functions that match the main components of seismic data \citep{Ibrahim2014SEG,Ibrahim2014EAGE,Ibrahim2015GEO,Ibrahim2015SEG,Ibrahim2015patent}.

\section{Contribution}

In this thesis, robust inversion is incorporated into the denoising-based source separation problem by replacing the conventional $\ell_2$ norm misfit with $\ell_1$ norm misfit. This is commonly known as robust inversion since the $\ell_1$ norm is robust with respect to large outliers than the conventional $\ell_2$ norm misfit. Robust inversion is more suitable for the denoising-based source separation problem than sparse inversion that uses sparse model constraint. The accuracy of sparse inversion is negatively impacted by the large fitting errors that arise from source inferences, especially for low amplitude reflections. The contamination of weak seismic reflections with strong source interferences is an unavoidable situation in simultaneous seismic sources. The rapid decay in the energy of seismic reflections combined with the time delays between simultaneous seismic sources produce these strong interferences. Even in the case of spatially separated and nearly simultaneous sources, strong source interferences will appear at far offset due to the hyperbolic moveout of seismic reflections \citep{Beasley2008LeadingEdge}. 

The robust inversion denoising was tested using numerically blended simple synthetics and field data from the Gulf of Mexico. Four possible inversion cost functions were formulated, two non-robust and two robust. In each case, the transform model was either constrained using sparsity ($\ell_1$ norm) or the conventional least-squares ($\ell_2$ norm) regularization. All cost functions were solved using iteratively re-weighted least-squares (IRLS) \nomenclature{IRLS}{Iteratively Re-weighted Least-Squares} solver \citep{Holland1977,Daubechies2010}.  The IRLS solver is used because it can be easily adapted to solve all the four inversion cost functions which facilitate a fair comparison. Comparison results showed that misfit robustness is more efficient than model sparsity in removing strong source interferences and preserving weak signals \citep{Ibrahim2013SEG,Ibrahim2014GEO}. This was especially true for the field data example due to the difficulty of imposing a strict sparsity constraint. This difficulty arises from the mismatch between the transform basis functions and the seismic reflections which reduce the model sparsity. For example, Radon transform basis functions do not account for amplitude variations with offset of seismic reflection hyperbolas. More significantly, the travel times used in Radon transform and the actual travel times of seismic reflections usually do not match. This mismatch arises from the coarse sampling of the Radon parameters, such a velocity and apex location. Radon transform parameters are coarsely sampled in practical implementations to decrease computational cost and facilitate the inversion by limiting the model space.  

In order to optimize the Radon operator focusing power, we used the apex shifted hyperbolic Radon transform (ASHRT) \nomenclature{ASHRT}{Apex Shifted Hyperbolic Radon Transform} for denoising \citep{Ibrahim2014GEO}. The ASHRT transform basis functions are tailored to match the reflection travel time hyperbolas in common receiver gathers. The ASHRT model is an extension to the conventional hyperbolic Radon transform (HRT) model \nomenclature{HRT}{Hyperbolic Radon Transform} commonly used in processing common mid gathers. The HRT transform basis functions are derived with the assumption that the minimum travel time for the seismic reflection hyperbola (the apex of the hyperbola) is located at zero offset. The zero offset apex assumption in common receiver gathers is only valid for subsurface structures that consist of horizontal layers. The ASHRT model uses hyperbolic basis functions with different apex locations in order to match seismic reflection travel time hyperbolas more closely.  Tailoring the transform basis functions to match the data will improve the transform ability to focus seismic reflections and attenuate noise. Source separation tests using the ASHRT transform showed its efficiency as a tool for focusing reflections and attenuating source interferences \citep{Ibrahim2013SEG,Ibrahim2014GEO}.  The main drawback for the ASHRT transform is its computational cost due to the extension of the model dimensions by scanning for the apex locations. Moreover, the ASHRT operator belongs to the time variant Radon operator category which cannot be computed efficiently in the frequency domain. 

In order to improve computationally efficiency, we designed a new ASHRT operator that use the Stolt migration/demigration operators as its kernel. The Stolt migration model \citep{Stolt1978} is derived by solving the seismic wave equation in the frequency-wavenumber ($f-k$) domain with a constant velocity subsurface assumption. The Stolt operator focuses seismic reflections back to its subsurface reflection points by mapping the seismic data in the $f-k$ domain. The Stolt migration operator is considered to be the most computationally efficient migration operator. The computational efficiency of Stolt operator results from employing Fast Fourier Transform (FFT) \nomenclature{FFT}{Fast Fourier Transform}operators. Moreover, seismic data have a band limited structure in the $f-k$ domain which speeds up the Stolt $f-k$ mapping considerably. However, the Stolt transform model is derived using a constant subsurface velocity assumption which limits its applications in seismic migration. The Stolt transform model can be viewed as a constant section across the ASHRT model. Therefore, we extend the single velocity Stolt operator to scan for multiple velocities and construct an ASHRT model. The Stolt-based ASHRT operator is equivalent to the time domain ASHRT operator but with higher computational efficiency \citep{Ibrahim2014SEG,Ibrahim2014EAGE}. Additionally, the Stolt-based operator improves the denoising accuracy since its basis functions match seismic reflections more closely than the time domain ASHRT operator. The Stolt-based ASHRT operator includes a scaling factor that accounts for the change in reflection hyperbola amplitude with incidence angle (obliquity factor). The new Stolt-based ASHRT operator is tested for the denoising-based source separation using robust inversion. Again, four different inversion scenarios are tested using numerically blended simple synthetics and field data from the Gulf of Mexico. Tests show that the Stolt-based ASHRT operator used in robust inversion can achieve both accurate and fast source separation \citep{Ibrahim2015GEO}. 

The computational efficiency of the Stolt-based ASHRT operator allows us to extend the model dimension to match seismic data more closely. Improving the transform operator ability to focus seismic data into sparse model is vital for many processing applications that rely on sparsity. For example, the accuracy of interpolation methods that use sparse inversion is related to the similarity between transform basis functions and seismic data.   In addition to seismic reflections, seismic data usually contain low amplitude seismic diffractions. Seismic diffractions have hyperbolic travel times similar to seismic reflections but with the hyperbola asymptote shifted in time. Unlike reflections, the seismic diffraction travel path in the subsurface is asymmetric, which causes the time delay of the diffraction hyperbola asymptote. Despite being weaker than seismic reflections, seismic diffractions are important part of seismic data. Seismic diffractions can be used to map the subsurface faults and increase the subsurface imaging resolution considerably \citep{Khaidukov2004}. In order to account for seismic diffractions, the Stolt-based ASHRT basis functions are extended to account for the asymptote time shift. We named this new transform Asymptote and Apex Shifted Hyperbolic Radon Transform (AASHRT) \nomenclature{AASHRT}{Asymptote and Apex Shifted Hyperbolic Radon Transform}. This transform is used to interpolate data which contain a significant amount of seismic diffractions. The results of the interpolation tests show that the AASHRT transform can be a powerful tool interpolating seismic diffractions \citep{Ibrahim2015SEG,Ibrahim2015patent}.

\section{Thesis outline}

{\bf Chapter 2} gives an overview of the motivations for developing simultaneous seismic sources and its potential benefits for wide azimuth surveys and time lapse seismic. The chapter introduces the background of simultaneous seismic development from land vibratory sources to marine impulsive sources applications.  It reviews the current research trends in processing and designing simultaneous seismic sources. An important part of this chapter is the description of the different source separation methods. It introduces the denoising-based source separation that utilizes the incoherency of source interferences in common receiver gathers. The denoising-based source separation will be used throughout this thesis.

{\bf Chapter 3} introduces different types of Radon transforms used for seismic data processing. 
It briefly reviews the development of Radon transforms and their applications in seismic data processing.
The low resolution of the adjoint operator associated with Radon transform non-orthogonality is demonstrated using a simple synthetic example. The same example is used to demonstrate the power of sparse Radon transforms for denoising and/or interpolation applications. 

{\bf Chapter 4} introduces the concept of robust Radon transform using the $\ell_1$ misfit function. 
The robust Radon transform is formulated as an inversion problem.
The inversion cost function is reformulated using weighted matrices. 
We examine four different inversion scenarios using misfit robustness and/or model sparsity. Numerically blended simple synthetic data and field data from the gulf of Mexico are used for separation tests. The simple synthetics example shows that adding misfit robustness into sparse inversion improves source separation by preserving weak reflections. The field data example shows that adding a robust misfit to the cost function is more critical than sparsity in suppressing inference and preserving weak reflections.

{\bf Chapter 5} introduces the Stolt migration operator which is derived from solving the seismic wave equation using constant subsurface assumption. The Stolt migration/demigration operators are used as the kernel of a Stolt-based ASHRT transform. The computational efficiency of the new Stolt-based ASHRT operator is compared with the conventional time domain ASHRT operator. Results show a significant increase in the computational efficiency using the Stolt-based ASHRT operator.  Again, four different inversion scenarios are formulated and solved using the IRLS algorithm. Results show that the Stolt-based operator can achieve both accurate and computationally efficient source separation.

{\bf Chapter 6} deals with extending the Stolt-based operator to account for the asymptote shift of seismic diffractions hyperbolas. The new ASSHRT transform is used for interpolating simple and complicated synthetic data that contain significant diffractions. Results show that the AASHRT transform can efficiently interpolate highly decimated data while preserving seismic diffractions.  Although this chapter can appear as a disconnected topic from the main purpose of this thesis, it is closely connected Radon transform and its applications. 

{\bf Chapter 7} contains the thesis work contributions, conclusions, limitations of the proposed methods, and suggestion for future work. 

\chapter[Simultaneous seismic sources ]
{
	Simultaneous seismic sources 
}

\section{Introduction} 

There are two possible approaches to reduce the total seismic survey time and thereby reduce its cost. The first approach reduces the total number of seismic sources used in the survey.  In order to achieve this, the spatial spacing between seismic sources is increased to cover the required survey area. This is considered to be the conventional approach to reduce survey cost. However, coarsely sampling the source grid can introduce unacceptable aliasing problems. Therefore, an interpolation processing step is usually required prior to imaging in order to estimate the data of missing sources. However, interpolation has its limitations and will not introduce any new information about the subsurface.   

A second and a more recent approach uses simultaneous seismic sources to reduce survey cost. This can be achieved by decreasing the time interval between sequential source firings for a single source acquisition (self-simultaneous source) \citep{Abma2013EAGE}. Alternatively, multiple sources can be fired simultaneously or near simultaneously to reduce the total survey time. Source interferences that contaminate recorded seismic data are the major drawback for simultaneous seismic sources. Source interferences can impact the accuracy of subsurface images estimated using simultaneous sources data. Therefore, source interferences must be removed prior to imaging or an appropriate imaging algorithm is used to handle them.  In recent years, the simultaneous seismic sources approach is becoming more accepted in the oil and gas industry \citep{Beasley2012,Abma2013EAGE,Ellis2013,Walker2014,Abma2015}. 

Simultaneous seismic sources are becoming more accepted since removing source interference is more preferable than interpolating missing sources data \citep{Berkhout2008,Berkhout2008SEG}. Unlike interpolation, increasing the total number of sources in a seismic survey will introduce new information about the subsurface. Therefore, images estimated with a dense source sampling grid should be more truthful to the subsurface structure than images estimated using interpolated sources. In addition to the cost reduction advantage, reducing the total survey time is beneficial for short acquisition time windows. Time limitations on seismic acquisitions commonly arise in oil and gas exploration due to safety or environmental restrictions. Also, simultaneous seismic sources introduce an additional degree of freedom in survey design that increases the operational flexibility of seismic surveys. For example, the speed of seismic streamer boats can be increased without compromising the density of the source sampling grid \citep{Berkhout2008SEG}.

Simultaneous seismic sources can be beneficial for time lapse (4D seismic) surveys to reduce both its acquisition cost and non-repeatability problems  \citep{Ayeni2009,Krupovnickas2012,Wason2014}. Time lapse surveys acquires and analyses repeated 3D seismic images of producing oil and gas fields to monitor changes in the rock properties of the reservoir. Changes in estimated subsurface images over time are assumed to be the result of changes in rock properties such as oil, water, or gas saturation \citep{Lumley2001}. Monitoring these changes accurately is important for oil production especially when enhanced oil recovery techniques are used. Therefore, subsurface imaging artefacts due to non-repeatable seismic acquisition conditions have to be kept to minimal.  Increasing the source sampling density using simultaneous seismic sources can lead to more accurate imaging and reduce the non-repeatability effects. Additionally, reducing the acquisition cost can reduce the time interval between acquiring repeated 3D seismic acquisitions. Increasing the acquisition for time-lapse images will provide more accurate and continuous reservoir monitoring \citep{Ayeni2009}.

Moreover, simultaneous seismic sources development aided the design of new acquisition geometries that increase the survey offset range.  In marine steamer acquisition, the offset range is usually limited by the operational conditions in the sea. The lateral deviation of the streamer from the towing direction due to sea currents (called feathering) restricts the maximum streamer length. This offset limitation has negative impact on imaging challenging subsurface targets such as sub basalt structures. Long offset data can be used to identify converted seismic waves \citep{Li1997} and record refracted seismic waves (such as head waves, diving waves) and wide angle reflections \citep{Fruehn1998,Wombell1999}. This information can e.g. improve the imaging accuracy of sub-basalt structures and full waveform inversion \citep{Herrmann2009,Plessix2012}. Long offset data can be recorded using a dual-boat operation such as continuous long offset (CLO) \nomenclature{CLO}{Continuous Long Offset} acquisition  \citep{Mastrigt2002}. In CLO acquisition, each boat tows a short streamer and long offset records are constructed by overlapping the two boats records. The dual-boat operation effectively doubles the offset range while avoiding operational complications such as streamer feathering.
However, the two boats sources need to be fired in alternative fashion with a long time interval between them to avoid source interference. Since seismic boats have minimum towing speed, the spatial spacing between seismic sources is effectively double with respect to that of a single boat acquisition. Simultaneous seismic sources used in CLO acquisition can reduce the source spacing by firing the two boat sources simultaneously.  This revised configuration is known as simultaneous long offset (SLO) \nomenclature{SLO}{Simultaneous Long Offset} acquisition \citep{Long2013,Kumar2015}.

\section{Background}

Firing seismic sources into the Earth simultaneously is similar to transmitting voice signals (sources) simultaneously over the same channel in cellular communications \citep{Ikelle_coding}. Channel sharing in cellular communications is achieved by coding each source signal into a specific frequency or phase band. The combined coded signal is then decoded (decomposed) back to its original constituent signals on the receiving end of the communication line. Research into signals mixing and separation have its early origins in the study of the "Cocktail Party Effect" in speech recognition \citep{Cherry1953}. The cocktail party effect is described as the human brain ability to focus on certain conversation while filtering other sounds in a noisy party. The human brain uses information like voice strength (amplitude), pitch (frequency) and timing (phase) to focus on (decode) the desired conversation. Modern technologies based on simultaneous sources such as multiple-input multiple-output (MIMO) \nomenclature{MIMO}{Multiple-Input Multiple-Output} have revolutionized the telecommunications industry. These technologies increased the amount of information that can be transmitted without the need to increase the infrastructure cost. 

Unlike cellular communications and human speech, seismic sources have relatively simple source signatures that are not easy to code. Moreover, seismic waves travel through the Earth subsurface which is a complex medium with unknown properties and structure. Therefore, transmitting seismic signals through the Earth can cause considerable change in the seismic wave properties and degrade the features used for coding. For this reason, early developments of simultaneous seismic sources were mainly focused on land vibratory (also called vibroseis) sources \citep{Silverman1979, Garotta1983, Landrum1987, Womack1990, Bagaini2014}. Land vibratory sources have long and complicated source signatures that are more suitable for coding than the short simple signatures of impulsive sources. The long vibratory source sweeps are needed to increase the signal to noise ratio of recorded seismic data. 
This long sweep requirement makes vibratory seismic sources less economic than the impulsive seismic sources.
Therefore, the simultaneous seismic sources technique was especially beneficial for vibratory source acquisition.
In simultaneous vibratory sources, each source frequency sweep can be coded in order to be orthogonal to other vibratory sources sweep. Then, each source reflections can be separated by cross-correlating the recorded data with this specific source sweep. An overview of the various simultaneous vibratory sources coding and decoding methods are listed in \cite{Bagaini2006,Howe2008,Bouska2013}. In addition to frequency coding, slip sweep techniques extend source coding to include source firing time delays \citep{Rozemond1996}. Simultaneous vibratory sources are becoming a mature technology that is commonly used in oil and gas industry. On the other hand, coding impulsive sources such as marine air guns is not an easy task. Despite recent research \citep{Abma2013,Mueller2015GEO}, coded impulsive sources is not as mature a technology as its land counterparts .  

\begin{figure}[htbp]
\centering
\includegraphics{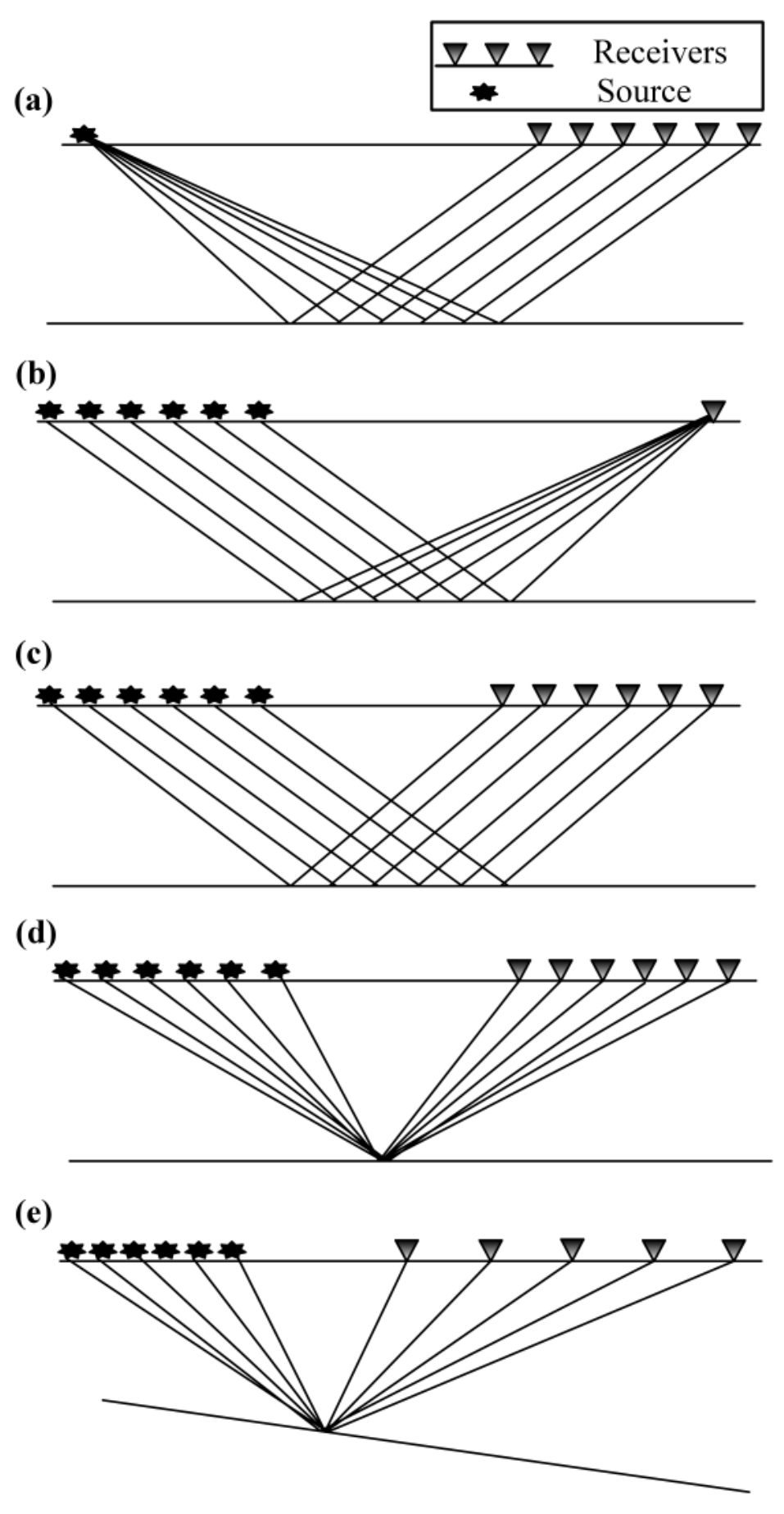}
\caption{Different seismic gathers used in data processing. (a) Common source gather (CSG). 
(b) Common receiver gather (CRG). (c) Common offset gather (COG). (d) Common mid-point gather (CMP). 
(e) Common depth-point gather (CDP).}
\label{ch2_gathers}
\end{figure}

The study of interfering seismic sources is not a completely new topic for impulsive seismic sources or marine seismic acquisition. However, early research was focused on avoiding or removing these unintended source interferences. These interferences were usually attributed to other seismic boats shooting in nearby areas and commonly known as crew noise or cross talk. \cite{Lynn1987} presented an extensive study of seismic crew noise by analyzing datasets from the Gulf of Mexico and North Sea. They concluded that source interferences can be treated as noise as long as the interfering seismic sources are not shooting in a synchronized fashion. In this case, the interfering signals will appear as incoherent noise in common mid-point seismic (CMP) \nomenclature{CMP}{Common Mid-Point gather} gather. Therefore, \cite{Lynn1987} suggested using weighted stacking of CMP gathers to attenuate these unsynchronized interferences.  If interfering sources are not firing with constant time delays, then only the primary source signal will be coherently aligned in gathers that contain traces that belong to different sources. Therefore, asynchronous sources interferences can be treated as incoherent noise in seismic gathers with traces generated from different shots. Seismic gathers such as common receiver gather\nomenclature{CRG}{Common receiver gather} (CRG), common offset gather (COG) and common depth-point gather (CDP) can be used to remove interferences. Figure \ref{ch2_gathers} shows schematics for the different seismic gathers commonly used in seismic data processing. 

In recent years, simultaneous seismic sources (also commonly known as blended sources) became an important research topic in seismic exploration. Simultaneous seismic sources can provide significant uplift in terms of both acquisition quality and economic efficiency. \cite{Beasley1998} were the first to study simultaneous seismic sources in marine acquisition to reduce survey cost \citep{Beasley1999patent}. They used two near simultaneous sources, positioned symmetrically at each end of a marine streamer cable. This approach uses the seismic source location as the coding information for source separation. Seismic sources were separated using geometry-based filtering. The filtering process assumes that the seismic data of each source have different dip. However, the dip discrimination assumption is only suitable when sources have significant spatial separation. Additionally, complicated subsurface geology can severely change the dip of recorded seismic waves. For example, seismic diffractions have significantly different dip than seismic reflections of the same source. These changes in dip coding information can severely degrade the quality of geometry based filtering.

\cite{De-Kok2002} suggested using random time delays and amplitude polarity reversal as coding information to improve source separation. \cite{Vaage2005} also explored using plurality reversal and tested different time delay schemes for simultaneous seismic sources coding. Introducing small random time delays into seismic sources firing is commonly known as source dithering. In dithered source acquisition, the source interferences appear as noise when traces are sorted into a different seismic gathers such as CMP gather. Casting the source separation problem as a denoising problem opened the door for using the fast arsenal of  well-developed denoising techniques for source separation.
 
\cite{Ikelle2007SEG} discussed using the source amplitude in addition to the firing time delays as coding information. Source separation is then achieved by higher order statistics and sparse inversion.  \cite{Stefani2007} showed that conventional imaging algorithms such as 3D pre-stack time migration (PSTM) \nomenclature{PSTM}{Pre-Stack Time Migration} can handle the asynchronous source interferences in some data sets. This approach is especially effective when the interfering seismic sources have significant spatial separation. 
\cite{Fromyr2008} and \cite{Dragoset2009} presented field tests of 3D wide azimuth surveys that used dithered simultaneous seismic sources in the Gulf of Mexico. They showed promising results for handling source interferences with conventional processing and migration algorithms. However, appropriate processing strategies were still needed to improve the quality of source separation.  

Current research into simultaneous seismic sources can be classified into three main topics. The first research topic is concerned with improving the accuracy and efficiency of source separation which is the topic of this thesis. The second research topic is focused on developing imaging algorithms that can image simultaneous seismic sources data directly without separation. The third research topic is concerned with improving source coding to facilitate accurate separation and/or imaging. In the following sections, the developments in each of these topics will be reviewed briefly. A special emphasis will be given to source separation method since it is the topic of this thesis. 

\section{Simultaneous source separation}

Current seismic sources separation methods (also known as deblending ) can be sorted into three main categories. The first category uses conventional seismic processing for simultaneous seismic source data without separation. These passive source separation methods use the power of stacking in the conventional processing sequence to suppress source interferences. Conventional stacking and migration algorithms can attenuate the incoherent source interferences enough to produce acceptable subsurface images. However, \cite{Stefani2007} and \citep{Spitz2008} noted that conventional processing and migration algorithms could be insufficient for some complicated datasets.  \cite{Aaron2009} used synthetic data for a controlled experiment that compare different source configurations for simultaneous seismic sources acquisition. They noted that source interferences can impact the accuracy of conventional seismic migration. Additionally, source interferences can impact the accuracy of many pre-imaging steps such as surface-related multiple elimination (SRME) \nomenclature{SRME}{Surface Related Multiple Elimination}. Additionally, \cite{Dragoset2009} and \cite{Abma2010} suggested that conventional processing cannot achieve the quality required for amplitude dependent analysis such as AVO, time lapse seismic and fracture analysis. \cite{Akerberg2008} attributed the inaccuracy of passive separation for some datasets to source non-orthogonality and the strong source interferences mixing with weak signals.  For all these reasons, more advanced source separation methods are needed prior to applying the conventional processing sequence. \cite{Berkhout2008} presented a theoretical framework for blended sources which is an extension to the simultaneous sources idea. In blended seismic sources framework, the time delay between successive sources is relatively long and the data is recorded continuously. This framework is built on the well-known data matrix notation that represent seismic data in frequency domain as a matrix \citep{Berkhout1982}.
 
\subsection{Denoising-based source separation}
\begin{sidewaysfigure}[htbp]
\centering
\includegraphics[angle=270]{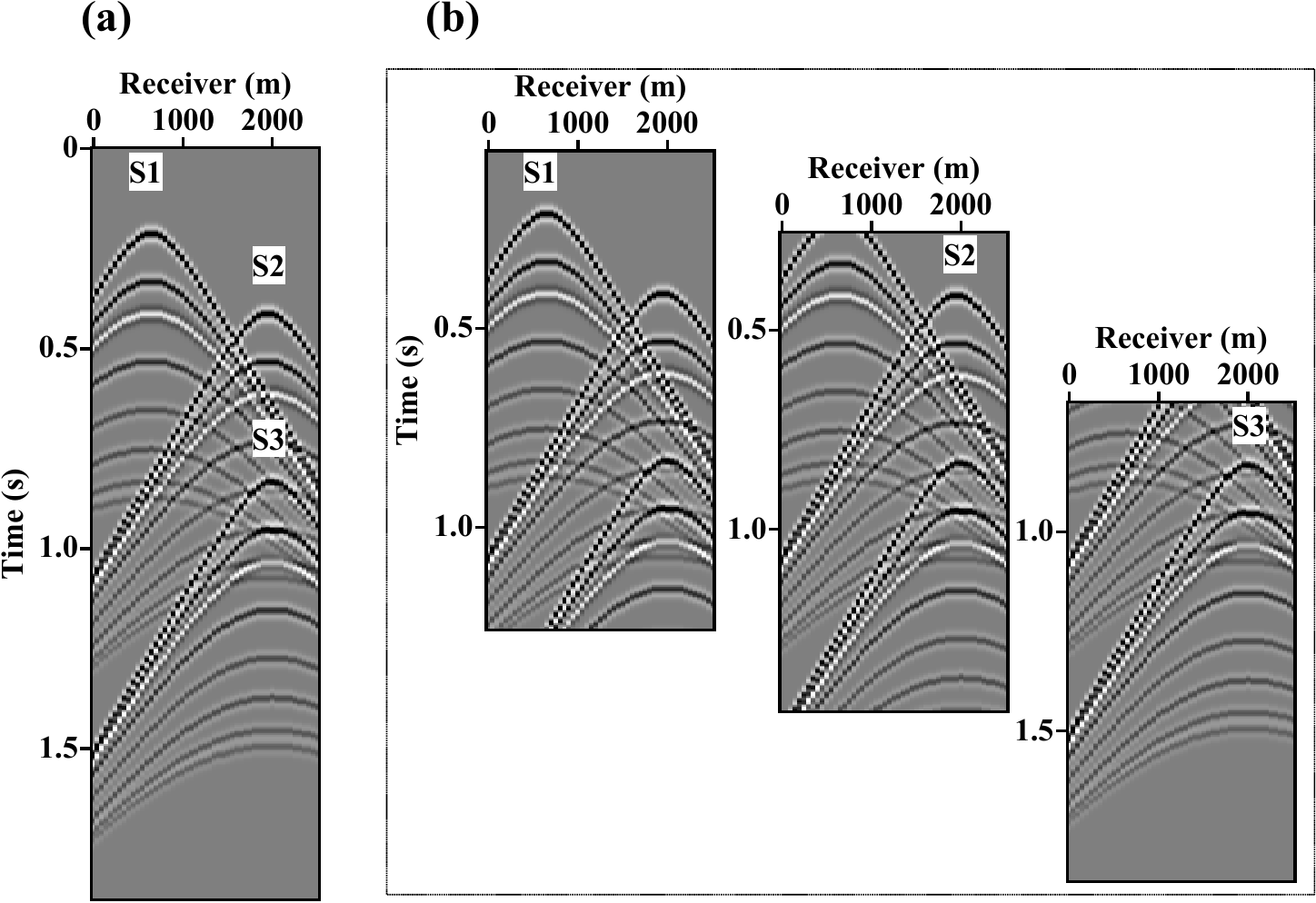}
\vspace*{5mm}
\caption{Schematics for the pseudo deblending  operator. 
				(a) Blended gather. 
				(b) Pseudo de-blended common shot gathers.}
\label{ch2_psuedo}
\end{sidewaysfigure}

\begin{sidewaysfigure}[htbp]
\centering
\includegraphics{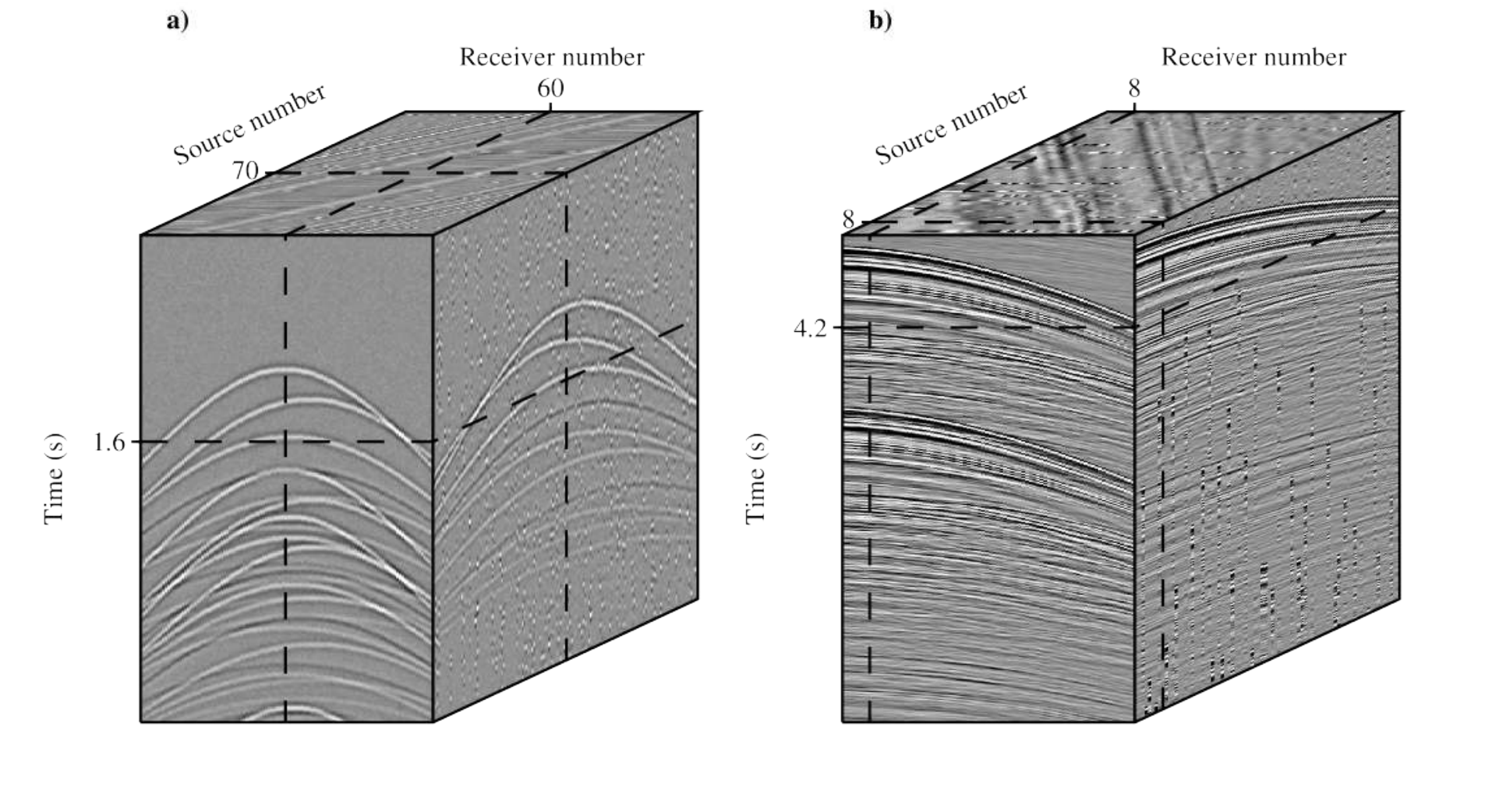}
\caption{Pseudo de-blended data cubes showing the different structures of source interferences.
			  (a) Synthetic example data cube. 
			  (b) Gulf of Mexico example data cube.}
\label{ch2_pseudo_cube}
\end{sidewaysfigure}

A second category of source separation methods treat source interferences as noise and remove using denoising algorithms. The denoising-based source separation methods utilize the incoherency of source interferences due to the spatial source separation and firing time dithering. Source separation assumes that simultaneous source data can be modelled from its single source components \citep{Berkhout2008}. For instance, if we let ${\bf D}$ represents the data of all seismic sources in time-space domain arranged into a data cube and ${\bf b}$ represents the two dimensional simultaneous seismic sources data, then  
\begin{equation}\label{ch2_blending}
{\bf b} ={ \Gamma}~{\bf D},
\end{equation}
where ${\Gamma}$ represents the blending operator that contains the source coding information (firing times and spatial locations) \citep{Berkhout2008}. Therefore, simultaneous seismic source data ${\bf b}$ can be separated by compensating for the source firing delays and subdividing of the data into single source segments. This operation is commonly known as pseudo deblending  and Figure \ref{ch2_psuedo} shows a simple schematic for pseudo deblending . Pseudo deblending  is equivalent to applying the adjoint blending operator ${\Gamma^{T}}$ to the simultaneous seismic source data ${\bf b}$ such that  
\begin{equation}\label{pseudodeblending_equ}
	\widetilde{\bf D}=\Gamma^{T}{\bf b},
\end{equation}
where $\widetilde{\bf D}$ represents the pseudo de-blended data cube. However, pseudo deblending  does not remove source interferences as shown in Figures \ref{ch2_psuedo}. Figure \ref{ch2_pseudo_cube} shows pseudo de-blended cubes for simple synthetics and for field data from the Gulf of Mexico. The front of the cubes represent common source gathers while the side shows common receiver gathers. Source interferences have an incoherent structure in common receiver gathers due to the time dithering of sources. Therefore, a denoising algorithm can be used to attenuate source interferences and achieve source separation. 
In recent years, several researchers suggested different denoising approaches to remove source interferences.  \cite{Moore2008} suggested using Radon transforms as a stacking tool to attenuate source interferences in common receiver gathers. \cite{Akerberg2008} also suggested Radon transform combined with a sparse inversion scheme to enhance the source separation quality. \cite{Spitz2008} suggested using a prediction error filter (PEF) \nomenclature{PEF}{Prediction Error Filter} based subtraction to attenuate source interferences.  \cite{Kim2009} suggested modelling the coherent interferences and adaptively subtracting them in order to preserve weak signals. \cite{Huo2009} suggested using a multi-dimensional vector-median filter to suppress the interferences in a robust fashion. \cite{Maraschini2012} used an iterative method based on rank reduction filtering to remove source interferences. \cite{Trad2012} suggested using the apex shifted Radon transform to geometrically filter seismic sources in common shot gathers. \cite{Ibrahim2013SEG,Ibrahim2014GEO} suggested using apex shifted Radon transform and robust inversion to remove source interferences and preserve weak signals. \cite{Sacchi2014IEEE} reformulated the sparse Radon inversion to move source interferences from the misfit term to the regularization term to remove source interferences and preserve weak signals.   \cite{Ibrahim2014SEG,Ibrahim2014EAGE,Ibrahim2015GEO} suggested using migration/demigration operators to speed up source separation using robust Radon transform. 
\subsection{Inversion-based source separation}

A third category of separation methods casts source separation as an inversion problem. In \cite{Berkhout2008} blended sources framework, the firing times are contained in a blending operator.  This operator is used to formulate an inversion problem for source separation. Similar to many inversion problems in geophysics, the source separation inversion problem is ill posed.  Therefore, an inversion constraint is added to the inversion cost function. Additionally, rather than inverting directly for the de-blended data, $\bf D$, one can invert for the representation of the de-blended data in terms of coefficients $\bf c$ in an auxiliary domain.  In other words, if the data are represented in terms of coefficients $\bf c$ in a basis $\Phi$, such that ${\bf D} = {\Phi} {\bf c}$, the aim is to estimate $\bf c$ by minimizing the following cost function
 \begin{equation}\label{deblending_inversion_equ}
	J=\|{\bf b} -{ \Gamma} {\Phi } {\bf c} \|_2^2 + \mu {\cal R} ({\bf c}), 
\end{equation}
where ${\cal R} ({\bf c})$ is a suitable constraint on the model of the de-blended data. If a suitable transform is chosen, the single source data can be represented by a sparse collection of coefficients in the transform domain. Therefore, a sparse model constraint can be used to estimate the coefficients that represent the unknown de-blended data. \cite{Moore2010} noted the connection between the efficiency of the sparsity based source separation and the simplicity of subsurface geology. A simple subsurface will have seismic data that can be easily decomposed into a sparse model using a suitable transformation. Therefore, including the sparsity constraint to the inversion problem will increase the quality of source separation. \cite{Mahdad2011} developed an iterative inversion algorithm that uses a coherency constraint in the Fourier domain to separate seismic sources. \cite{vanBorselen2012} developed an inversion based separation method that uses the constraint that nearby sources produce similar records. \cite{Wason2011} combined the inversion based source separation problem with the compressive sensing problem to increase the survey efficiency. In this approach, the source separation and the data interpolation problems can be solved simultaneously.  \cite{Wason2011} used the curvelet transform domain to impose the sparsity constrain to the inversion problem. \cite{Kontakis2015b} suggested combining coherency-based separation and sparse inversion-based source separation in the focal transform domain to improve the separation quality and reduce the dependency on model sparsity. Similarly, \cite{Cheng2015} used rank reduction to solve for source separation and interpolation simultaneously.

\section{Simultaneous source imaging}
A third approach for processing simultaneous seismic sources is to use it for for imaging directly.  The \cite{Berkhout2008} framework for blended seismic sources suggested two approaches for processing blended seismic data. In the first approach, sources are de-blended and then conventional processing and imaging sequence is used.  In the second approach, simultaneous sources data is used for imaging directly without source separation.  In the imaging approach, the blending information is included in the imaging algorithm to handle source interferences. Earlier research on multiple source imaging was concerned with reducing the computational cost of imaging conventional single source data \citep{Romero2000}. \cite{Tang2009} suggested using least-squares wave equation migration to image simultaneous seismic sources directly. \cite{Dai2009} incorporated a de-blurring filter into least-squares migration to improve the convergence of simultaneous seismic sources imaging \citep{Dai2011}. \cite{Verschuur2009SEG} suggested using a target-oriented migration algorithm with multi-shift least-squares imaging condition to image simultaneous seismic sources data. \cite{Godwin2010} formulated both the conventional seismic imaging and blended seismic imaging in the context of matrix operation and used singular value decomposition (SVD) \nomenclature{SVD}{Singular Value Decomposition} to optimize the sources amplitude coding. \cite{Verschuur2011GEO} suggested incorporating the multiples into simultaneous sources imaging to increase illumination (double illumination).  \cite{Soni2015} used full wave-field migration algorithm to image simultaneous source VSP data.
 
\section{Simultaneous source coding}

\citet{Berkhout2009} suggested extending the blending concept to include both sources and receivers in order to compress the recorded seismic traces. In this approach, the continuous recording of blended source acquisition is divided between different receivers that have different time delays. \cite{Berkhout2012} suggested using source arrays with each source element operate in different frequency bandwidths (dispersed source array) and called this approach inhomogeneous blending. Dispersed source array (DSA) \nomenclature{DSA}{Dispersed Source Array} is designed such that the combined wavefield have the desired temporal and angular spectral properties at each point in the subsurface.  DSA blended acquisition will allow the source elements to be designed without the low versus high frequency compromise.  Additionally, the spatial sampling intervals can be optimized for sources with different frequency bands.  \cite{Jiang2010} showed that the source separability depends on the firing time delays and the data frequency limits. They performed a series of separation experiment to define the limit for high quality separation (Jiang limit) \citep{Abma2012}. \cite{Abma2012} adopted the independent simultaneous source (ISS) \nomenclature{ISS}{Independent Simultaneous source} survey \citep{Howe2008} into marine OBS acquisition. In this method, sources fire independently from each other and the source dithering results from natural randomness that arise from source movement. Unlike land sources, marine sources have regularity in their movement and additional randomness is included in source firing times. \cite{Abma2013} suggested coding the individual airguns in marine sources to fire with different time delays (popcorn shooting) to improve sources efficiency and achieve greater flexibility.  \cite{Mueller2015GEO} used time dithering to design source firing sequences that are close to orthogonal. This coding approach is used to enhance source separation using multi-frequency separation methods. \cite{Wu2015} proposed an alternative approach to blending, called shot repetition, which activates a broadband source more than once at the same location. The shot-repetition codes are designed such that the sources at different locations are uncorrelated to each other.

\chapter[Radon transforms]
{
Radon transforms  
}

\section{Introduction}

Radon transforms were first introduced by the Austrian mathematician Johann Radon in 1917 \citep{Radon1917, Radon1983}. Radon mathematically proved that the integrals of a function along arbitrary geometrical projections can form a complete basis for this function. This basis functions can be used to form a transformation similar to the sinusoidal basis functions of the Fourier transform. Radon transforms are popular in many applications such as medical imaging \citep{Kuchment2013}, remote sensing \citep{Copeland1995} and seismic data processing \citep{Thorson1985}. In these applications, the recorded data consists of signals that are reflected and/or transmitted through the desired object. Therefore, Radon transforms that use appropriate geometrical paths that resemble these signals can focus them efficiently \citep{Deans1983,Kuchment2013}. Designing and improving mathematical transforms that focus seismic data is a topic of ongoing and extensive research \citep{Berkhout2006,Fomel2010,Jones2013,Kutscha2014}. Transforms that focus seismic signals can be used as powerful processing tool. Therefore, Radon transforms have been widely used in seismic data processing for many applications such as interpolation \citep{Kabir1995,Sacchi1995,Trad2002,Ibrahim2015SEG}, multiple separation \citep{Hampson1986b,Foster1992,Landa2015}, noise removal \citep{Russell1990a,Russell1990b} and micro-seismic signal detection \citep{Sabbione2013,Sabbione2015a}.

\section{Background}

Radon transforms are not orthogonal transforms for signals with finite spatial length. Therefore, seismic data cannot be recovered accurately from estimated Radon models. This inherent deficiency combined with the fact that seismic data are usually incomplete and have noise contamination complicates the data recovery.  Incomplete data situations arise from limited seismic aperture, sparse sampling and/or data gaps. Limited seismic aperture is a direct consequence of the limited length of receiver array used in the field. Cost reduction in data acquisition usually results in sparsely sampled data, especially for land acquisitions. Data gaps or dead traces result from human or geographical obstacles to data acquisition. \cite{Thorson1985} were the first to suggest casting the problem of the Radon transformation as an inversion problem. They included a sparse model constraint in the inversion cost function to increase Radon transform focusing power. \cite{Hampson1986} suggested using least-squares inversion and the parabolic Radon approximation in the $f-x$ domain to remove multiples in CMP gathers after NMO \nomenclature{NMO}{Normal Move Out correction} correction. \cite{Yilmaz1989} used a $t^2$ stretching transformation of CMP gathers to improve the reflections approximation by the parabolic Radon transform. \cite{Kostov1990} and \cite{Gulunay1990} utilized the Toeplitz structure of the linear and parabolic Radon operators in the $f-x$ domain to speed up inversion using fast solvers like Levinson recursion. \cite{Sacchi1995,Sacchi1995b} incorporated the sparse model constraint into the $f-x$ domain parabolic Radon transform to reconstruct missing traces. \cite{Sacchi1999} introduced a fast sparse parabolic Radon transform that use conjugate gradient inversion and efficient matrix multiplications.  \cite{Cary1998} observed that the time domain Radon operators can estimate a sparser Radon model than the frequency domain operators. Time domain Radon operators can achieve better sparsity since one can impose sparsity in both time and Radon parameter simultaneously. However, frequency domain operators are more computationally efficient \citep{Trad2003}. Moreover, \cite{Cary1998} proposed to use frequency domain operators to compute the time invariant Radon model in time domain. In this approach, the frequency domain operator is inserted between the forward and the inverse Fourier transform operators. Therefore, the Radon model is estimated in time domain while the Radon operator is computed in the frequency domain. This approach will achieve computational efficiency without compromising  model sparsity in Radon space \citep{Trad2003}.  \cite{Hermann2000} proposed to redesign the frequency domain Radon algorithm of  \cite{Sacchi1995} to deal with aliasing problems. They suggested using model weights derived from low non-aliased frequencies to regularize the aliased high frequencies. \cite{vanDedem2000,vanDedem2005} introduced apex shifting into Radon transform to predict 3D surface related multiples. \cite{Trad2003} proposed implementing a single velocity apex shifted Radon using Stolt migration operator for multiples removal and interpolation. \cite{Abbad2011} proposed a modified fast parabolic Radon transform that uses complex singular value decomposition (SVD).  \cite{Trad2012} used the apex shifted to Radon transform to geometrically filter interfering sources in common shot gathers. \cite{Sabbione2015a} and \cite{Sabbione2015b} used apex shifted parabolic and hyperbolic Radon transforms to detect and de-noise micro-seismic signals. \cite{Guitton2003,Ji2006,Ji2012} suggested using both Radon sparsity and robust misfit function in Radon inversion to process data with erratic noise. \cite{Hu2013} introduced a fast implementation of Radon transform that uses a low rank approximation of the transform kernel and butterfly algorithm to compute the Fourier integral operator. \cite{Lu2013} suggested using iterative shrinkage solver to accelerate the sparse inversion of the time invariant Radon transform. \cite{Zhang2014} expanded the iterative shrinkage solver to accelerate 3D time invariant Radon transforms. \cite{Ibrahim2013SEG,Ibrahim2014GEO} used robust inversion and apex shifted Radon to remove simultaneous sources interferences in common receiver gathers. \cite{Ibrahim2015GEO,Ibrahim2014SEG,Ibrahim2014EAGE} used a multi-velocity Stolt-based apex shifted Radon transform for simultaneous source separation. \cite{Ibrahim2015SEG} proposed apex and asymptote shifted Radon transform to interpolate seismic diffractions.

\section{Radon operators}

In order to derive Radon operators, let us assume that $d(t,x)$ denote the two dimensional seismic data and $m(\tau,\xi)$ denote the Radon model \citep{Sacchi1995b}. We first define the forward Radon operator, $\mathscr{L}$, and its adjoint operator $\mathscr{L}^{\dag}$ as follow
\begin{align} \label{ch3_eq1}
d(t,x)&=\mathscr{L}~m(\tau,\xi)=\int^{\infty}_{-\infty}~m(\tau={\phi}\,(t,x,\xi),\xi)~d\xi, \\ 
\widetilde{m}(\tau,\xi)&=\mathscr{L}^{\dag}~d(t,x)= \int^{\infty}_{-\infty}~d(t=\widetilde{\phi}\,(\tau,x,\xi),x)~dx, 
\end{align}
where $\widetilde{m}(\tau,\xi)$ is the Radon model estimated by the adjoint operator. The parameter $\xi$ is the Radon parameter (or parameters) that define the Radon operator integration path by the function $\phi(\tau,x,\xi)$. In the operator format, the forward and adjoint Radon transforms can be expressed as
\begin{eqnarray}
	{\bf d} &={\bf Lm} \label{ch3_2}\\ 
	\widetilde{{\bf m}} &= {\bf L}^{T} {\bf d}. \label{ch3_3}
\end{eqnarray}
In practice, we estimate the Radon model using inversion of equation \ref{ch3_2} instead of the adjoint operator in equation \ref{ch3_3}. 
Popular variants of the Radon transform operators are linear (LRT), parabolic (PRT), apex shifted parabolic (APRT), hyperbolic (HRT), apex shifted hyperbolic (ASHRT) and asymptote and apex shifted hyperbolic (AASHRT) operators. Table \ref{ch3_table_Radon_operators} lists the different forward and adjoint Radon operators used in seismic data processing. Radon operators can be sorted into the time-invariant and time-variant operators according transform integration path function $\phi(\tau,x,\xi)$. Time invariant Radon operators have integration functions that can be computed efficiently in frequency domain using Fourier time shift property. On the other hand, time variant integration path functions can only be computed in time domain which results in slow computation \citep{Trad2002}. Improving both the accuracy and efficiency of Radon operators is the subject of ongoing research \citep{Abbad2011,Hu2013,Lu2013,Zhang2014,Ibrahim2014GEO,Ibrahim2015GEO,Ibrahim2015SEG}. 
\setlength{\tabcolsep}{3.0pt}
\renewcommand{\arraystretch}{3.0}
\begin{table}[htp] 
\begin{center}
\caption{Radon transform operators used in seismic processing.}
\label{ch3_table_Radon_operators} 
\begin{tabular}{ | c |  c | c | c | c |}
\hline 
Type			     &Operator 	  		&$\xi$			       			&$\widetilde{\phi}\,(\tau,x,\xi)$         			   	  			&${\phi}\,(t,x,\xi)$\\ \hhline{|=|=|=|=|=|}
Time				&{\sc LRT} 			&$\xi=p$					    &$t=\tau+p~x$ 															&$\tau=t-p~x$ \\ \cline{2-5} 
 invariant	 	&{\sc PRT}     		&$\xi=q$       	   			&$t=\tau+q~{x^2}$                            							&$\tau=t-q~{x^2}$	\\ \cline{2-5} 
					&{\sc ASPRT}   	&$\xi=[q,x_0]$ 	   		&$t=\tau+q~{(x-x_0)^2}$                      						&$\tau=t-q~{(x-x_0)^2}$	\\  \hhline{|=|=|=|=|=|}
Time				&{\sc HRT}  	  	&$\xi=v$      		   		&$t=\sqrt{\tau^2+\frac{x^2}{v^2}}$          					&$\tau=\sqrt{t^2-\frac{x^2}{v^2}}$ \\ \cline{2-5}  
variant			&{\sc ASHRT}	&$\xi=[v,x_0]$ 	   		&$t=\sqrt{\tau^2+\frac{(x-x_0)^2}{v^2}}$    				&$\tau=\sqrt{t^2-\frac{(x-x_0)^2}{v^2}}$\\  \cline{2-5}  
				   &{\sc AASHRT}  &$\xi=[v,x_0,\tau_0]$ 	&$t=\tau_0+\sqrt{\tau^2+\frac{(x-x_0)^2}{v^2}}$ 		&$\tau=-\tau_0+\sqrt{t^2-\frac{(x-x_0)^2}{v^2}}$\\\hline	
\end{tabular}
\end{center}
\end{table} 

\section{Sparse Radon}

The main reason for Radon transform popularity in seismic data processing is its ability to focus seismic signals. This focusing property of Radon transforms justifies the inclusion of sparse model constraint in Radon inversion. The sparsity model constraint is a very powerful tool that can achieve significant increase the accuracy and resolution of the transform. The sparsity constraint is built on the priori assumption that the data can be represented by few coefficients in the Radon model. Since the Radon operator is usually chosen to resemble seismic signals, this assumption is generally accepted. The flexibility of designing Radon operators to match seismic data makes it a very versatile tool in seismic data processing.   

\begin{figure}[htbp]
 		\includegraphics{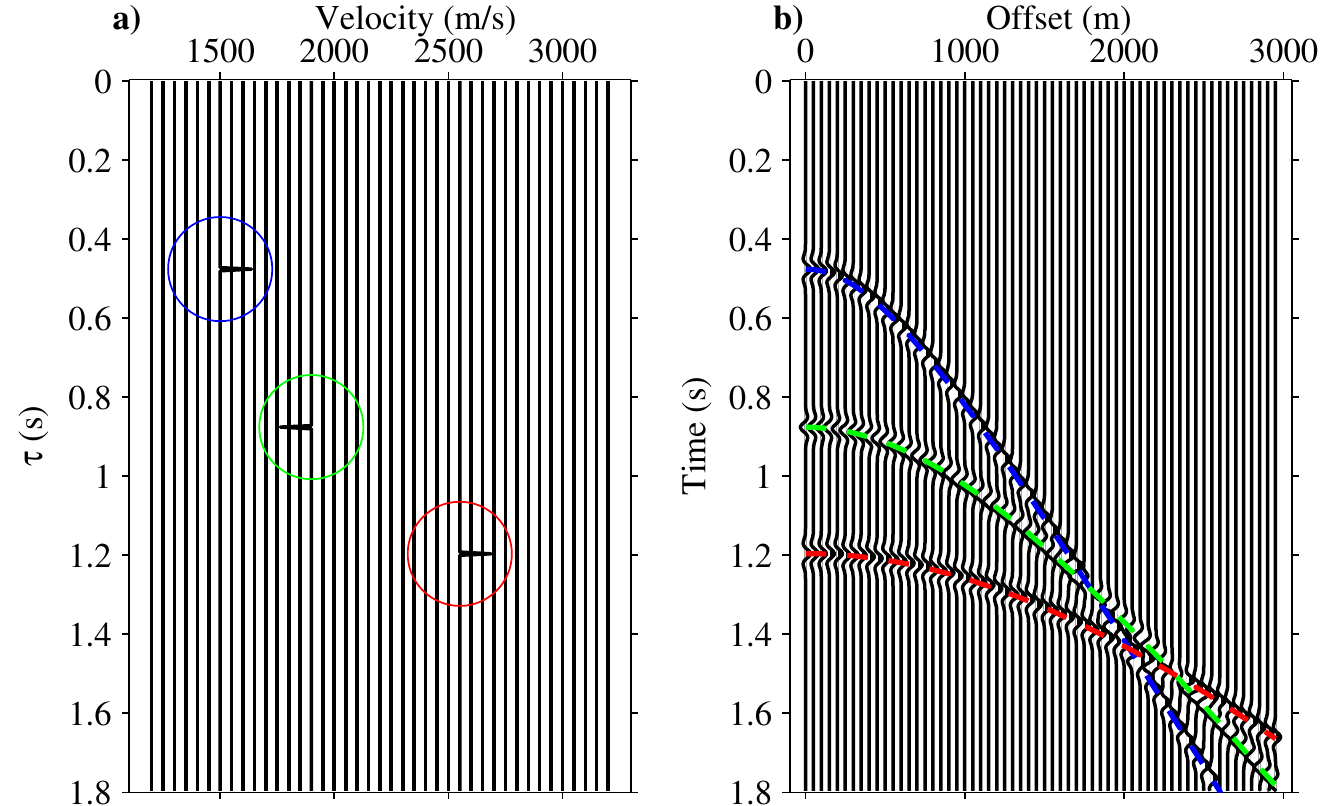}
 		\vspace*{5mm}
 		\caption{Hyperbolic Radon transform example. 
   		(a) Radon model.
   		(b) Data modelled using Radon transform. }
		\label{ch3_modelling}
\end{figure}

\begin{figure}[htbp]
 		\includegraphics{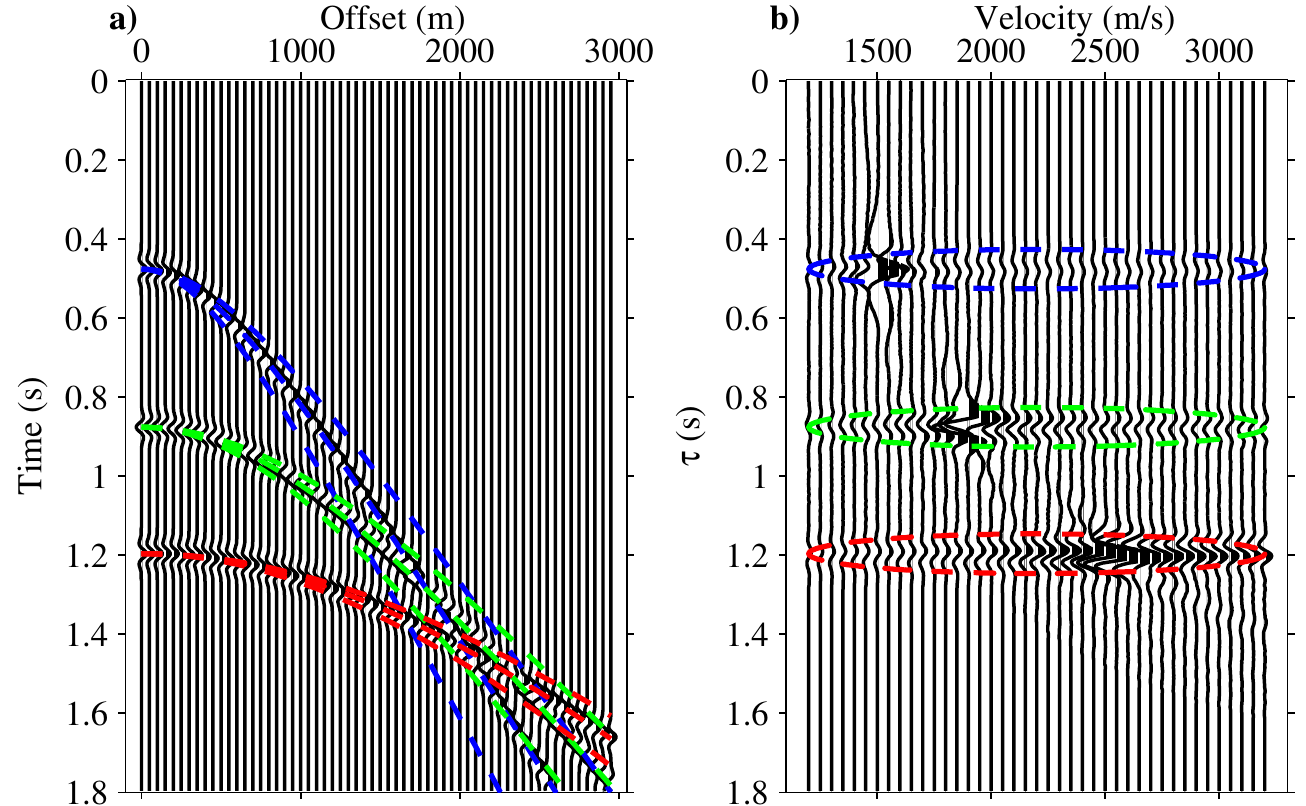}
 		\vspace*{5mm}
 		\caption{Smearing artefacts due to similarity in apex time. 
   		(a) Input data.
   		(b) Adjoint Radon model showing smearing along Radon $v$ parameter.  }
		\label{ch3_v_smearing}	
\end{figure}

\begin{figure}[htbp]
 		\includegraphics{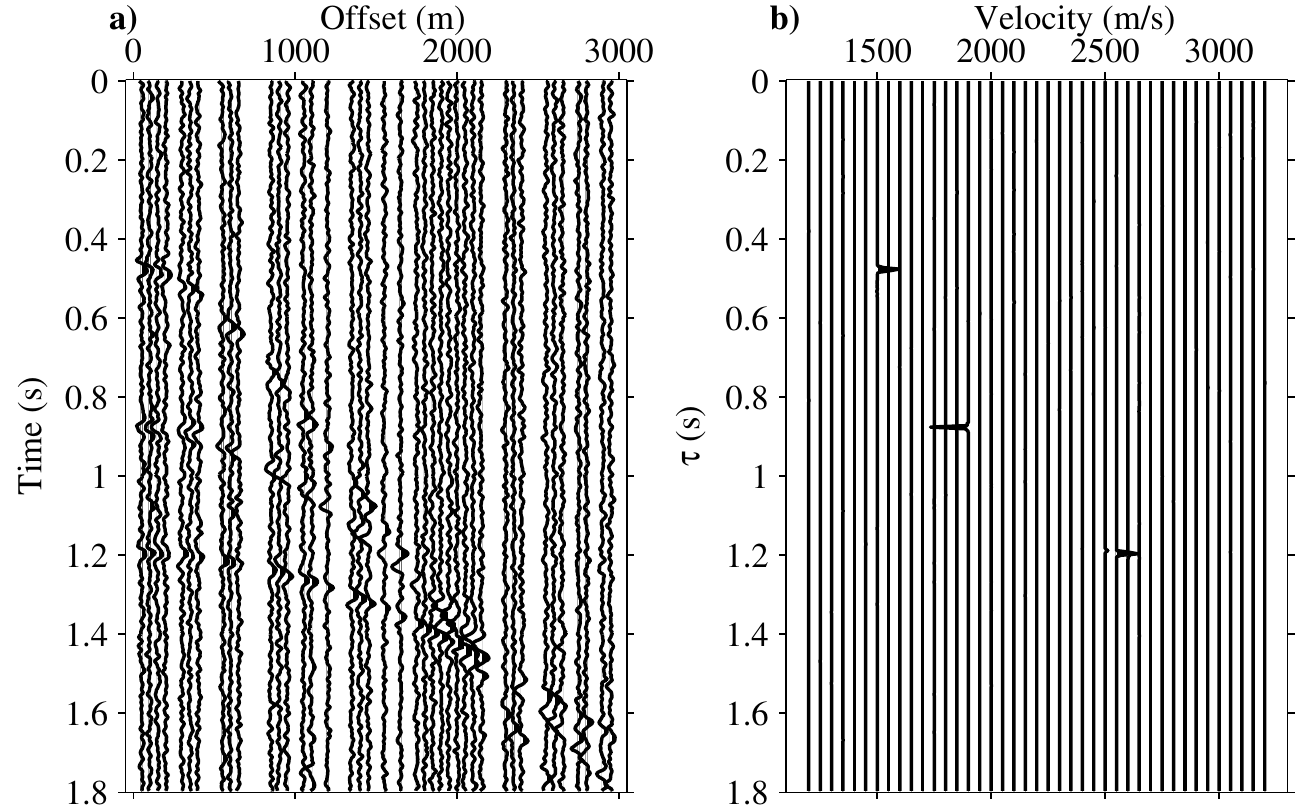}
 		\vspace*{5mm}
 		\caption{Sparse inversion of Radon model. 
   		(a) Data with missing traces and contaminated with noise.
   		(b) Radon model estimated by sparse inversion. }
		\label{ch3_sparse}
\end{figure}

\begin{figure}[htbp]
 		\includegraphics{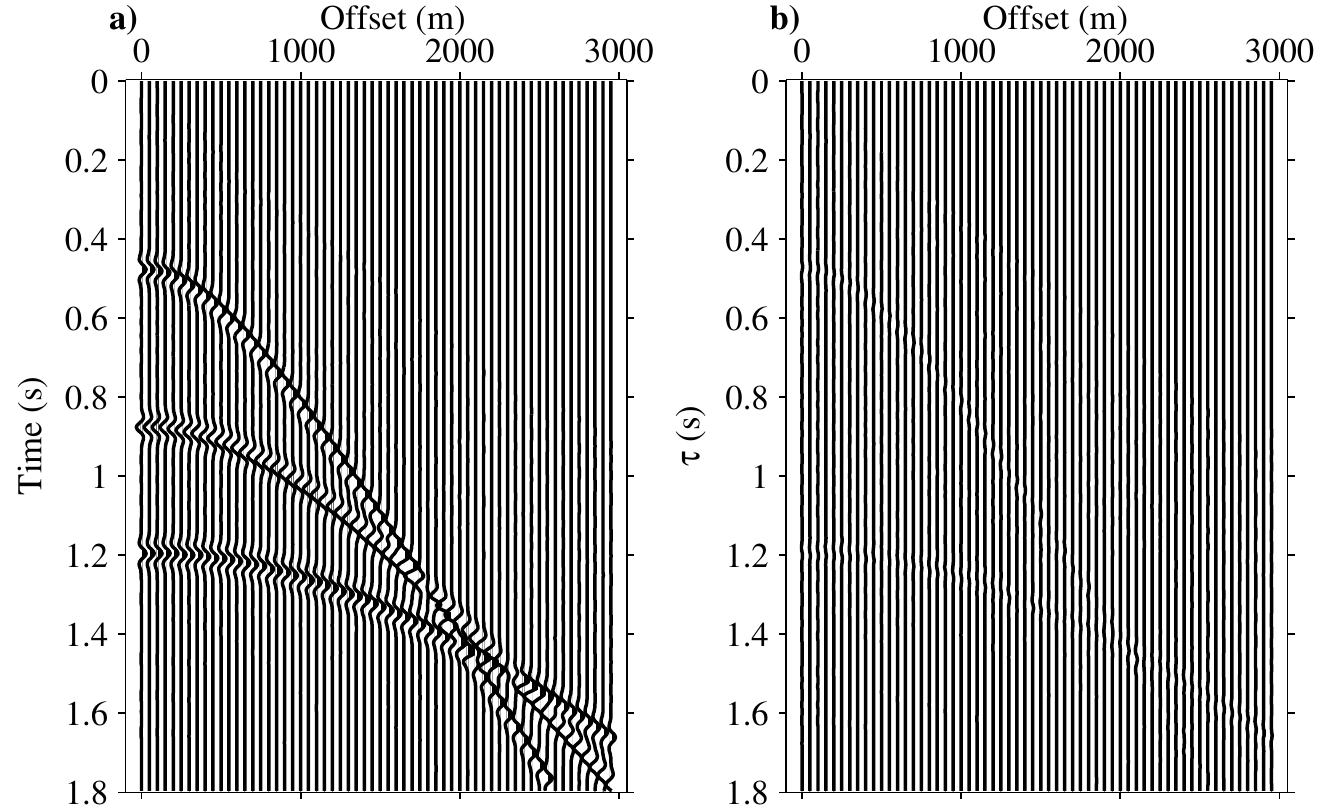}
 		\vspace*{5mm}
 		\caption{Data recovered from model estimate by sparse inversion. 
   		(a) Recovered data.
   		(b) Error in recovered data. }
		\label{ch3_sparse2}
\end{figure}

In order to demonstrate the deficiency of the Radon adjoint operator and the power of sparse Radon transform, we construct a simple synthetic example.  Figure \ref{ch3_modelling}a shows a synthetic hyperbolic Radon transform model that consists of three events. Every event represents a hyperbolic reflection defined by two parameters, velocity $v$ and apex time $\tau$. This model resembles seismic reflections from horizontal layers with 1500, 1900 and 2550 m/s velocities.  The forward Radon modelling operator is used to generate synthetic seismic data from this model. Using the modelling operator to generate the data is commonly known as the inverse problem crime. However, using the modelling operator to generate the data guarantees that the transform basis functions and the signal match closely. This allow use to test the efficiency of our inversion algorithms and prove the main concepts. The modelling operator use the velocity and apex time information to define the hyperbolic travel time curve for each event. Then, the modelling transform generates the data by repeating the event wavelet along the travel time hyperbola as shown in Figure \ref{ch3_modelling}b.
If the adjoint Radon transform is used to estimate the Radon model for the data in Figure \ref{ch3_modelling}b, 
the estimated model will not match the original model used to generate the data. This indicates that the Radon transform is not orthogonal transformation (${\bf L}^T {\bf L} \neq 1$). The transform non-orthogonality results in smearing artefacts in the Radon model estimated using the adjoint operator. For example, Figure \ref{ch3_v_smearing}a shows different theoretical travel time hyperbolas that are used by the transform.  These theoretical travel time hyperbolas are generated using the same apex times but with different velocities that are close the true event velocity. The overlap between these theoretical travel time hyperbolas (transform basis functions) generate the smearing effects along the model $v$ dimension that are marked in Figure \ref{ch3_v_smearing}b. The smearing artefacts problem increases for real data with missing traces and contaminated with noise.   
Fortunately, the similarity between the transform basis functions and the signals can be used to impose sparsity on the model inversion. 
The transform model can be assumed to have a sparse representation for the data. Therefore, an accurate Radon model can be estimated using sparse inversion. For example, data in Figure \ref{ch3_modelling} have three reflections only and it should be represented by three events in the Radon model. This sparse representation assumption should not be affected by noise contamination or missing traces. For example, Figure \ref{ch3_sparse}a shows the same synthetic data in Figure \ref{ch3_modelling}b with added random noise (signal to noise ratio SNR=2) and missing traces (30\% missing). Using this noisy and irregularly sampled data as input, the Radon model in Figure \ref{ch3_sparse}b is estimated using sparse inversion. The sparse Radon model is estimated by minimizing the following cost function 
\begin{equation}\label{ch3_cost_fn}
J=\|{ {\bf d} -{\bf L}\, {\bf m}}\|_2^2+\mu \| {\bf m} \|_1^1,
\end{equation}
where ${\bf d}$ is the observed data (data with missing traces and noise), ${\bf L}$ is the forward Radon operator,
${\bf m}$ is the sparse Radon model. The first term in equation \ref{ch3_cost_fn} represents the misfit between observed and modelled data and the second term represent the model constraint. This cost function estimate a sparse Radon model by imposing a constraint on the model size in $\ell_1$ norm sense. The trade-off parameter $\mu$ represents the trade-off between the misfit term and the model constraint. The estimated sparse Radon model is shown in Figure \ref{ch3_sparse}b which matches the original model closely. 
In order to recover a noise free and regularly sampled data, the forward modelling is used to model the data using the model estimated by sparse inversion. Figure \ref{ch3_sparse2}a shows the recovered seismic data and \ref{ch3_sparse2}b shows the error in the recovered data compared to the true, noise free data from Figure \ref{ch3_modelling}b. The denoising/interpolation quality can be measured using the following equation
\begin{equation}
Q =10\log_{10}\left(\frac{\| {\bf d}_{original} \|_2^2}{\| {\bf d}_{original}- {\bf d}_{recovered} \|_2^2}\right).
\end{equation}
where ${\bf d}_{original}$ is the original noise free and complete data and ${\bf d}_{recovered}$ is the data recovered form sparse Radon model. The $Q$ value for denoising/interpolating example shown in Figure \ref{ch3_sparse2} is $22.472$ dB which demonstrates the high efficiency of sparse Radon inversion.  

\chapter[Robust Radon Transform]
{ Robust Radon Transform
\footnote{A version of this chapter has been published in Ibrahim, A. and M. D. Sacchi, 2014, Simultaneous source separation using a robust Radon transform, Geophysics 79(1): V1-V11.}
}
\section{Motivation}
\begin{figure}[htbp]
\centering
\includegraphics{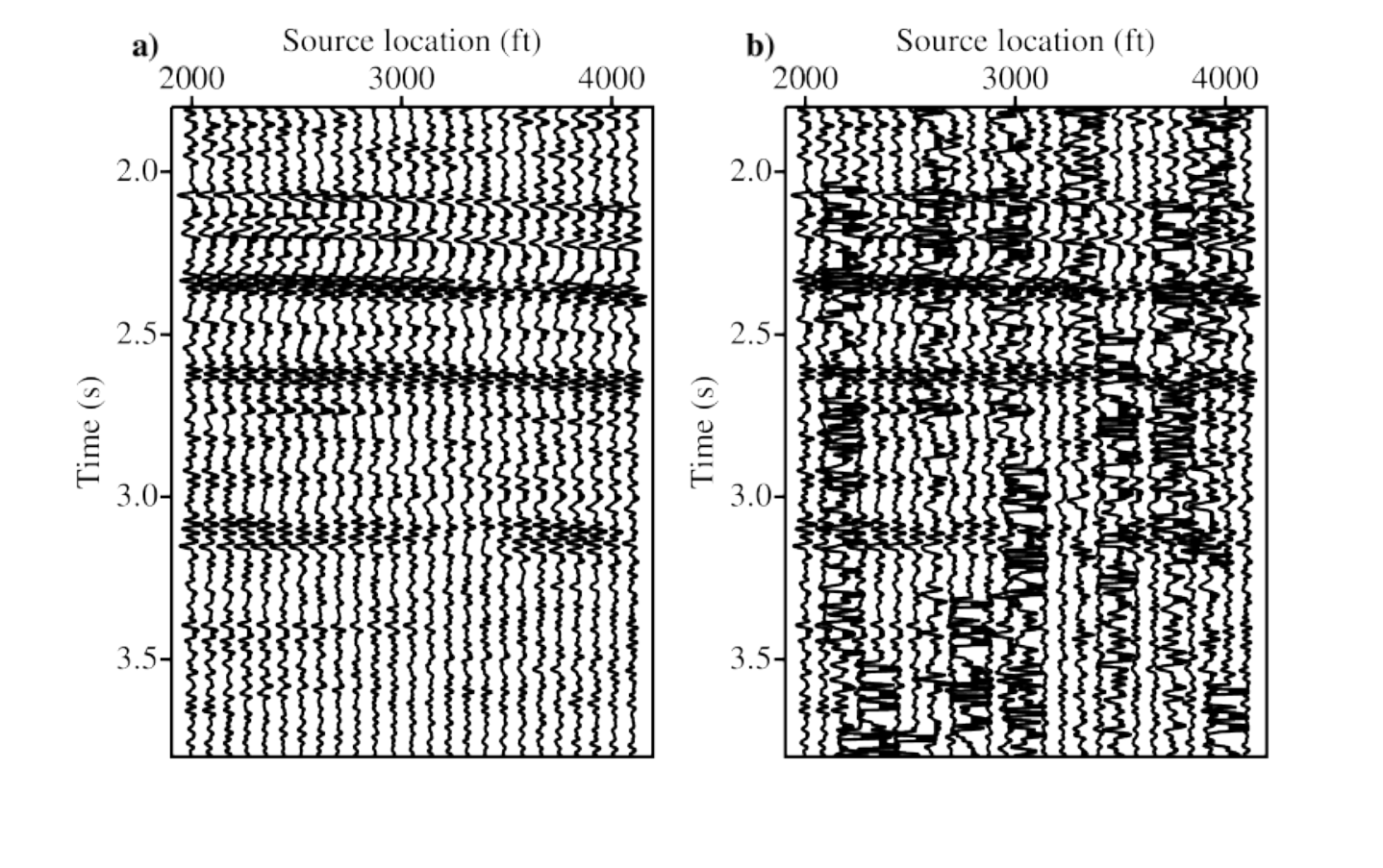}
\vspace*{-12mm}
\caption{A close-up of numerically blended common receiver gather from the Gulf of Mexico field data 
showing source interferences. (a) Original gather. (b) Blended gather. }
\label{ch4_erraric_noise}
\end{figure}

R
Source interferences with amplitudes much larger than the primary signal usually appear in simultaneous seismic sources data. Interferences have large amplitude due to the time delay between sources and the rapid decay of seismic reflections with arrival time. For example, Figure \ref{ch4_erraric_noise} shows a numerically blended field data example from the Gulf of Mexico. The weak late reflections of the primary source are almost obscured by the much stronger early reflections of the interfering source. These strong interferences will appear as erratic noise in common receiver gathers. The susceptibility of the $\ell_2$ norm misfit function to outliers that result from source interferences can reduce the accuracy of the model estimation. Therefore, the conventional $\ell_2$ norm misfit should be replaced by a misfit function that is robust to outliers. \cite{Claerbout1973} suggested using the $\ell_1$ norm misfit as a robust alternative to the conventional $\ell_2$ norm. In this chapter, we test the effect of incorporating robust misfit into the Radon transform inversion in relation to suppressing source interferences. We use simple synthetic examples and a numerically blended field data from the Gulf of Mexico to test robust and non-robust Radon transforms. 

\section{Robust Radon transform}
We assume that the common receiver gather is contaminated with noise (source interferences). Therefore, we pose the problem of estimating a noise free Radon model $\bf m$ of this gather as an inversion problem. This problem minimize the residual between the observed data and the modelled data,  
\begin{equation} \label{ch4_eq1}
{\bf r } ={\bf d}  - {\bf L}\, {\bf m},
\end{equation}
\noindent
where ${\bf d}$ is the common receiver gather data contaminated with interferences, ${\bf L}$ is the Radon modelling operator and ${\bf m}$ is the estimated Radon model. This problem is an ill-posed inversion \citep{Sacchi1995}. Therefore, a regularization term must be included in the inversion cost function to estimate a unique and stable model.  Therefore, the Radon transform inversion problem can be formulated by minimizing the following general cost function 
\begin{align}\label{ch4_cost_fn}
J=&\|{\bf r }\|_p^p+\mu \| {\bf m} \|_q^q \nonumber \\
  =&\|{ {\bf d} -{\bf L}\, {\bf m}}\|_p^p+\mu \| {\bf m} \|_q^q,
\end{align}
where the first term on the right hand side is the misfit term and the second term is the regularization term.  
In both terms we assume that $\ell_p$ and $\ell_q$ norms are given by the general expressions $\ell_p = \sum_i |r_i|^p$ and $\ell_q  = \sum_i |m_i|^q$. Through the minimization of this cost function with respect to the unknown vector of Radon coefficients $\bf m$ one finds a solution that honors the observed data $\bf d$.

The parameters $p$ and $q$ represent the exponent of the $p$-norm of the misfit and the $q-$norm of the model regularization term, respectively. The $\ell_2$ norm ($p = 2$) for the misfit is conventionally used with the implicit assumption that the data contains Gaussian additive noise. This Gaussian additive noise will result in fitting errors that follow a Gaussian probability distribution. The $\ell_2$ norm misfit is very efficient in handling small fitting errors that follow a Gaussian probability distribution. However, if the data are contaminated with strong noise that does not follow a Gaussian probability distribution, this noise will produce unrealistic models.  The $\ell_2$ norm misfit is sensitive to outliers since minimizing the $\ell_2$ norm misfit is equivalent to using the mean of the misfits. In order to show the relationship between the $\ell_2$ misfit and the mean, let us define $r_2$ as the point with the minimum sum of squared differences ($\ell_2$ norm) to the misfit vector ${\bf r}$, then we have 
\begin{equation}
{ r_2} := {x} \mid \sum^{N}_{i=1} (x -r_i)^2 ~~is~min. 
\end{equation} 
To find the minimum, we set the partial derivative with respect to ${\bf x}$ equal to zero and set $x=r_2$, then we obtain 
\begin{equation}
0 = \sum^{N}_{i=1} 2( r_2  -  r_i) 
\end{equation}
which is equivalent to  
\begin{equation}
r_2 = \frac{1}{N}\sum^{N}_{i=1} r_i 
\end{equation}
Clearly, $r_2$ is the definition of the mean. Therefore, large fitting errors (outliers) will bias the misfit $\ell_2$ norm and produce inaccurate mean. Next, let us define $r_1$ as the value that minimize the absolute differences ($\ell_1$ norm) to the misfit vector ${\bf r}$, then we have 
\begin{equation}
{ r_1} := {x} \mid \sum^{N}_{i=1} |x -r_i| ~~is~min 
\end{equation} 
Again to find the minimum, we set the partial derivative with respect to ${\bf x}$ equal to zero. We obtain 
\begin{equation}
0 = \sum^{N}_{i=1} sgn( r_1  -  r_i) 
\end{equation} 
The $sgn$ function is defined as equal to $+1$ when the difference is positive and equal to $-1$ when the difference is negative and undefined for zero difference. This equation indicates that $r_1$ should be chosen so that $N/2$ of the differences are positive and $N/2$ of the difference are negative which defines $r_1$ as the median. Thus, the $\ell_1$ norm misfit is equivalent to using the misfit median.  
Therefore, \cite{Claerbout1973} suggested using the $\ell_1$ norm ($p = 1$) for the misfit when the data contains erratic and large amplitude noise. We will name Radon transforms that use the $\ell_1$ norm misfit as robust Radon transforms. 
For the model regularization, the $\ell_2$ norm ($q=2$) model regularization induces the estimation of Radon models that are smooth.  On the other hand, the $\ell_1$ model regularization induces the estimation of Radon models that are sparse (high resolution).

In this thesis, we test four Radon transforms to remove source interferences in common receiver gathers. 
The first transform is the classical least-squares (non-robust) Radon transform obtained via quadratic regularization that corresponds to $p=2$ and $q=2$ \citep{Hampson1986b}.
The second transform is the classical high resolution (non-robust) Radon transform \citep{Sacchi1995b,Trad2003} that corresponds to $p=2$ and $q=1$. The third transform is the robust Radon transform with quadratic regularization ( $p=1$ and $q=2$). 
Finally, the fourth transform is the robust Radon transform with sparse regularization ( $p=1$ and $q=1$). 
The Iteratively Reweighed Least Squares (IRLS) algorithm can be used to estimate $\bf m$ for all four transforms \citep{Holland1977}. 
Our implementation of the IRLS algorithm starts by defining  either the $p$ or $q$ norms by the following expression
\begin{align}
\| {\bf x}\|_p^p &=\sum\limits_{i}|x_i||x_i|^{p-2}|x_i|= {\bf x}^{T}\, {\bf Q}\, {\bf x}  \nonumber \\
						   &= {\bf x}^{T} {\bf Q}{\bf x} ={\bf x}^{T}\,{\bf W_x}^{T}\,{\bf W_x}\, {\bf x}=\| {\bf W_x} \,{\bf x} \|_2^2
\end{align}
\noindent 
where
\begin{equation}
[{\bf Q}]_{ii}=\frac{1}{|x_i|^{2-p}} ~~and ~~[{\bf W_x}]_{ii}=\frac{1}{\sqrt{|x_i|^{2-p}}} ~~~~ 0<p\le 2.
\end{equation}  

The weighting ${\bf W_x}$ matrix cannot be computed for $x_i=0$. 
Therefore, the weighting matrix for $\bf m$ will be redefined as follows 
\[
 [{{\bf W}_{m}}]_{ii}  =
   \begin{dcases}
		\frac{1}{\sqrt{|m_i|^{2-q}}}     & \mbox{if } m_i > \epsilon_m  \, ,\\ 
		\frac{1}{\sqrt{\epsilon_{m}^{2-q}}}  & \mbox{if } m_i \leq \epsilon_m \,.
   \end{dcases}
\]

Similarly, we redefine the weighting matrix for the residuals as follows 
\[
[{{\bf W}_{r}}]_{ii} =
   \begin{dcases}
		\frac{1}{\sqrt{|r_i|^{2-p}}}     & \mbox{if } r_i > \epsilon_r \, ,\\
		\frac{1}{\sqrt{\epsilon_{r}^{2-p}}}  & \mbox{if } r_i \leq \epsilon_r \,.
   \end{dcases}
\]
Both $\epsilon_r$ and $\epsilon_m$ represent small numbers to avoid the singularity at  $r=0$ and $m=0$.
\citet{Holland1977} defined the value of $\epsilon_{r}$ as
\begin{equation} \label{ch4_epcilon_r}
\epsilon_{r}=b_r ~~ \frac{\operatorname{MAD}({\bf r})}{0.6745},
\end{equation}
where $\operatorname{MAD}$ indicates the median absolute deviation of the residuals $\bf r$. 
The parameter $b_r$ is a tuning parameter and \citet{Holland1977} recommended using $b_r=1.345$.
The estimator $\operatorname{MAD}$ use the median twice, first to estimate the center of the residuals in order to form the set of absolute residual about the median, and then compute the median of these absolute residuals \citep{Maronna2006} 
\begin{equation}
\operatorname{MAD}({\bf r})=\operatorname{Med}({{\bf r} - \operatorname{Med}({\bf r})}).
\end{equation}
The reason for $0.6745$ in equation \label{ch4_epcilon_r} is to make $\operatorname{MAD}$ comparable with standard deviation of a normal random variable \citep{Maronna2006}.

The parameter $\epsilon_m$ is computed via the following expression
 $$\epsilon_m=b_m~~\frac{\operatorname{max}|{\bf m}|}{100}$$,   
where $b_m$ is a tuning parameter that in our simulations was selected in an heuristic fashion \citep{Trad2003, Ji2006}. 
We can now represent the inversion by the following new cost function 
\begin{equation}\label{equ:10}
J=\|{\bf W}_{r}\, {\bf r}\|_2^2+\mu \| {\bf W}_{m} {\bf m} \|_2^2. 
\end{equation}
\noindent 
Thus, we have turned the non-quadratic problem into a sequence of quadratic minimization problems for fixed weighting matrices ${\bf W}_r$ and ${\bf W}_m$ that depends on unknowns $\bf r$ and $\bf m$, respectively. 
Equation \ref{equ:10} can be written in its standard form \citep{Hansen1998} by a simple change of variable ${\bf u} = {\bf W}_{m} {\bf m} $
\begin{equation}\label{equ:11}
J=\|{\bf W}_{r}\, [{\bf L} {({\bf W}_m)}^{-1} {\bf u}-{\bf d}] \|_2^2+\mu \| {\bf u} \|_2^2. 
\end{equation}
\noindent
Equation \ref{equ:11} is minimized via the method of conjugate gradients \citep{Hestenes1952,Scales1987,Shewchuk1994} followed by an update of the matrices of weights ${\bf W_r} $ and ${\bf W_m}$. We follow the method described by \cite{Trad2003} where  the regularization term in equation  \ref{equ:11} is omitted by setting $\mu=0$ and the number of iterations of the method of conjugate gradients replaces the trade-off parameter $\mu$ \citep{Hansen1998}. 
In essence, we have an internal iteration to minimize equation (\ref{equ:11}) via the method of Conjugate Gradients and an external iteration to update the weights. 
The algorithm is stopped when the misfit change between iterations is less than a defined tolerance value ($tolerance=0.01$) or when it reaches a maximum number of iterations.   
Finally, we want to clarify that the forward Radon operator is also convolved with a wavelet.
This permits representing a constant amplitude hyperbola via a single coefficient in the Radon space.  
Consequently, the adjoint operator requires cross-correlation with the wavelet. 
In other words, in our algorithm we have replaced the operators $\bf L$ by $\bf C \bf L$ and ${\bf L}^T$  by ${\bf L}^T {\bf C}^T$. 
The operators $\bf C$ and ${\bf C}^T$ correspond to the convolution and cross-correlation with a known wavelet, respectively \citep{Claerbout1992}. 
We have selected a zero phase wavelet with an amplitude spectrum similar to the amplitude spectrum of the wavelet in the data.

\section{Synthetic data example}

\begin{figure}[htbp]
		\includegraphics{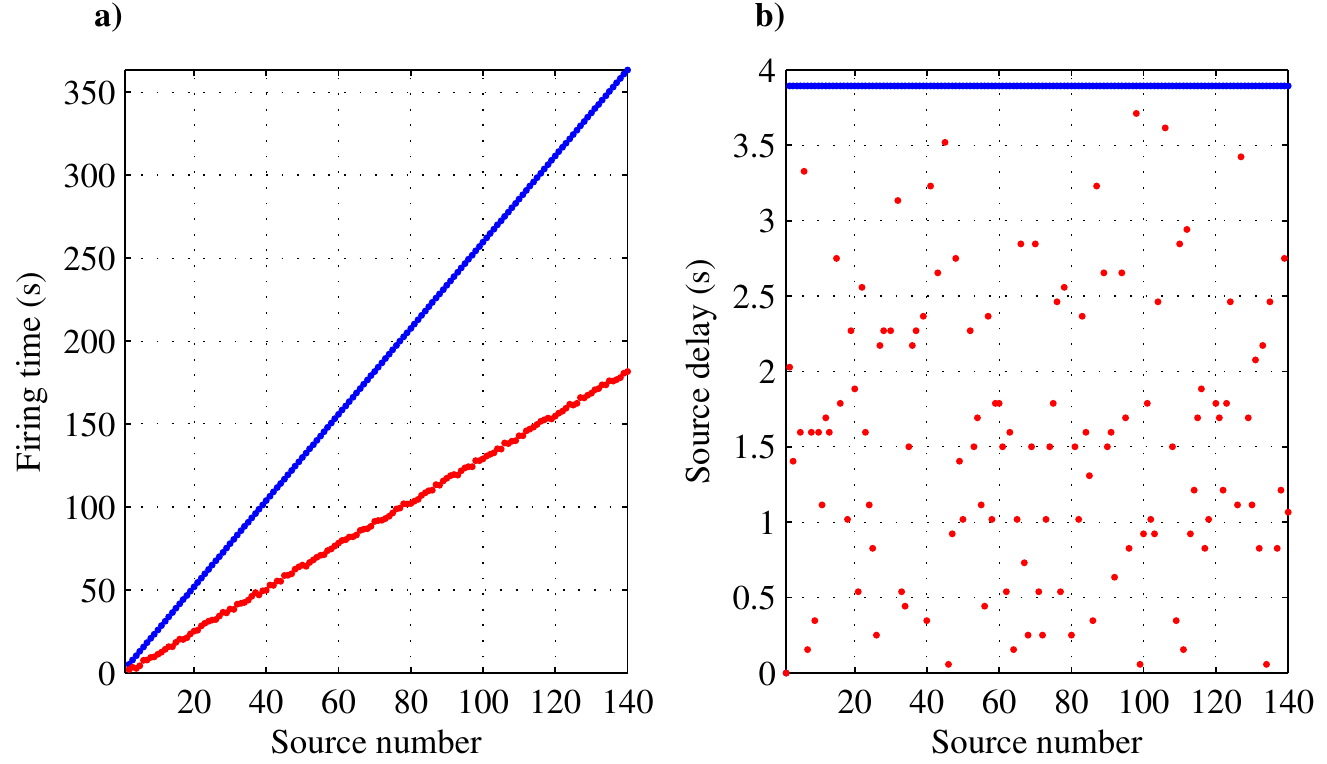}
		\vspace*{2mm}
		\caption{Seismic sources firing times for synthetic data example.
			(a) Firing times of conventional (blue) and simultaneous seismic sources (red). 
			(b) Time delay between successive sources for conventional (blue) and simultaneous sources (red).}
		\label{ch4_synth_firing_times}
\end{figure}

We tested the robust Radon transform with a numerically blended synthetic data set. The single shot gathers of the synthetic data were modelled using the forward apex shifted hyperbolic Radon modelling operator. This guarantees that the main components of the data are composed of reflection hyperbolas similar to the transform basis. This synthetic data resembles a subsurface model of dipping reflectors. However, the synthetic data are composed of 10 seismic reflections with the velocities $1500, 1560, 1590, 1600, 1655, 1700, 1750, 1780, 1850$ and $1975$ m/s, while, the transform basis use only 5 velocities $1500, 1600, 1700, 1800$ and $1900$ m/s. This situation is similar to field data examples where the true velocity is unknown and the Radon transform use coarse velocity sampling.   The data are numerically blended with a $50\%$ reduction in time compared to conventional acquisition. 
The acquisition scenario represents a single source boat with the time interval between successive sources is nearly half of the conventional acquisition. In order to make the source interferences appear incoherent, the source firing times are dithered using random time delays.  The firing times of the sources for both conventional and blended acquisition sources are shown in Figure \ref{ch4_synth_firing_times}.
\begin{figure}[htbp] 
	\centering
	\includegraphics{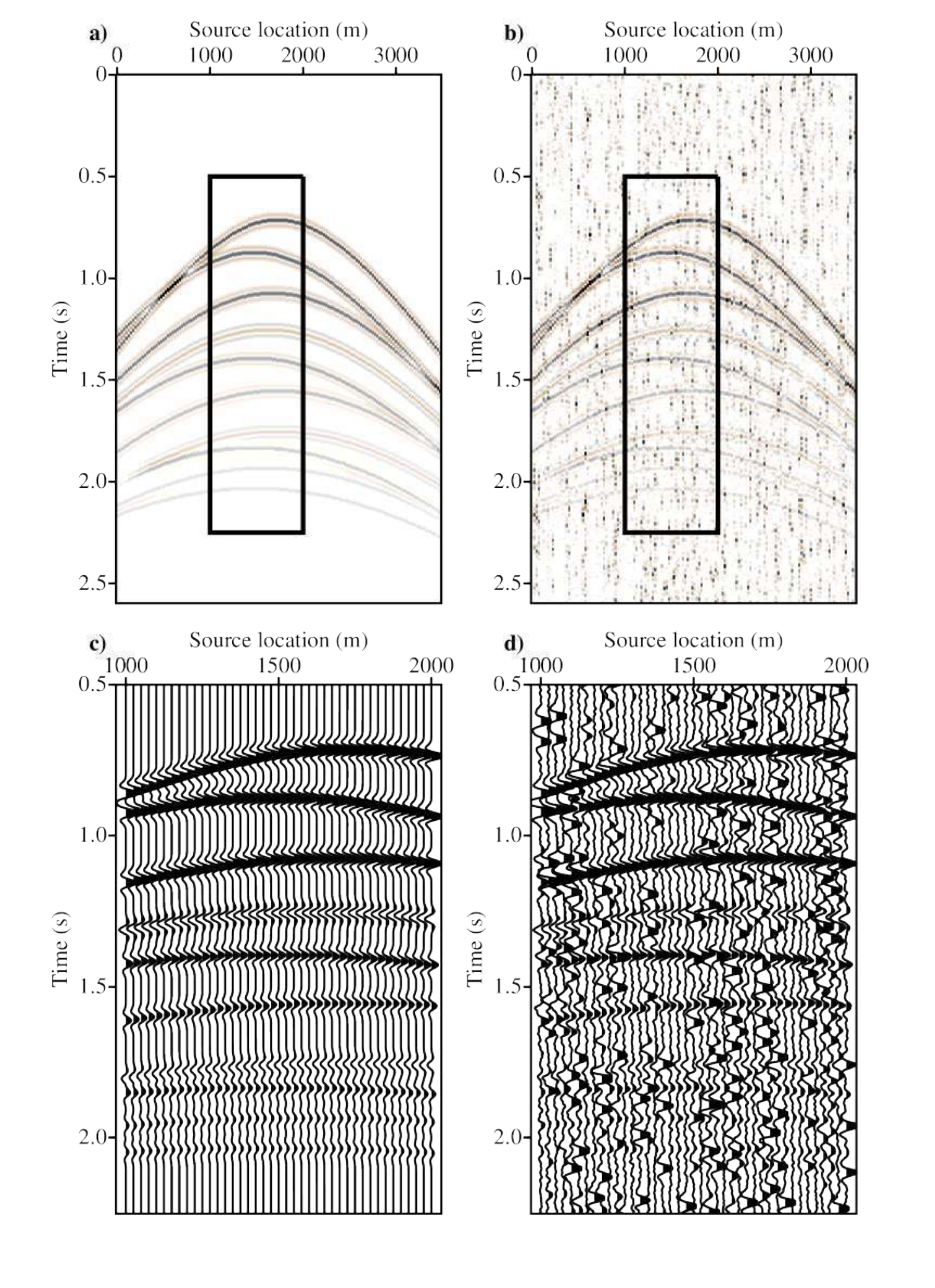}
	\reduceVspace
	\caption{A numerically blended common receiver gather from the synthetic data example.
	(a) Original gather.   
	(b) Pseudo-deblended gather .  
	(c) Close-up of the original gather.   
	(d) Close-up of the pseudo-deblended gather. }
	\label{ch4_synth_CRG_example}
\end{figure}
Figure \ref{ch4_synth_CRG_example}a shows an original common receiver gather of the simple synthetic data example and Figure \ref{ch4_synth_CRG_example}b shows the same gather after blending and pseudo-deblending.
Figures \ref{ch4_synth_CRG_example}c and d are close-up wiggle plots of the areas marked on Figures \ref{ch4_synth_CRG_example}a and b, respectively. These close-up figures show more clearly the high amplitude source interferences mixing with the low amplitude late arrivals of the primary source (notice the reflections at 2 sec). 
   
\begin{sidewaysfigure}[htbp]
 		\includegraphics{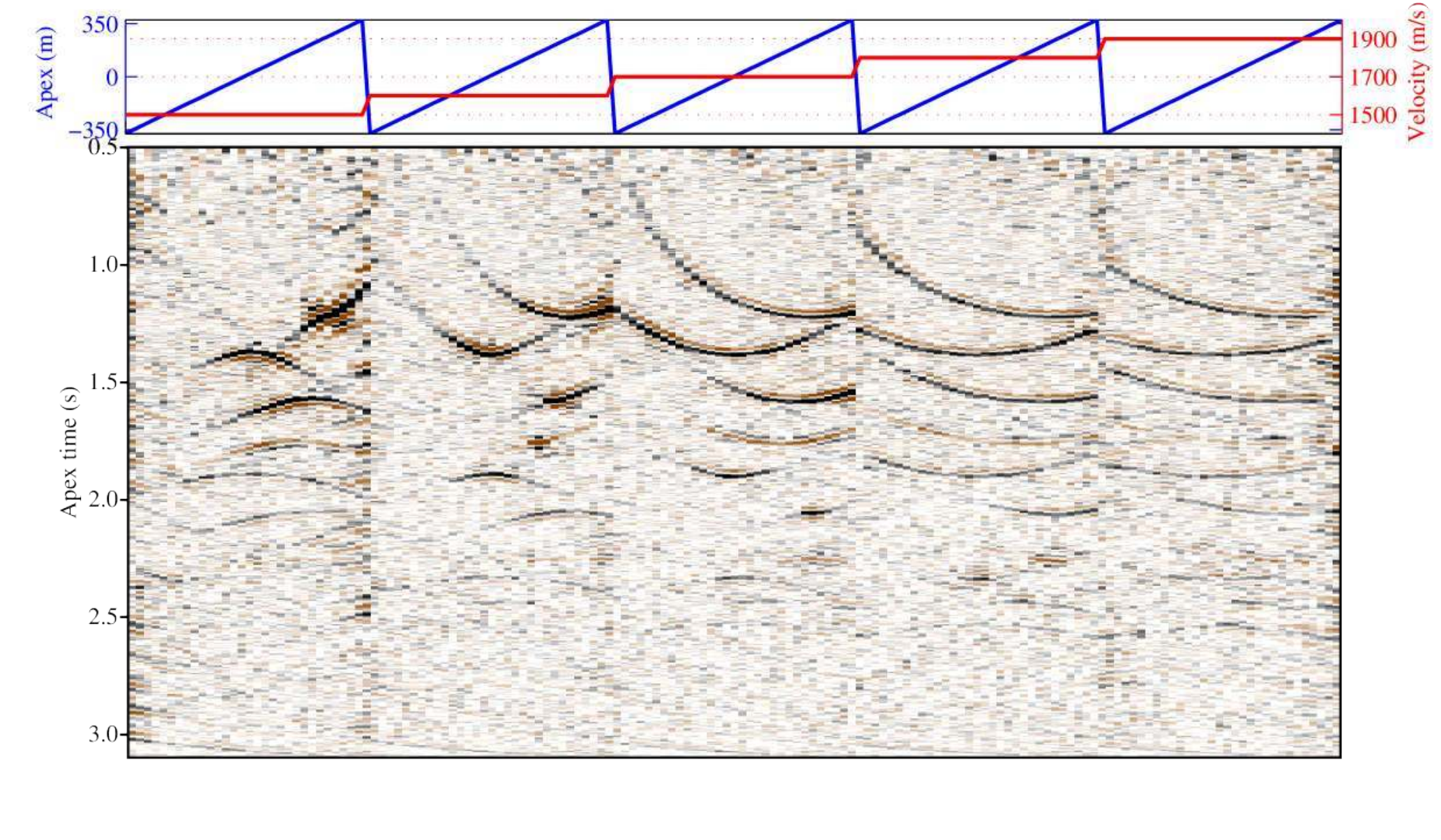}
 		\reduceVspace
 		\caption{Apex-shifted Radon model for the common receiver gather from the synthetic data example estimated using $p=2$ and $q=2$ inversion.}
		\label{ch4_synth_model_L2L2}
\end{sidewaysfigure}

\begin{sidewaysfigure}[htbp]
 		\includegraphics{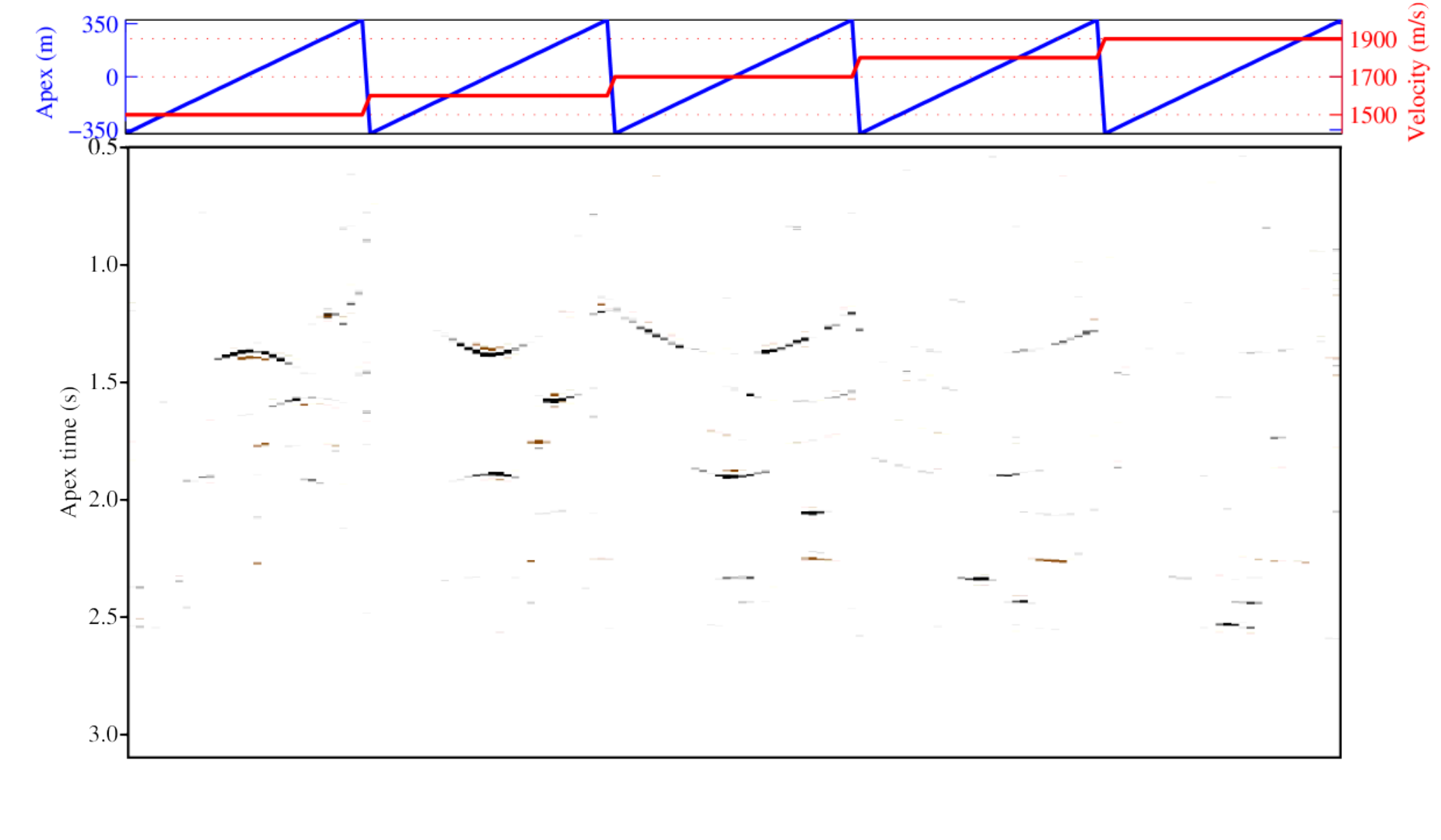}
 		\reduceVspace
 		\caption{Apex-shifted Radon model for the common receiver gather from the synthetic data example estimated using $p=2$ and $q=1$ inversion.}
		\label{ch4_synth_model_L2L1}
\end{sidewaysfigure}

\begin{sidewaysfigure}[htbp]
 		\includegraphics{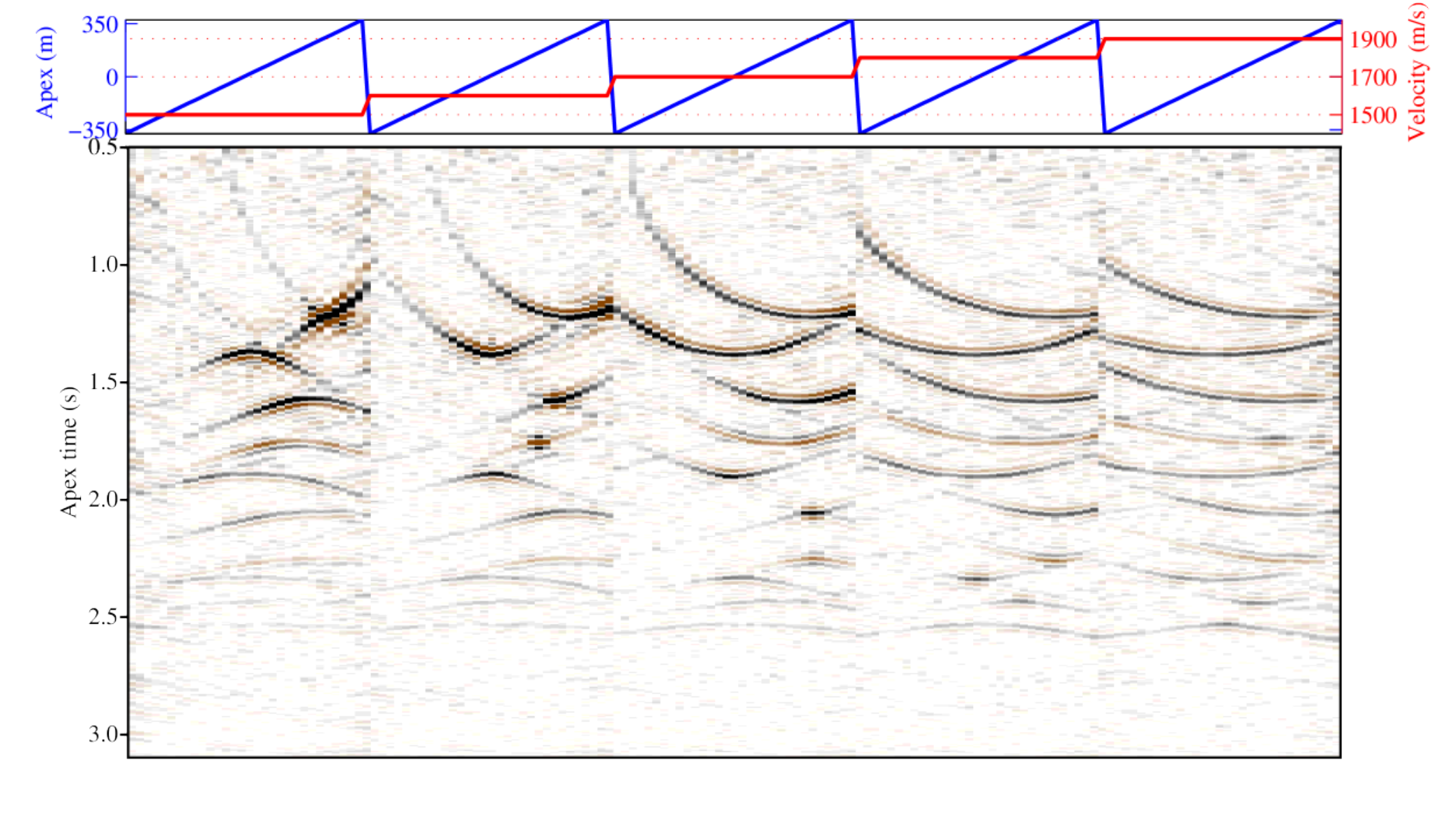}
 		\reduceVspace
 		\caption{Apex-shifted Radon model for the common receiver gather from the synthetic data example estimated using $p=1$ and $q=2$ inversion.}
		\label{ch4_synth_model_L1L2}
\end{sidewaysfigure}

\begin{sidewaysfigure}[htbp]
 		\includegraphics{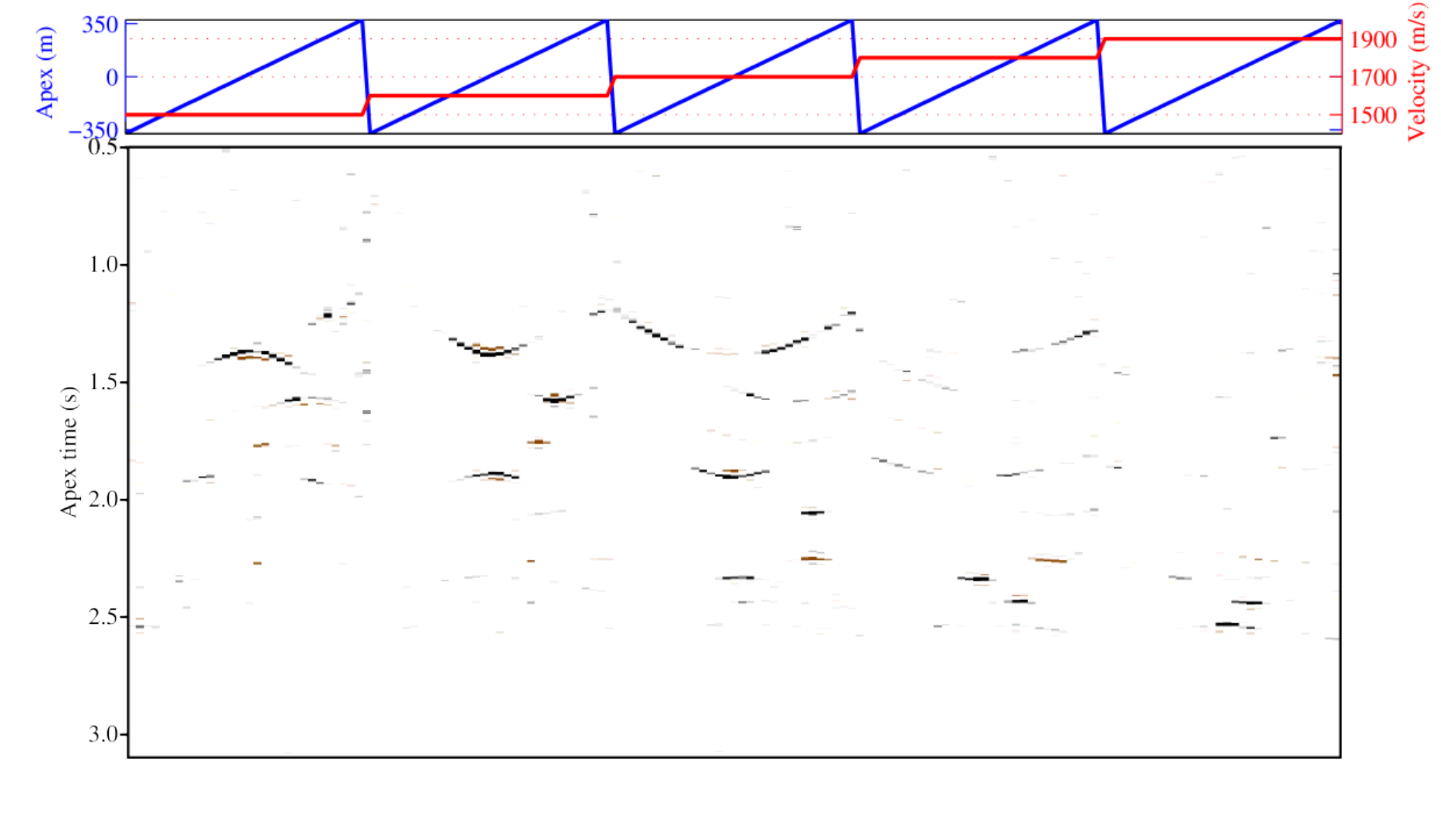}
 		\reduceVspace
 		\caption{Apex-shifted Radon model for the common receiver gather from the synthetic data example estimated using $p=1$ and $q=1$ inversion.}
		\label{ch4_synth_model_L1L1}
\end{sidewaysfigure}

\begin{figure}[htbp]
 		\includegraphics{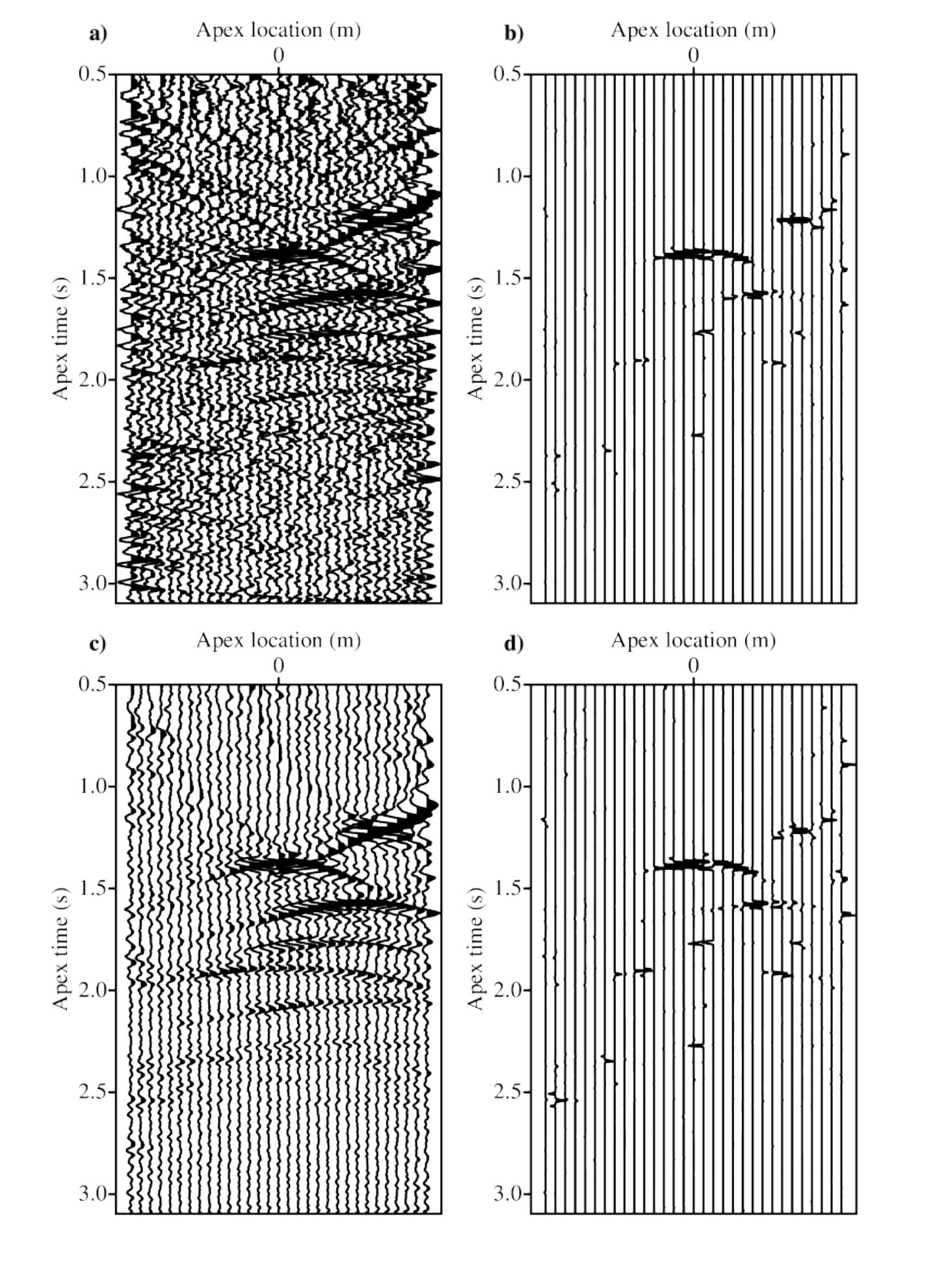}
 		\reduceVspace
   		\caption{One velocity panel ($v=1500~m/s$) of the ASHRT models for the simple synthetic data estimated by inversion with 
   		(a) $p=2$ and $q=2$, (b) $p=2$ and $q=1$, (c) $p=1$ and $q=2$, and (d) $p=1$ and $q=1$. }
		\label{ch4_synth_modeld_closeup}
\end{figure}

\begin{figure}[htp] 
	\centering
	\includegraphics{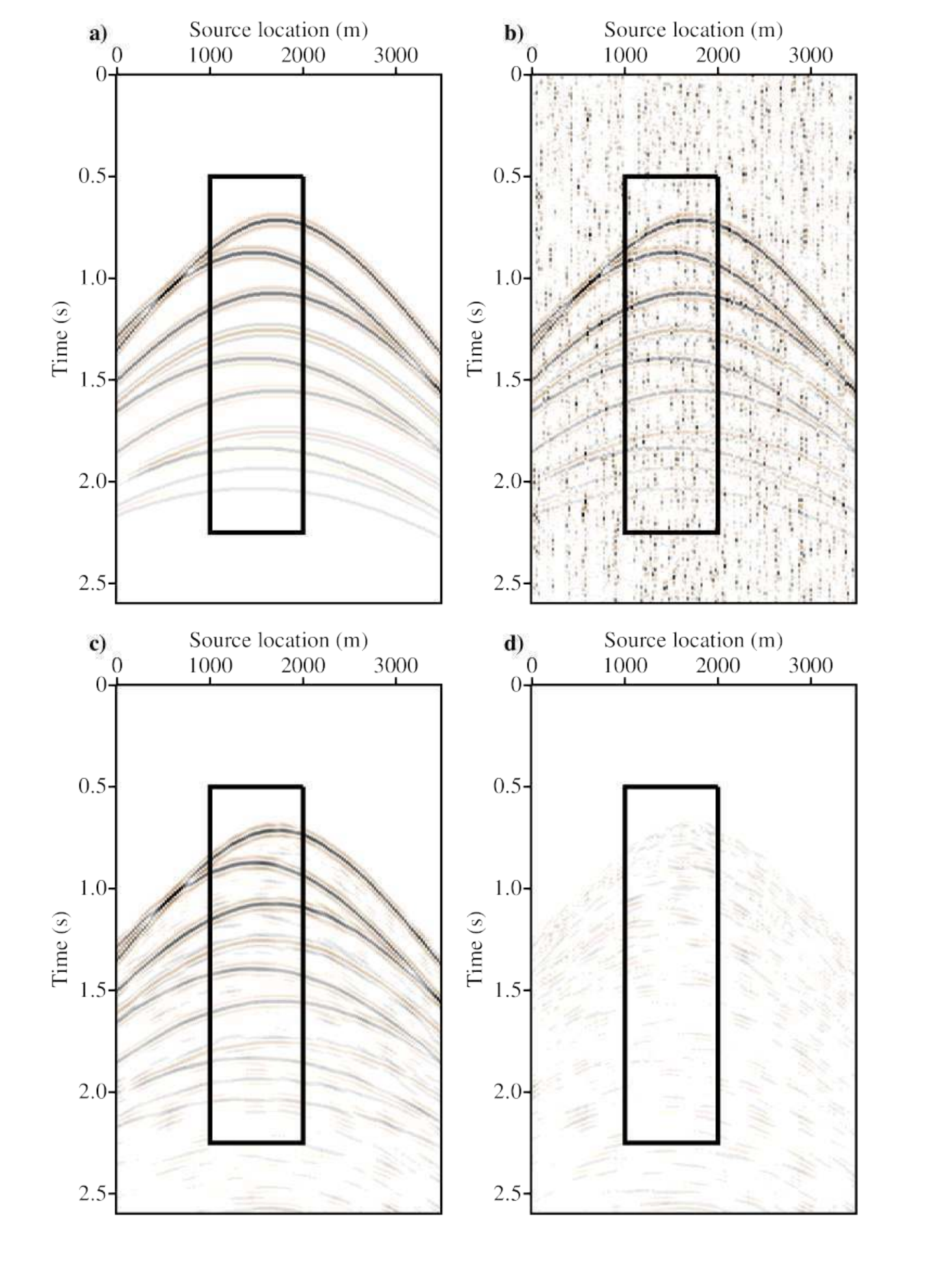}
	\reduceVspace
	\caption{Synthetic data common receiver gather recovered from Radon model estimated using $p=2$ and $q=2$ inversion.
	(a) Original gather.   (b) Pseudo-deblended gather.  (c) Recovered gather.   (d) Recovered gather error.}
	\label{ch4_synth_CRG_L2L2}
\end{figure}

\begin{figure}[htp] 
	\centering
	\includegraphics{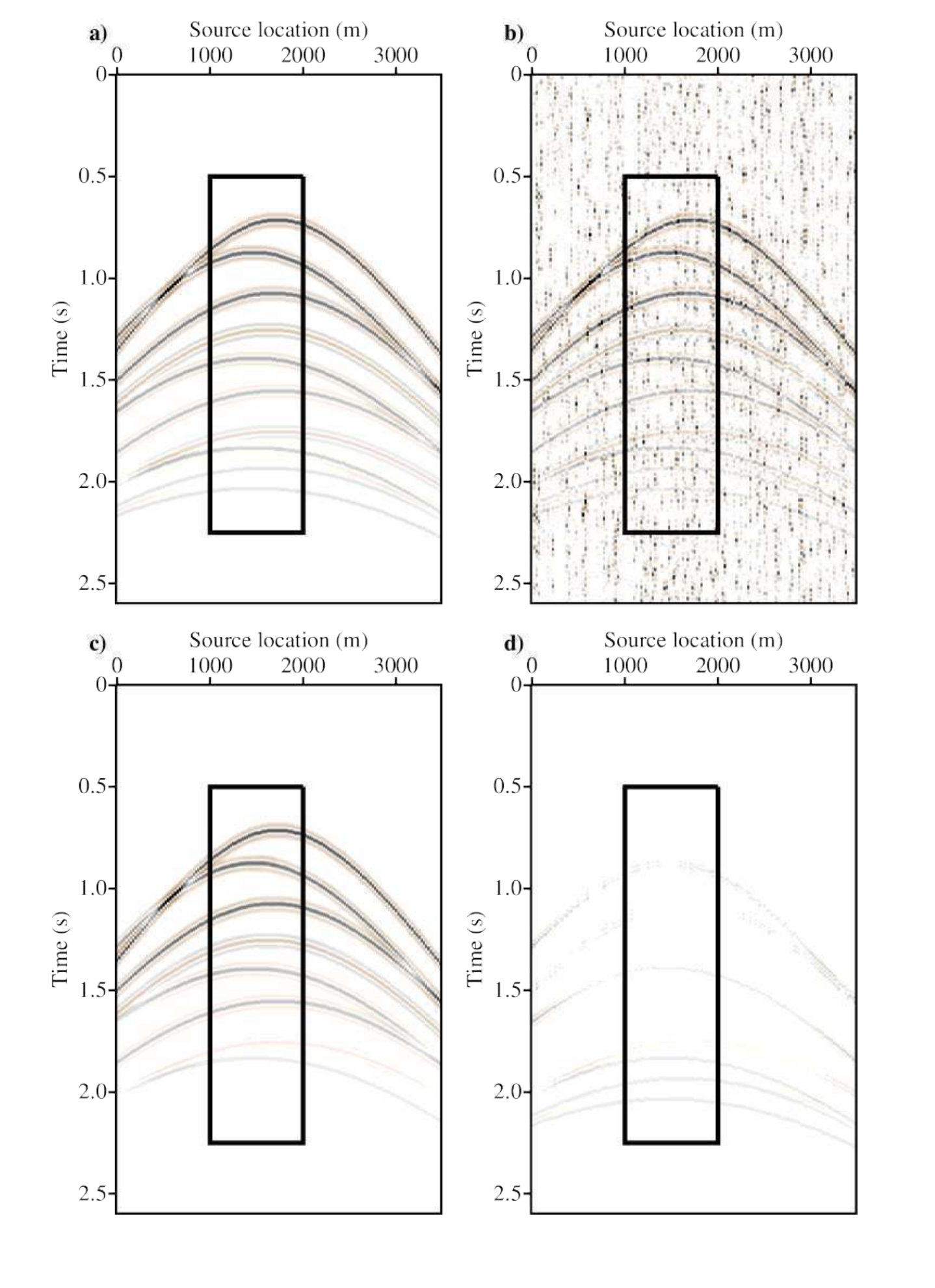}
	\reduceVspace
	\caption{Synthetic data common receiver gather recovered from Radon model estimated using $p=2$ and $q=1$ inversion.
	(a) Original gather.   (b) Pseudo-deblended gather.  (c) Recovered gather.   (d) Recovered gather error.}
	\label{ch4_synth_CRG_L2L1}
\end{figure}

\begin{figure}[htp] 
	\centering
	\includegraphics{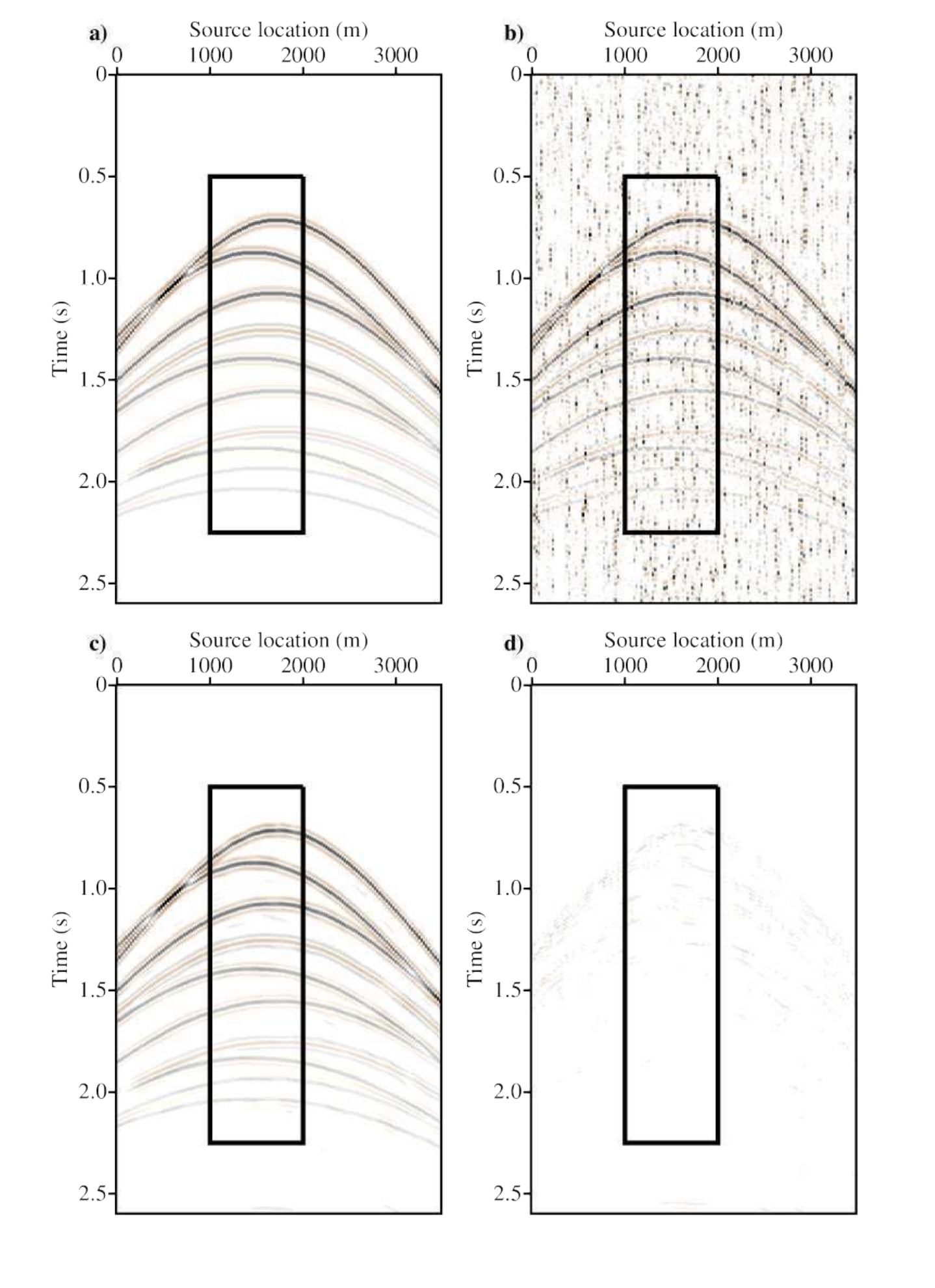}
	\reduceVspace
	\caption{Synthetic data common receiver gather recovered from Radon model estimated using $p=1$ and $q=2$ inversion.
	(a) Original gather.   (b) Pseudo-deblended gather.  (c) Recovered gather.   (d) Recovered gather error.}
	\label{ch4_synth_CRG_L1L2}
\end{figure}

\begin{figure}[htp] 
	\centering
	\includegraphics{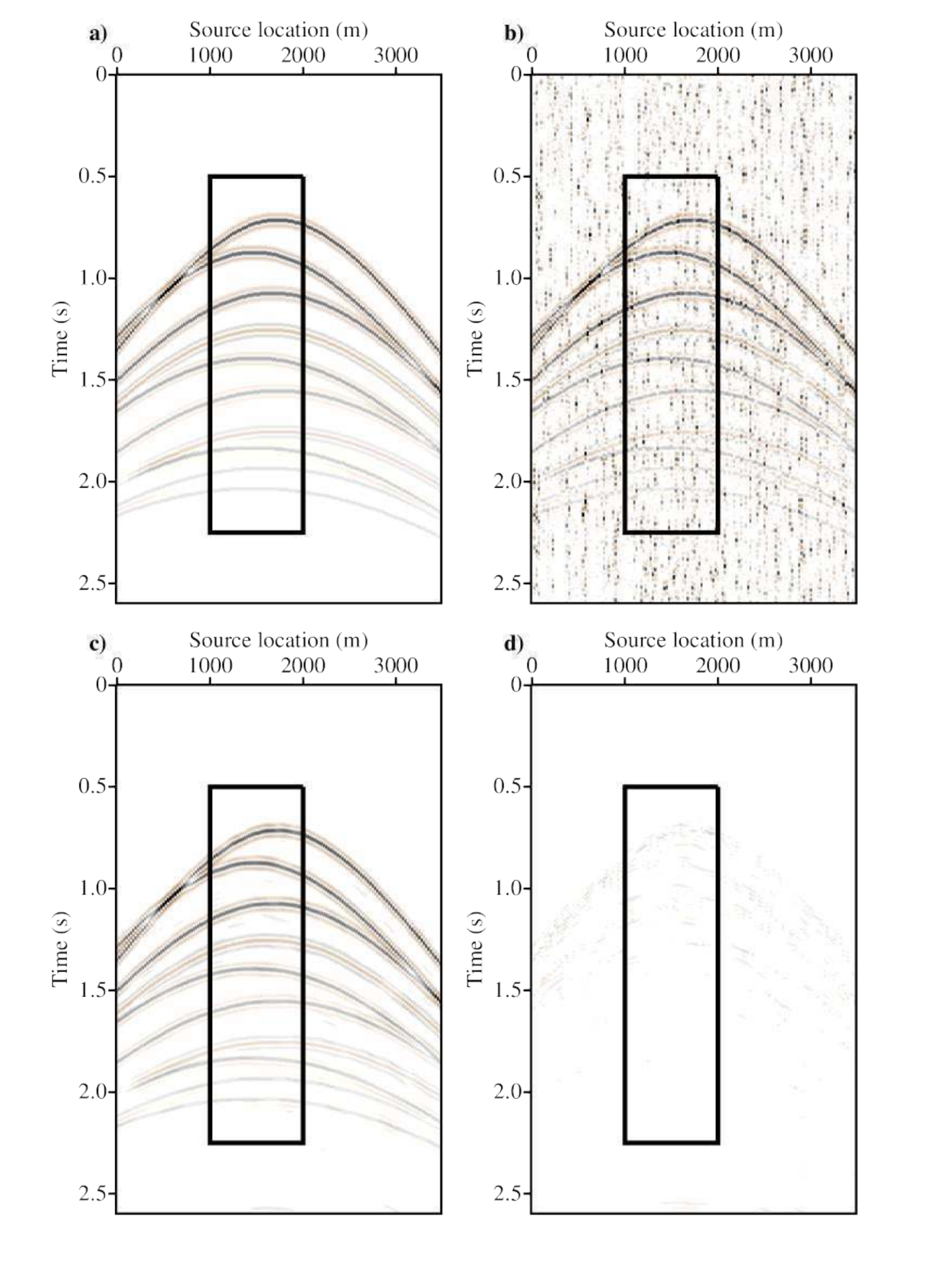}
	\reduceVspace
	\caption{Synthetic data common receiver gather recovered from Radon model estimated using $p=1$ and $q=1$ inversion.
	(a) Original gather.   (b) Pseudo-deblended gather.  (c) Recovered gather.   (d) Recovered gather error.}
	\label{ch4_synth_CRG_L1L1}
\end{figure}

\begin{figure}[htp] 
	\centering
	\includegraphics{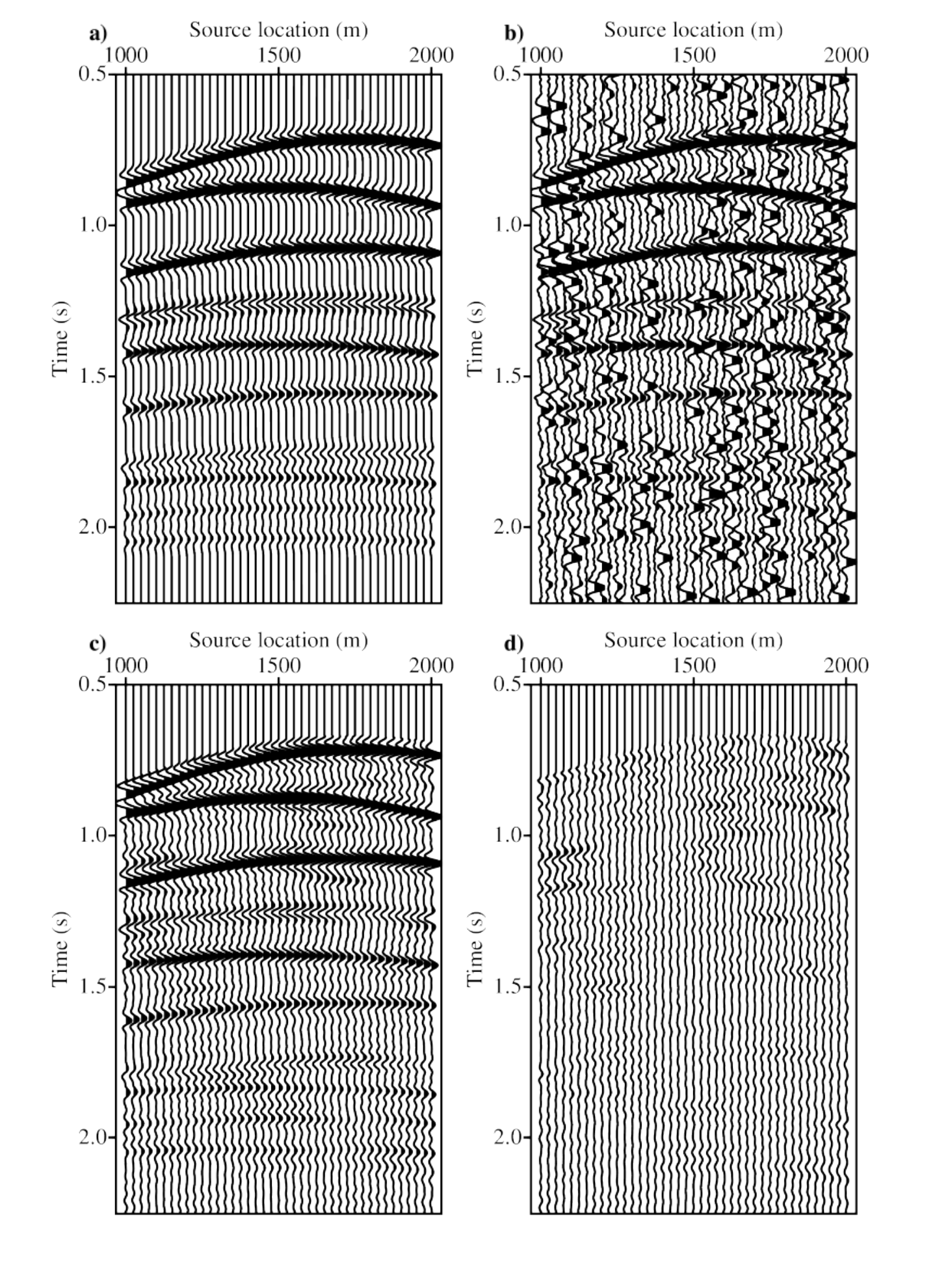}
	\reduceVspace
	\caption{Close-up of synthetic data common receiver gather recovered from Radon model estimated using $p=2$ and $q=2$ inversion.
	(a) Original gather.   (b) Pseudo-deblended gather.  (c) Recovered gather.   (d) Recovered gather error.}
	\label{ch4_synth_CRG_L2L2_closeup}
\end{figure}

\begin{figure}[htp] 
	\centering
	\includegraphics{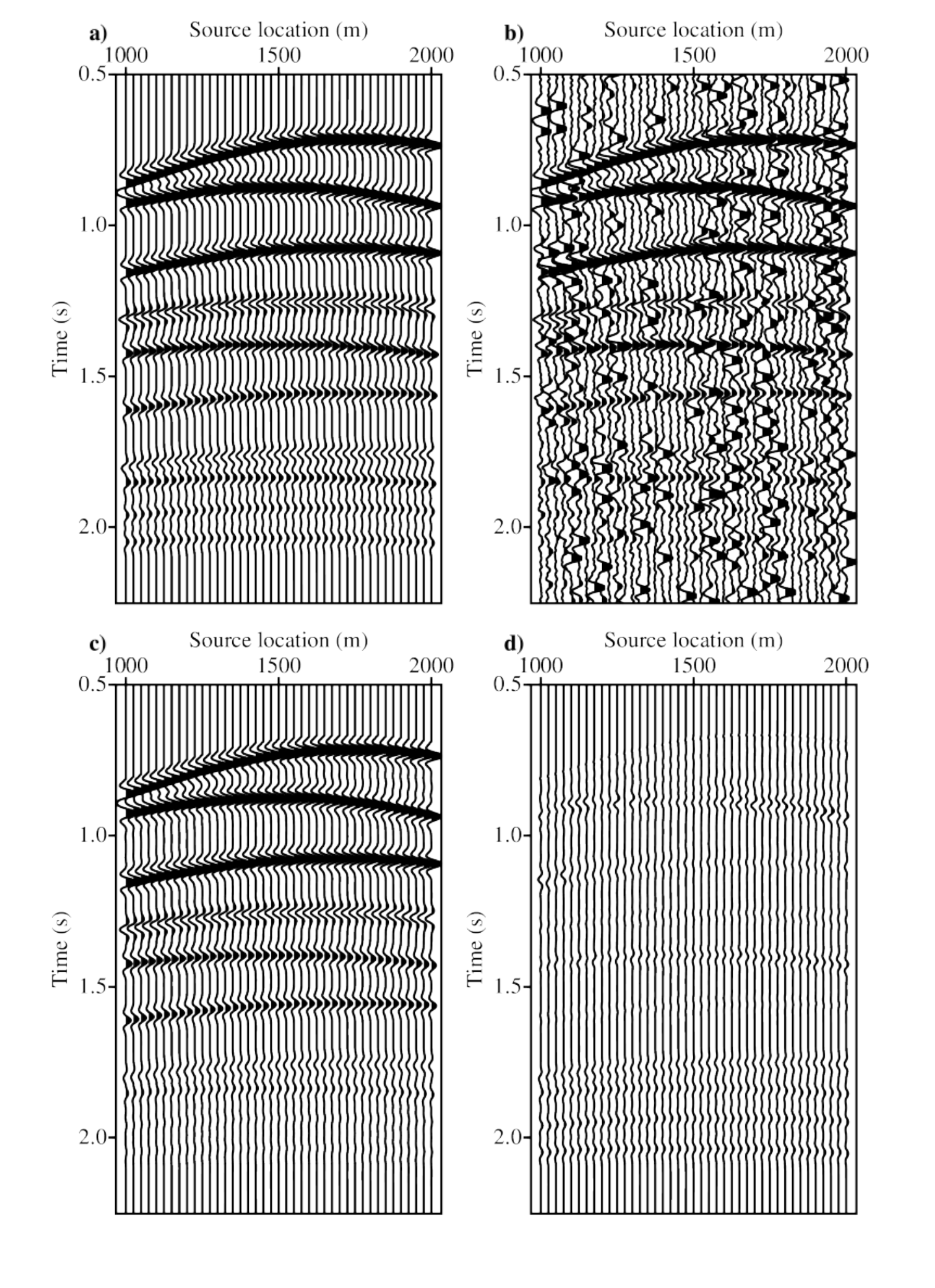}
	\reduceVspace
	\caption{Close-up of synthetic data common receiver gather recovered from Radon model estimated using $p=2$ and $q=1$ inversion.
	(a) Original gather.   (b) Pseudo-deblended gather.  (c) Recovered gather.   (d) Recovered gather error.}
	\label{ch4_synth_CRG_L2L1_closeup}
\end{figure}

\begin{figure}[htp] 
	\centering
	\includegraphics{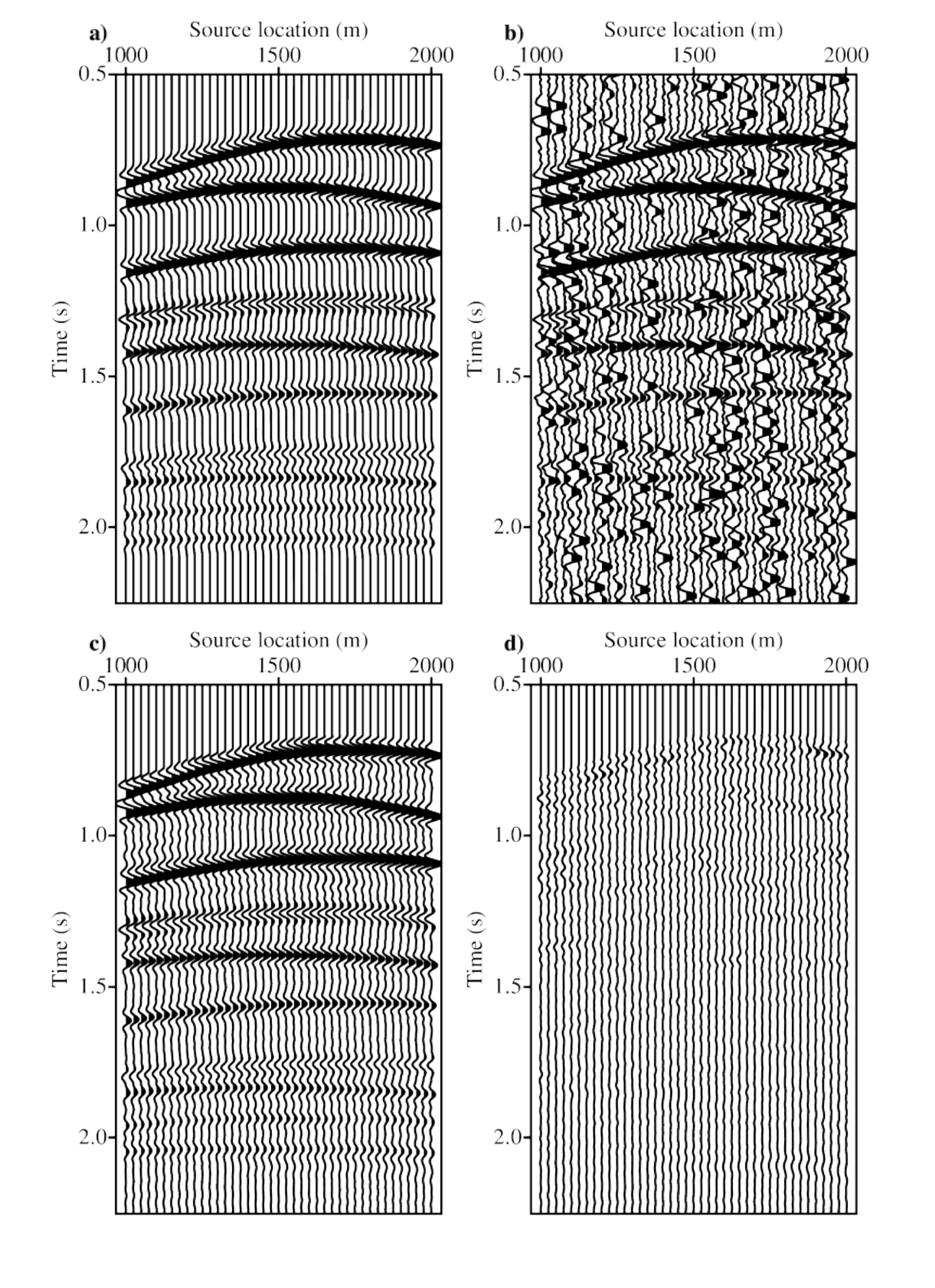}
	\reduceVspace
	\caption{Close-up of synthetic data common receiver gather recovered from Radon model estimated using $p=1$ and $q=2$ inversion.
	(a) Original gather.   (b) Pseudo-deblended gather.  (c) Recovered gather.   (d) Recovered gather error.}
	\label{ch4_synth_CRG_L1L2_closeup}
\end{figure}

\begin{figure}[htp] 
	\centering
	\includegraphics{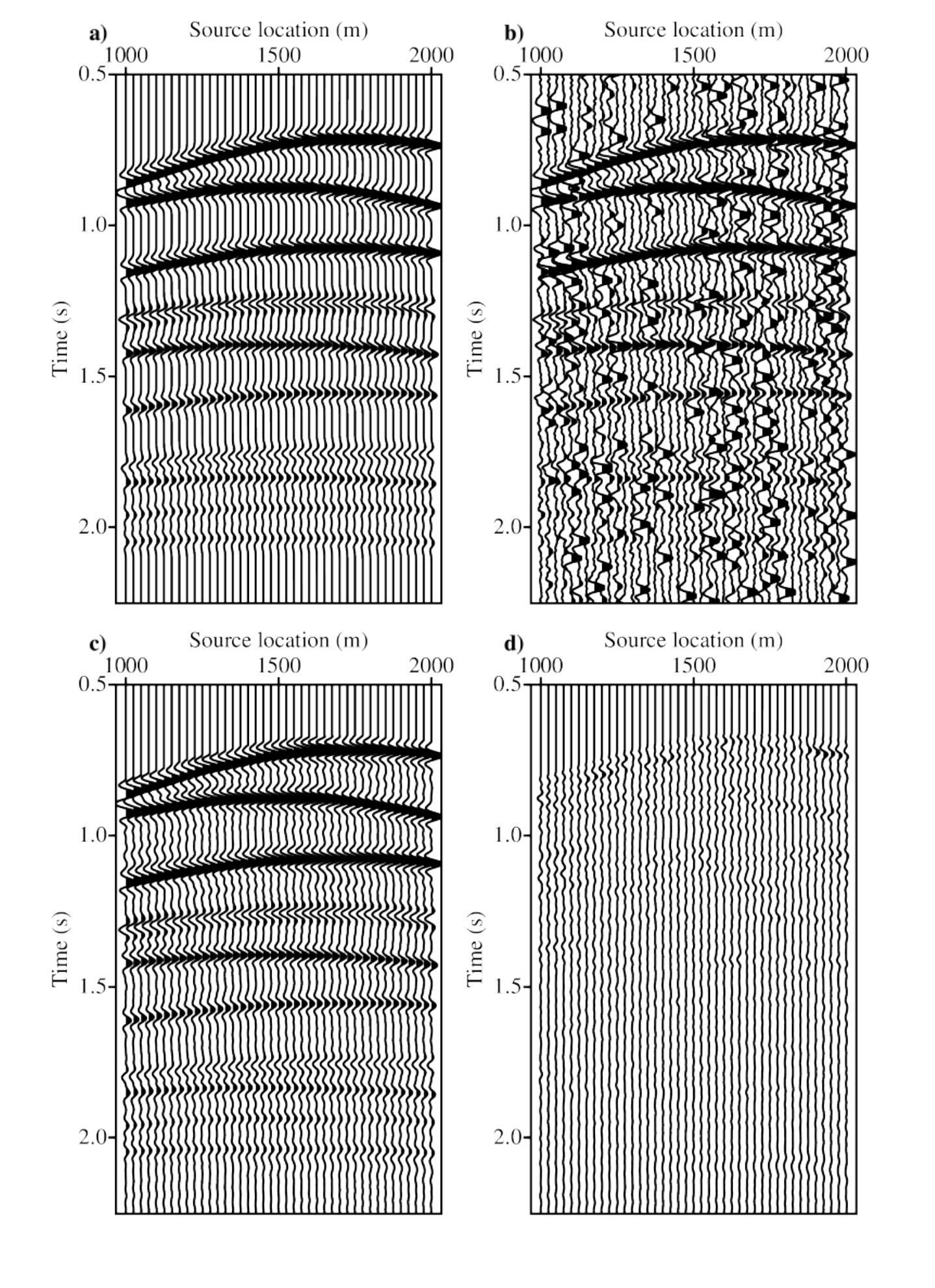}
	\reduceVspace
	\caption{Close-up of synthetic data common receiver gather recovered from the Radon model estimated using $p=1$ and $q=1$ inversion.
	(a) Original gather.   (b) Pseudo-deblended gather.  (c) Recovered gather.   (d) Recovered gather error.}
	\label{ch4_synth_CRG_L1L1_closeup}
\end{figure}

\begin{figure}[htp] 
	\centering
	\includegraphics{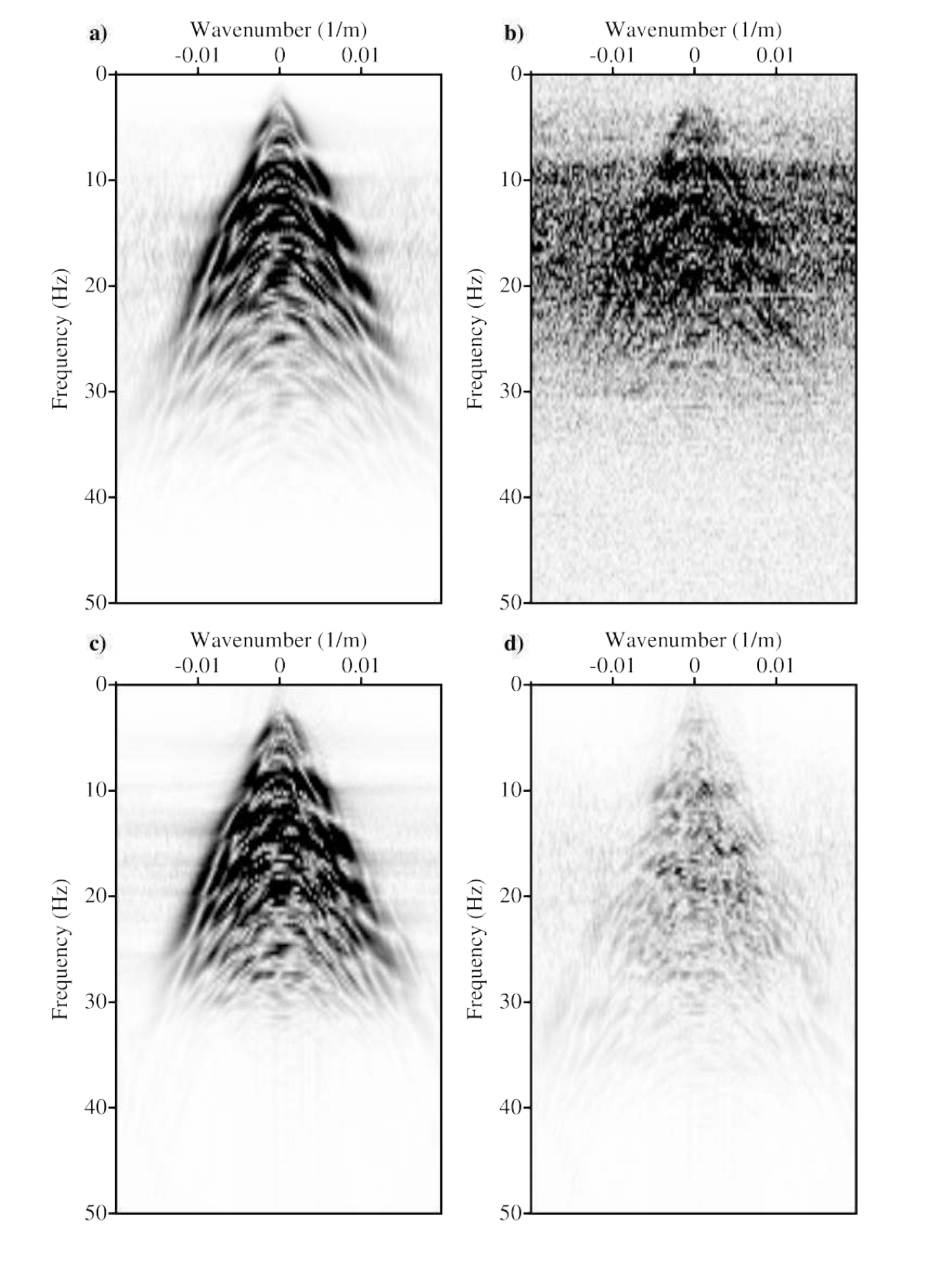}
	\reduceVspace
	\caption{The $f-k$ spectra of the synthetic data common receiver gather recovered from the estimated Radon model using $p=2$ and $q=2$ inversion.
	(a) Original gather.   (b) Pseudo-deblended gather.  (c) Recovered gather.   (d) Recovered gather error.}
	\label{ch4_synth_CRG_L2L2_fk}
\end{figure}

\begin{figure}[htp] 
	\centering
	\includegraphics{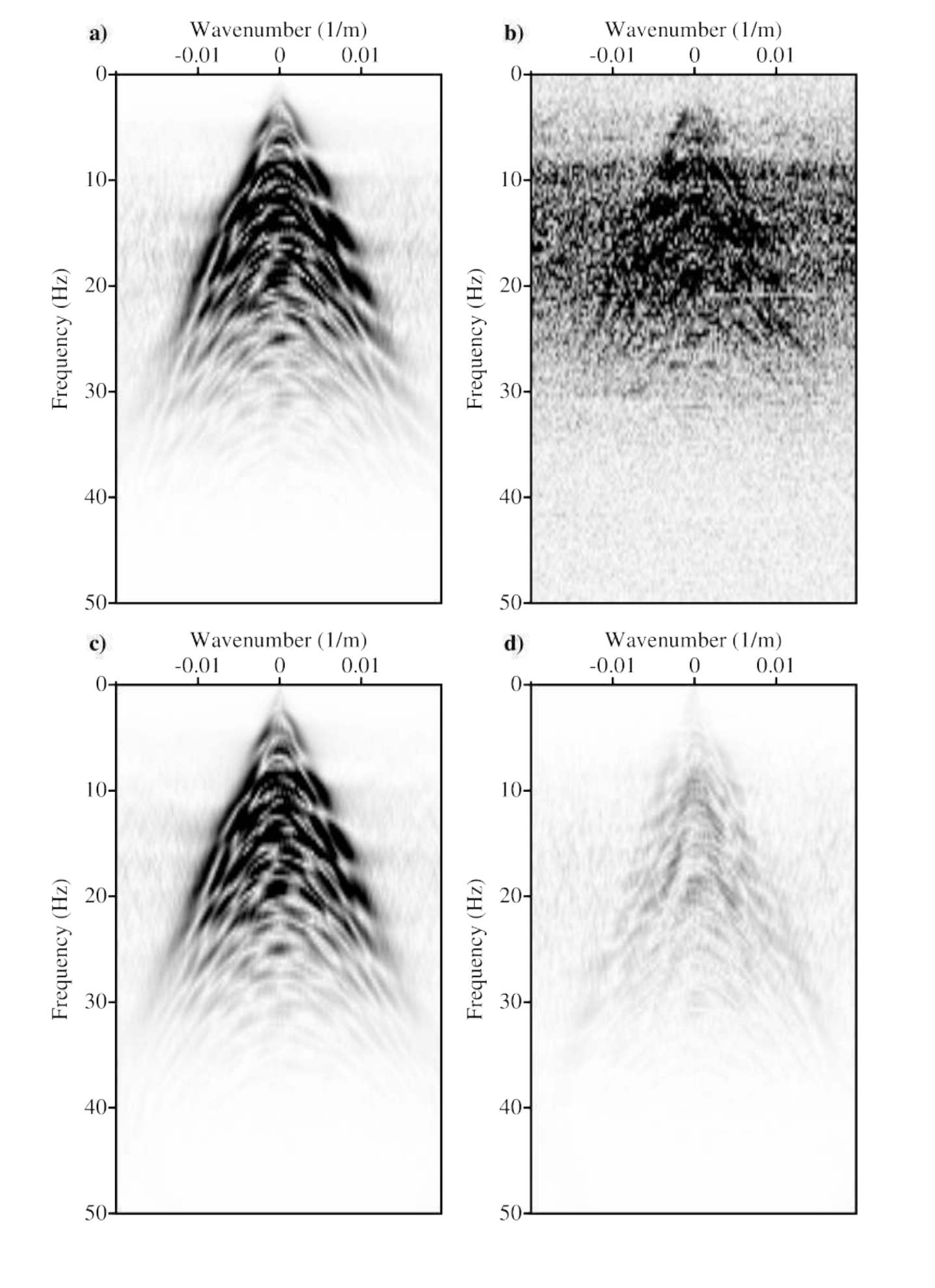}
	\reduceVspace
	\caption{The $f-k$ spectra of the synthetic data common receiver gather recovered from the estimated Radon model using $p=2$ and $q=1$ inversion.
	(a) Original gather.   (b) Pseudo-deblended gather.  (c) Recovered gather.   (d) Recovered gather error.}
	\label{ch4_synth_CRG_L2L1_fk}
\end{figure}

\begin{figure}[htp] 
	\centering
	\includegraphics{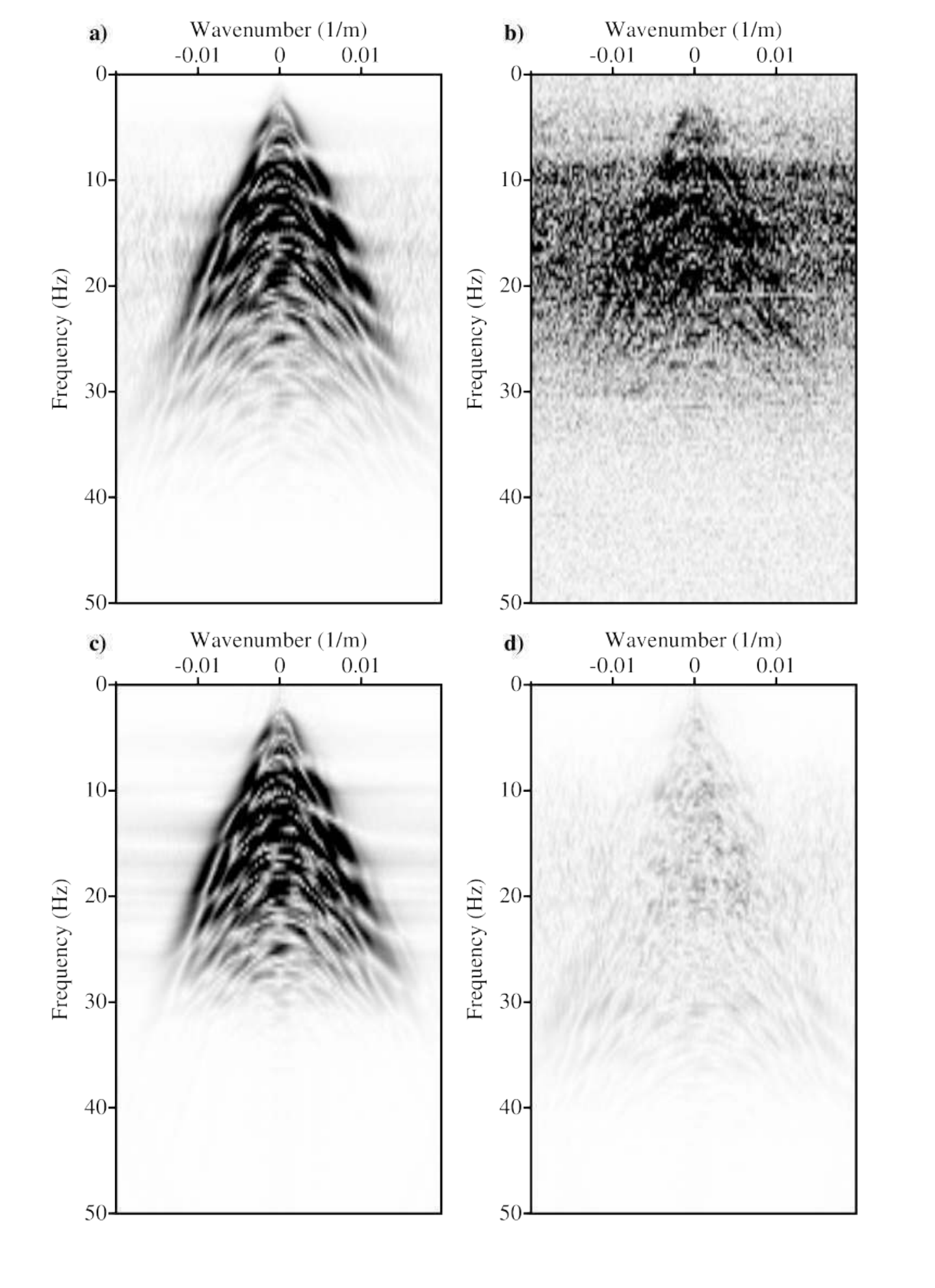}
	\reduceVspace
	\caption{The $f-k$  spectra of the synthetic data common receiver gather recovered from the estimated Radon model using $p=1$ and $q=2$ inversion.
	(a) Original gather.   (b) Pseudo-deblended gather.  (c) Recovered gather.   (d) Recovered gather error.}
	\label{ch4_synth_CRG_L1L2_fk}
\end{figure}

\begin{figure}[htp] 
	\centering
	\includegraphics{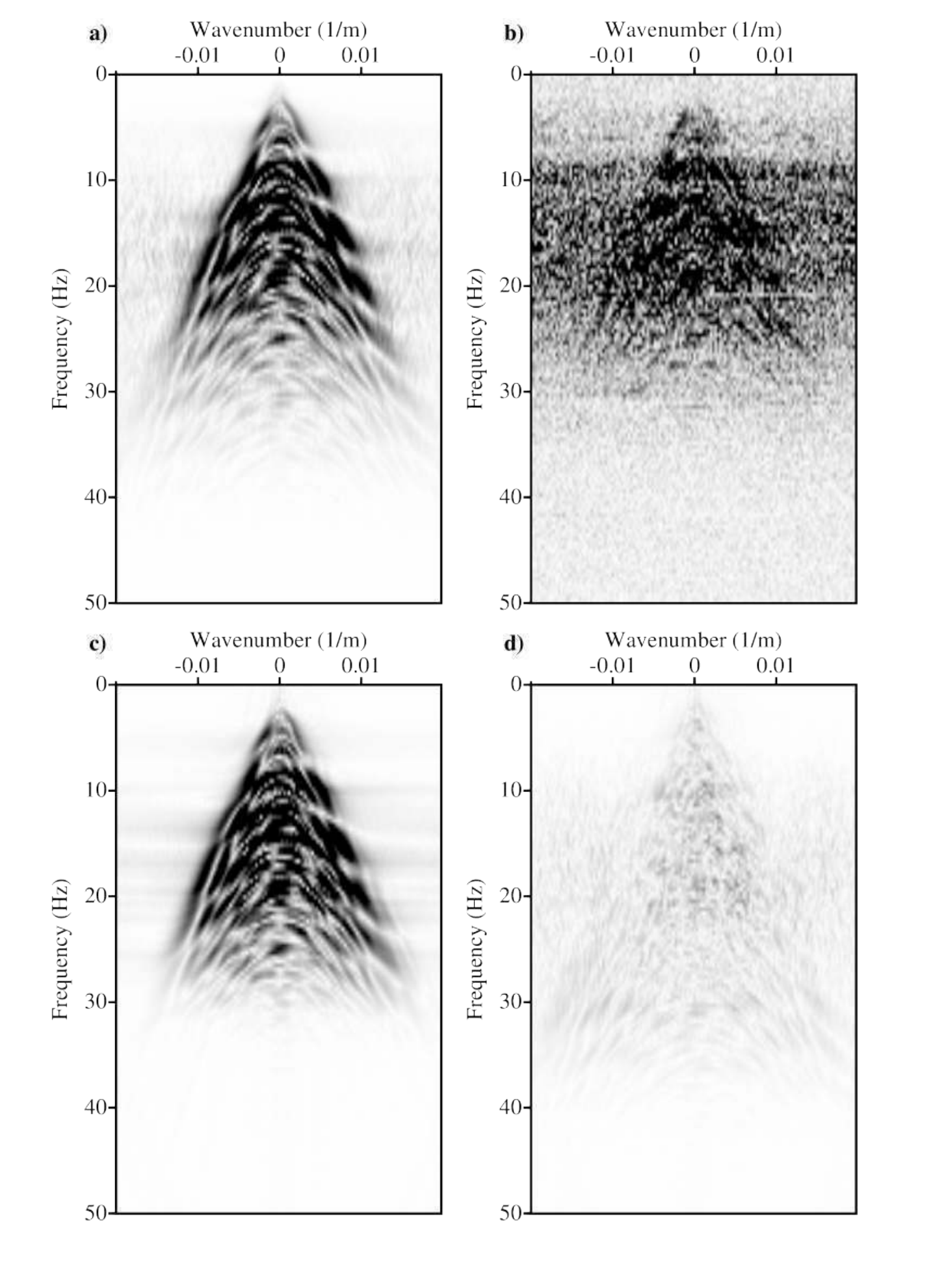}
	\reduceVspace
	\caption{The $f-k$ spectra of the synthetic data common receiver gather recovered from the estimated Radon model using $p=1$ and $q=1$ inversion.
	(a) Original gather.   (b) Pseudo-deblended gather.  (c) Recovered gather.   (d) Recovered gather error.}
	\label{ch4_synth_CRG_L1L1_fk}
\end{figure}

\begin{figure}[htp] 
	\centering
	\includegraphics{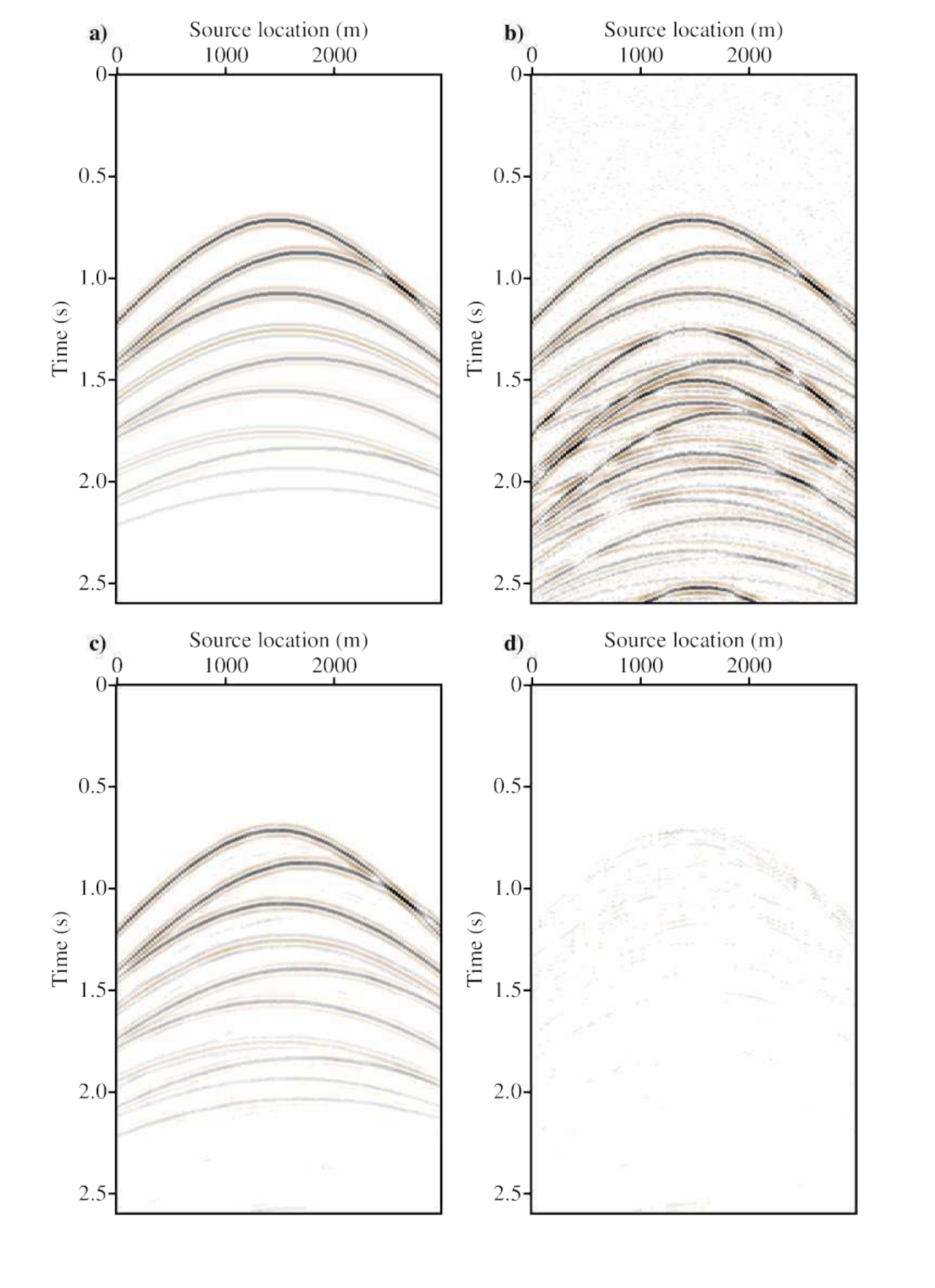}
	\vspace*{2mm}
	\caption{Synthetic data common source gather recovered from the Radon model estimated using $p=1$ and $q=1$ inversion.
	(a) Original gather.   (b) Pseudo-deblended gather.  (c) Recovered gather.   (d) Recovered gather error.}
	\label{ch4_synth_CSG_L1L1}
\end{figure}

\begin{sidewaysfigure}[htp] 
 		\includegraphics{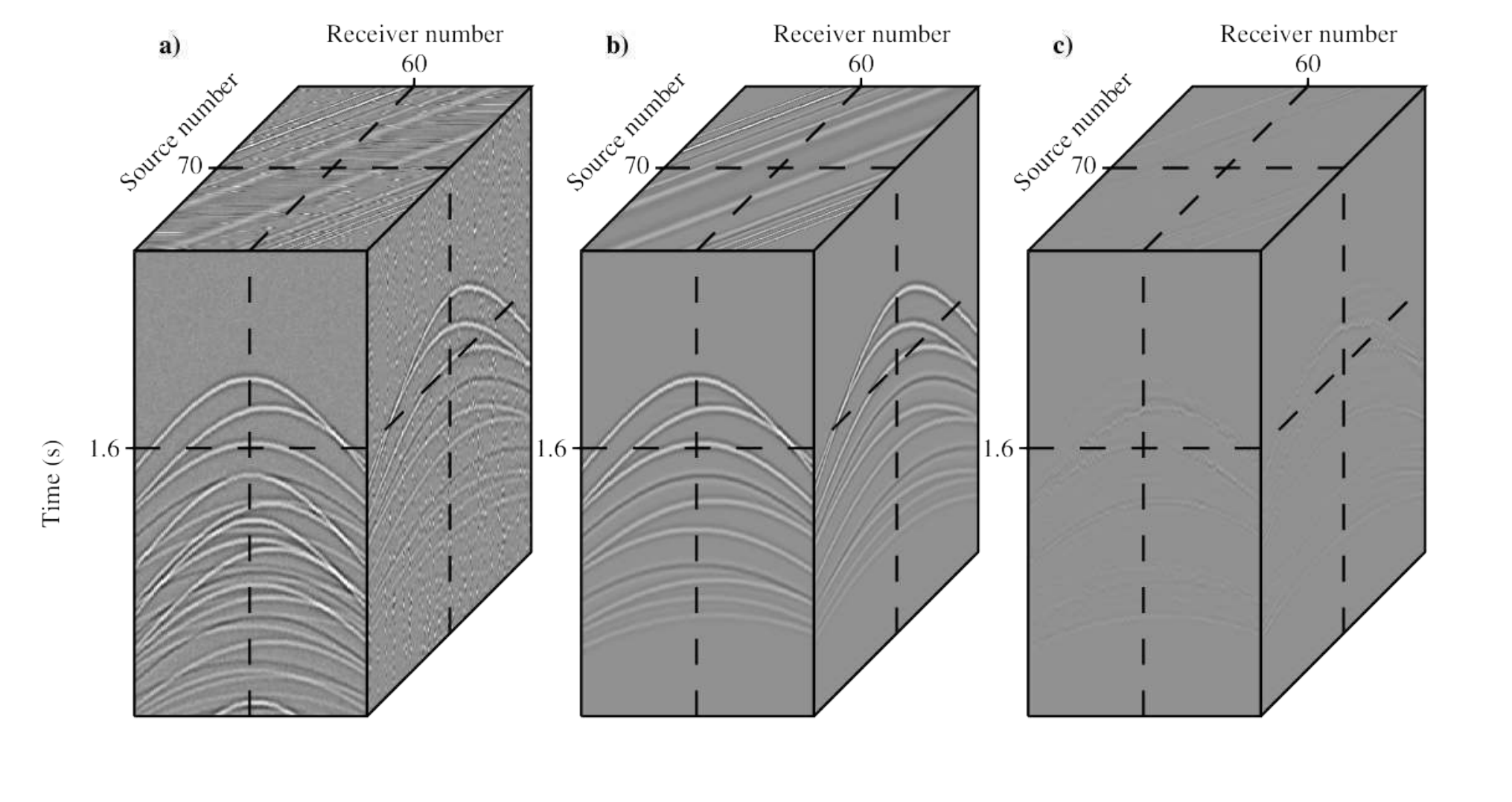}
 		\reduceVspace
 		\caption{Synthetic data example cubes. 
   		(a) Pseudo-deblended data cube. 
   		(b) Data cube recovered by forward modelling ASHRT models estimated via $p=1$ and $q=1$ inversion.
       (c) Difference between recovered and original data cubes. }
		\label{ch4_synth_cubes_L1L1}
\end{sidewaysfigure}

Four different Radon models for the pseudo-deblended common receiver gather in Figure \ref{ch4_synth_CRG_example}b are estimated using four different inversion scenarios. The ASHRT Radon models for ($p=2,q=2$), ($p=2,q=1$), ($p=1,q=2$) and ($p=1,q=1$) inversions are shown in Figures \ref{ch4_synth_model_L2L2}, \ref{ch4_synth_model_L2L1}, \ref{ch4_synth_model_L1L2} and \ref{ch4_synth_model_L1L1}, respectively. 
These models are easy to interpret since the synthetic data is simple. 
The coarse velocities used in the transform reduced the focusing power of the transform as shown in Figures \ref{ch4_synth_model_L2L1} and \ref{ch4_synth_model_L1L1}. 
However, the effect of including a robust misfit is clear in comparing the robust models in Figures \ref{ch4_synth_model_L1L2} and \ref{ch4_synth_model_L1L1} versus the non-robust models in Figures \ref{ch4_synth_model_L2L2} and \ref{ch4_synth_model_L2L1}. 
Figure \ref{ch4_synth_modeld_closeup} shows one velocity panel wiggle plot from each of the four estimated Radon models. These wiggle plots show the effects of the robust misfit and model sparsity on the model estimation more clearly.
In order to evaluate the efficiency of removing incoherent source interference and preserving the coherent signals, the four estimated Radon models were used to recover the de-noised common receiver gathers.
Figures \ref{ch4_synth_CRG_L2L2}, \ref{ch4_synth_CRG_L2L1}, \ref{ch4_synth_CRG_L1L2} and \ref{ch4_synth_CRG_L1L1} show the common receiver gathers recovered from Radon models estimated using ($p=2,q=2$), ($p=2,q=1$), ($p=1,q=2$) and ($p=1,q=1$) inversion, respectively. 
These figures clearly show that the robust Radon transforms using ($p=1,q=2$) and ($p=1,q=1$) inversion were able to attenuate interference while preserving the signals better than the non-robust transforms using ($p=2,q=2$) and ($p=2,q=1$). The sparse Radon transform ($p=2,q=1$) was able to remove all the interferences, however, it attenuated the weak reflections at $2.0$ seconds. The best results were obtained using both sparsity and robustness ($p=1,q=1$) as shown in Figure \ref{ch4_synth_CRG_L1L1}. 
\clearpage
\clearpage
\clearpage
In order to examine the recovered common receiver gathers more closely, Figures \ref{ch4_synth_CRG_L2L2_closeup}, \ref{ch4_synth_CRG_L2L1_closeup}, \ref{ch4_synth_CRG_L1L2_closeup} and \ref{ch4_synth_CRG_L1L1_closeup} show the close-up gathers recovered from Radon models estimated using ($p=2,q=2$), ($p=2,q=1$), ($p=1,q=2$) and ($p=1,q=1$) inversion, respectively. A close-up of the data recovered by sparse Radon model ($p=2,q=1$) in Figure \ref{ch4_synth_CRG_L2L1_closeup} shows that the sparse Radon transform caused considerable amplitude losses for the weak signals. On the other hand, using both sparsity and robustness ($p=1,q=1$) achieves better preservation of the weak signals as shown in Figure \ref{ch4_synth_CRG_L2L1_closeup}.  The quality for removing the incoherent source interferences and preserving the coherent signals can also be evaluated in the $f-k$ spectra of the common receiver gathers. Figures \ref{ch4_synth_CRG_L2L2_fk}, \ref{ch4_synth_CRG_L2L1_fk}, \ref{ch4_synth_CRG_L1L2_fk} and \ref{ch4_synth_CRG_L1L1_fk} show the $f-k$ spectra for the common receiver gathers recovered from their Radon models estimated using ($p=2,q=2$), ($p=2,q=1$), ($p=1,q=2$) and ($p=1,q=1$) inversion, respectively. These figures also confirm that common receiver gathers recovered from Radon models estimated using ($p=1,q=1$) inversion attenuate the incoherent interferences without distorting the coherent signals. 
The quality of the pseudo-deblended common receiver gather and the recovered common receiver gather data from Radon models can be evaluated using the following expressions
\begin{equation}  \label{ch4_Qpd}
Q_{PD} =10~log_{10}\left(\frac{\| {\bf d}_{original} \|_2^2}{\| {\bf d}_{original}-{\bf d}_{PD} \|_2^2} \right), \\
\end{equation}
\begin{equation}\label{ch4_Qr}
Q_{R} =10~log_{10}\left(\frac{\| {\bf d}_{original} \|_2^2}{\| {\bf d}_{original}-{\bf d}_{recovered} \|_2^2} \right), 
\end{equation}
where is ${\bf d}_{original}$ is the original common receiver gather, ${\bf d}_{PD}$ is the pseudo-deblended common receiver gather and ${\bf d}_{recovered}$ is the common receiver gather recovered from Radon models estimated by inversion.  Therefore, the improvement due to the Radon transform can be calculated using the following expression
\begin{align}
Q =Q_{R} - Q_{PD}.
\end{align}
Note that if $Q=10$ dB, then the $\ell_2$ norm of the original data is 10 times larger than the error of common receiver gather recovered from the Radon model. Generally, values of Q above $10$ dB are considered acceptable deblending results. The $Q$ values for all the four recovered common receiver gathers are listed in Table \ref{table:ch4_quality}. As expected, the highest $Q$ value for recovered synthetic data common receiver gather is $18.65$ dB using $p=1,q=1$ inversion.  Source separation can be achieved by denoising all common receiver gathers in the pseudo-deblended data cube. Figure \ref{ch4_synth_CSG_L1L1} shows the common source gather recovered after denoising all common receiver gathers using $p=1,q=1$ inversion.  This figure shows that coherent interferences in common source gather were effectively removed after denoising all common receiver gathers. Figure \ref{ch4_synth_cubes_L1L1} shows the data cubes recovered using Radon models estimated using $p=1,q=1$ inversion.  The top of the cube shows a time slice through the data. Comparing the time slice of the pseudo-deblended cube and the de-noised cube, we can observe that all source interferences were removed. 
\clearpage
\clearpage
\section{Field data example}
\begin{figure}[htbp]
		\includegraphics{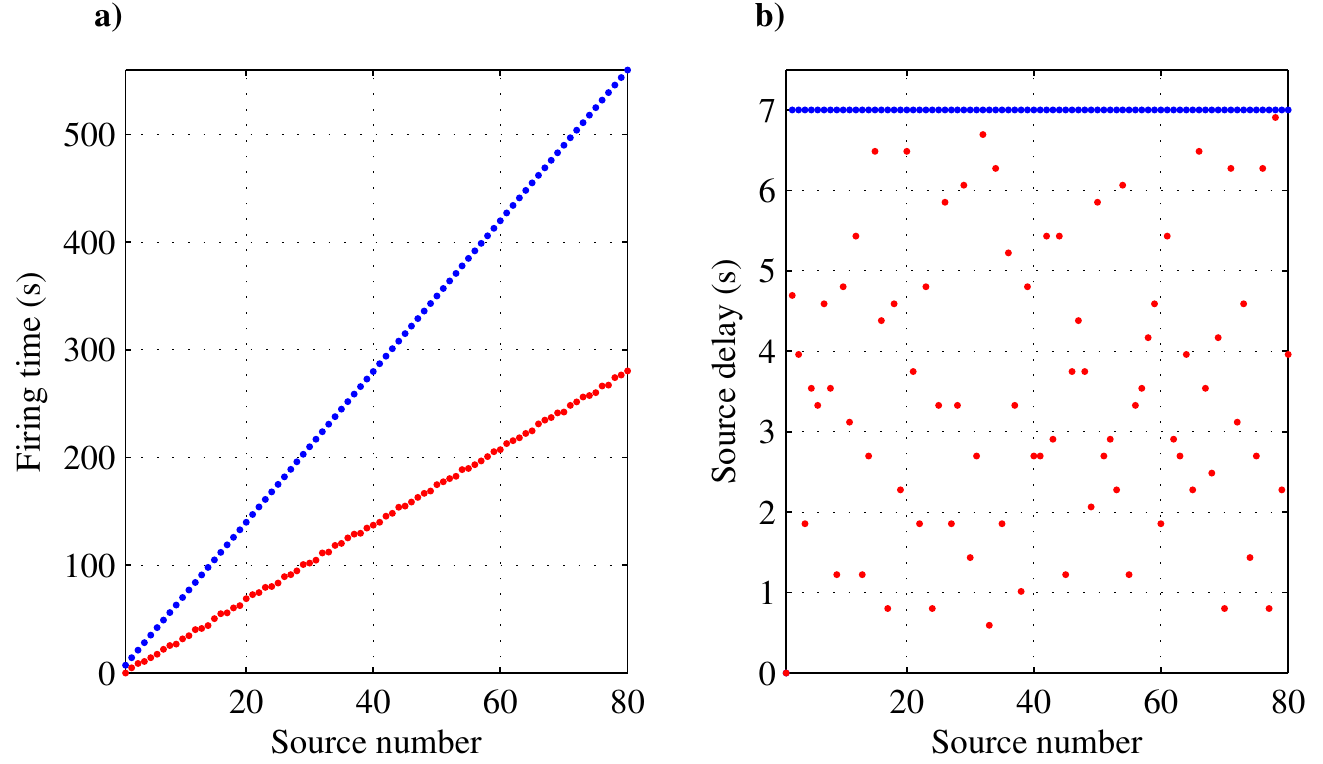}
		\increaseVspace
		\caption{Seismic sources firing times for numerically blended Gulf of Mexico data.
			(a) Firing times of conventional (blue) and simultaneous seismic sources (red). 
			(b) Time delay between successive sources for conventional (blue) and simultaneous sources (red).}
		\label{ch4_GOM_firing_times}
\end{figure}

We also tested the four different Radon transforms using a numerically blended marine data from the Gulf of Mexico. The acquisition scenario represents a single source boat with the time interval between successive sources is nearly half of the conventional acquisition. In order to make the source interferences appear incoherent, the source firing times are dithered using random time delays. The firing times of the sources for both conventional and blended acquisition sources are shown in Figure \ref{ch4_GOM_firing_times}.

\begin{figure}[htbp]
	\centering
	\includegraphics{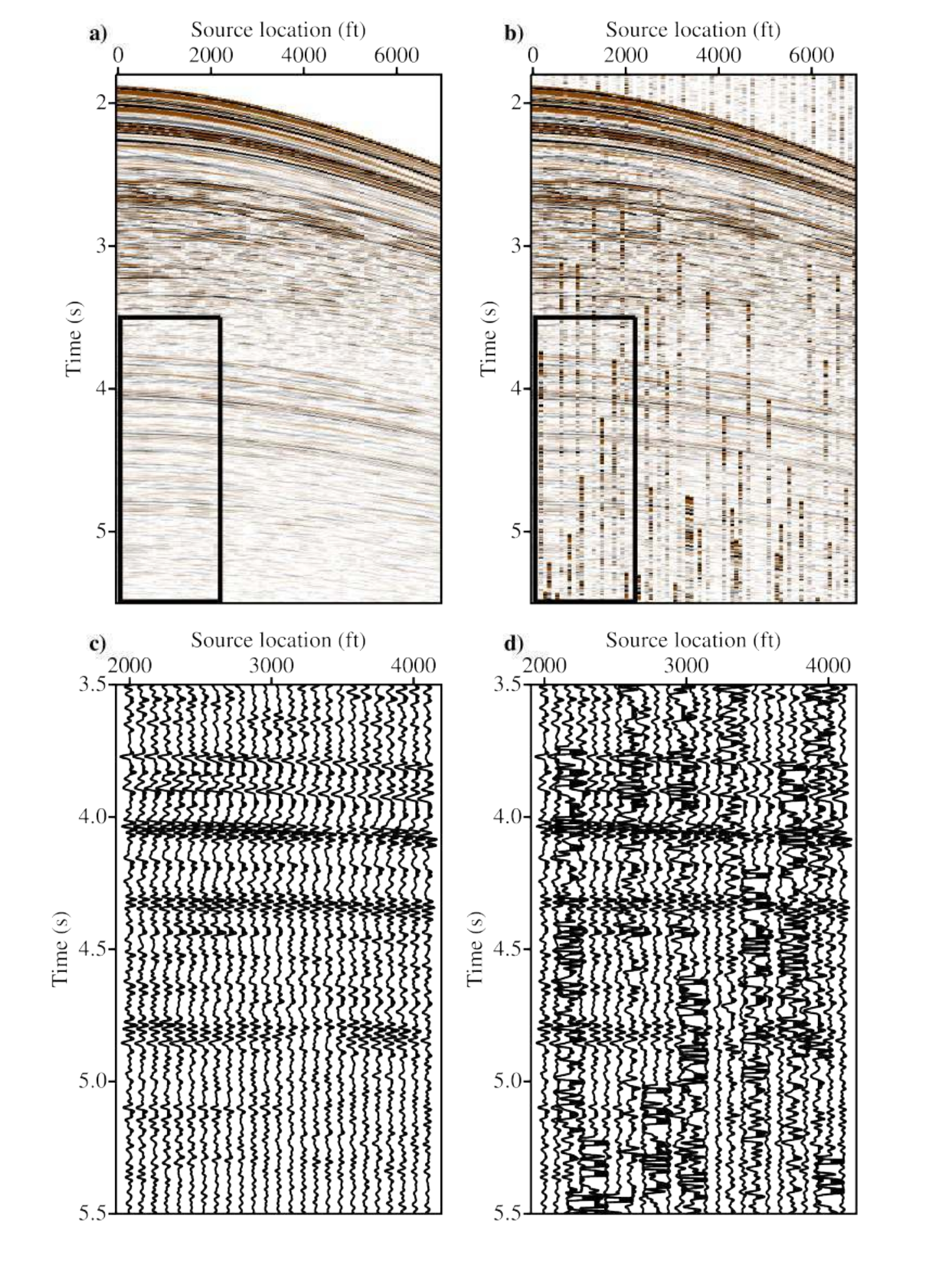}
	\reduceVspace
	\caption{A numerically blended common receiver gather from Gulf of Mexico field data example.
	(a) Original gather.   
	(b) Pseudo-deblended gather .  
	(c) Close-up of the original gather.   
	(d) Close-up of the pseudo-deblended gather. }
	\label{ch4_GOM_CRG_example}
\end{figure}

\begin{sidewaysfigure}[htbp]
 		\includegraphics{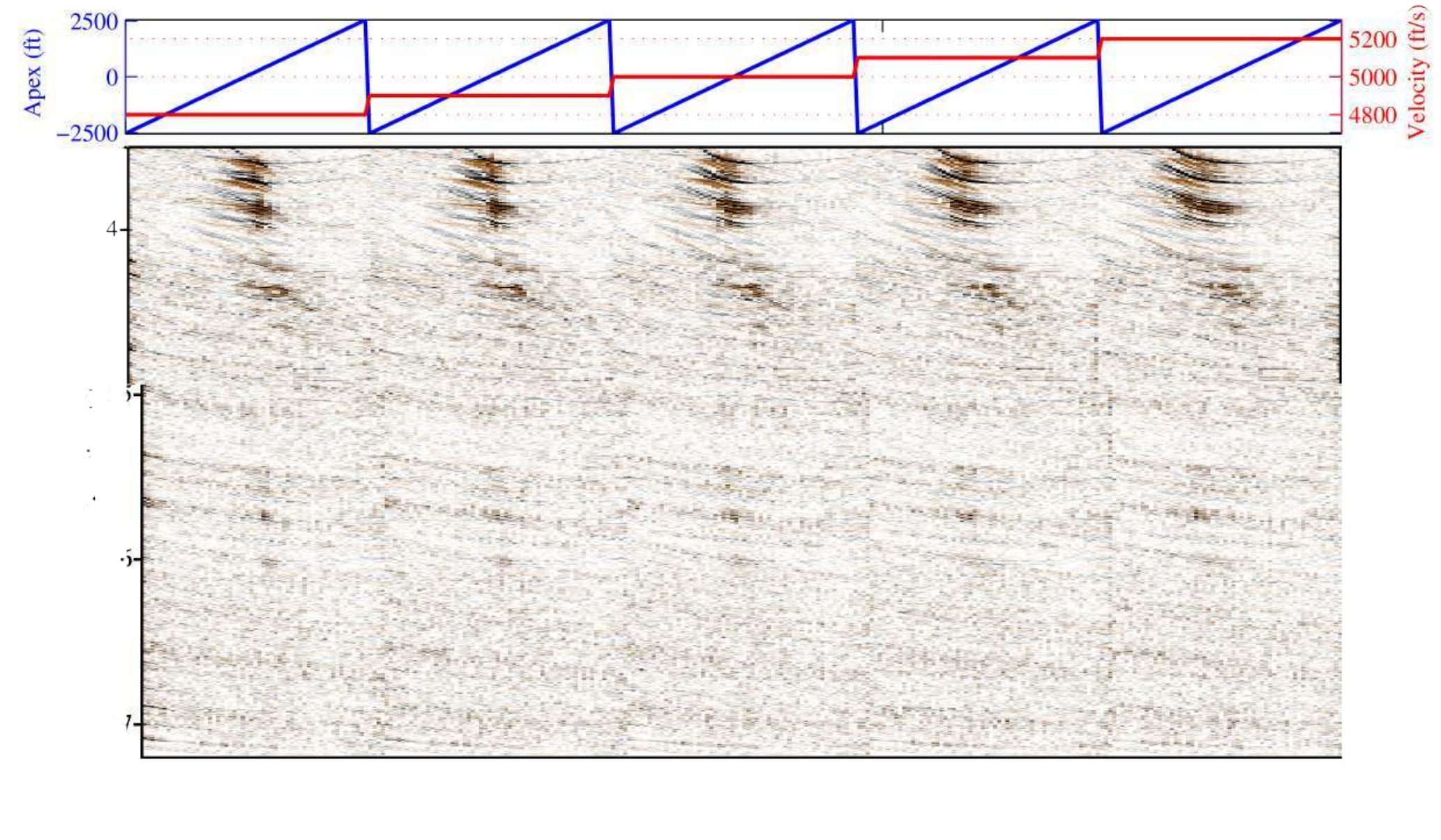}
 		\reduceVspace
 		\caption{Estimated apex-shifted Radon model for one Gulf of Mexico common receiver gather using $p=2$ and $q=2$ inversion.}
		\label{ch4_GOM_model_L2L2}
\end{sidewaysfigure}

\begin{sidewaysfigure}[htbp]
 		\includegraphics{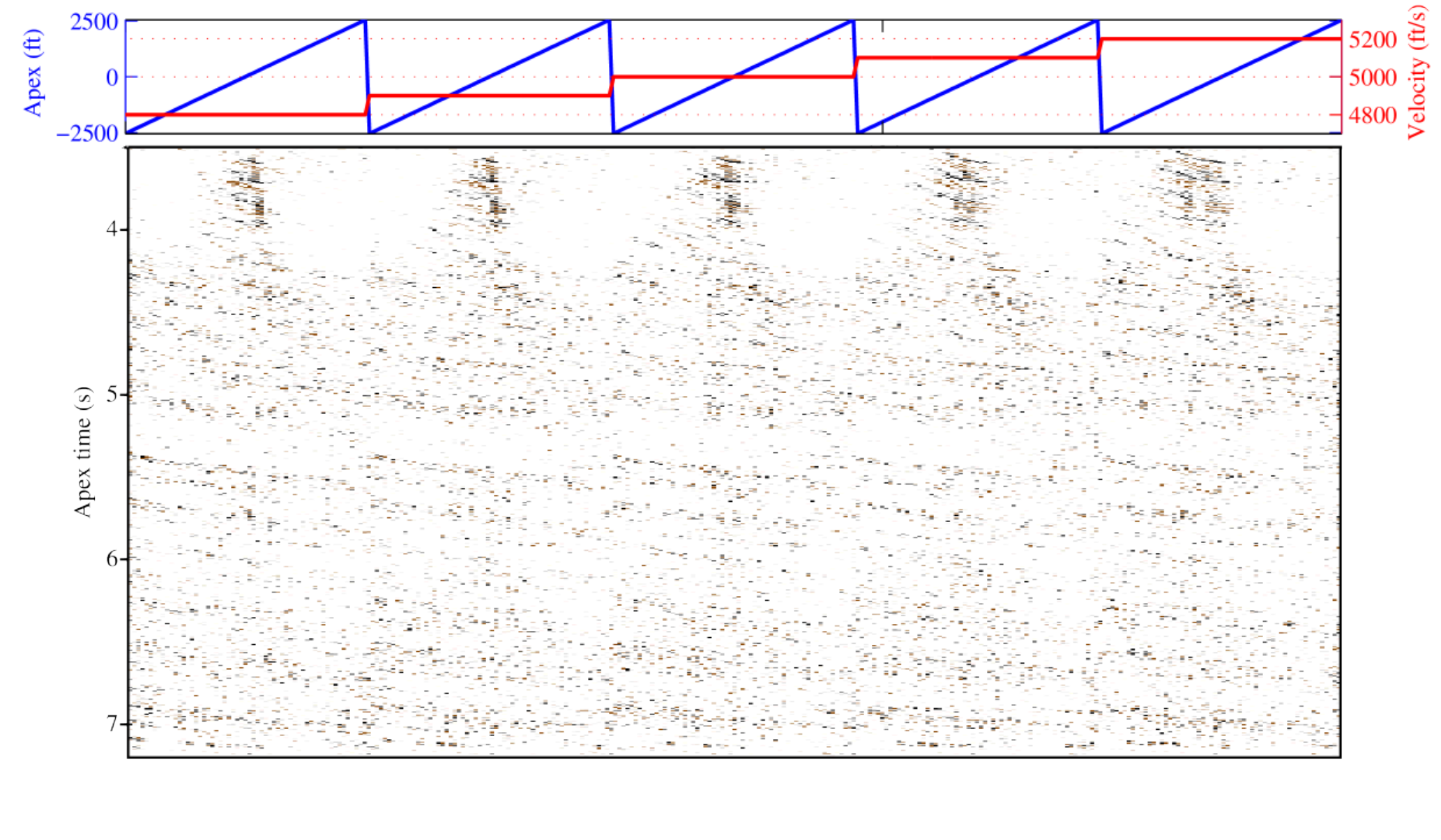}
 		\reduceVspace
 		\caption{Estimated apex-shifted Radon model for one Gulf of Mexico common receiver gather using $p=2$ and $q=1$ inversion.}
		\label{ch4_GOM_model_L2L1}
\end{sidewaysfigure}

\begin{sidewaysfigure}[htbp]
 		\includegraphics{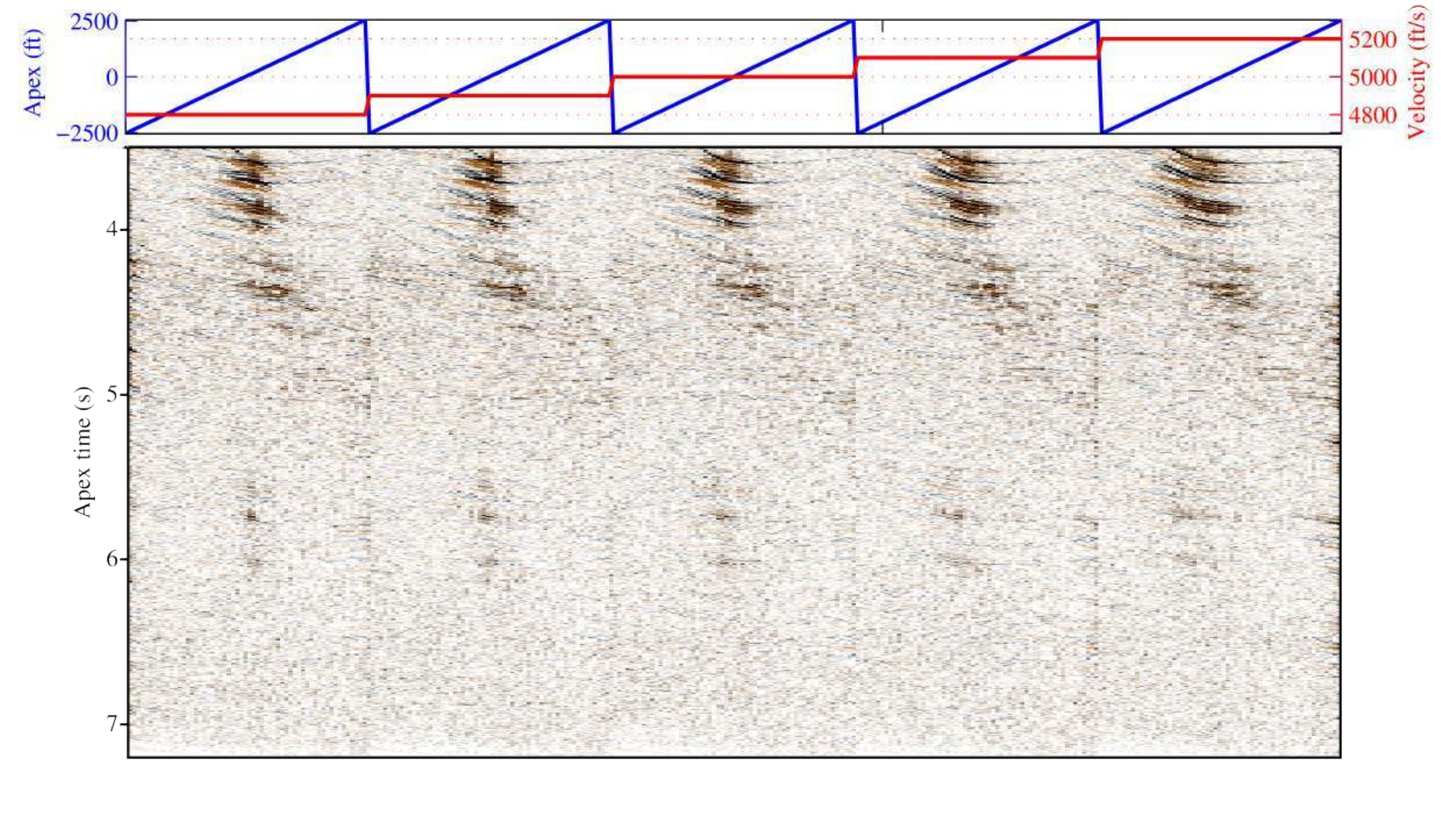}
 		\reduceVspace
 		\caption{Estimated apex-shifted Radon model for one Gulf of Mexico common receiver gather using $p=1$ and $q=2$ inversion.}
		\label{ch4_GOM_model_L1L2}
\end{sidewaysfigure}

\begin{sidewaysfigure}[htbp]
 		\includegraphics{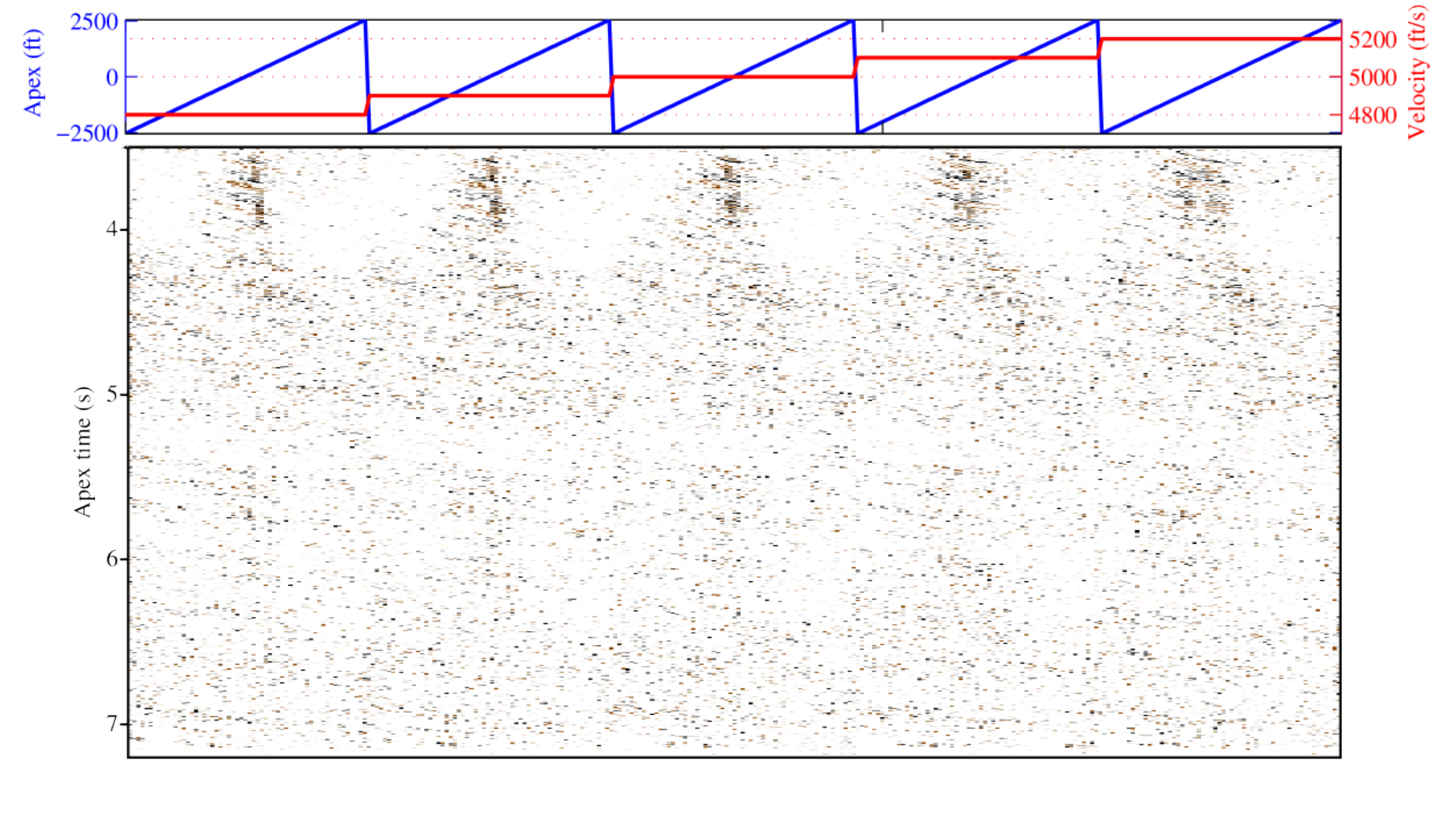}
 		\reduceVspace
 		\caption{Estimated apex-shifted Radon model for one Gulf of Mexico common receiver gather using $p=1$ and $q=1$ inversion.}
		\label{ch4_GOM_model_L1L1}
\end{sidewaysfigure}

Figure \ref{ch4_GOM_CRG_example}a shows an original common receiver gather from the Gulf of Mexico data example and Figure \ref{ch4_GOM_CRG_example}b shows the same gather after blending and pseudo-deblending. Figures \ref{ch4_GOM_CRG_example}c and d show a close-up of the areas marked on Figures \ref{ch4_GOM_CRG_example}a and b, respectively. We have chosen a close-up window between 3.5 and 5.5 seconds to show the effect of the strong incoherent source interferences and weak coherent signals. Four different Radon models for the pseudo-deblended common receiver gather in Figure \ref{ch4_GOM_CRG_example}b were estimated using four different inversion scenarios. The ASHRT Radon models for ($p=2,q=2$), ($p=2,q=1$), ($p=1,q=2$) and ($p=1,q=1$) inversions are shown in Figures \ref{ch4_GOM_model_L2L2}, \ref{ch4_GOM_model_L2L1}, \ref{ch4_GOM_model_L1L2} and \ref{ch4_GOM_model_L1L1}, respectively. The Radon transform scans the following five velocities ($4800, 4900, 5000, 5100,$ and $5200$ ft/s) and scans 59 apex locations from $2537.5$ ft to $-2537.5$ ft. This coarse sampling of the Radon parameters makes it difficult to impose a strict sparsity constraint on the Radon model. The sparsity constraint ideally requires representing a single reflection hyperbola with a single Radon coefficient. This is often not feasible with field data due to the mismatch between the theoretical travel-time hyperbolas used by the Radon transform and the travel-times of the actual reflections. Additionally, the amplitude variation of seismic reflection with offset further complicates this mismatch. For these reasons, the choice of optimal regularization parameters for robust inversion with sparse regularization ($p=1,q=1$) is rather difficult. For the field data example, we opted to use low sparsity to preserve the coherent signal rather than using strong sparsity and risk signal loss.  

\begin{figure}[htbp]
 		\includegraphics{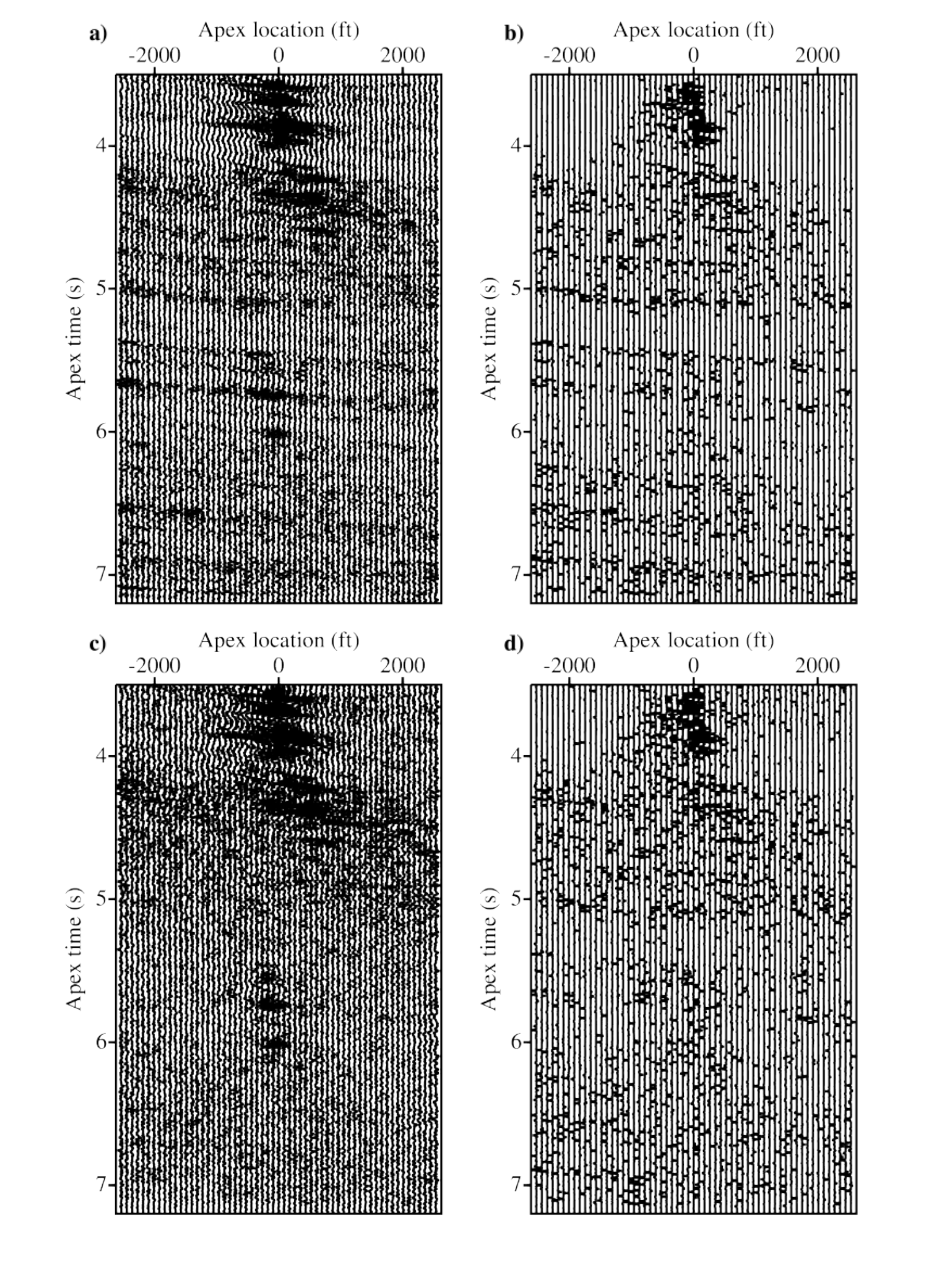}
 		\reduceVspace
   		\caption{One velocity panel ($v=4800~ft/s$) of the ASHRT models for Gulf of Mexico data estimated by inversion with 
   		(a) $p=2$ and $q=2$, (b) $p=2$ and $q=1$, (c) $p=1$ and $q=2$, and (d) $p=1$ and $q=1$. }
		\label{ch4_GOM_modeld_closeup}
\end{figure}

\begin{figure}[htp] 
	\centering
	\includegraphics{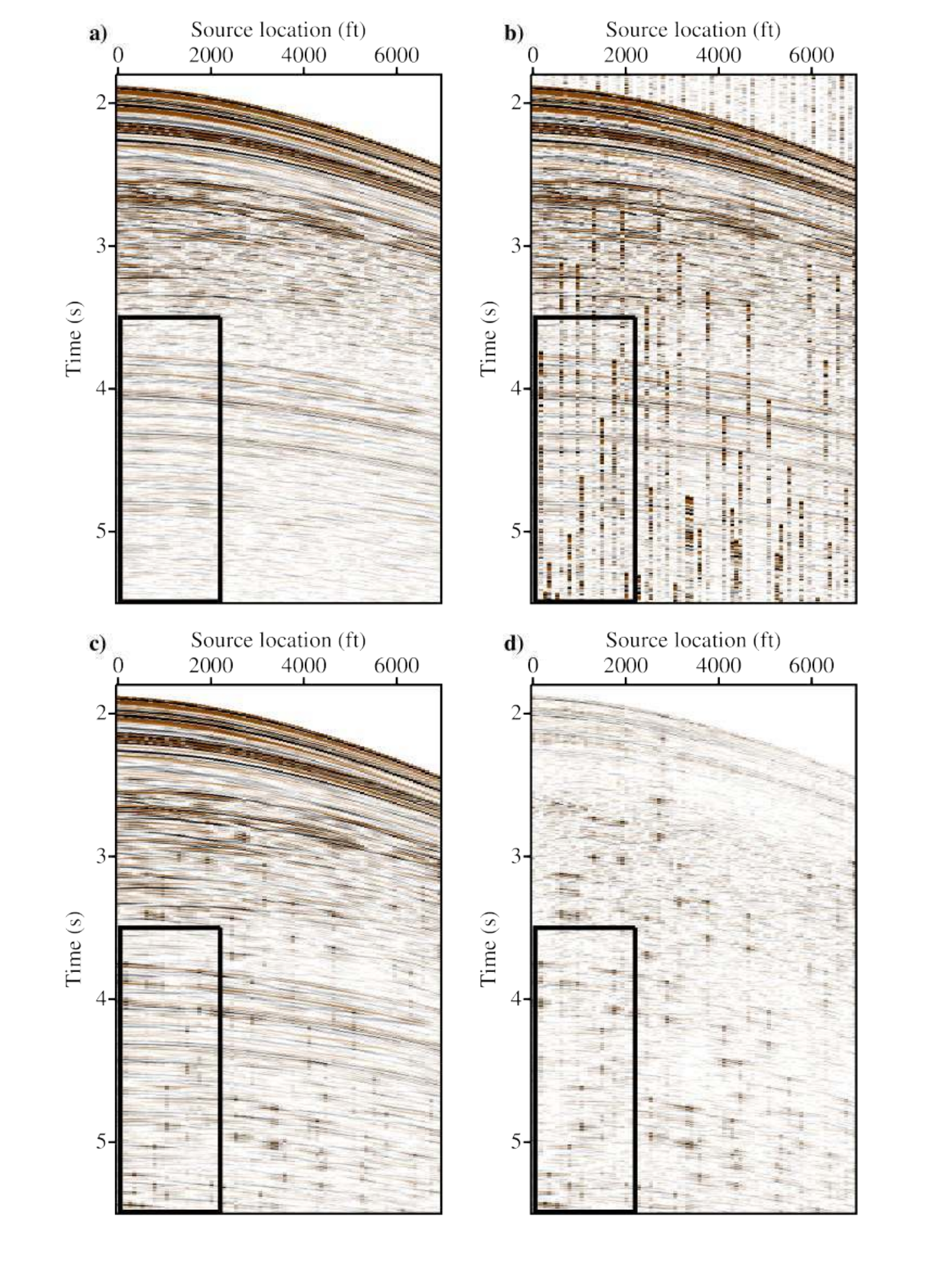}
	\reduceVspace
	\caption{Common receiver gather from the Gulf of Mexico data recovered from the Radon model estimated using $p=2$ and $q=2$ inversion.
	(a) Original gather.   (b) Pseudo-deblended gather.  (c) Recovered gather.   (d) Recovered gather error.}
	\label{ch4_GOM_CRG_L2L2}
\end{figure}

\begin{figure}[htp] 
	\centering
	\includegraphics{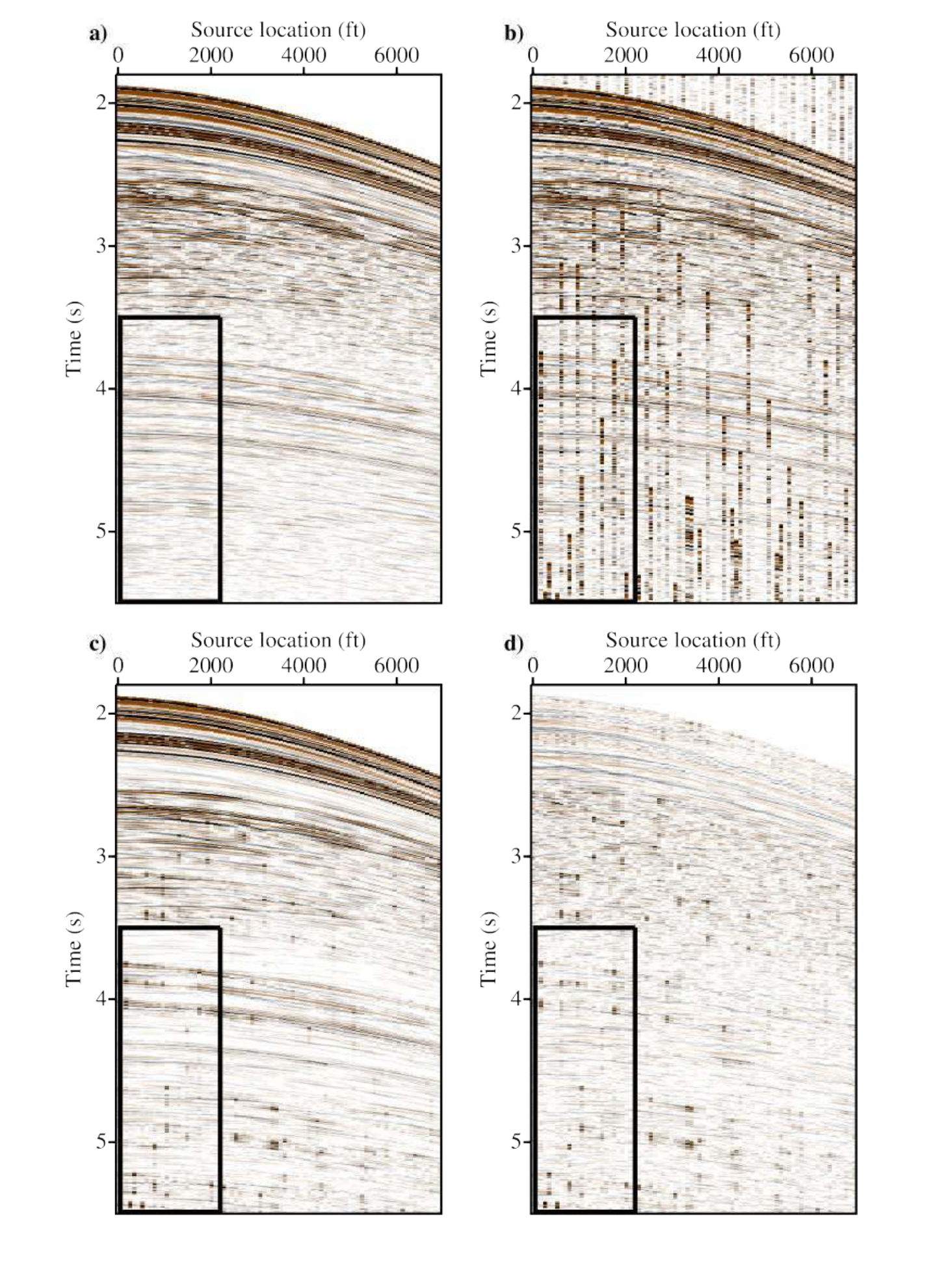}
	\reduceVspace
	\caption{Common receiver gather from the Gulf of Mexico data recovered from the Radon model estimated using $p=2$ and $q=1$ inversion.
	(a) Original gather.   (b) Pseudo-deblended gather.  (c) Recovered gather.   (d) Recovered gather error.}
	\label{ch4_GOM_CRG_L2L1}
\end{figure}

\begin{figure}[htp] 
	\centering
	\includegraphics{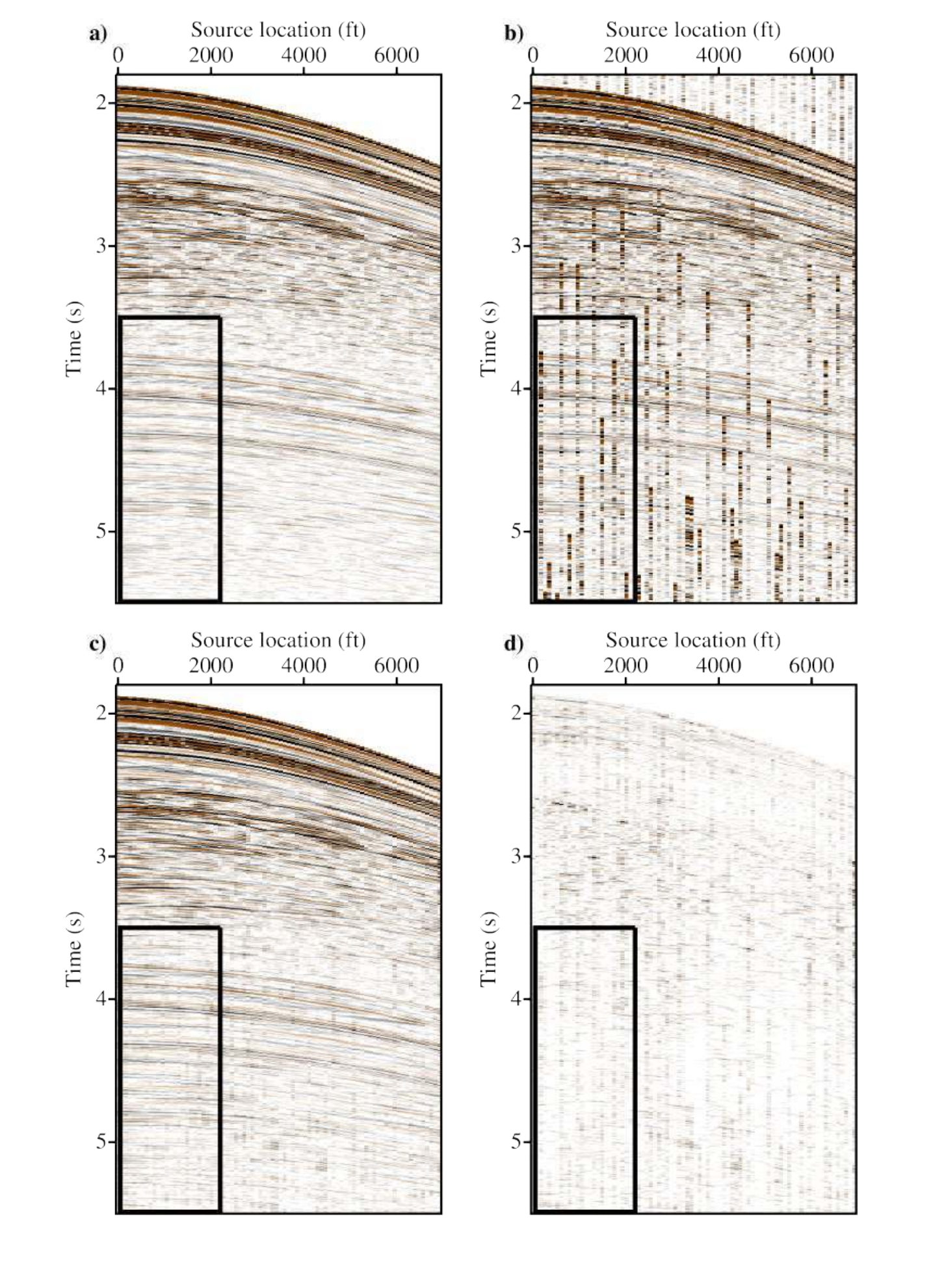}
	\reduceVspace
	\caption{Common receiver gather from the Gulf of Mexico data recovered from the Radon model estimated using $p=1$ and $q=2$ inversion.
	(a) Original gather.   (b) Pseudo-deblended gather.  (c) Recovered gather.   (d) Recovered gather error.}
	\label{ch4_GOM_CRG_L1L2}
\end{figure}

\begin{figure}[htp] 
	\centering
	\includegraphics{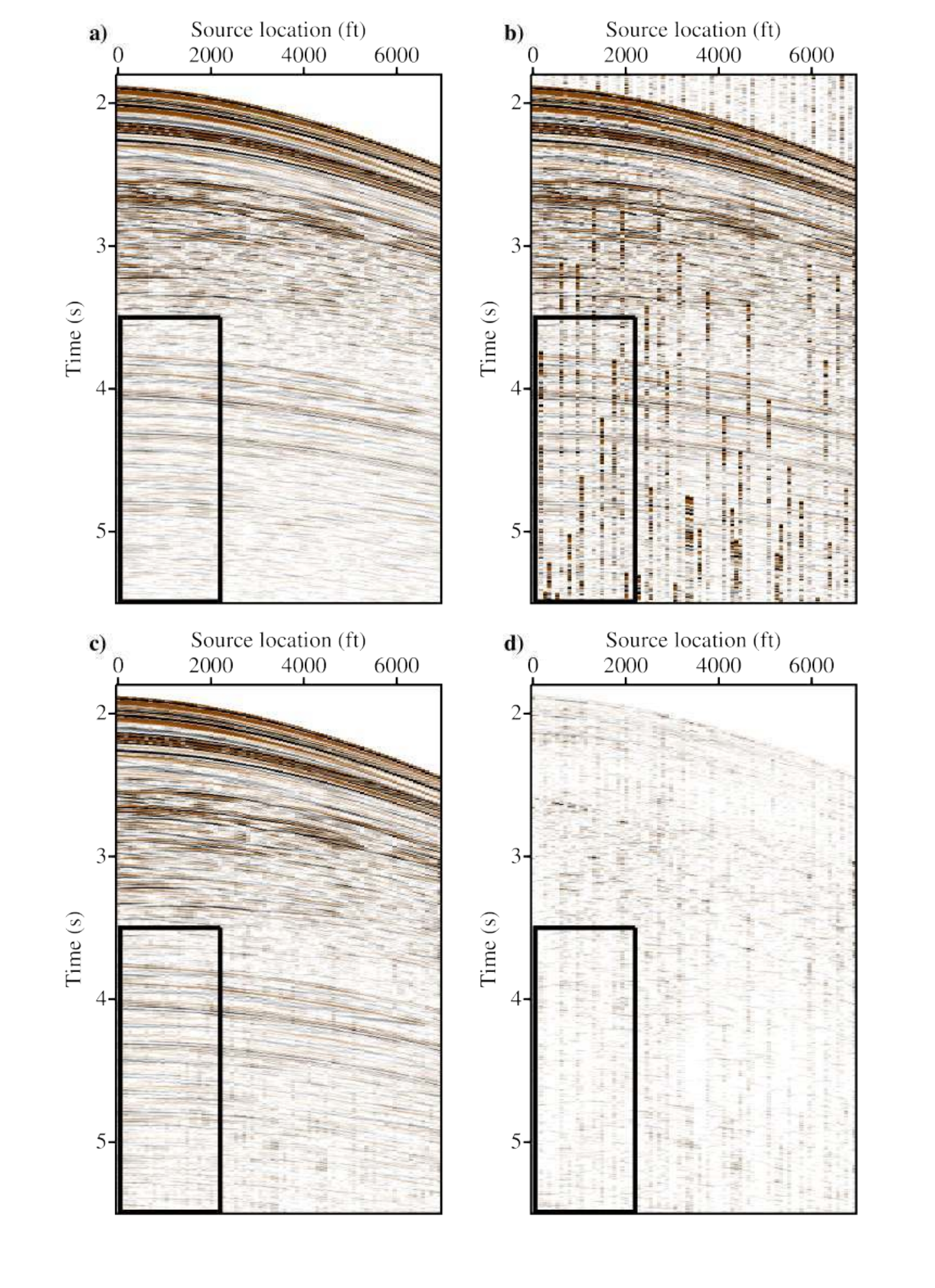}
	\reduceVspace
	\caption{Common receiver gather from the Gulf of Mexico data recovered from the Radon model estimated using $p=1$ and $q=1$ inversion.
	(a) Original gather.   (b) Pseudo-deblended gather.  (c) Recovered gather.   (d) Recovered gather error.}
	\label{ch4_GOM_CRG_L1L1}
\end{figure}

\begin{figure}[htp] 
	\centering
	\includegraphics{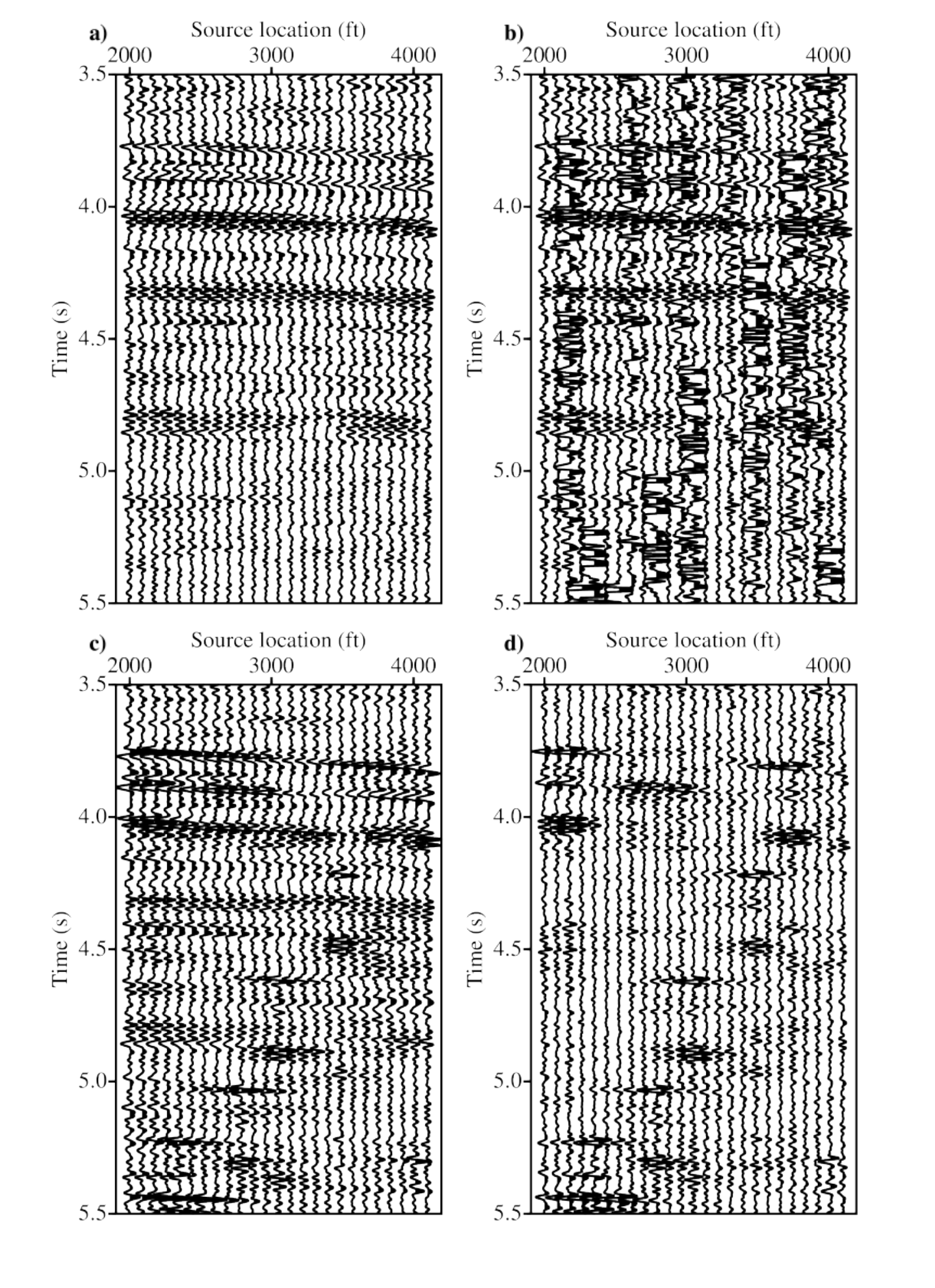}
	\reduceVspace
	\caption{Close-up of one common receiver gather from the Gulf of Mexico data recovered from the Radon model estimated using $p=2$ and $q=2$ inversion.
	(a) Original gather.   (b) Pseudo-deblended gather.  (c) Recovered gather.   (d) Recovered gather error.}
	\label{ch4_GOM_CRG_L2L2_closeup}
\end{figure}

\begin{figure}[htp] 
	\centering
	\includegraphics{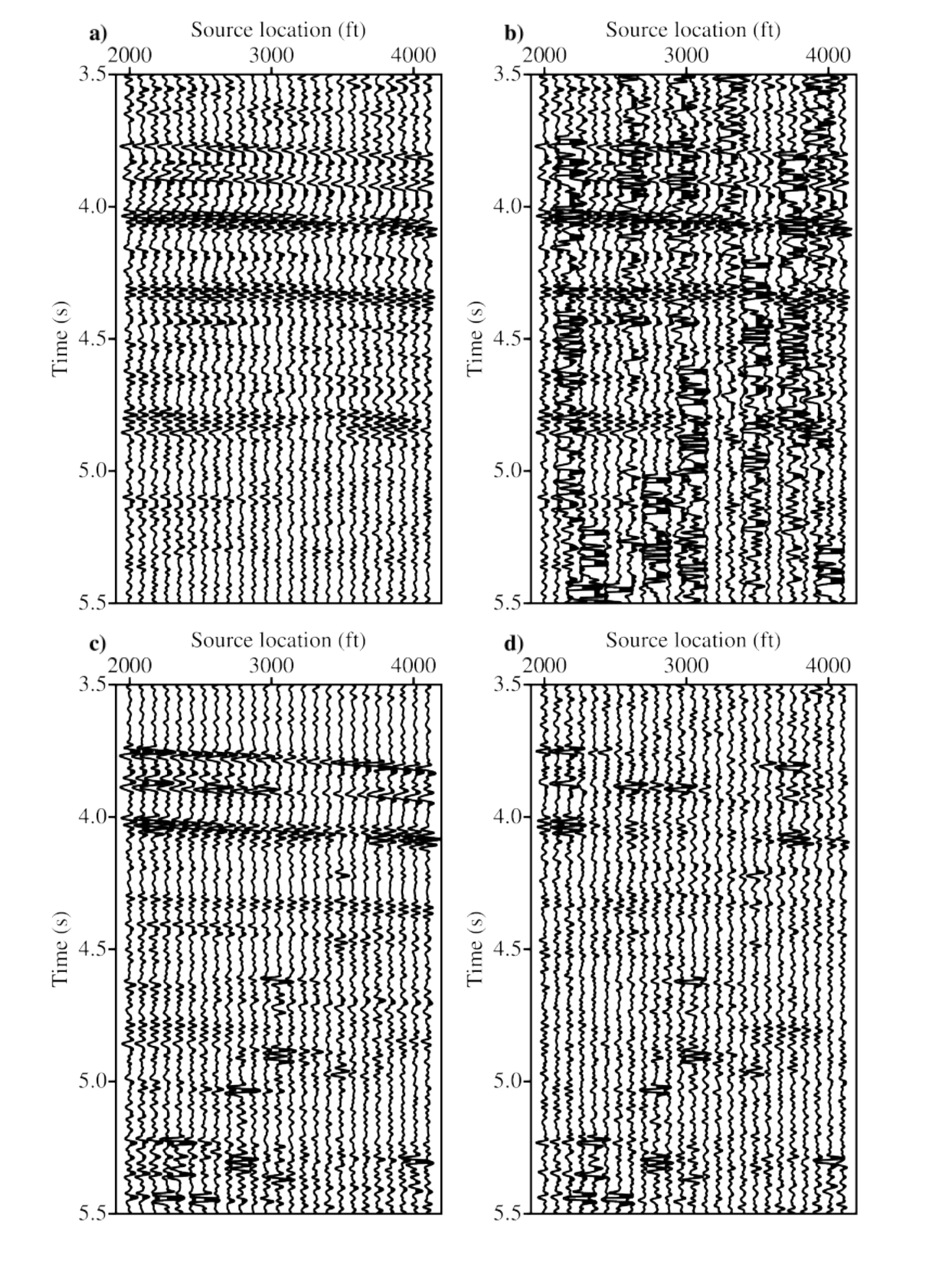}
	\reduceVspace
	\caption{Close-up of one common receiver gather from the Gulf of Mexico data recovered from the Radon model estimated using $p=2$ and $q=1$ inversion.
	(a) Original gather.   (b) Pseudo-deblended gather.  (c) Recovered gather.   (d) Recovered gather error.}
	\label{ch4_GOM_CRG_L2L1_closeup}
\end{figure}

\begin{figure}[htp] 
	\centering
	\includegraphics{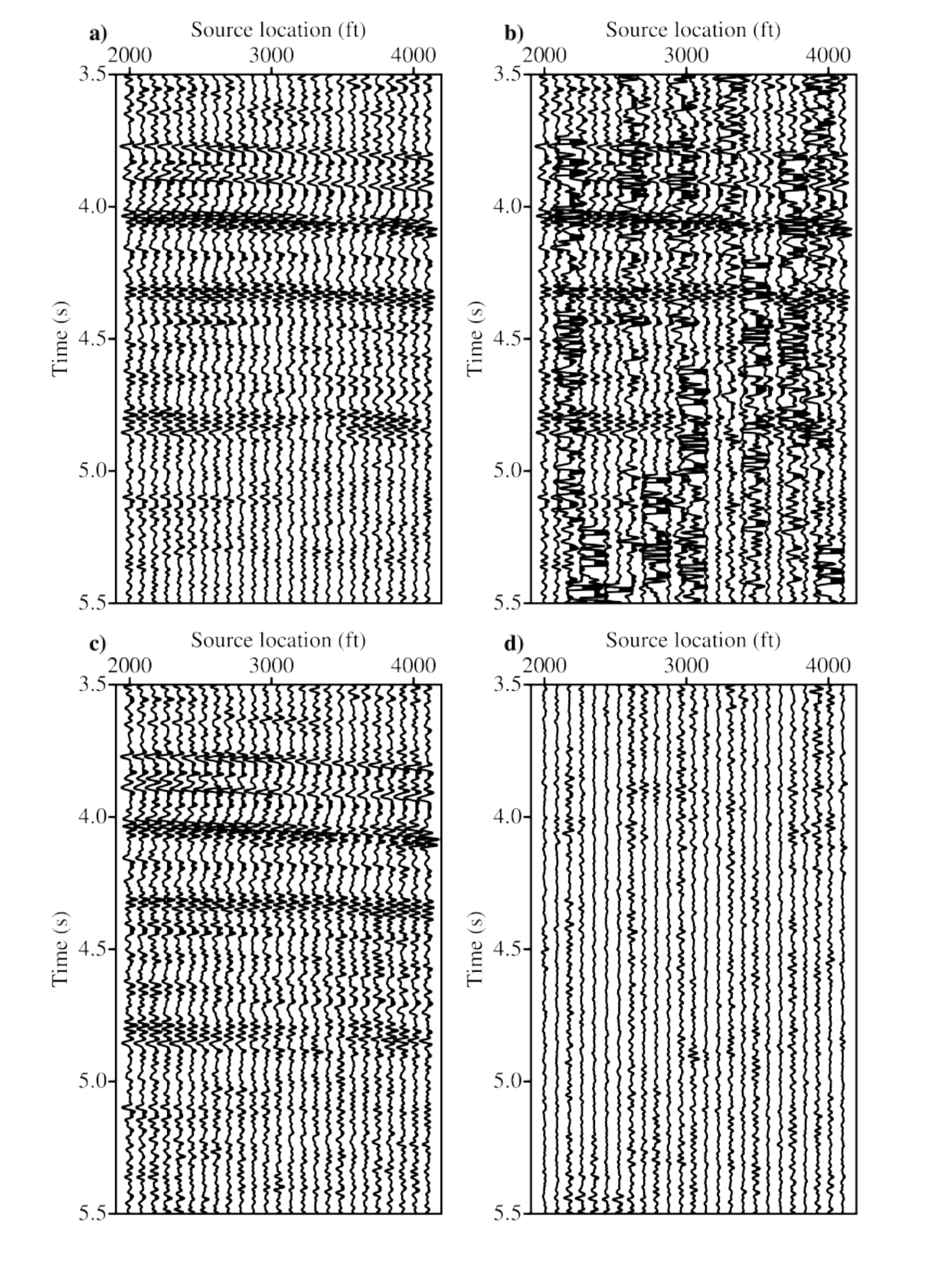}
	\reduceVspace
	\caption{Close-up of one common receiver gather from the Gulf of Mexico data recovered from the Radon model estimated using $p=1$ and $q=2$ inversion.
	(a) Original gather.   (b) Pseudo-deblended gather.  (c) Recovered gather.   (d) Recovered gather error.}
	\label{ch4_GOM_CRG_L1L2_closeup}
\end{figure}

\begin{figure}[htp] 
	\centering
	\includegraphics{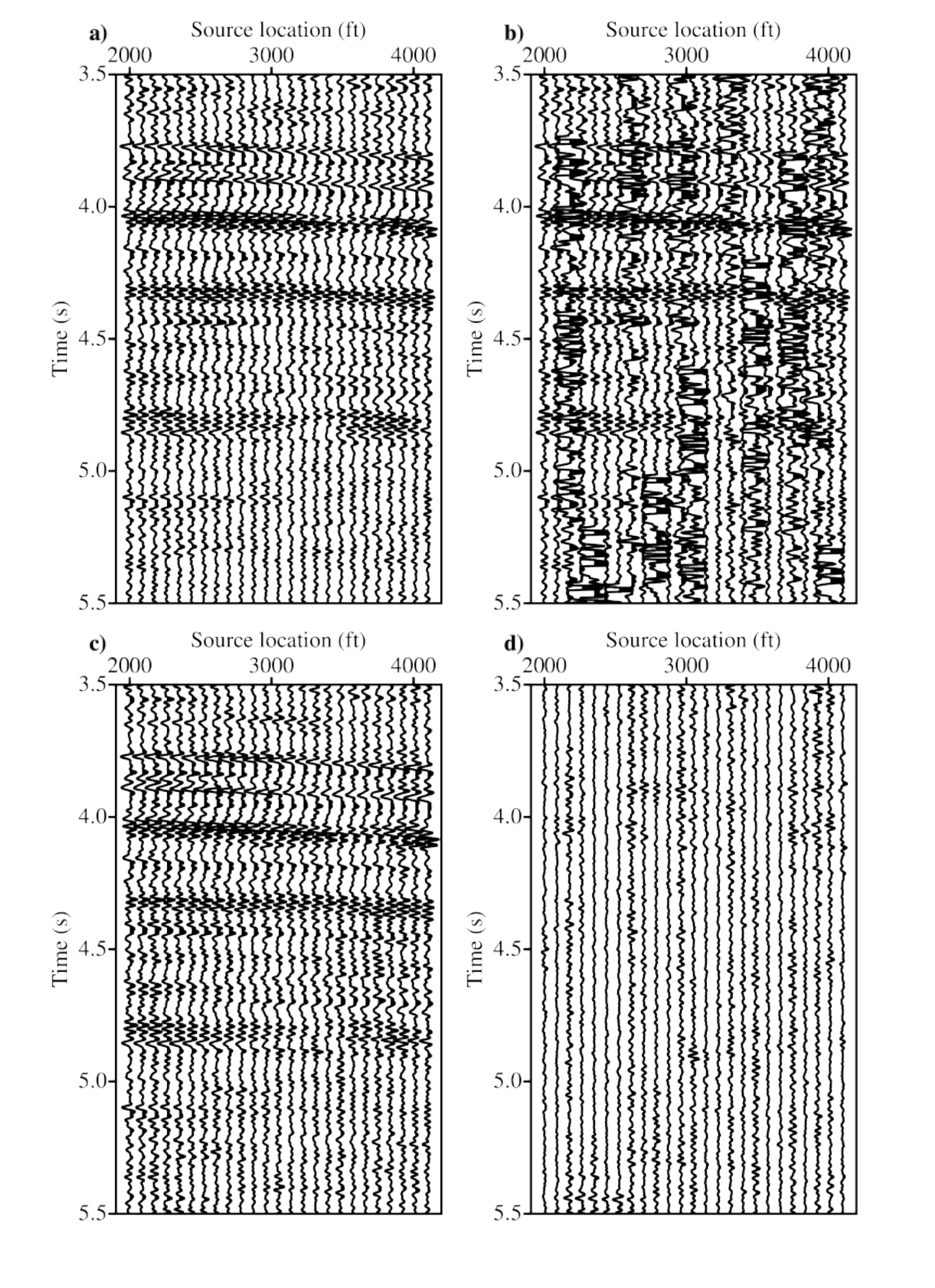}
	\reduceVspace
	\caption{Close-up of one common receiver gather from the Gulf of Mexico data recovered from the Radon model estimated using $p=1$ and $q=1$ inversion.
	(a) Original gather.   (b) Pseudo-deblended gather.  (c) Recovered gather.   (d) Recovered gather error.}
	\label{ch4_GOM_CRG_L1L1_closeup}
\end{figure}

\begin{figure}[htp] 
	\centering
	\includegraphics{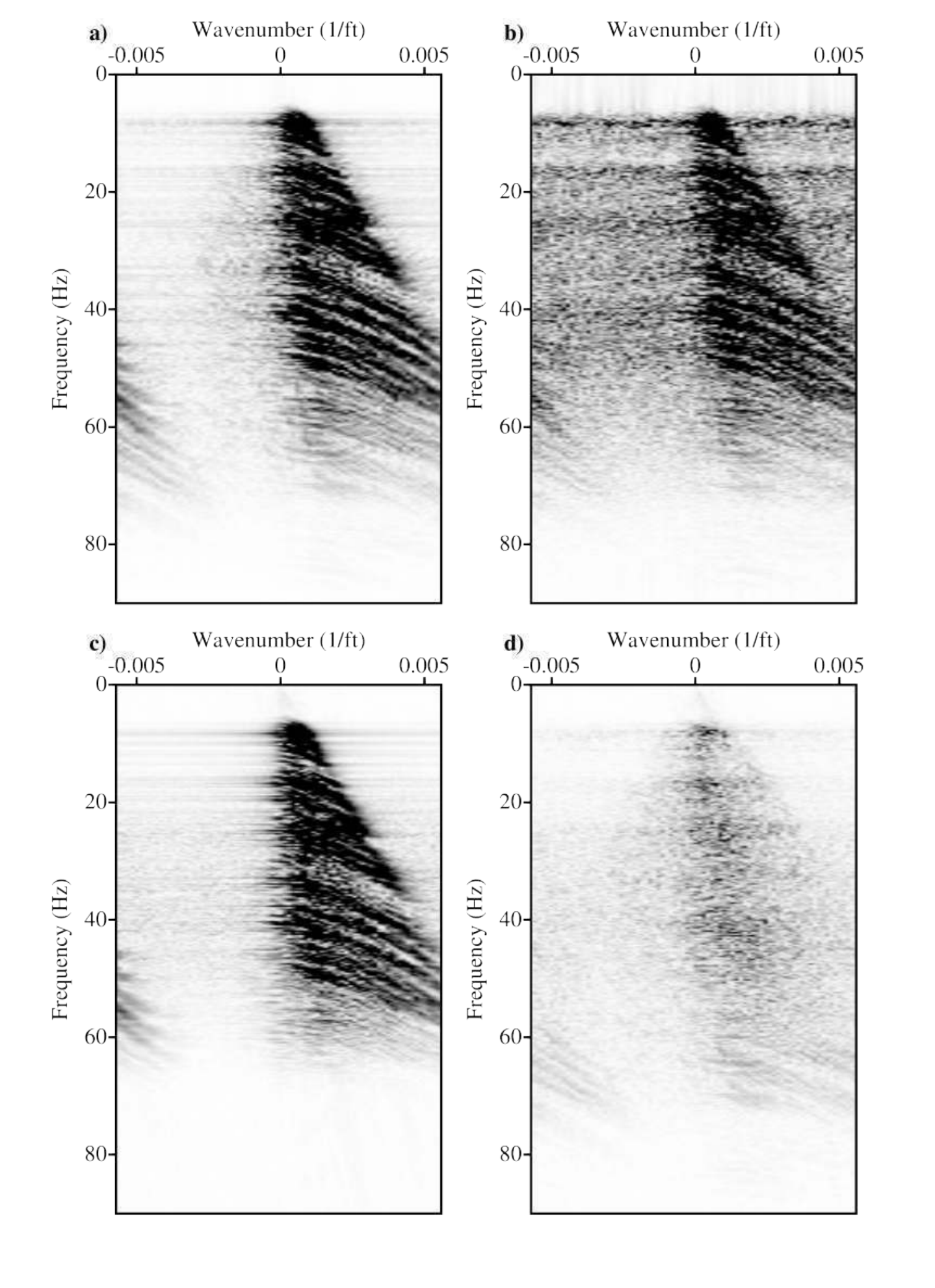}
	\reduceVspace
	\caption{The $f-k$ spectra of one common receiver gather from the Gulf of Mexico data recovered from the estimated Radon model using $p=2$ and $q=2$ inversion.
	(a) Original gather.   (b) Pseudo-deblended gather.  (c) Recovered gather.   (d) Recovered gather error.}
	\label{ch4_GOM_CRG_L2L2_fk}
\end{figure}

\begin{figure}[htp] 
	\centering
	\includegraphics{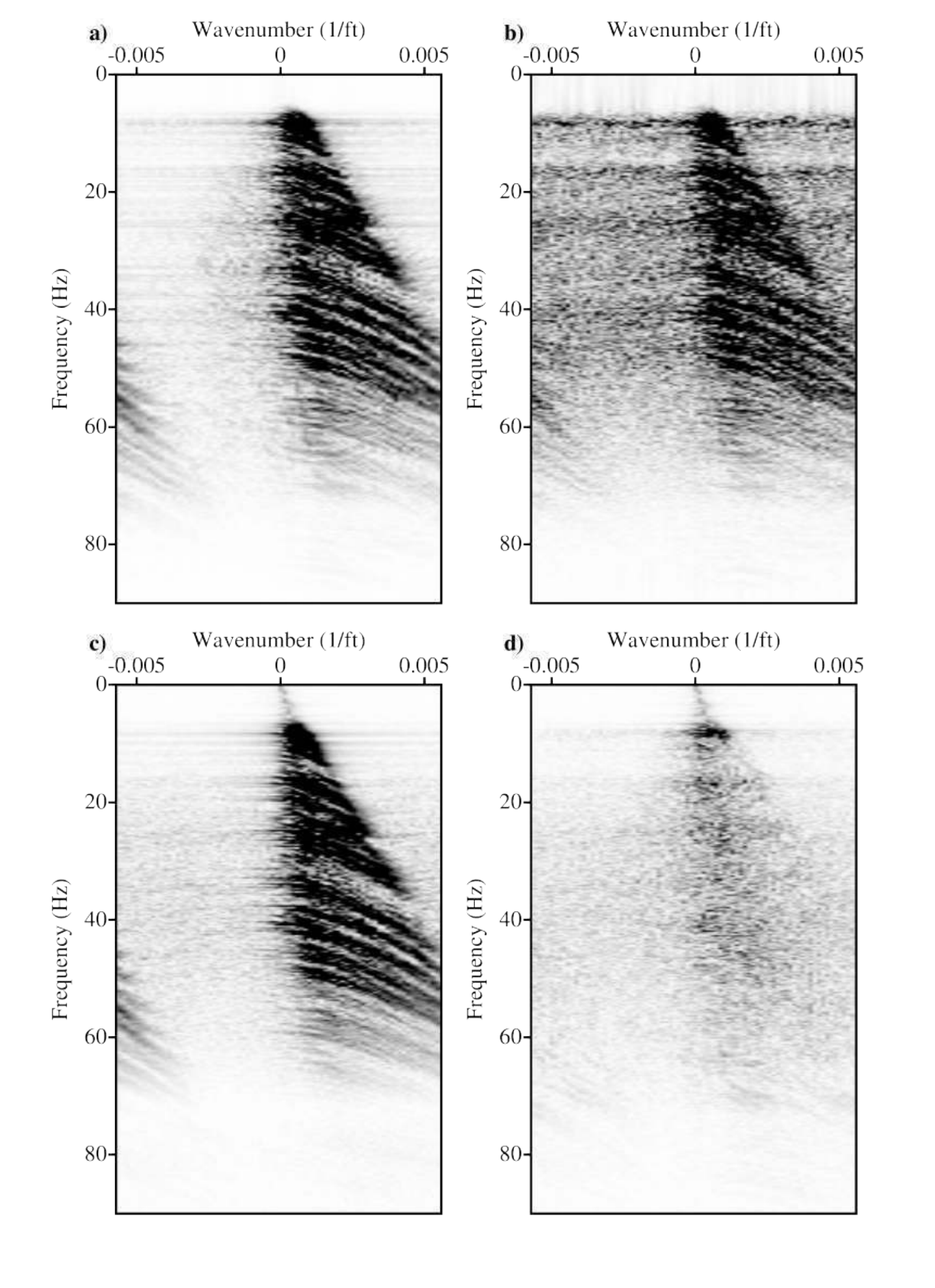}
	\reduceVspace
	\caption{The $f-k$ spectra of one common receiver gather from the Gulf of Mexico data recovered from the estimated Radon model using $p=2$ and $q=1$ inversion.
	(a) Original gather.   (b) Pseudo-deblended gather.  (c) Recovered gather.   (d) Recovered gather error.}
	\label{ch4_GOM_CRG_L2L1_fk}
\end{figure}

\begin{figure}[htp] 
	\centering
	\includegraphics{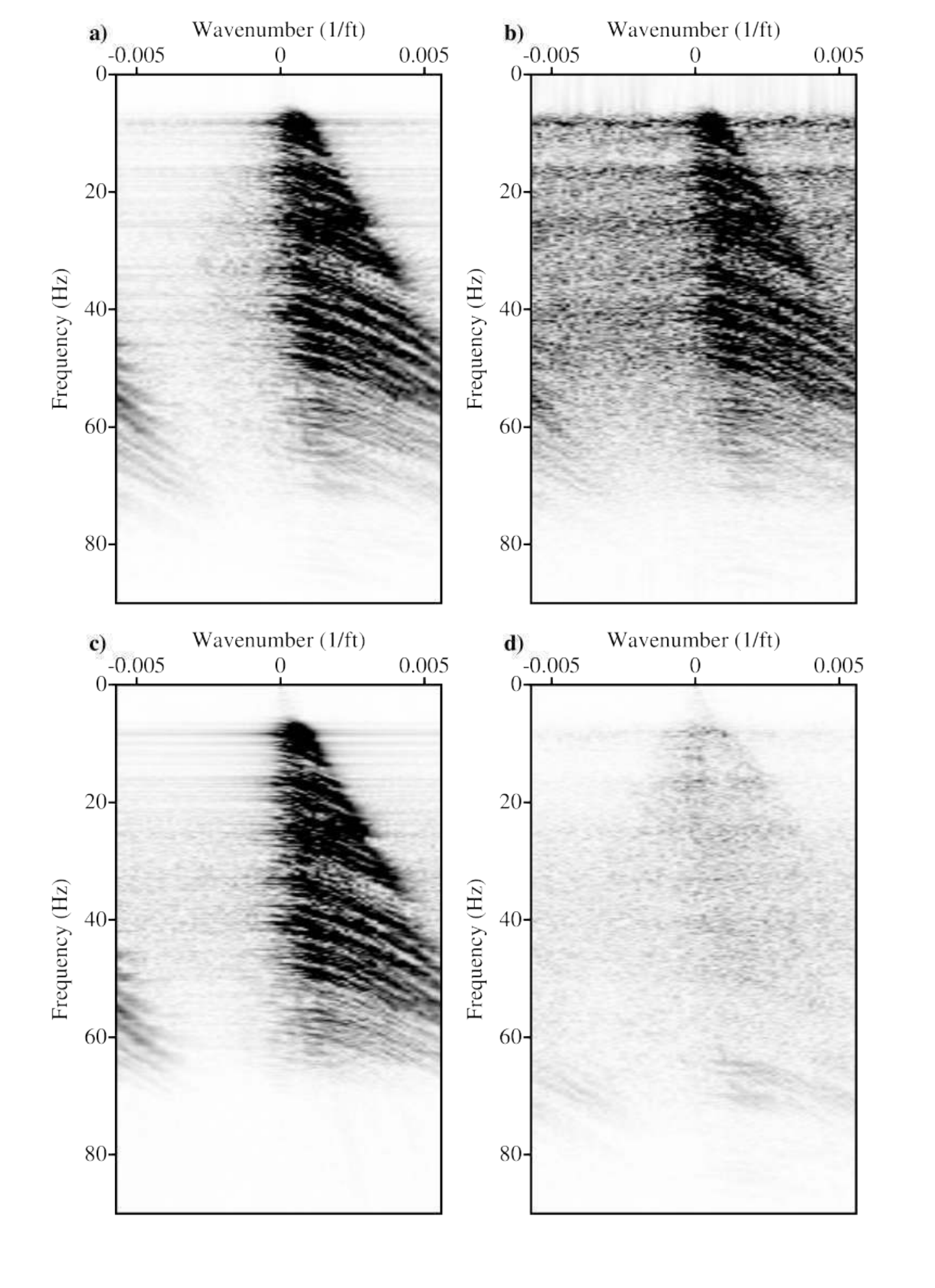}
	\reduceVspace
	\caption{The $f-k$ spectra of one common receiver gather from the Gulf of Mexico data recovered from the estimated Radon model using $p=1$ and $q=2$ inversion.
	(a) Original gather.   (b) Pseudo-deblended gather.  (c) Recovered gather.   (d) Recovered gather error.}
	\label{ch4_GOM_CRG_L1L2_fk}
\end{figure}

\begin{figure}[htp] 
	\centering
	\includegraphics{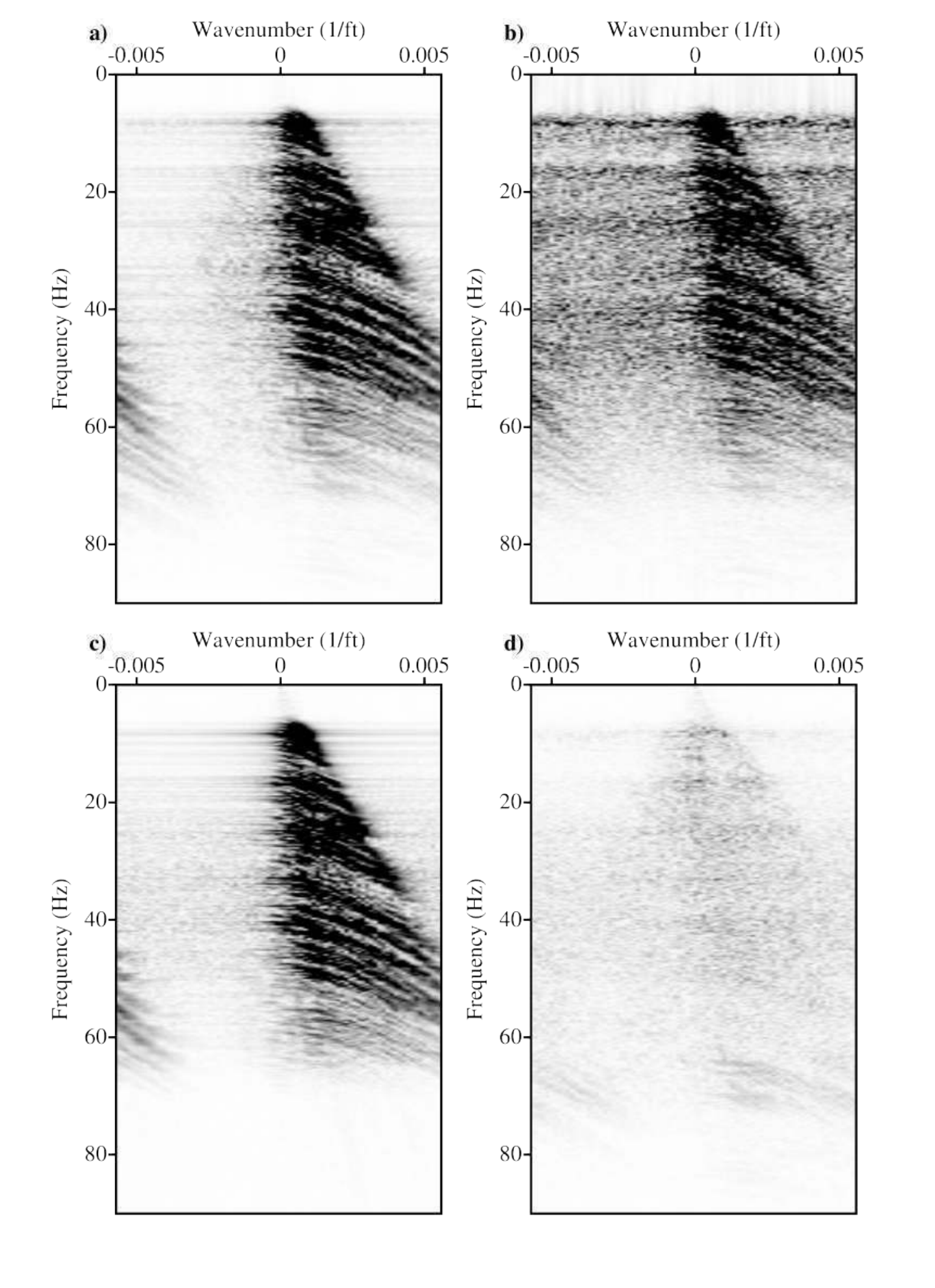}
	\reduceVspace
	\caption{The $f-k$ spectra of one common receiver gather from the Gulf of Mexico data recovered from the estimated Radon model using $p=1$ and $q=1$ inversion.
	(a) Original gather.   (b) Pseudo-deblended gather.  (c) Recovered gather.   (d) Recovered gather error.}
	\label{ch4_GOM_CRG_L1L1_fk}
\end{figure}

The estimated Radon models are used to recover the common receiver gathers.  Figures \ref{ch4_GOM_CRG_L2L2}, \ref{ch4_GOM_CRG_L2L1}, \ref{ch4_GOM_CRG_L1L2} and \ref{ch4_GOM_CRG_L1L1} show data recovered from Radon models estimated using ($p=2,q=2$), ($p=2,q=1$), ($p=1,q=2$) and ($p=1,q=1$) inversion, respectively. These figures clearly show that the two robust Radon transforms using ($p=1,q=2$) and ($p=1,q=1$) inversion were able to attenuate interference while preserving the weak signals better than the non-robust transforms using ($p=2,q=2$) and ($p=2,q=1$) inversions. The ability of robust Radon transforms to preserve weak reflections is quite clear in the close-up in 
Figures \ref{ch4_GOM_CRG_L1L2_closeup} and \ref{ch4_GOM_CRG_L1L1_closeup}. These figures show that the data recovered from Radon models estimated using ($p=1,q=2$) and ($p=1,q=1$) robust inversion, respectively. The error of the recovered common receiver gathers in Figures \ref{ch4_GOM_CRG_L1L2_closeup}d and \ref{ch4_GOM_CRG_L1L1_closeup}d confirm that the robust inversion represents an effective method for removing source interferences. On the other hand, Figures \ref{ch4_GOM_CRG_L2L2_closeup} and \ref{ch4_GOM_CRG_L2L1_closeup} shows a close-up of common receiver gather recovered from Radon models estimated using ($p=2,q=2$) and ($p=2,q=1$) non-robust inversion, respectively. The recovered common receiver gathers error in Figures \ref{ch4_GOM_CRG_L2L2_closeup}d and \ref{ch4_GOM_CRG_L2L1_closeup}d show clearly that non-robust inversions cannot remove source interferences effectively. Figures \ref{ch4_GOM_CRG_L2L2_fk}, \ref{ch4_GOM_CRG_L2L1_fk}, \ref{ch4_GOM_CRG_L1L2_fk} and \ref{ch4_GOM_CRG_L1L1_fk} show the $f-k$ spectra of the common receiver gathers recovered from Radon models estimated using ($p=2,q=2$), ($p=2,q=1$), ($p=1,q=2$) and ($p=1,q=1$) inversion, respectively. The $f-k$ spectra of the recovered common receiver gathers confirm that the robust inversion results in more accurate data recovery than the non-robust inversion. 

\begin{figure}[htp] 
	\centering
	\includegraphics{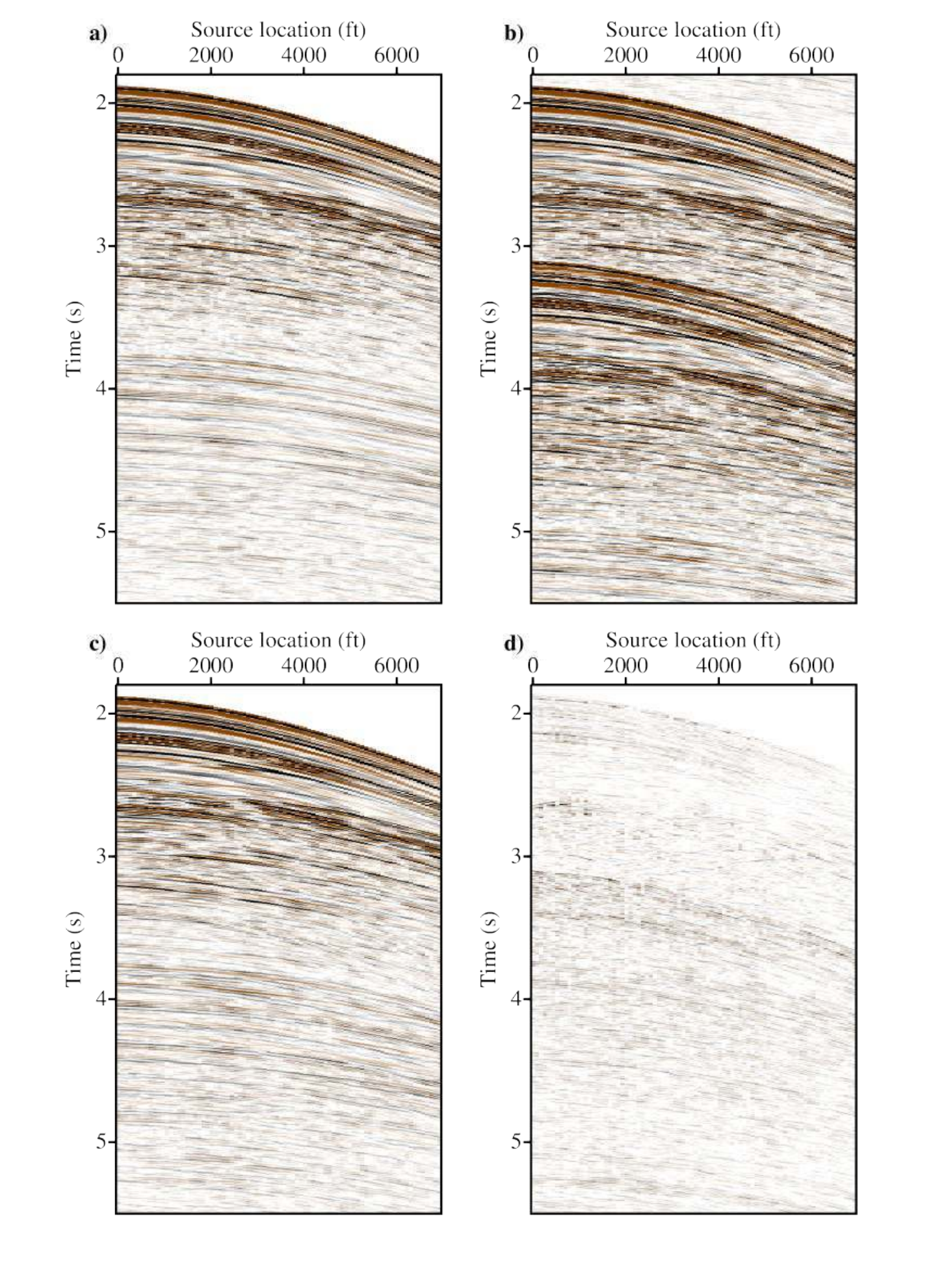}
	\reduceVspace
	\caption{Common source gather from the Gulf of Mexico data separated using $p=1$ and $q=2$ Radon model inversion.
	(a) Original gather.   (b) Pseudo-deblended gather.  (c) Recovered gather.   (d) Recovered gather error.}
	\label{ch4_GOM_CSG_L1L2}
\end{figure}

\begin{sidewaysfigure}[htp] 
 		\includegraphics{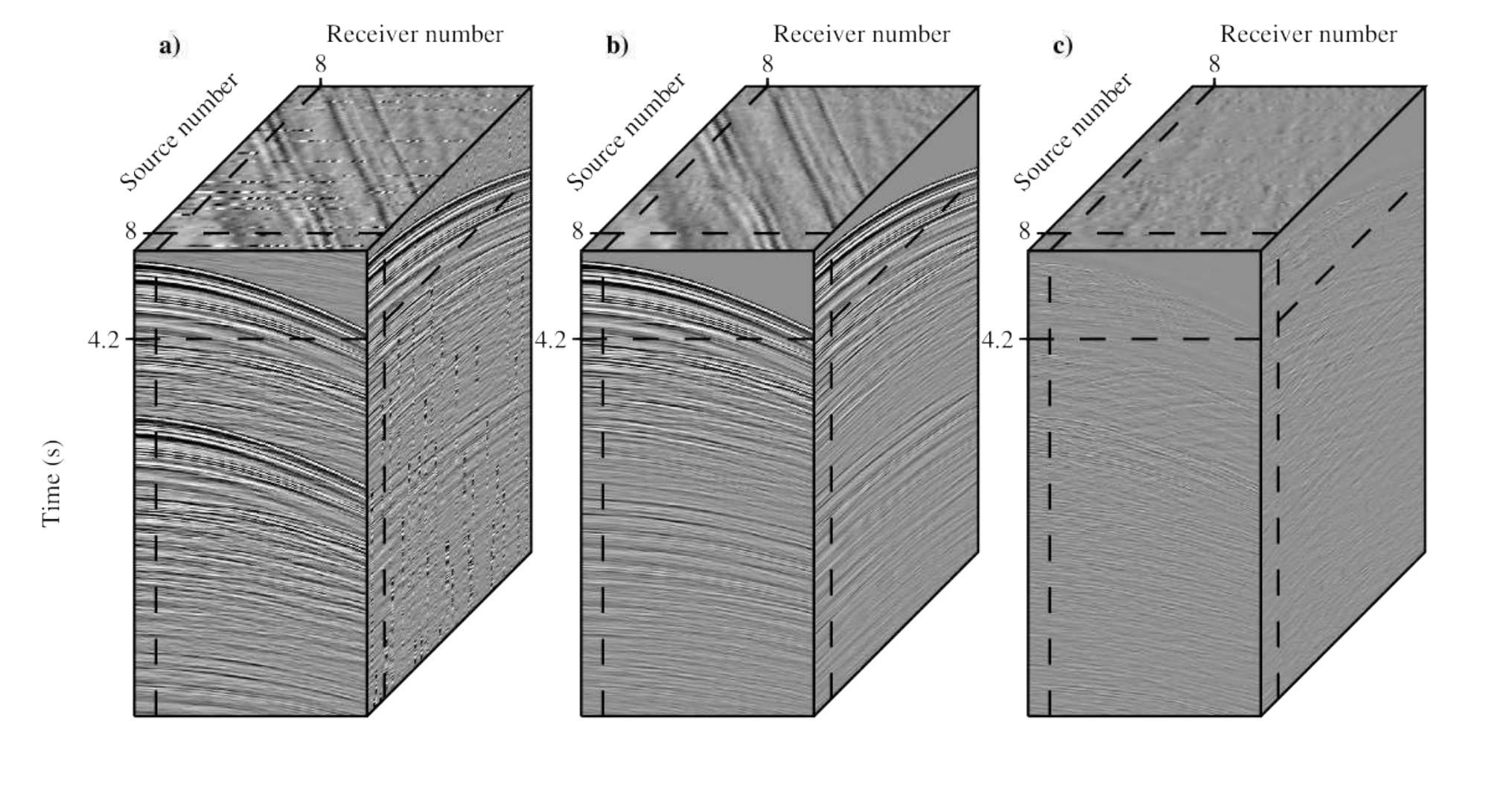}
 		\reduceVspace
 		\caption{Gulf of Mexico data cubes. 
   		(a) Pseudo-deblended data.
   		(b) Data recovered by forward modelling using the ASHRT model estimated via $p=1$ and $q=2$ inversion.
   		(c) Difference between recovered and original data cubes. }
		\label{ch4_GOM_cubes_L1L2}
\end{sidewaysfigure}

The $Q$ values for the recovered Gulf of Mexico common receiver gathers are listed in Table \ref{table:ch4_quality}. Our tests show that inversion using both robustness and quadratic regularization ($p=1$, $q=2$) produce the best results ($Q=11.4833$ dB). For the Gulf of Mexico data, we also measure the quality for recovering the weak events showed in the close-up window. The best quality for recovered weak events window was achieved using $p=1$, $q=2$ model inversion with $Q=13.12$ dB. This illustrates quantitatively that the robust Radon transform can remove source interferences effectively while preserving weak signals. Figure \ref{ch4_GOM_CSG_L1L2} shows the common source gather recovered after denoising all common receiver gathers using $p=1,q=2$ inversion.  This figure shows that coherent interferences in common source gather were removed effectively after denoising all common receiver gathers. Figure \ref{ch4_GOM_cubes_L1L2} shows the data cube recovered using Radon models estimated using $p=1,q=2$ inversion.   
Regarding the Gulf of Mexico field data example, imposing both robustness and sparsity  ($p=q=1$) is not as simple as one might think. In this case, the algorithm becomes sensitive to the selection of $\epsilon_r$, $\epsilon_m$ and the number of internal and external iterations of the IRLS method. Recent results in the area of robust deconvolution that include sparsity constraints \citep{Ali2012} suggest that more sophisticated algorithms are needed to obtain sparse and robust solutions that are not prone to failure due to incorrect parameter selection \citep{Li2012}.  

\setlength{\tabcolsep}{3.5pt}
\renewcommand{\arraystretch}{3.0}
\begin{table}[htp] 
\small{
\begin{center}
\caption{Quality of denoising CRG using ASHRT transform.}
\label{table:ch4_quality}
\begin{tabular}{| c | c | c | c | c |}
\hline 
{Model inversion} 	    							&$p=2$, $q=2$	     &$p=2$, $q=1$         &$p=1$, $q=2$	&$p=1$, $q=1$ \\ \hline  
{Synthetic data }   								& $10.7831$      	    & $14.7742$               &	$15.1325$	&	$18.6502$\\  \hline                    
{Field data  }										& $7.7762$			   &  $7.0399$      			  &   $11.4833$     &	$10.3468$	\\  \hline 
{Field data (weak events window)}		& $7.4004$          	   &  $6.6095$     			  &	$13.123$		  &$11.5293$	\\ \hline    
\end{tabular}
\end{center}}
\end{table} 

\section{Conclusions}
We have implemented robust Radon transforms to eliminate erratic incoherent noise that arises in common receiver gathers of simultaneous source data.  We showed that source interferences in common receiver gathers can be removed via a  robust Radon transform.  Our synthetic data example shows superior results when a sparse ($q=1$) and robust ($p=1$) Radon transform is adopted. It is well known that the stringent requirement of sparsity  can be easily satisfied with simulated data. Conversely, imposing sparsity in the Radon coefficients is not an easy task when there is a  mismatch between the travel-times and amplitudes of the data and those modeled by the transform. The latter limits the benefit of the  sparsity constraint  for real data applications.  However, our real data tests show that  Radon transforms with a robust misfit ($p=1$) and a simple quadratic regularization ($q=2$) provides an effective algorithm to eliminate source interferences in common receiver gathers.

\chapter[Stolt-based Radon Transforms]
{
Stolt-based Radon Transforms \footnote{A version of this chapter has been published. Ibrahim and Sacchi, 2015, Simultaneous source separation using migration and demigration operators, Geophysics 80(6), WD27 - WD36.}
}

\section{Motivation}

The ASHRT transform has a high computational cost due to the extension of the model dimensions by scanning  the apex locations. Additionally, the ASHRT operator is a time variant Radon operator that cannot be computed efficiently in the frequency domain. Therefore, we need to design a new operator that is computationally efficient and has basis functions that match seismic data in common receiver gathers. The Stolt migration/demigration operators are computationally efficient and they can focus hyperbolic reflections with shifted apexes. The Stolt operator can focus seismic reflections back to its subsurface reflection points by mapping the seismic data in the $f-k$ domain. The computational efficiency of the Stolt operator results from employing Fast Fourier Transforms (FFT) and mapping the band-limited data in the $f-k$ domain.  However, the Stolt migration model is derived using a constant velocity subsurface assumption. Thus, the Stolt transform model can be viewed as a constant velocity cross section of the ASHRT model. Therefore, we extend the single velocity Stolt operator to scan for multiple velocities and construct an ASHRT model. The Stolt-based ASHRT operator is equivalent to the time domain ASHRT operator but with higher computational efficiency.

\section{Stolt-based Radon Transform}
\begin{sidewaysfigure}[htbp]
	\includegraphics{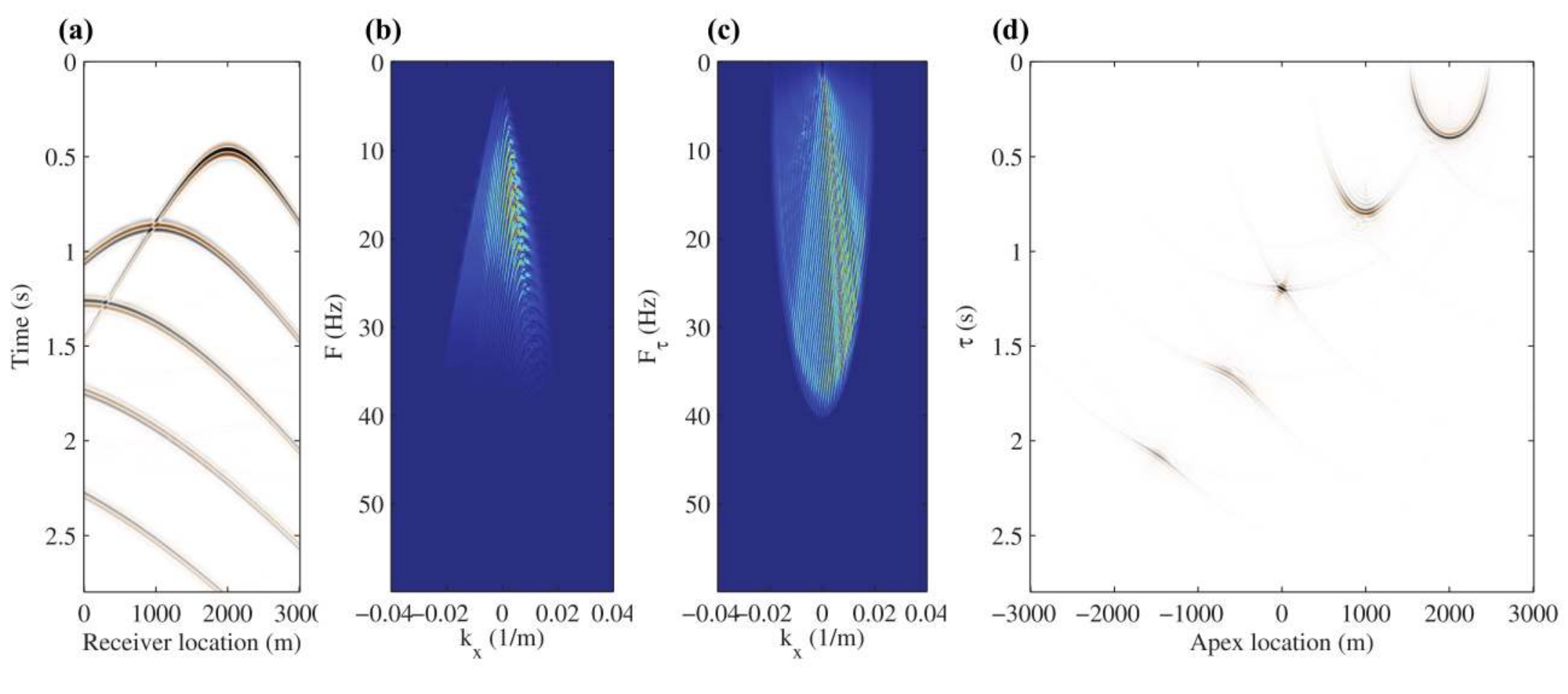}
	\caption{Computing apex shift hyperbolic Radon transform (ASHRT) using Stolt migration operator. 
		(a) Input data. (b) Data in the $\omega-k_x$ domain after 2D FFT. 	
   		(c) Data after mapping to $\omega_{\tau}-k_x$ domain. 
   		(d) Stolt model in the time domain.}
	\label{ASHRT_Stolt_schematics}
\end{sidewaysfigure}
\begin{sidewaysfigure}[htbp]
	\includegraphics{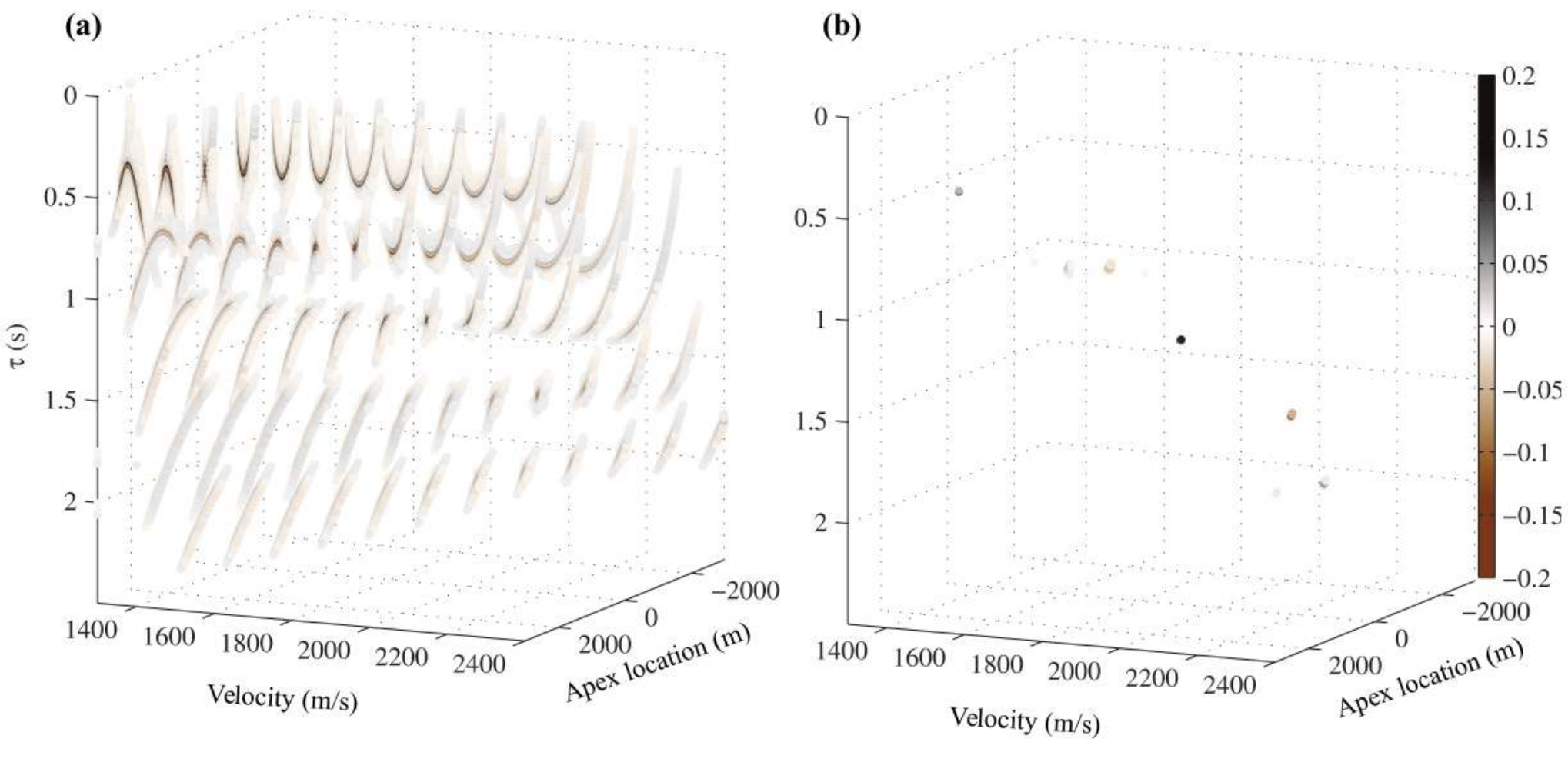}
	\caption{Apex shift hyperbolic Radon models. 
   		(a) Model estimated using least-squares inversion.
   		(b) Model estimated using sparse inversion.}
  	\label{ASHRT_Stolt_models}
\end{sidewaysfigure}

The Stolt migration method is considered to be the fastest migration algorithm \citep{Margrave2001}. 
This operator performs migration by mapping data in the $\omega-k$ domain to vertical wavenumber $k_z$ for a constant subsurface velocity.
Despite of not being widely used any more in seismic imaging due to its constant velocity limitation, the low computational cost of the Stolt operator made it a useful tool in other fields such as medical imaging \citep{Bamler1992,Garcia2013} and synthetic aperture radar imaging \citep{Cafforio1991,Li2014, Wu2014}. 

Using the exploding reflector principle \citep{Claerbout1985} and a constant velocity assumption, the Stolt operator can be used to estimate the subsurface model from zero offset data.
This estimated model ${\widetilde{m}}(\tau,v,x)$ is related to the data recorded at the surface ${d}(t,x)$ by the following relationship \citep{Yilmaz2001} 
\begin{equation}\label{ch5_Stolt_adj1}
\widetilde{m}(\tau,v,x)= \int\int {d}(\omega,k_x)~ \exp[{-ik_xx-i\omega_{\tau}(v)\tau }]~d\omega~dk_x,
\end{equation}
where $x$ represents the horizontal axis and $\omega_{\tau}$ is the Fourier dual of the apex time $\tau$ which is a function of the velocity through the modified dispersion relationship \citep{Yilmaz2001}
\begin{equation}
\omega_{\tau} = \sqrt{\omega^2-(vk_x)^2}. 
\end{equation}
Equation \ref{ch5_Stolt_adj1} can be rewritten by changing  the integration variable from $\omega$ to $\omega_{\tau}$ 
\begin{align}\label{ch5_Stolt_adj}
\widetilde{m}(\tau,v,x)=& \int\int C~{ d}{(\omega=\sqrt{\omega_{\tau}^2+(vk_x)^2},k_x)}& \nonumber \\ 
& \times \exp{[-ik_xx-i\omega_{\tau}(v)\tau]}~d\omega_{\tau}~dk_x,  
\end{align} 
where $C=\omega_{\tau}/\omega$ is a scaling factor resulting from the change of variables.

Figure \ref{ASHRT_Stolt_schematics} shows the steps required to image with the Stolt operator.
Similarly, the forward Stolt modelling operator can be written as 
\begin{align}\label{ch5_Stolt_forward}
{d}(t,x)=& \int\int\int {m} (\omega_{\tau}=\sqrt{\omega^2-(vk_x)^2},v,k_x) \nonumber\\
& \times \exp[{ik_xx+i\omega t}]~ d\omega~dk_x~dv.
\end{align}
The forward and adjoint transforms in equations (\ref{ch5_Stolt_adj}) and (\ref{ch5_Stolt_forward}) can be written in operator form as follows 
\begin{equation} \label{ch5_adj_operator}
{\bf L}^T={\bf F}_{\omega_{\tau},k_x}^{-1}~{\bf M}^{T}_{\omega,v,k_x}~{\bf F}_{t,x}~{\bf S}^{T},
\end{equation} 
\begin{equation} \label{ch5_forward_operator}
{\bf L}={\bf S}~{\bf F}_{\omega,k_x}^{-1}~{\bf M}_{\omega_{\tau},v,k_x}~{\bf F}_{\tau,x},
\end{equation} 
where, ${\bf F}$ is the Fourier transform, ${\bf M}_{\omega_{\tau},v,k_x}$ is the Stolt mapping operator and ${\bf S}$ is a summation operator and its adjoint is a spraying operator \citep{Claerbout1985}.
Although the Stolt operator is derived with a constant velocity assumption, it can be used to construct an equivalent of the ASHRT model with multiple velocities. 
Since each  image  represents one plane inside the ASHRT cube at constant velocity, the ASHRT model is a collection of all these images. 
Therefore, the adjoint Stolt operator in equation \ref{ch5_adj_operator} includes a spreading operator ${\bf S}^{T}$ that computes several images with different velocities from the same data while the forward Stolt operator in equation \ref{ch5_forward_operator} uses a summation operator to model the data.

The classical ASHRT operator has a computational cost of $O(n_{a} \times n_{\tau} \times n_{v} \times n_x)$, where $n_{a}, n_{\tau}, n_{v}$ and $n_x$ are the number of  apex locations, zero offset times, velocities and offsets, respectively.  
Assuming that we scan for all possible apex locations and times, then  $n_{a}=n_{x}$ and $n_{\tau}=n_{t}$.
Therefore, the ASHRT operator cost is $ O( n_x^2 \times n_t \times n_v)$. 
On the other hand, the Stolt based ASHRT (without FFT zero padding) operator has a cost that is of the 2D FFT of the data with size $n_t \times n_x$ followed by $\omega-k$ mapping and inverse 2D FFT of the model with size $n_t \times n_v \times n_x$. 
Therefore, the total computational cost of an ASHRT implemented via the Stolt operator is  $O( [n_t \log_2(n_t)+n_x \log_2(n_x) ][n_v+1]+ n_v \times n_{kx} \times n_{\omega})$, where $n_{k_x}$ and $n_{\omega}$ are the number of horizontal wavenumbers and temporal frequencies, respectively. 
In the previous analysis, we have assumed that both the data and the model  are regularly sampled in space and apex, respectively. The latter permits us to adopt the computationally efficient FFT operators. 
However, if the data and/or the model are not regularly sampled, the FFT operator will be replaced by the less efficient discrete Fourier transform (DFT). The cost of the $\omega-k$ mapping is proportional to $n_v \times n_{k_x} \times n_{\omega}$ and we stress that the latter is an upper limit, since in practice we only scan for a limited group of positive frequencies and use the Fourier domain symmetries to compute the negative frequencies.
\begin{figure}[htbp]
\centering
\includegraphics{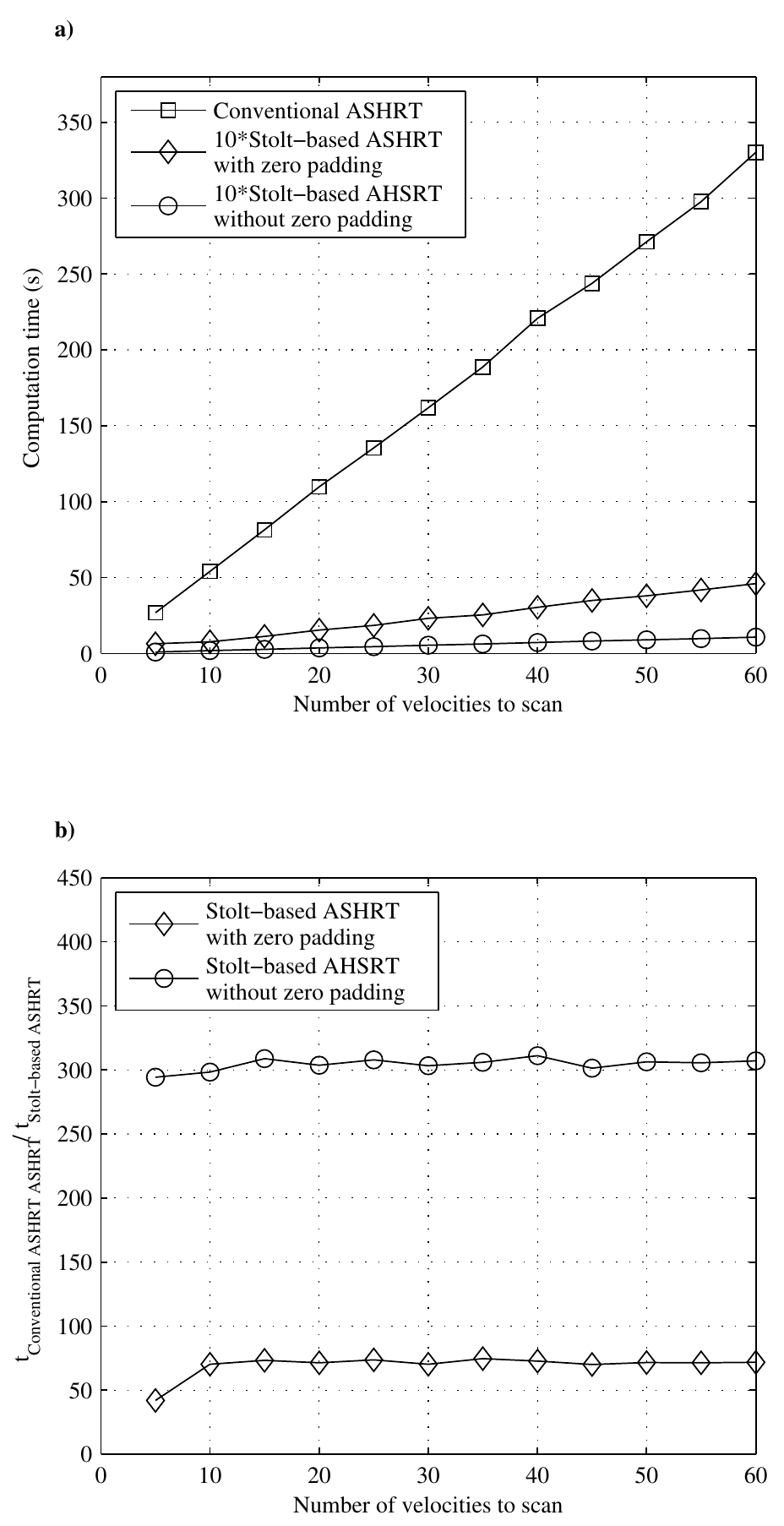}
\caption{Comparing operators (a) Computation times of conventional and Stolt-based ASHRT operators (note that the time for the Stolt-based operators are scaled by a factor of 10). (b) Ratio of the computation times for the Stolt-based ASHRT operators to the conventional ASHRT operator.}
\label{speed}
\end{figure}
Figure \ref{speed}a shows the computational times of the conventional ASHRT and the Stolt-based ASHRT with and without zero padding (note that the Stolt-based results are plotted with a scaling factor of 10). 
Zero padding is required to reduce artifacts associated with $\omega-k$ interpolation and improve Stolt's  mapping precision.  
Figure \ref{speed}b shows the improvement in the computational time by using the Stolt-based ASHRT operator with and without zero padding compared to the conventional ASHRT. 
It is clear that using the Stolt operator can lead to significant savings in computational cost.
Therefore, this operator can be useful in early stages of processing for quality control and velocity analysis. 
This is very important for processing large data sets that contain large number of common receiver gathers.

\section{Synthetic Example}
We tested the Stolt-based Radon transform with a numerically blended synthetic data set. 
The single shot gathers of the synthetic data were modelled using the forward modelling operator. 
This guarantees that the main components of the data are composed of reflection hyperbolas similar to the transform basis. 
However, the synthetic data are composed from 10 seismic reflections with the velocities $1500,1560,1590,1600,1655,1700,1750,1780,1850$ and $1975$ m/s, while, the transform basis use only 5 velocities $1500, 1600, 1700, 1800$ and $1900$ m/s. 
The data example is the same as the example in chapter 4. 
This situation is similar to field data examples where the true velocity is unknown and the Radon transform use coarse velocity sampling.   
The data are numerically blended with a $50\%$ reduction in time compared to conventional acquisition. 
The acquisition scenario represent a single source boat with the time interval between successive sources which is nearly half of the conventional acquisition. 
In order to make the source interferences appear incoherent, the source firing times are dithered using random time delays.  
The firing times of the sources are the same as the example in chapter 4 and it is shown in Figure 4.2.

\begin{figure}[htp] 
	\centering
	\includegraphics{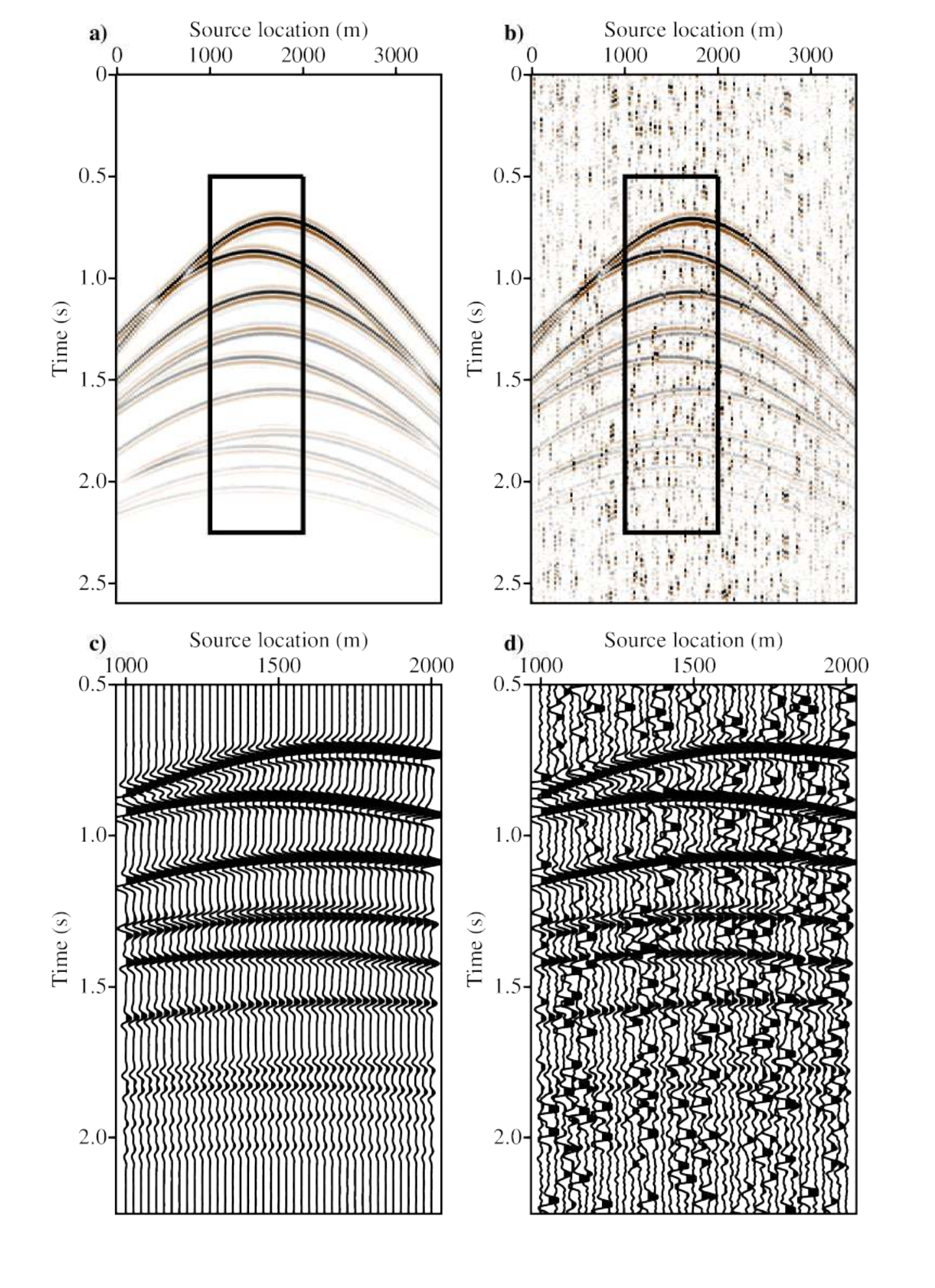}
	\reduceVspace
	\caption{A numerically blended common receiver gather from the synthetic data example.
	(a) Original gather.   
	(b) Pseudo-deblended gather .  
	(c) Close up of the original gather.   
	(d) Close up of the pseudo-deblended gather. }
	\label{ch5_synth_CRG_example}
\end{figure}

Figure \ref{ch5_synth_CRG_example}a shows an original common receiver gather of the simple synthetic example 
and Figure \ref{ch5_synth_CRG_example}b shows the same gather after blending and pseudo-deblending.
Figures \ref{ch5_synth_CRG_example}c and d are the close up wiggle plots of the areas marked on Figures \ref{ch5_synth_CRG_example}a and b, respectively. 
These close up figures show the high amplitude source interferences mixing with the low amplitude late arrivals of the primary source (notice the reflections at 2 sec). 
   
\begin{sidewaysfigure}[htbp]
 		\includegraphics{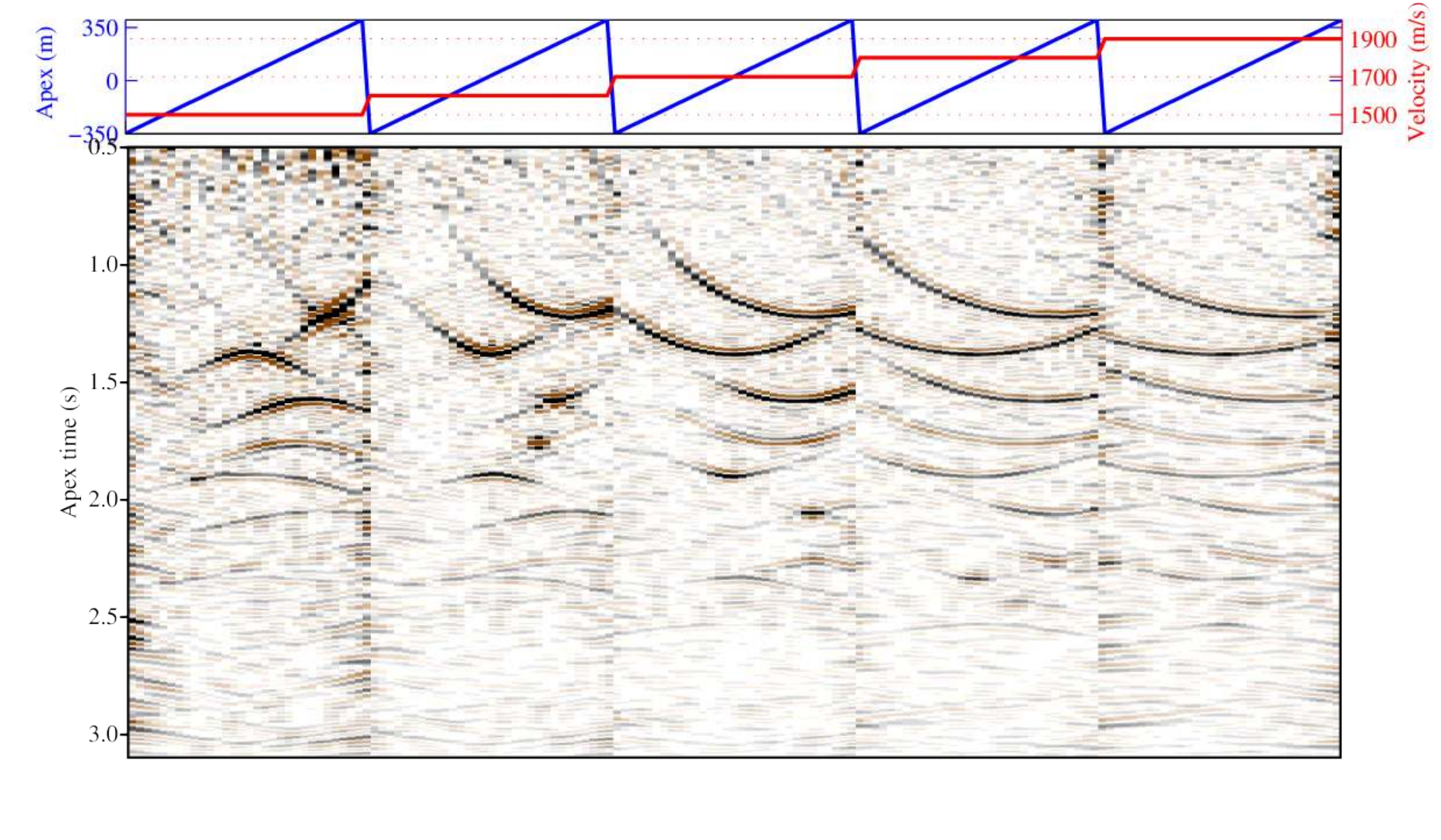}
 		\reduceVspace
 		\caption{ Stolt-based ASHRT model for one common receiver gather from the synthetic data example, estimated using $p=2$ and $q=2$ inversion.}
		\label{ch5_synth_model_L2L2}
\end{sidewaysfigure}

\begin{sidewaysfigure}[htbp]
 		\includegraphics{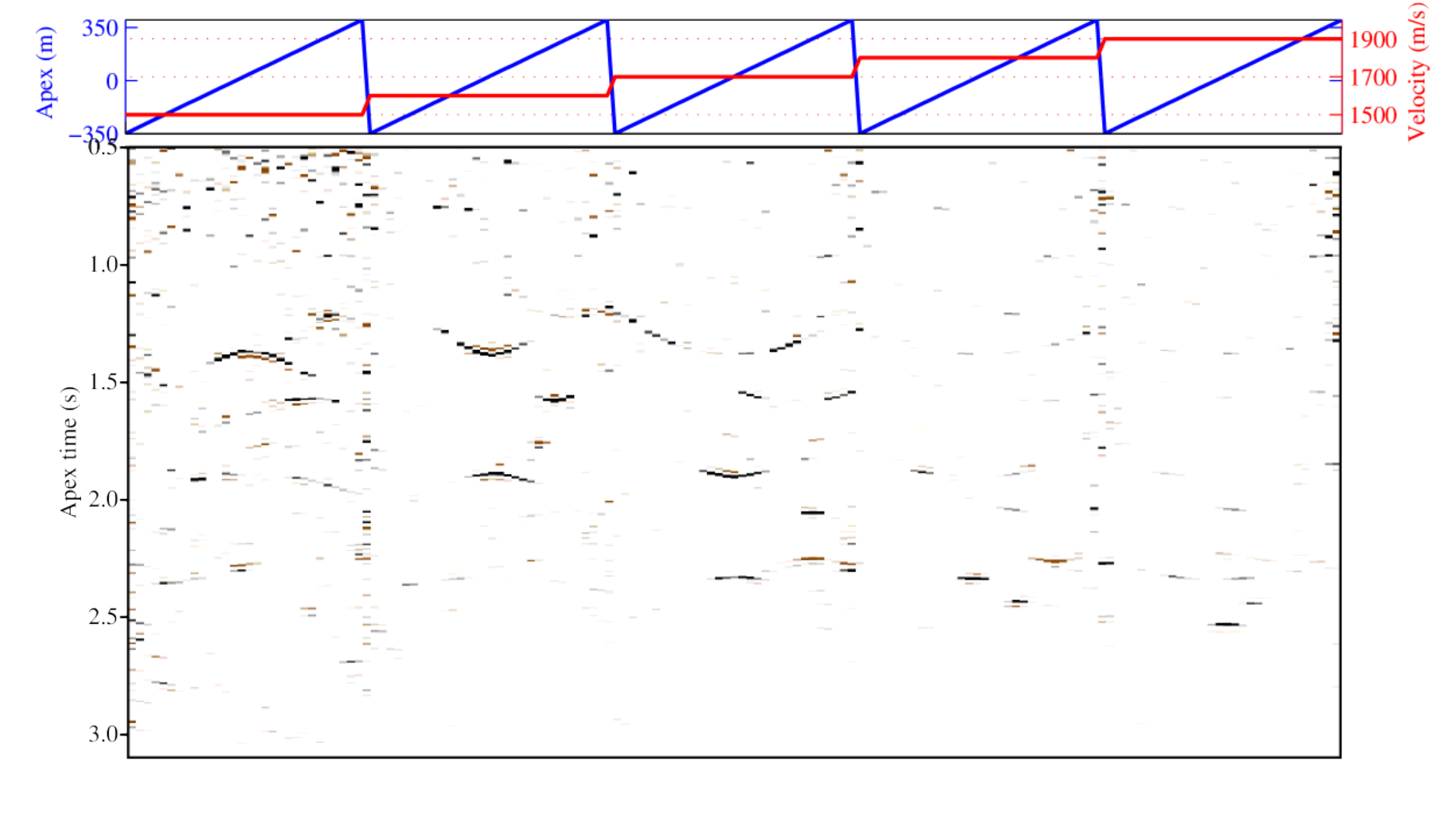}
 		\reduceVspace
 		\caption{ Stolt-based ASHRT model for one common receiver gather from the synthetic data example, estimated using $p=2$ and $q=1$ inversion.}
		\label{ch5_synth_model_L2L1}
\end{sidewaysfigure}

\begin{sidewaysfigure}[htbp]
 		\includegraphics{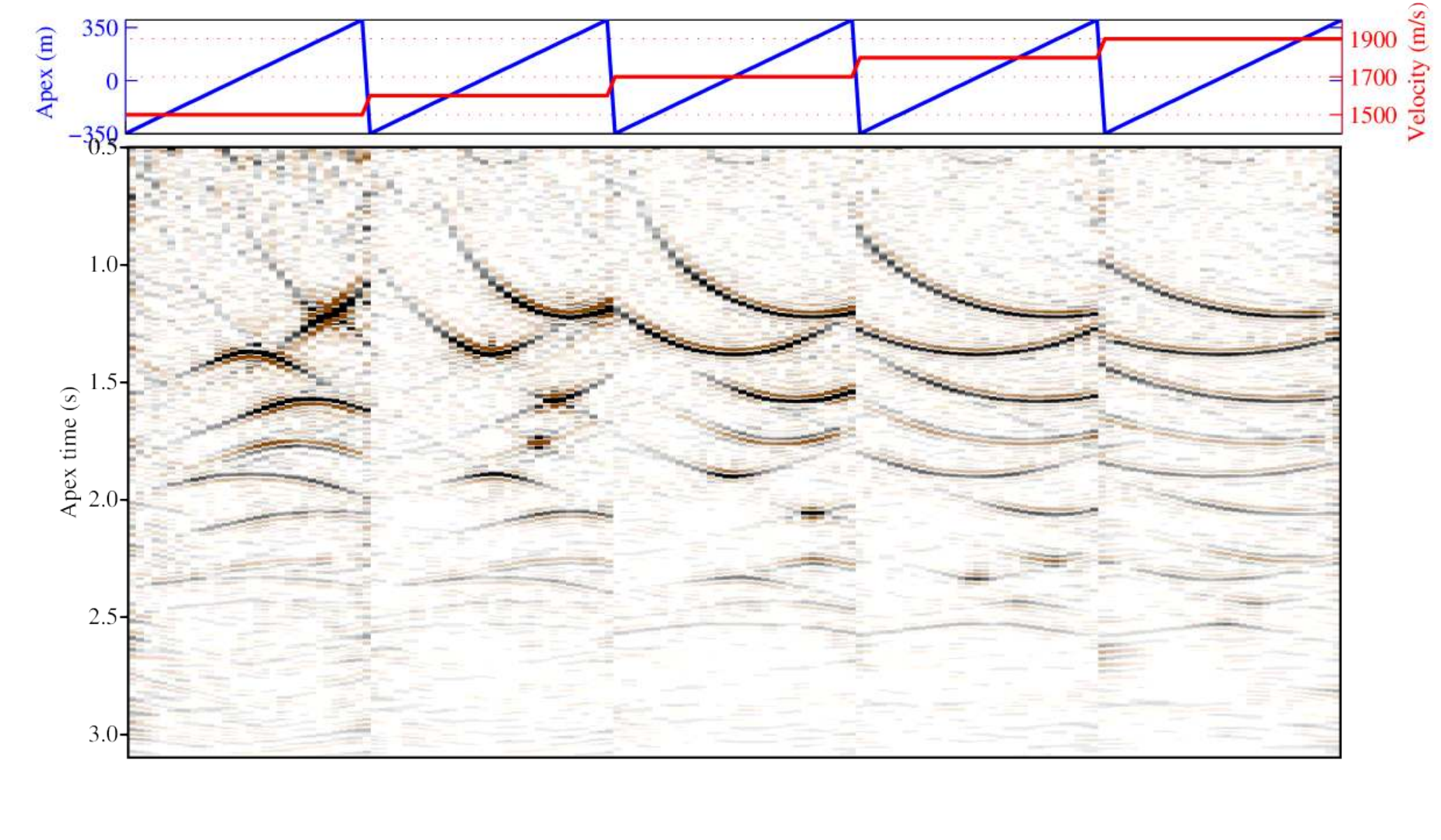}
 		\reduceVspace
 		\caption{ Stolt-based ASHRT model for one common receiver gather from the synthetic data example, estimated using $p=1$ and $q=2$ inversion.}
		\label{ch5_synth_model_L1L2}
\end{sidewaysfigure}

\begin{sidewaysfigure}[htbp]
 		\includegraphics{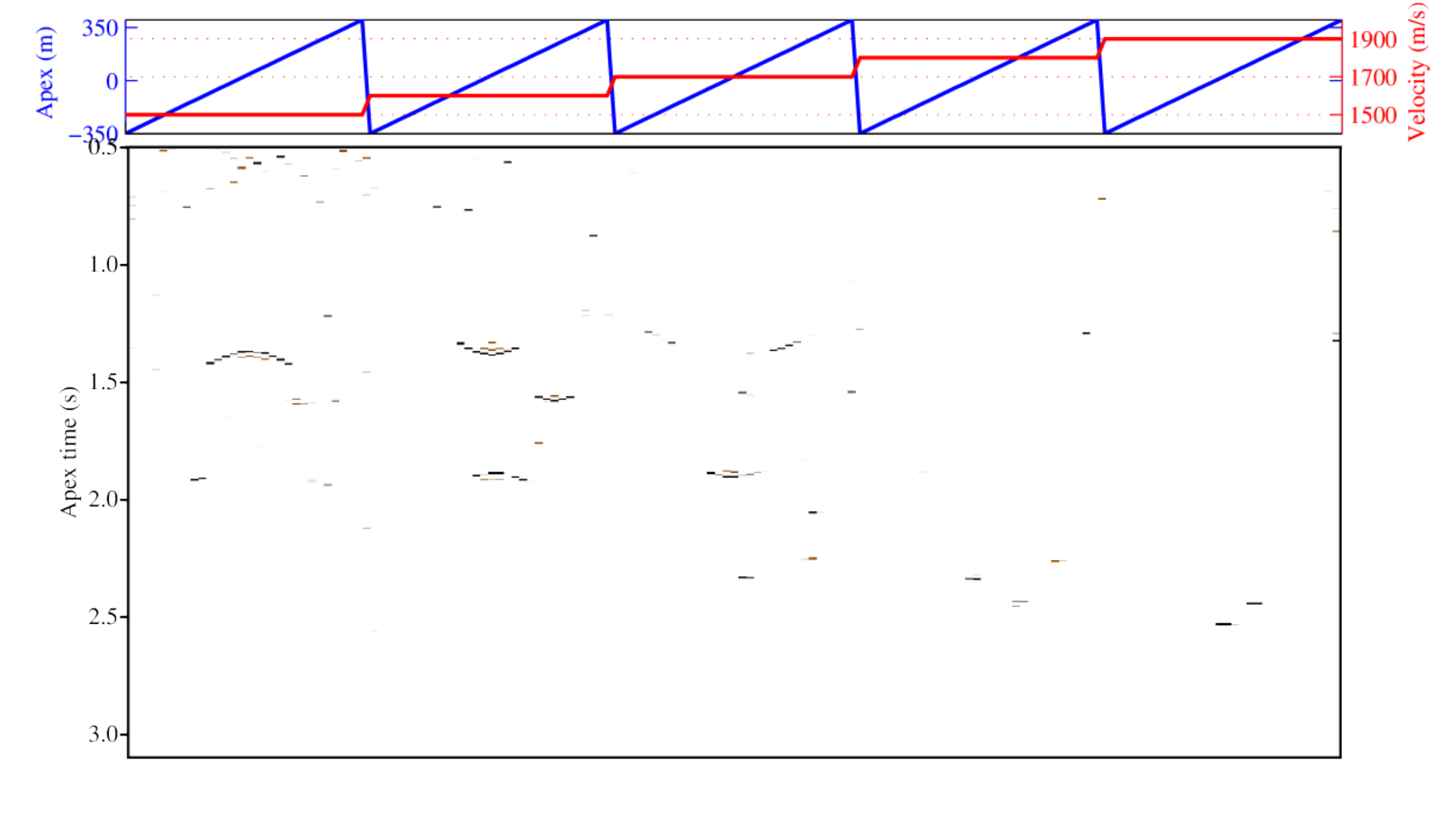}
 		\reduceVspace
 		\caption{ Stolt-based ASHRT model for one common receiver gather from the synthetic data example, estimated using $p=1$ and $q=1$ inversion.}
		\label{ch5_synth_model_L1L1}
\end{sidewaysfigure}

\begin{figure}[htbp]
 		\includegraphics{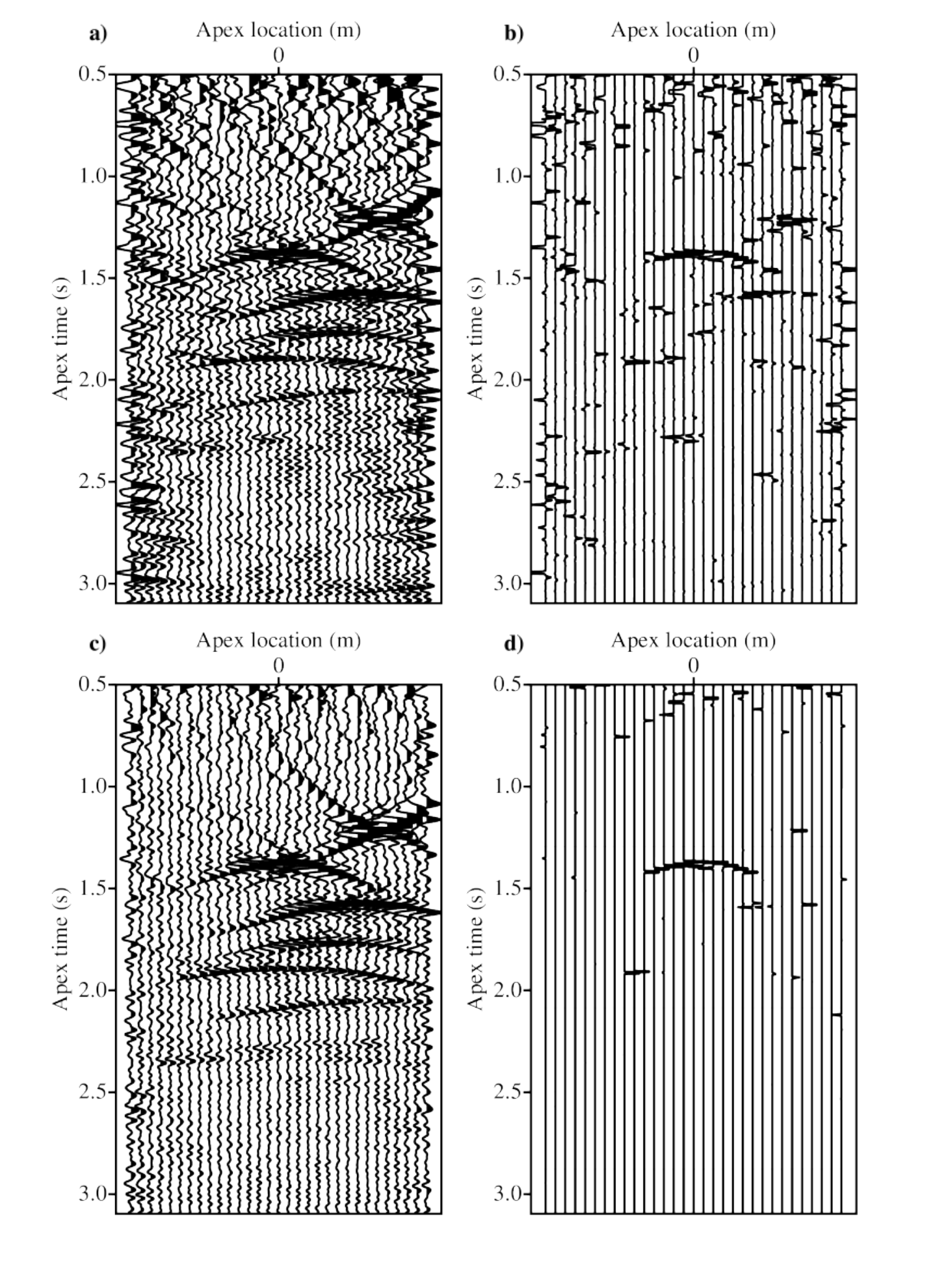}
 		\reduceVspace
   		\caption{One velocity panel ($v=1500~m/s$) of the Stolt-based ASHRT models for the simple synthetic data estimated by inversion with 
	    (a) $p=2$ and $q=2$, (b) $p=2$ and $q=1$, (c) $p=1$ and $q=2$, and (d) $p=1$ and $q=1$. }
		\label{ch5_synth_modeld_closeup}
\end{figure}

Four different Radon models for the pseudo-deblended common receiver gather in Figure \ref{ch5_synth_CRG_example}b are estimated using four different inversion scenarios.
The  Radon models for ($p=2,q=2$), ($p=2,q=1$), ($p=1,q=2$) and ($p=1,q=1$) inversions are shown in Figures \ref{ch5_synth_model_L2L2}, \ref{ch5_synth_model_L2L1}, \ref{ch5_synth_model_L1L2} and \ref{ch5_synth_model_L1L1}, respectively. 
These models are easy to interpret since the synthetic data are simple. 
The coarse velocities used in the transform clearly reduced the focusing power of the transform as shown in Figures \ref{ch5_synth_model_L2L1} and \ref{ch5_synth_model_L1L1}. 
The effect of robust misfit was clear in comparing the robust models in Figures \ref{ch5_synth_model_L1L2} and \ref{ch5_synth_model_L1L1}
versus the non robust models in Figures \ref{ch5_synth_model_L2L2} and \ref{ch5_synth_model_L2L1}. 
Figure \ref{ch5_synth_modeld_closeup} shows one velocity panel from each of the four estimated Radon models. 
This figures shows the effects of robustness and sparsity on the model estimation more clearly. 

\begin{figure}[htp] 
	\centering
	\includegraphics{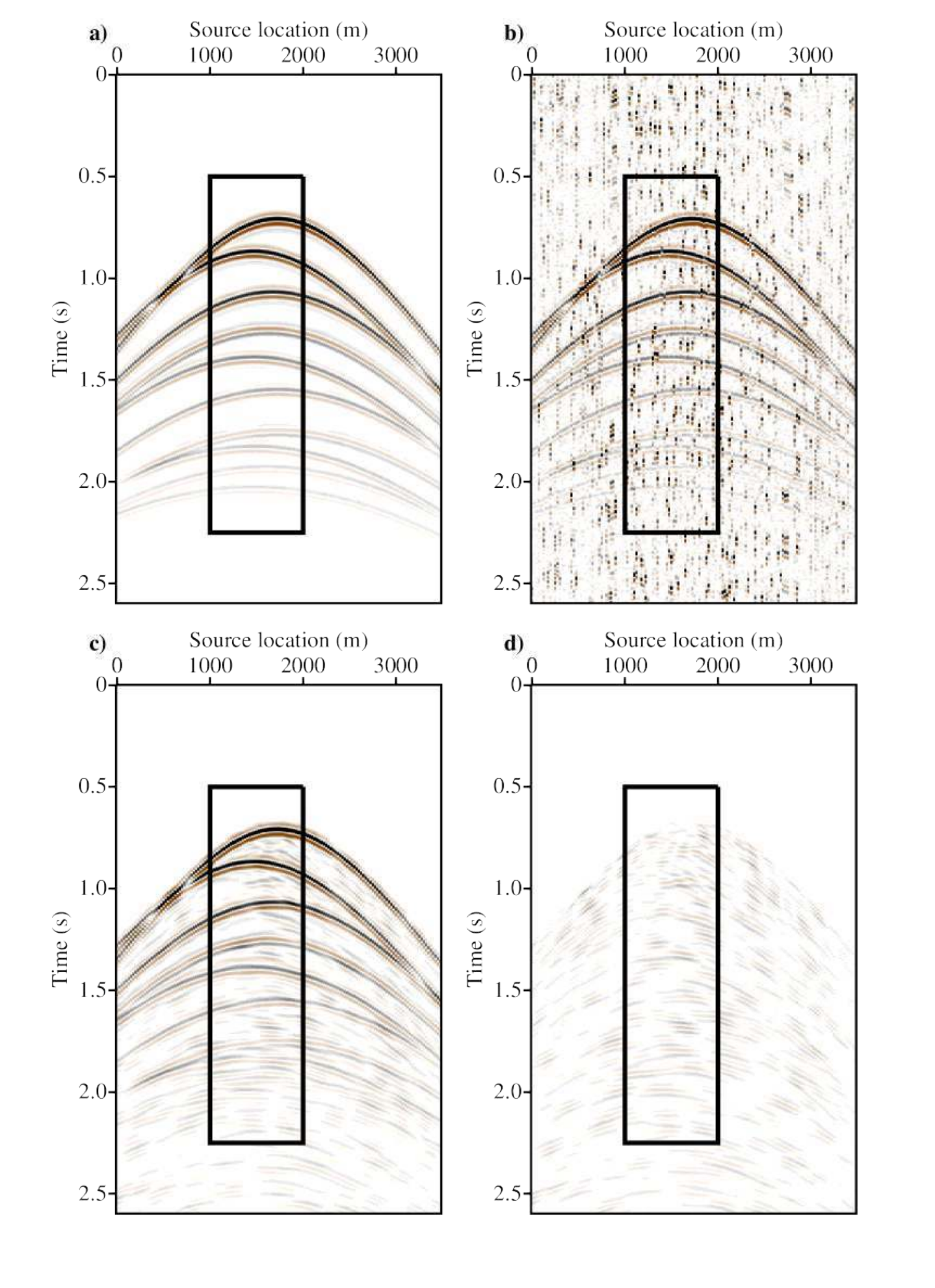}
	\reduceVspace
	\caption{Synthetic data common receiver gather recovered from the Radon model estimated using $p=2$ and $q=2$ inversion.
	(a) Original gather.   (b) Pseudo-deblended gather.  (c) Recovered gather.   (d) Recovered gather error.}
	\label{ch5_synth_CRG_L2L2}
\end{figure}

\begin{figure}[htp] 
	\centering
	\includegraphics{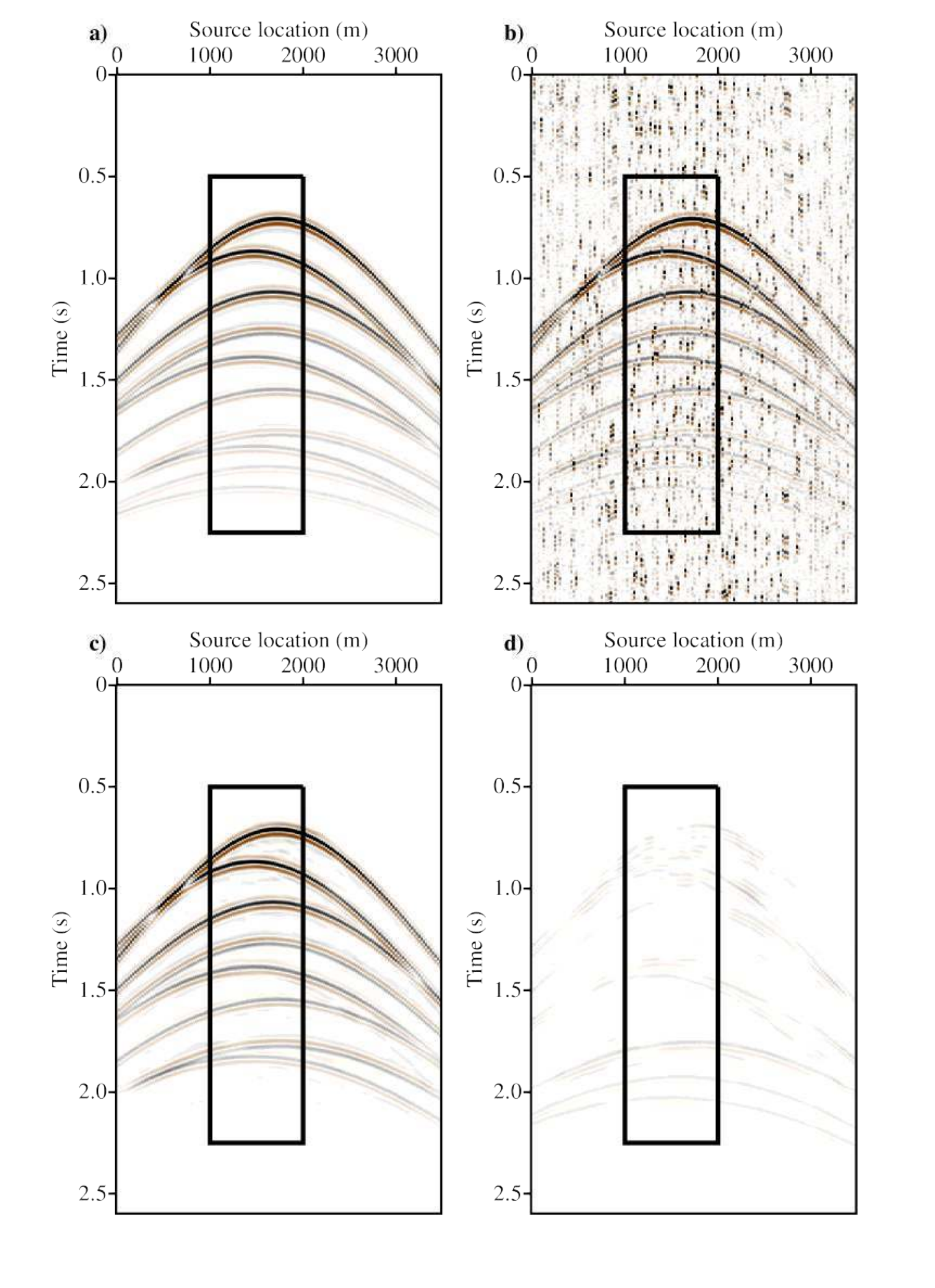}
	\reduceVspace
	\caption{Synthetic data common receiver gather recovered from the Radon model estimated using $p=2$ and $q=1$ inversion.
	(a) Original gather.   (b) Pseudo-deblended gather.  (c) Recovered gather.   (d) Recovered gather error.}
	\label{ch5_synth_CRG_L2L1}
\end{figure}

\begin{figure}[htp] 
	\centering
	\includegraphics{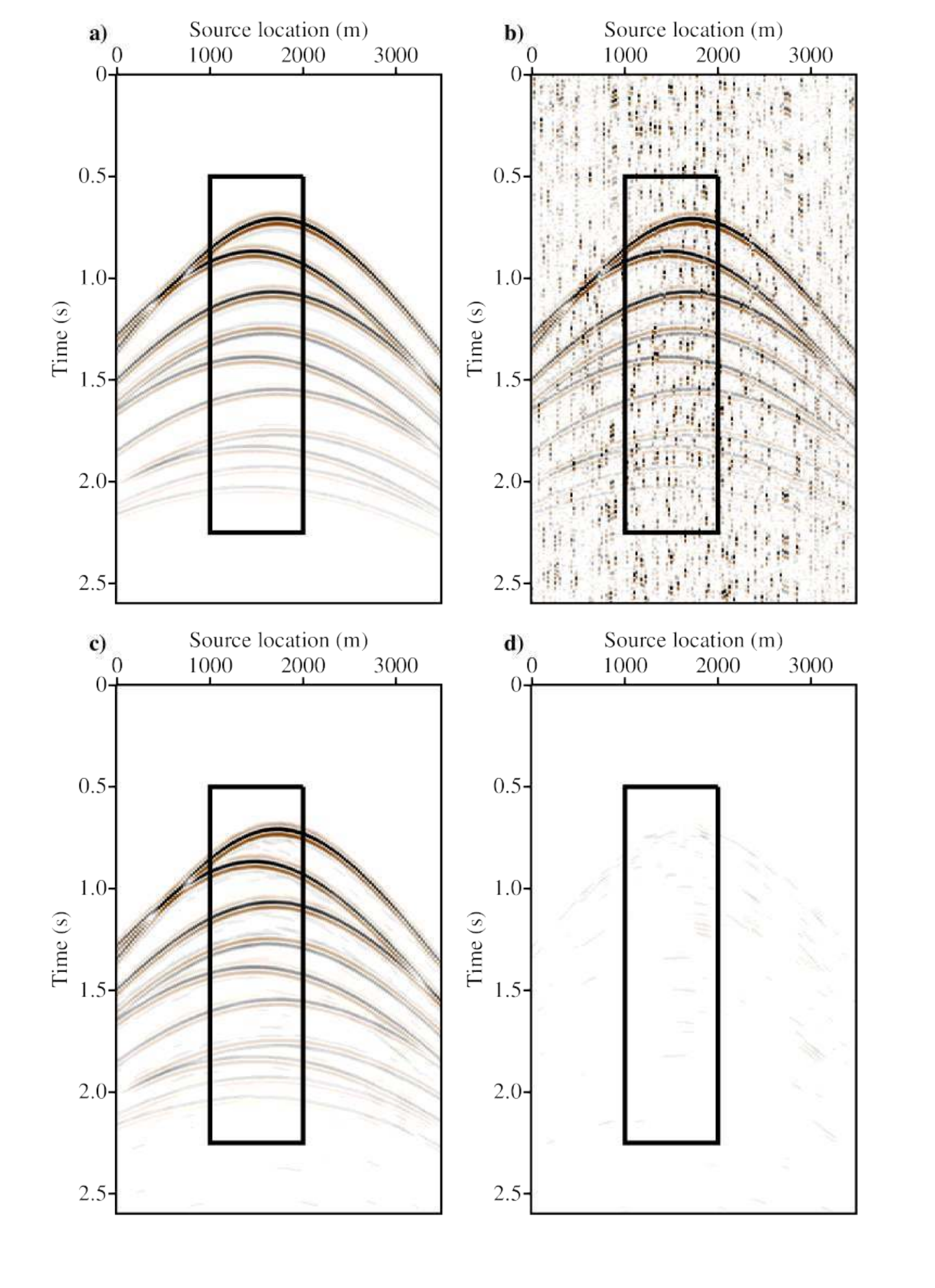}
	\reduceVspace
	\caption{Synthetic data common receiver gather recovered from the Radon model estimated using $p=1$ and $q=2$ inversion.
	(a) Original gather.   (b) Pseudo-deblended gather.  (c) Recovered gather.   (d) Recovered gather error.}
	\label{ch5_synth_CRG_L1L2}
\end{figure}

\begin{figure}[htp] 
	\centering
	\includegraphics{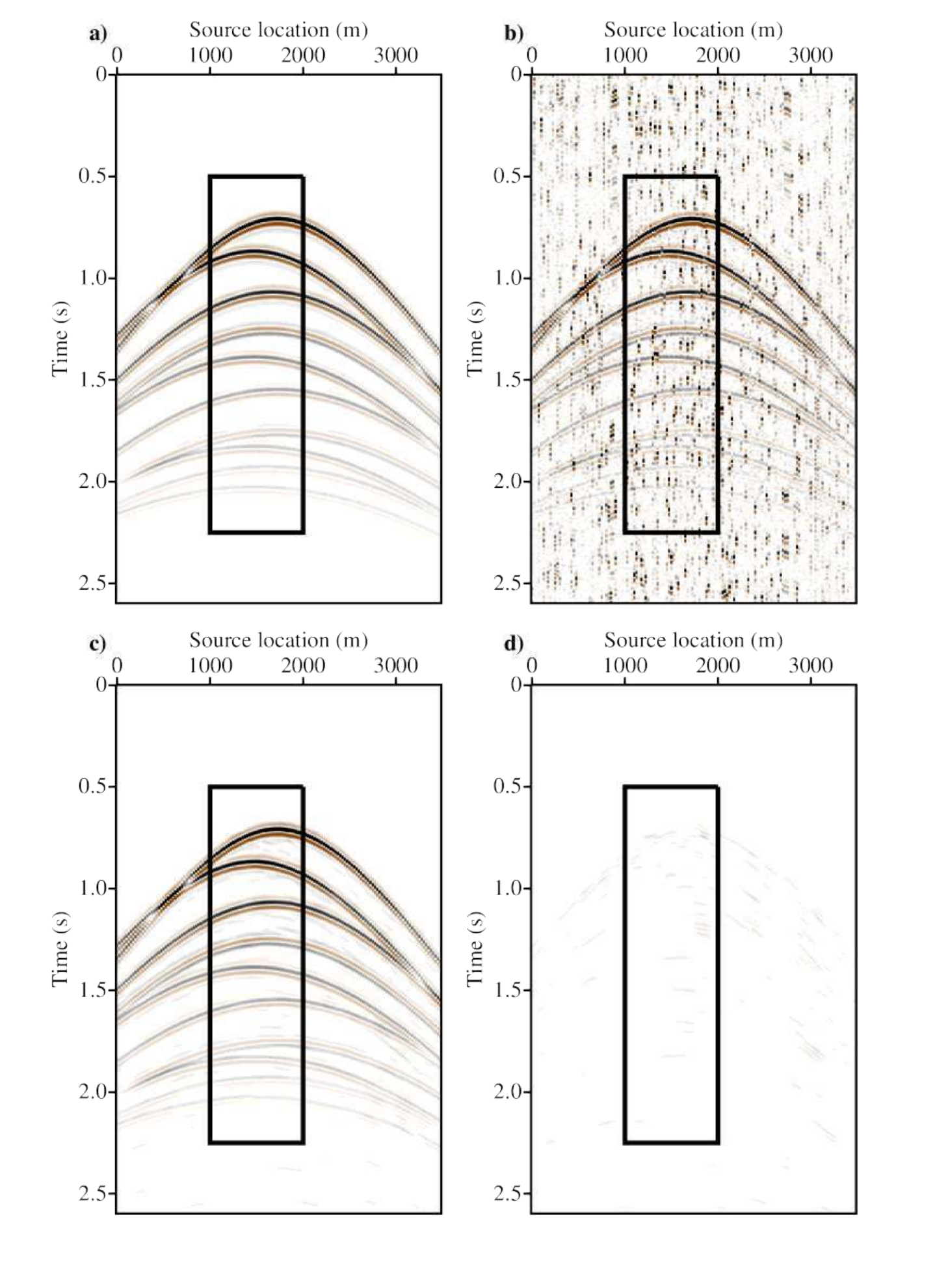}
	\reduceVspace
	\caption{Synthetic data common receiver gather recovered from the Radon model estimated using $p=1$ and $q=1$ inversion.
	(a) Original gather.   (b) Pseudo-deblended gather.  (c) Recovered gather.   (d) Recovered gather error.}
	\label{ch5_synth_CRG_L1L1}
\end{figure}

In order to evaluate the efficiency of removing source interference without distorting signals, the four estimated Radon models were used to recover the de-noised common receiver gathers.
Figures \ref{ch5_synth_CRG_L2L2}, \ref{ch5_synth_CRG_L2L1}, \ref{ch5_synth_CRG_L1L2} and \ref{ch5_synth_CRG_L1L1} show data recovered from 
Radon models estimated using ($p=2,q=2$), ($p=2,q=1$), ($p=1,q=2$) and ($p=1,q=1$) inversion, respectively. 
These figures clearly shows that the two robust Radon transforms using ($p=1,q=2$) and ($p=1,q=1$) inversion were able to attenuate interference while preserving the signals better than the non-robust transforms using ($p=2,q=2$) and ($p=2,q=1$). 


\begin{figure}[htp] 
	\centering
	\includegraphics{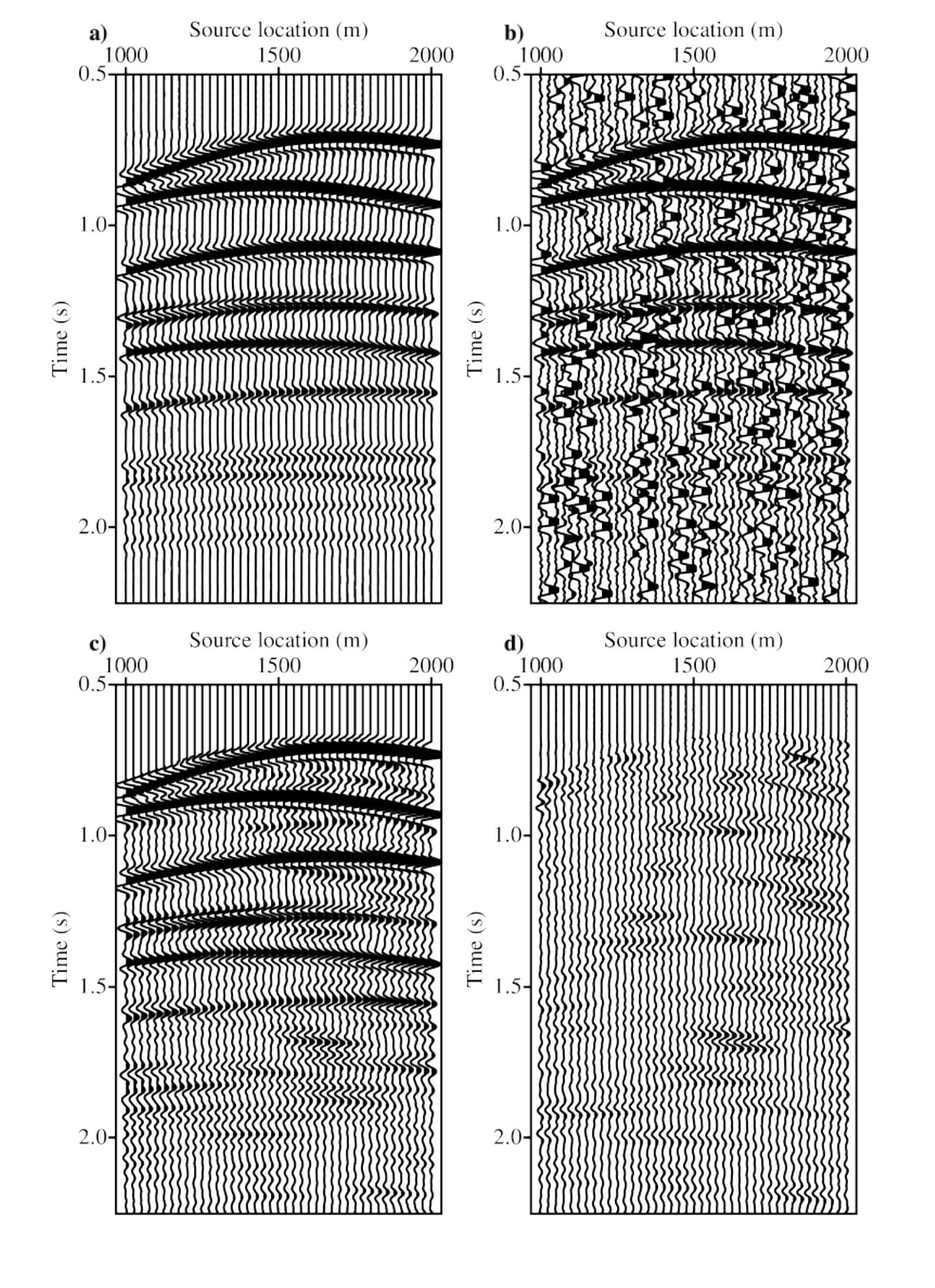}
	\reduceVspace
	\caption{Close-up of the synthetic data common receiver gather recovered from the Radon model estimated using $p=2$ and $q=2$ inversion.
	(a) Original gather.   (b) Pseudo-deblended gather.  (c) Recovered gather.   (d) Recovered gather error.}
	\label{ch5_synth_CRG_L2L2_closeup}
\end{figure}

\begin{figure}[htp] 
	\centering
	\includegraphics{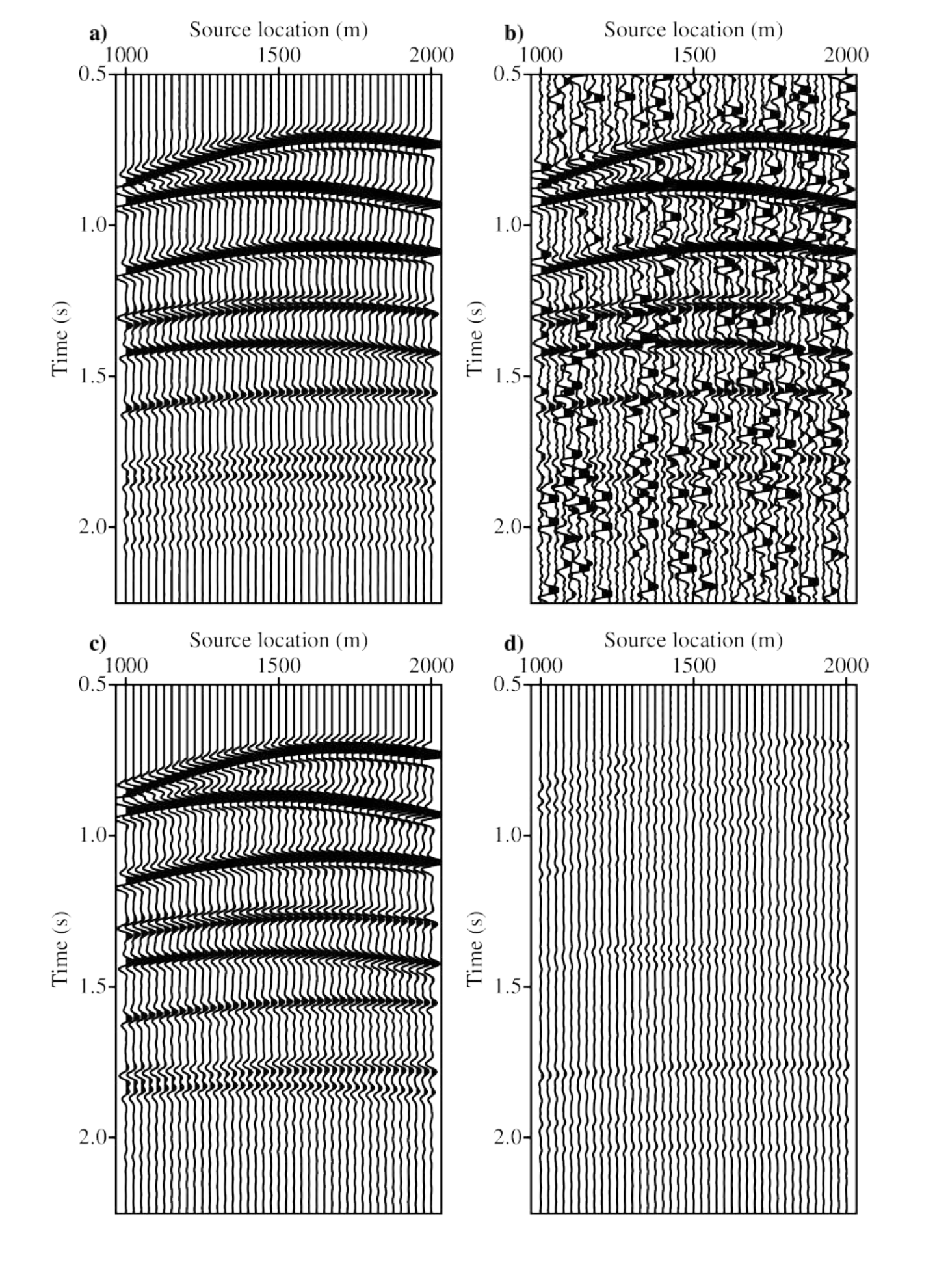}
	\reduceVspace
	\caption{Close-up of the synthetic data common receiver gather recovered from the Radon model estimated using $p=2$ and $q=1$ inversion.
	(a) Original gather.   (b) Pseudo-deblended gather.  (c) Recovered gather.   (d) Recovered gather error.}
	\label{ch5_synth_CRG_L2L1_closeup}
\end{figure}

\begin{figure}[htp] 
	\centering
	\includegraphics{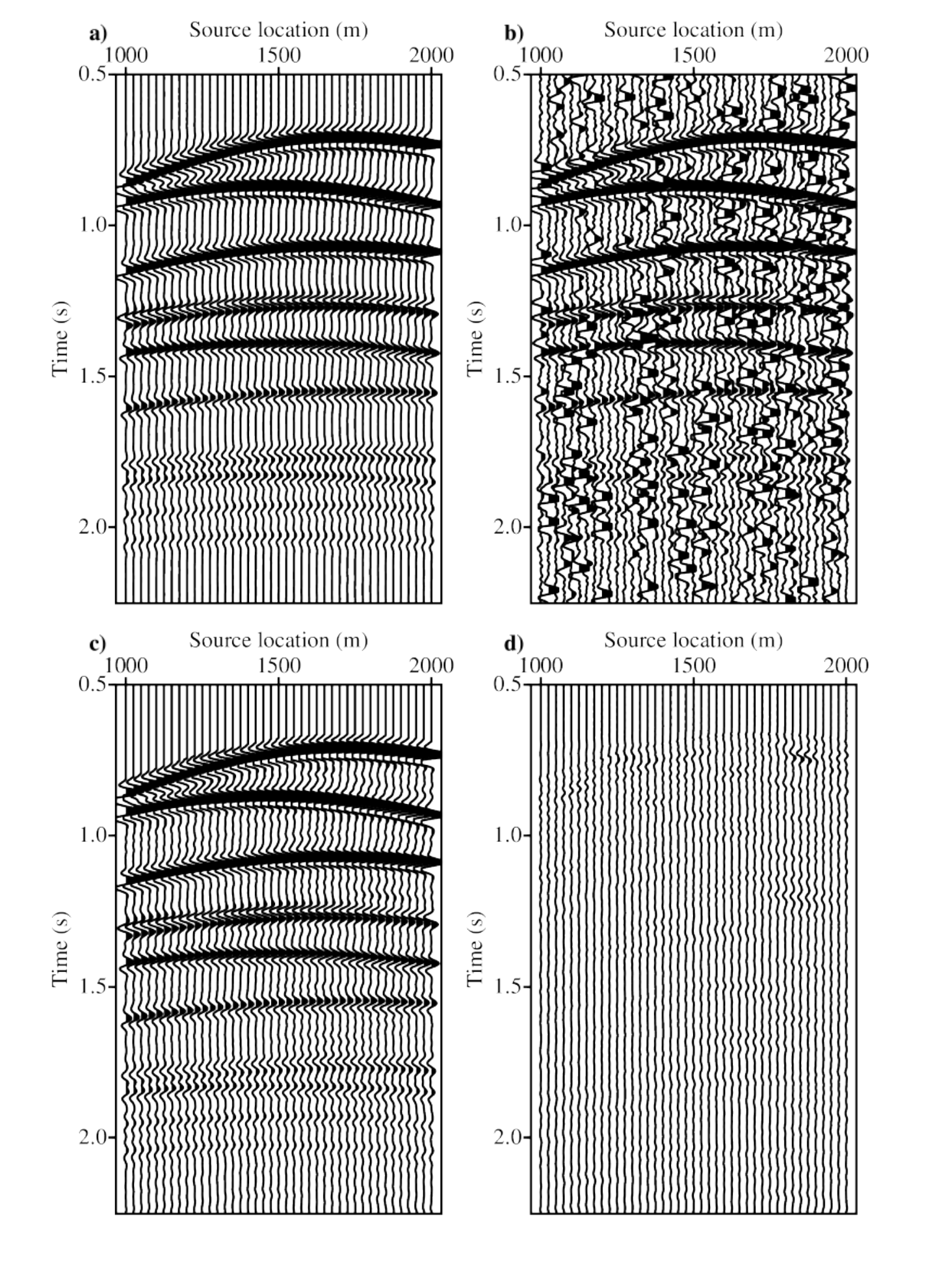}
	\reduceVspace
	\caption{Close-up of the synthetic data common receiver gather recovered from the Radon model estimated using $p=1$ and $q=2$ inversion.
	(a) Original gather.   (b) Pseudo-deblended gather.  (c) Recovered gather.   (d) Recovered gather error.}
	\label{ch5_synth_CRG_L1L2_closeup}
\end{figure}

\begin{figure}[htp] 
	\centering
	\includegraphics{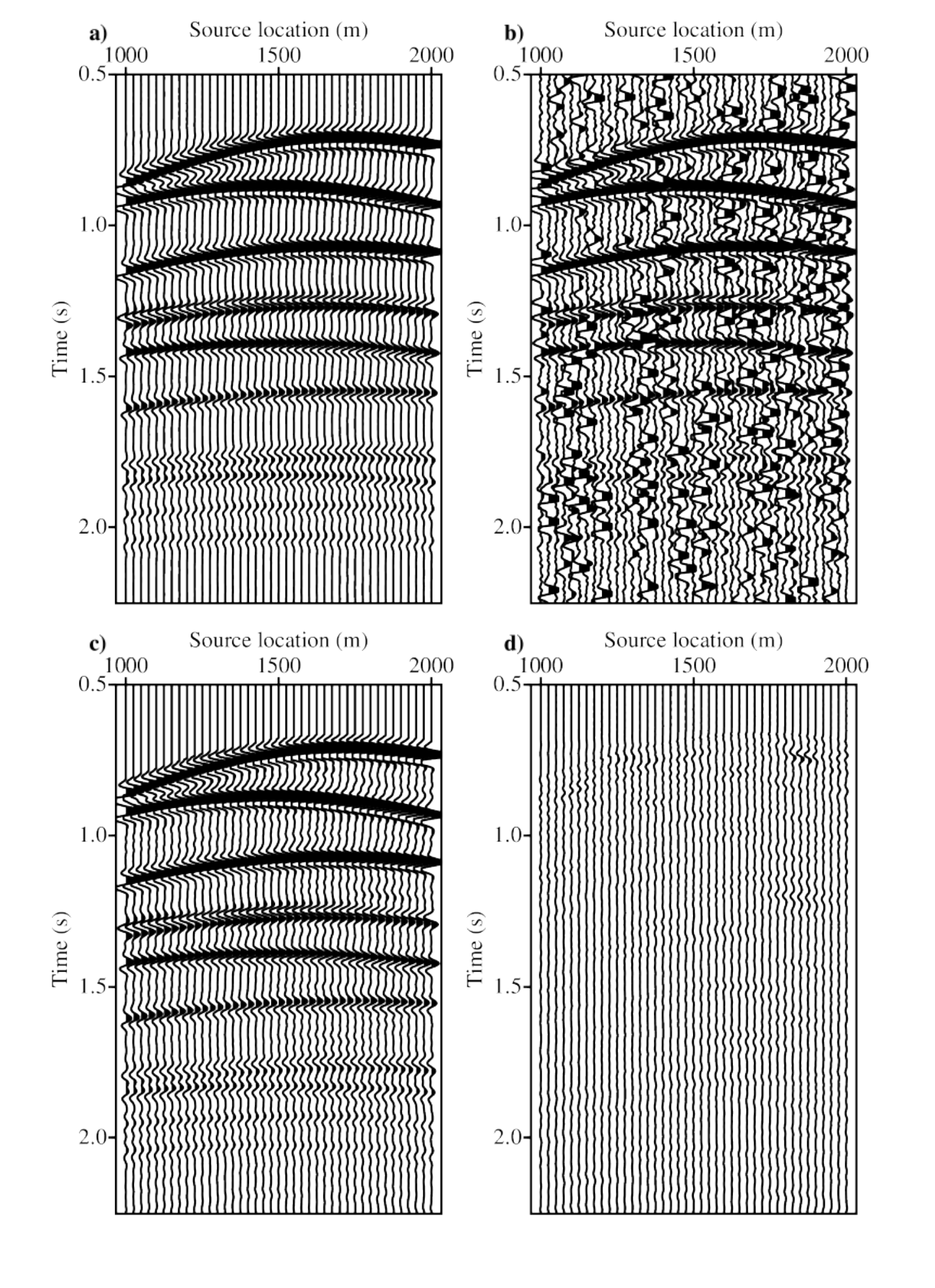}
	\reduceVspace
	\caption{Close-up of the synthetic data common receiver gather recovered from the Radon model estimated using $p=1$ and $q=1$ inversion.
	(a) Original gather.   (b) Pseudo-deblended gather.  (c) Recovered gather.   (d) Recovered gather error.}
	\label{ch5_synth_CRG_L1L1_closeup}
\end{figure}

\begin{figure}[htp] 
	\centering
	\includegraphics{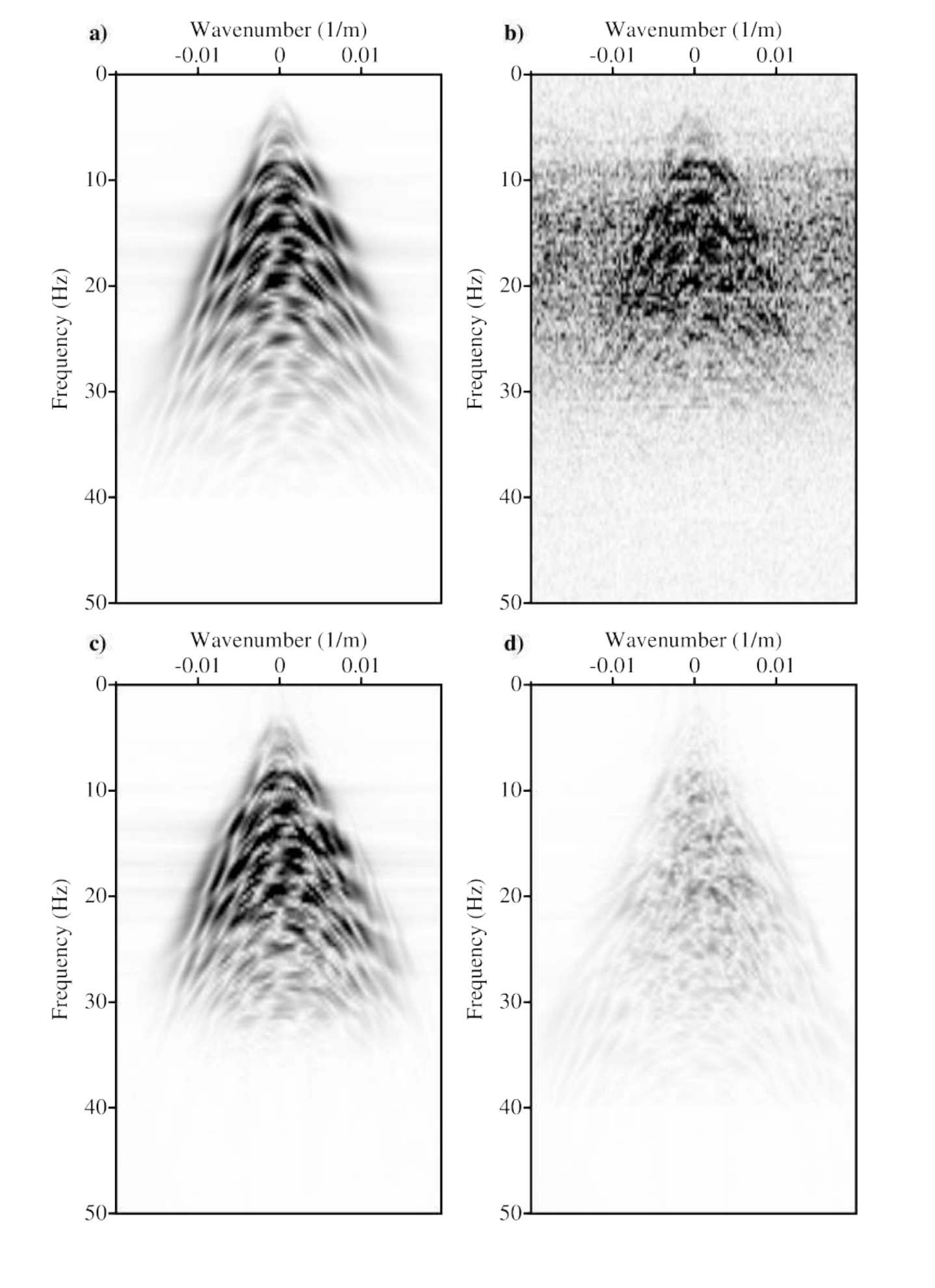}
	\reduceVspace
	\caption{The $f-k$ spectra of the synthetic data common receiver gather recovered from the Radon model estimated using $p=2$ and $q=2$ inversion.
	(a) Original gather.   
	(b) Pseudo-deblended gather .  
	(c) Recovered gather.   
	(d) Error of recovered gather.}
	\label{ch5_synth_CRG_L2L2_fk}
\end{figure}

\begin{figure}[htp] 
	\centering
	\includegraphics{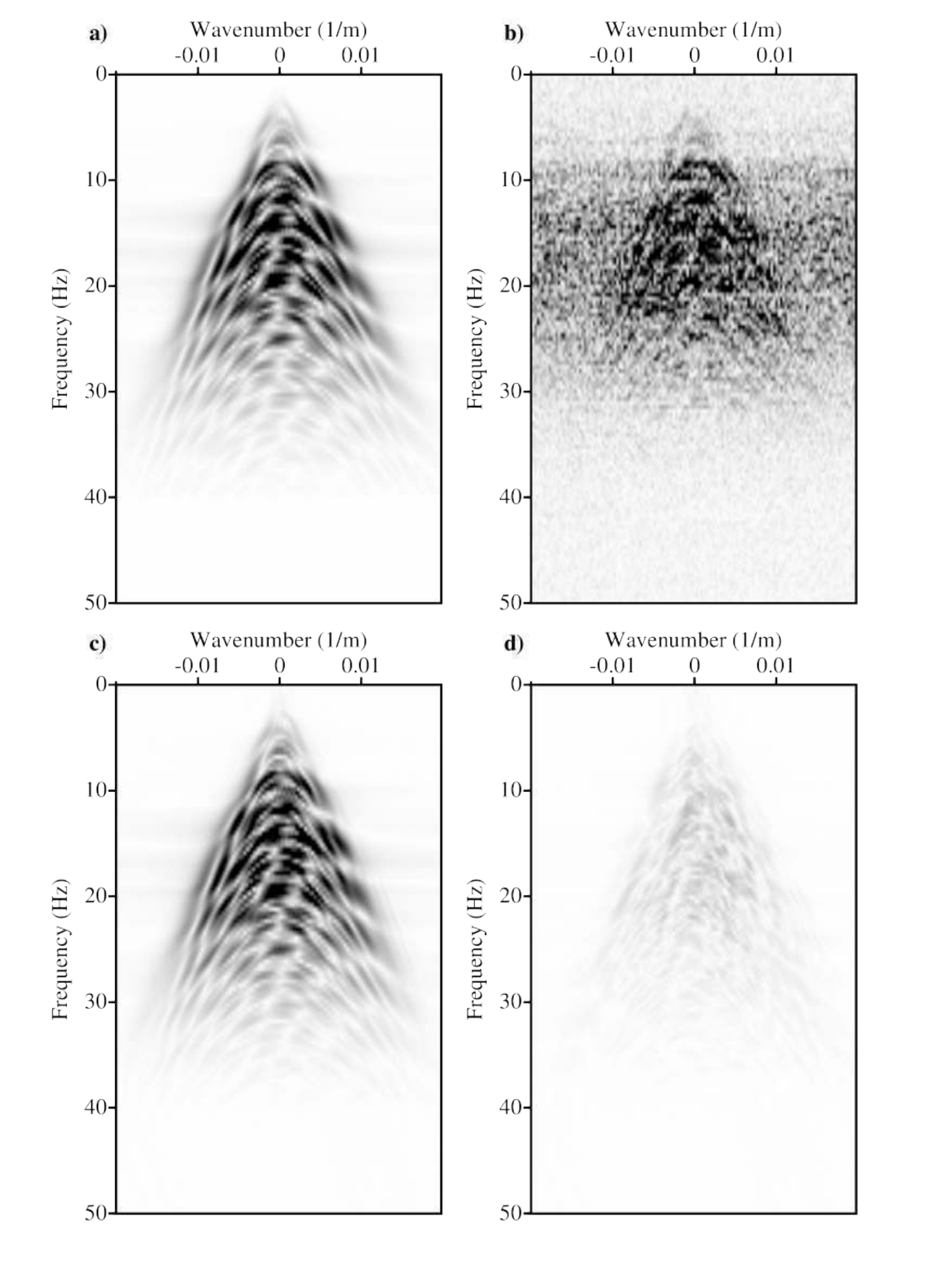}
	\reduceVspace
	\caption{The $f-k$ spectra of the synthetic data common receiver gather recovered from the Radon model estimated using $p=2$ and $q=1$ inversion.
	(a) Original gather.   
	(b) Pseudo-deblended gather .  
	(c) Recovered gather.   
	(d) Error of recovered gather.}
	\label{ch5_synth_CRG_L2L1_fk}
\end{figure}

\begin{figure}[htp] 
	\centering
	\includegraphics{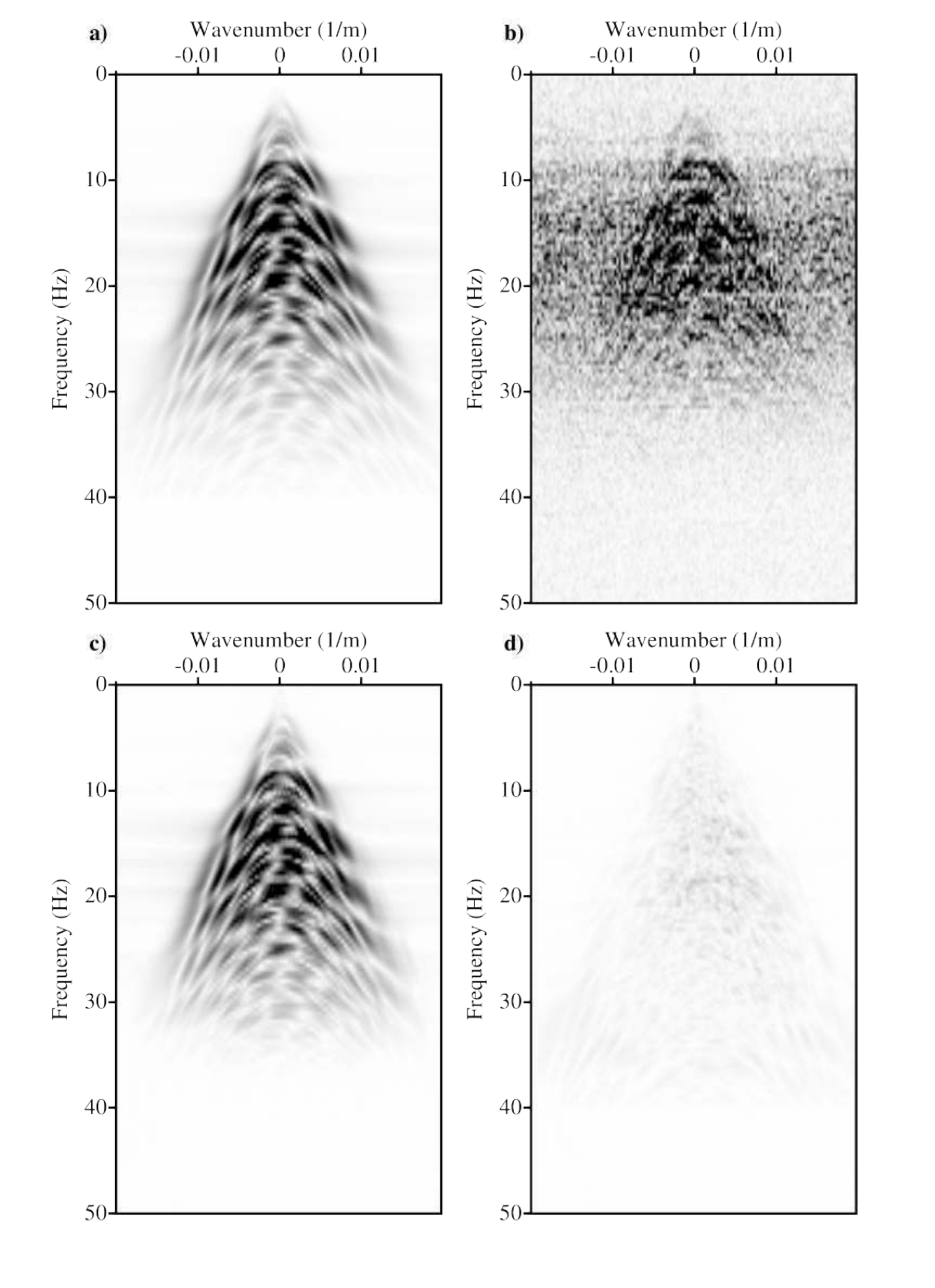}
	\reduceVspace
	\caption{The $f-k$ spectra of the synthetic data common receiver gather recovered from the Radon model estimated using $p=1$ and $q=2$ inversion.
	(a) Original gather.   
	(b) Pseudo-deblended gather .  
	(c) Recovered gather.   
	(d) Error of recovered gather.}
	\label{ch5_synth_CRG_L1L2_fk}
\end{figure}

\begin{figure}[htp] 
	\centering
	\includegraphics{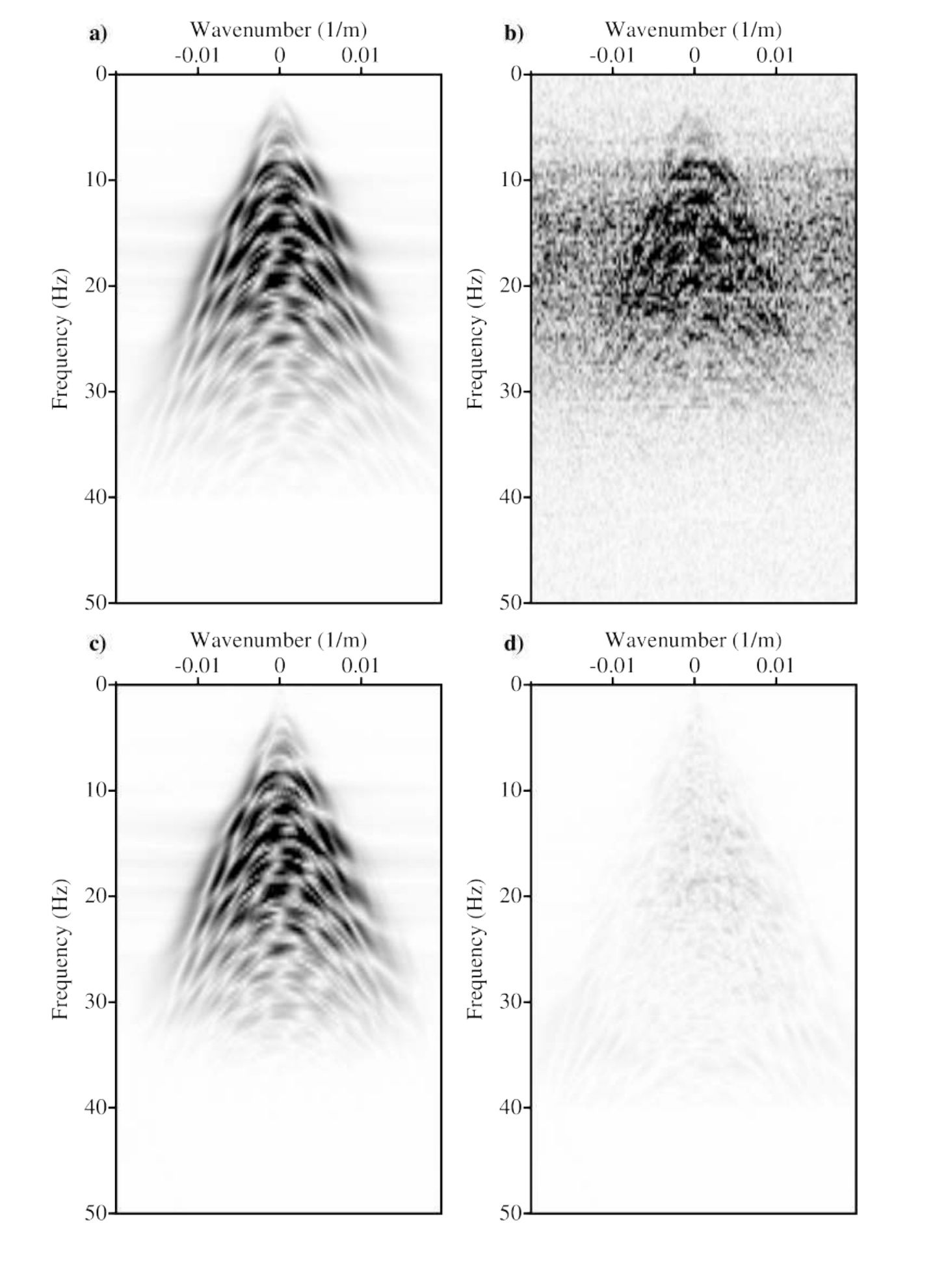}
	\reduceVspace
	\caption{The $f-k$ spectra of the synthetic data common receiver gather recovered from the Radon model estimated using $p=1$ and $q=1$ inversion.
	(a) Original gather.   
	(b) Pseudo-deblended gather .  
	(c) Recovered gather.   
	(d) Error of recovered gather.}
	\label{ch5_synth_CRG_L1L1_fk}
\end{figure}

In order to examine the results more closely, Figures \ref{ch5_synth_CRG_L2L2_closeup}, \ref{ch5_synth_CRG_L2L1_closeup}, \ref{ch5_synth_CRG_L1L2_closeup} and \ref{ch5_synth_CRG_L1L1_closeup} show close up of common receiver gathers recovered recovered from 
Radon models estimated using ($p=2,q=2$), ($p=2,q=1$), ($p=1,q=2$) and ($p=1,q=1$) inversion, respectively. 
The close-up of the data recovered from sparse Radon model ($p=2,q=1$)in Figure \ref{ch5_synth_CRG_L2L1_closeup} shows the difficulty of using sparsity to remove source interferences without losing the weak signals. 
On the other hand, using both sparsity and robustness ($p=1,q=1$) achieves better preservation of the weak signals as shown in Figure \ref{ch5_synth_CRG_L2L1_closeup}.   
The quality removing the incoherent source interference and preserving the coherent signals can also be evaluated in the $f-k$ spectra of the common receiver gathers. 
Figures \ref{ch5_synth_CRG_L2L2_fk}, \ref{ch5_synth_CRG_L2L1_fk}, \ref{ch5_synth_CRG_L1L2_fk} and \ref{ch5_synth_CRG_L1L1_fk} shows the $f-k$ spectra of common receiver gathers recovered from Radon models estimated using ($p=2,q=2$), ($p=2,q=1$), ($p=1,q=2$) and ($p=1,q=1$) inversion, respectively. 

\begin{figure}[htp] 
	\centering
	\includegraphics{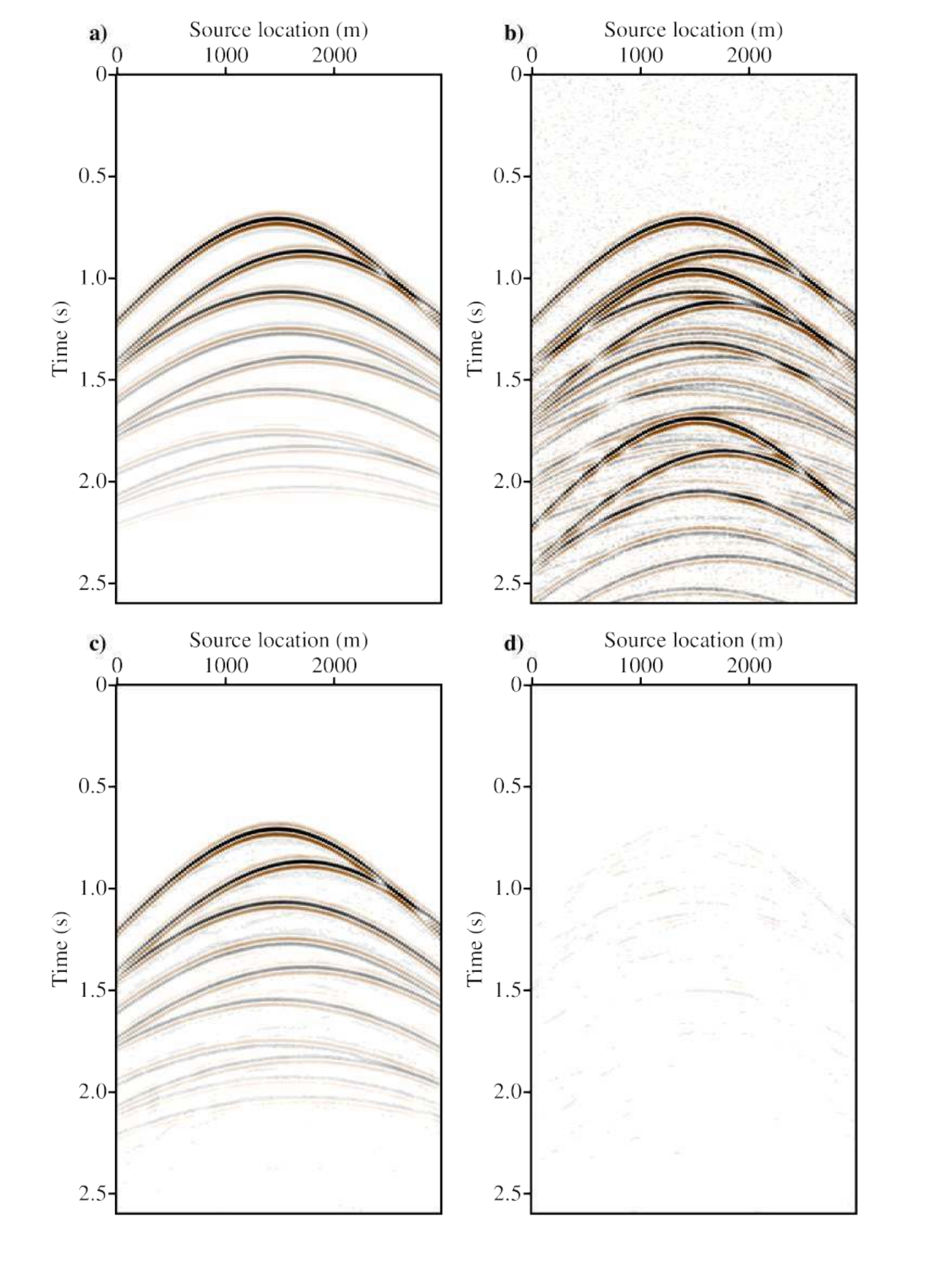}
	\reduceVspace
	\caption{Synthetic data common source gather recovered from the Radon model estimated using $p=1$ and $q=1$ inversion.
	(a) Original gather.   (b) Pseudo-deblended gather.  (c) Recovered gather.   (d) Recovered gather error.}
	\label{ch5_synth_CSG_L1L1}
\end{figure}

\begin{sidewaysfigure}[htp] 
 		\includegraphics{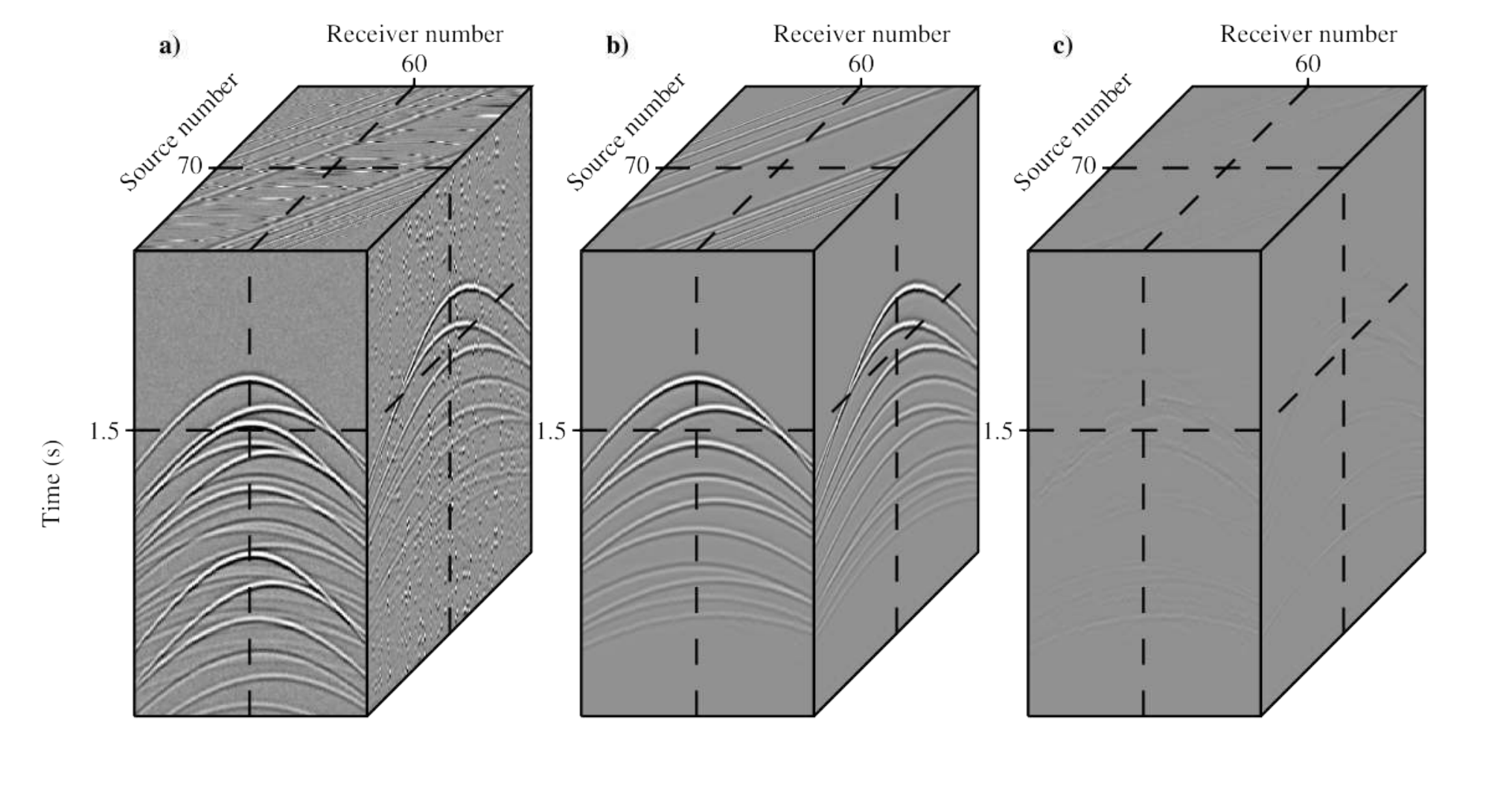}
 		\reduceVspace
 		\caption{Synthetic data example cubes. 
   		(a) Pseudo-deblended data cube. 
   		(b) Data cube recovered by forward modeling from the Stolt-based ASHRT model estimated via $p=1$ and $q=1$ inversion.
        (c) Difference between recovered and original data cubes. }
		\label{ch5_synth_cubes_L1L1}
\end{sidewaysfigure}

The quality improvement using Radon transform is calculated again using the following expression
\begin{align}
Q =Q_{R} - Q_{PD}, 
\end{align}
where $Q_{R}$ is the quality of the data recovered from the estimated Radon model and $Q_{PD}$ is the quality of pseudo-deblended gather.  
The values of $Q_{PD}$ and $Q_{R}$ are calculated using the equations 4.14 and 4.15, respectively. 
The $Q$ values for the recovered synthetic data common receiver gathers are listed in Table \ref{table:ch5_quality}.
The highest $Q$ value for recovered synthetic data common receiver gather is $22.32$ dB using $p=1,q=1$. 
Figure \ref{ch5_synth_CSG_L1L1} shows the common source gather recovered after denoising all common receiver gathers using $p=1,q=1$ inversion.  
This figure shows that coherent interferences in the common source gather were effectively removed after denoising all common receiver gathers. 
Figure \ref{ch5_synth_cubes_L1L1} shows the data cubes recovered using Radon models estimated using $p=1,q=1$ inversion.   


\section{Field Data Example}

\begin{figure}[htp] 
	\centering
	\includegraphics{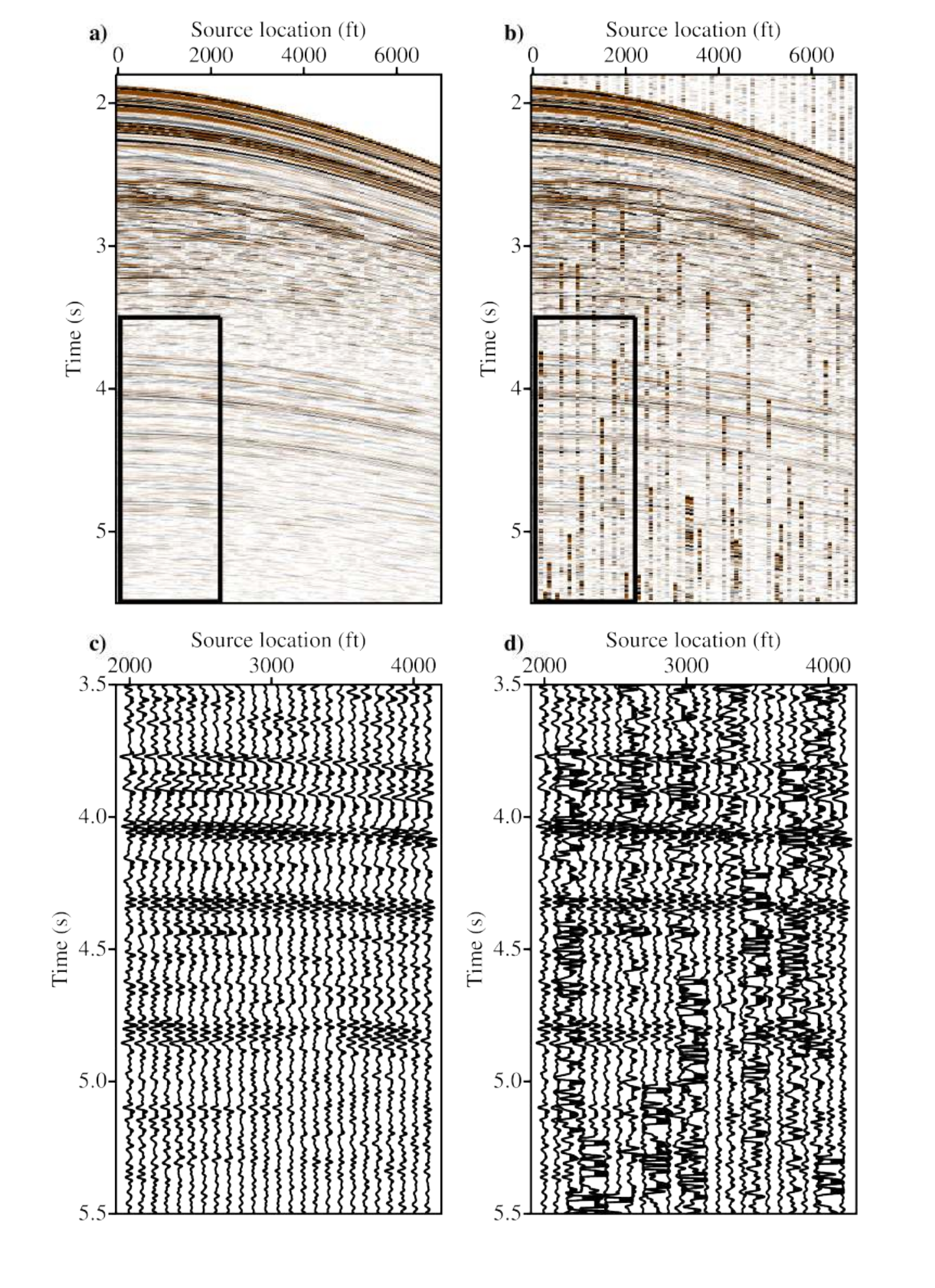}
	\reduceVspace
	\caption{A numerically blended common receiver gather from the Gulf of Mexico field data.
	(a) Original gather.   
	(b) Pseudo-deblended gather .  
	(c) Close up of the original gather.   
	(d) Close up of the pseudo-deblended gather. }
	\label{ch5_GOM_CRG_example}
\end{figure}

We also tested the four different Radon transforms using a numerically blended marine data from the Gulf of Mexico. The acquisition scenario represents a single source boat with the time interval between successive sources is nearly half of the conventional acquisition. 
In order to make the source interferences appear incoherent, the source firing times are dithered using random time delays.  
This example is the same that was used in chapter 4 and the source firing times are shown in Figure 4.23.
Figure \ref{ch5_GOM_CRG_example}a shows an original common receiver gather from the Gulf of Mexico data 
and Figure \ref{ch5_GOM_CRG_example}b shows the same gather after blending and pseudo-deblending.
Figures \ref{ch5_GOM_CRG_example}c and d show close up of the areas marked on Figures \ref{ch5_GOM_CRG_example}a and b, respectively. 
We have chosen a close up window between $3.5$  and $5.5$ seconds to show the effects of the strong source interferences mixing with weak signals. 

\begin{sidewaysfigure}[htbp]
 		\includegraphics{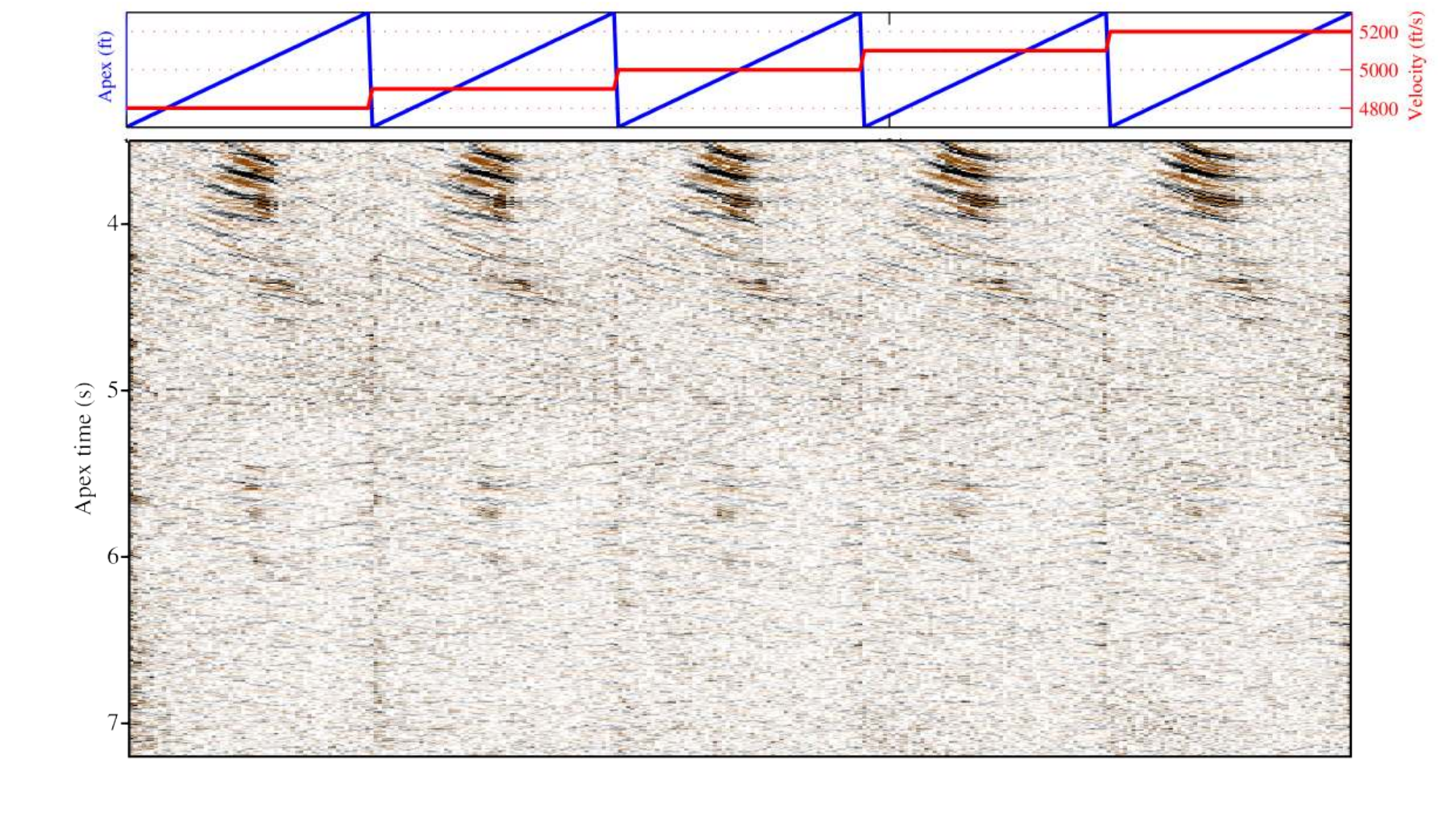}
 		\reduceVspace
 		\caption{ Stolt-based ASHRT model for one Gulf of Mexico common receiver gather estimated using $p=2$ and $q=2$ inversion.}
		\label{ch5_GOM_model_L2L2}
\end{sidewaysfigure}

\begin{sidewaysfigure}[htbp]
 		\includegraphics{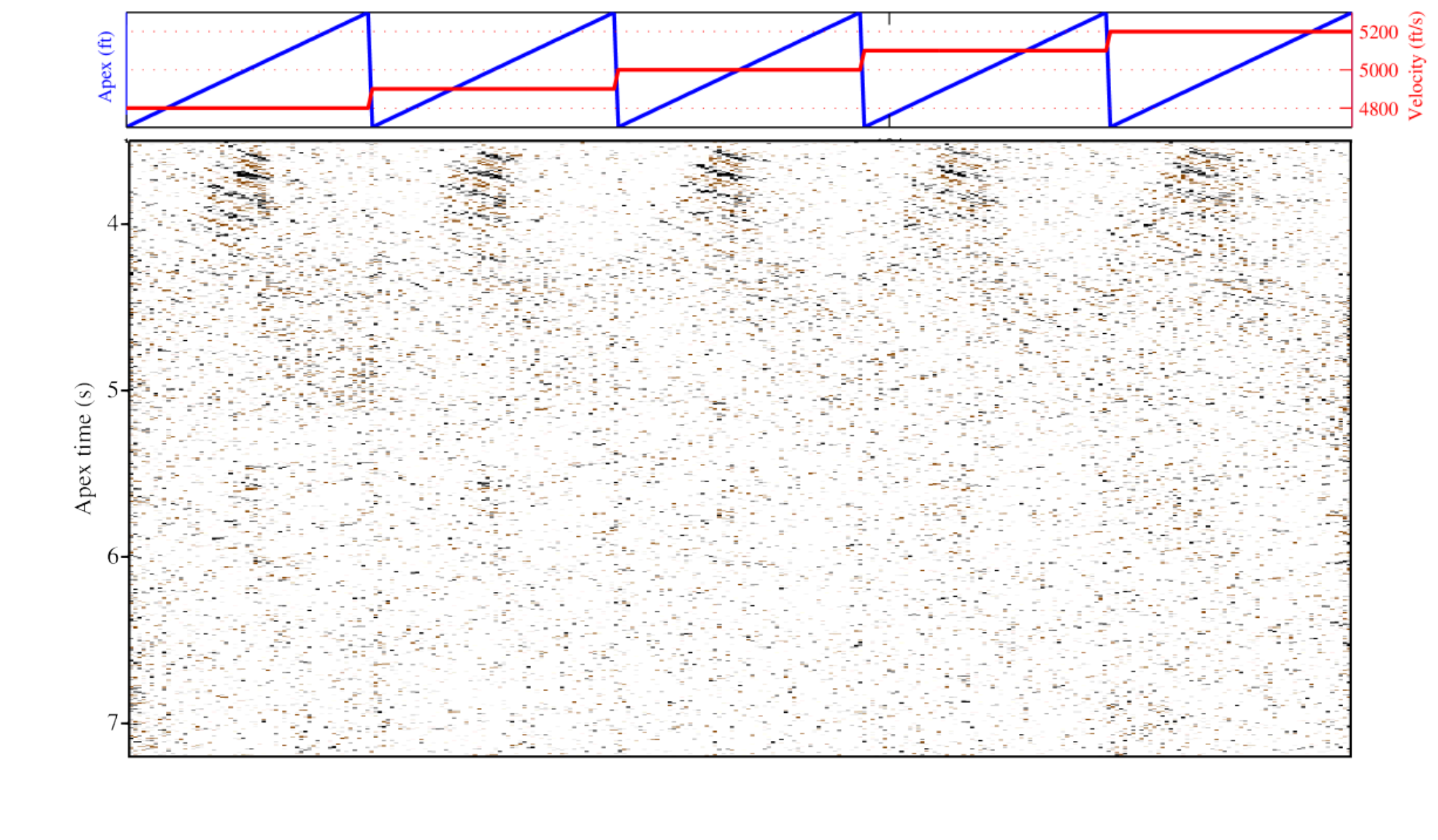}
 		\reduceVspace
 		\caption{ Stolt-based ASHRT model for one Gulf of Mexico common receiver gather estimated using $p=2$ and $q=1$ inversion.}
		\label{ch5_GOM_model_L2L1}
\end{sidewaysfigure}

\begin{sidewaysfigure}[htbp]
 		\includegraphics{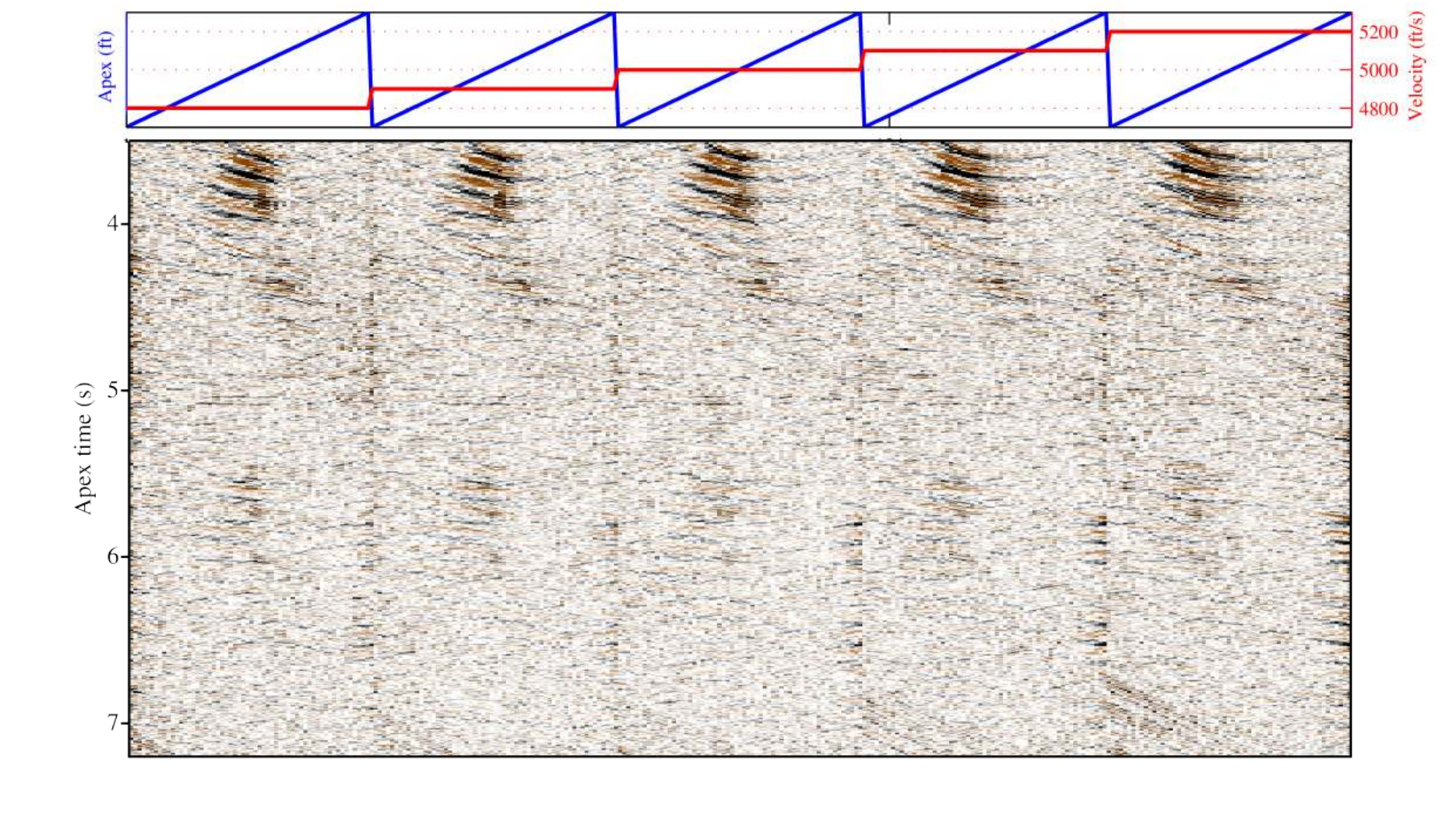}
 		\reduceVspace
 		\caption{ Stolt-based ASHRT model for one Gulf of Mexico common receiver gather estimated using $p=1$ and $q=2$ inversion.}
		\label{ch5_GOM_model_L1L2}
\end{sidewaysfigure}

\begin{sidewaysfigure}[htbp]
 		\includegraphics{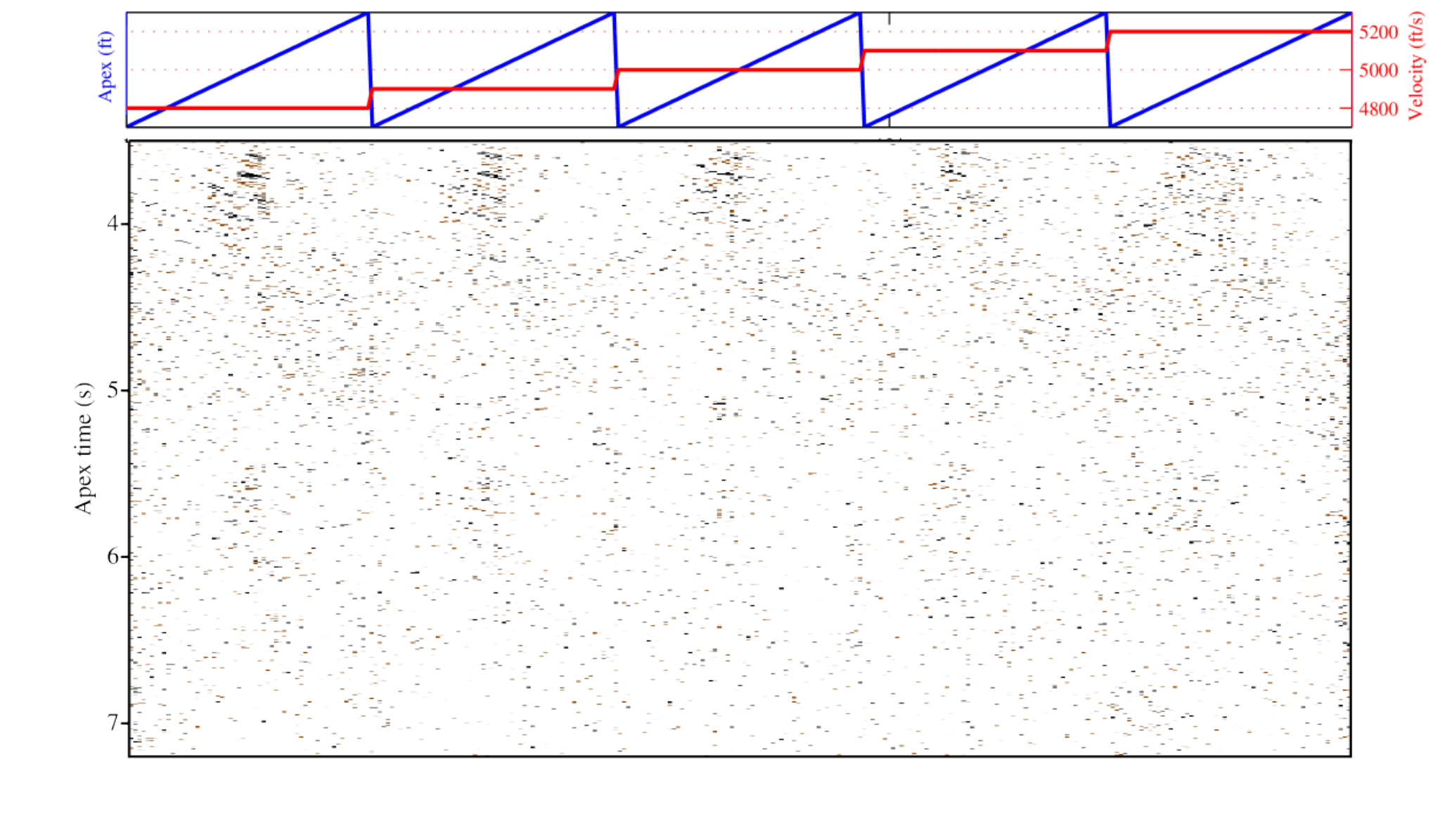}
 		\reduceVspace
 		\caption{ Stolt-based ASHRT model for one Gulf of Mexico common receiver gather estimated using $p=1$ and $q=1$ inversion.}
		\label{ch5_GOM_model_L1L1}
\end{sidewaysfigure}

Four different Radon models for the pseudo-deblended common receiver gather in Figure 4.24b are estimated using four different inversion scenarios.
The ASHRT Radon models for ($p=2,q=2$), ($p=2,q=1$), ($p=1,q=2$) and ($p=1,q=1$) inversions are shown in Figures \ref{ch5_GOM_model_L2L2}, \ref{ch5_GOM_model_L2L1}, \ref{ch5_GOM_model_L1L2} and \ref{ch5_GOM_model_L1L1}, respectively. 
The Radon transform scans five velocities  (4800, 4900, 5000, 5100, and 5200 ft/s) and 59 apexes from $2537.5$ ft to $-2537.5$ ft. 
Note that the same Radon parameters were used for the time domain ASHRT  shown in chapter 4 to facilitate comparisons. 
Note again that, this coarse sampling of the Radon parameters makes it difficult to impose a strict sparsity constraint on Radon model. 
For these reasons, the choice of optimal regularization parameters for robust inversion with sparse regularization $p=1,q=1$ is rather difficult. 

\begin{figure}[htbp]
 		\includegraphics{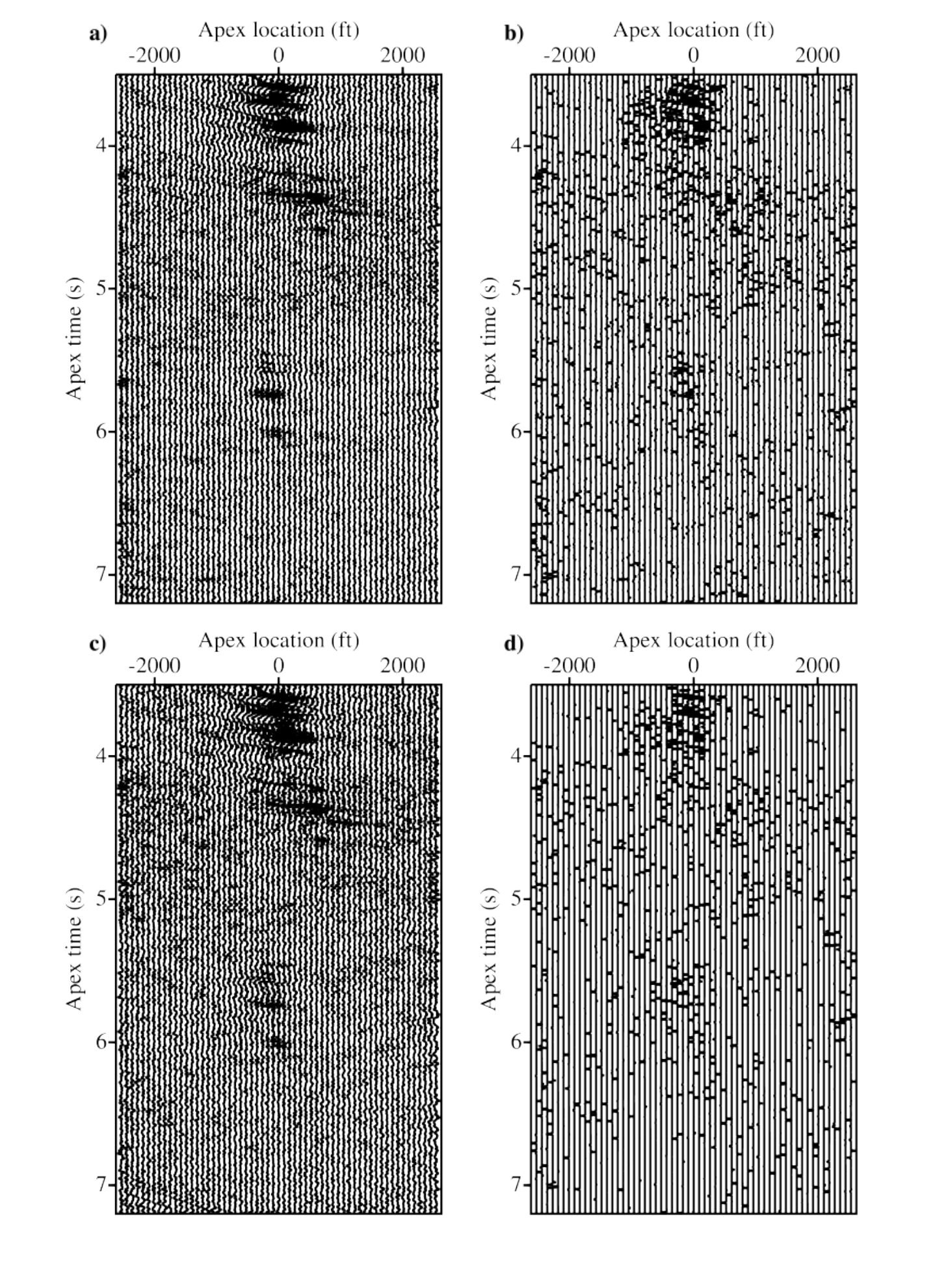}
 		\reduceVspace
   		\caption{One velocity panel ($v=4300ft/s$) of the Stolt-based ASHRT models for the Gulf of Mexico data estimated by inversion with 
	    (a) $p=2$ and $q=2$, (b) $p=2$ and $q=1$, (c) $p=1$ and $q=2$, and (d) $p=1$ and $q=1$. }
		\label{ch5_GOM_modeld_closeup}
\end{figure}

\begin{figure}[htp] 
	\centering
	\includegraphics{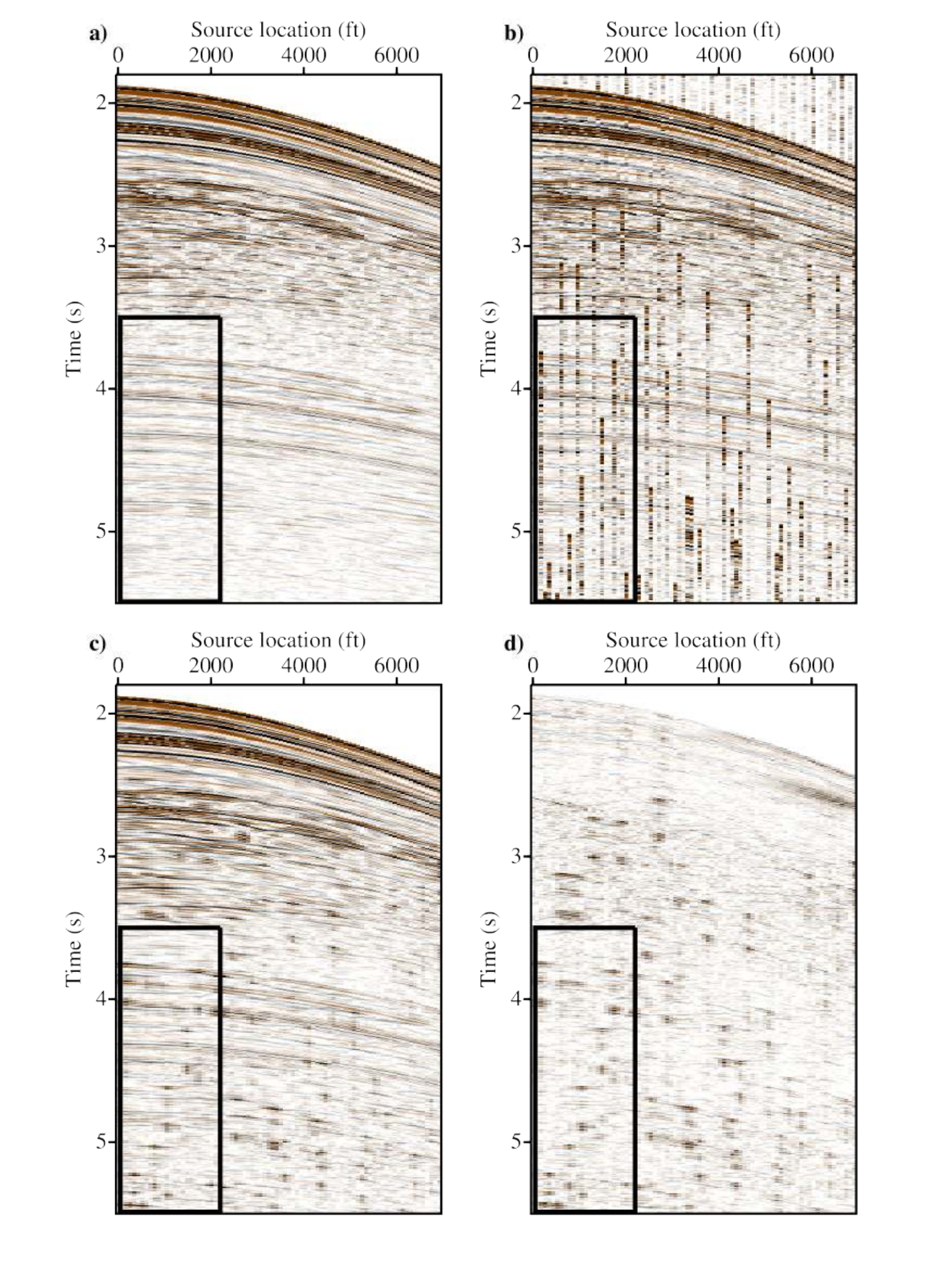}
	\reduceVspace
	\caption{Common receiver gather from the Gulf of Mexico data recovered from the Stolt-based ASHRT model estimated using $p=2$ and $q=2$ inversion.
	(a) Original gather.   (b) Pseudo-deblended gather.  (c) Recovered gather.   (d) Recovered gather error.}
	\label{ch5_GOM_CRG_L2L2}
\end{figure}

\begin{figure}[htp] 
	\centering
	\includegraphics{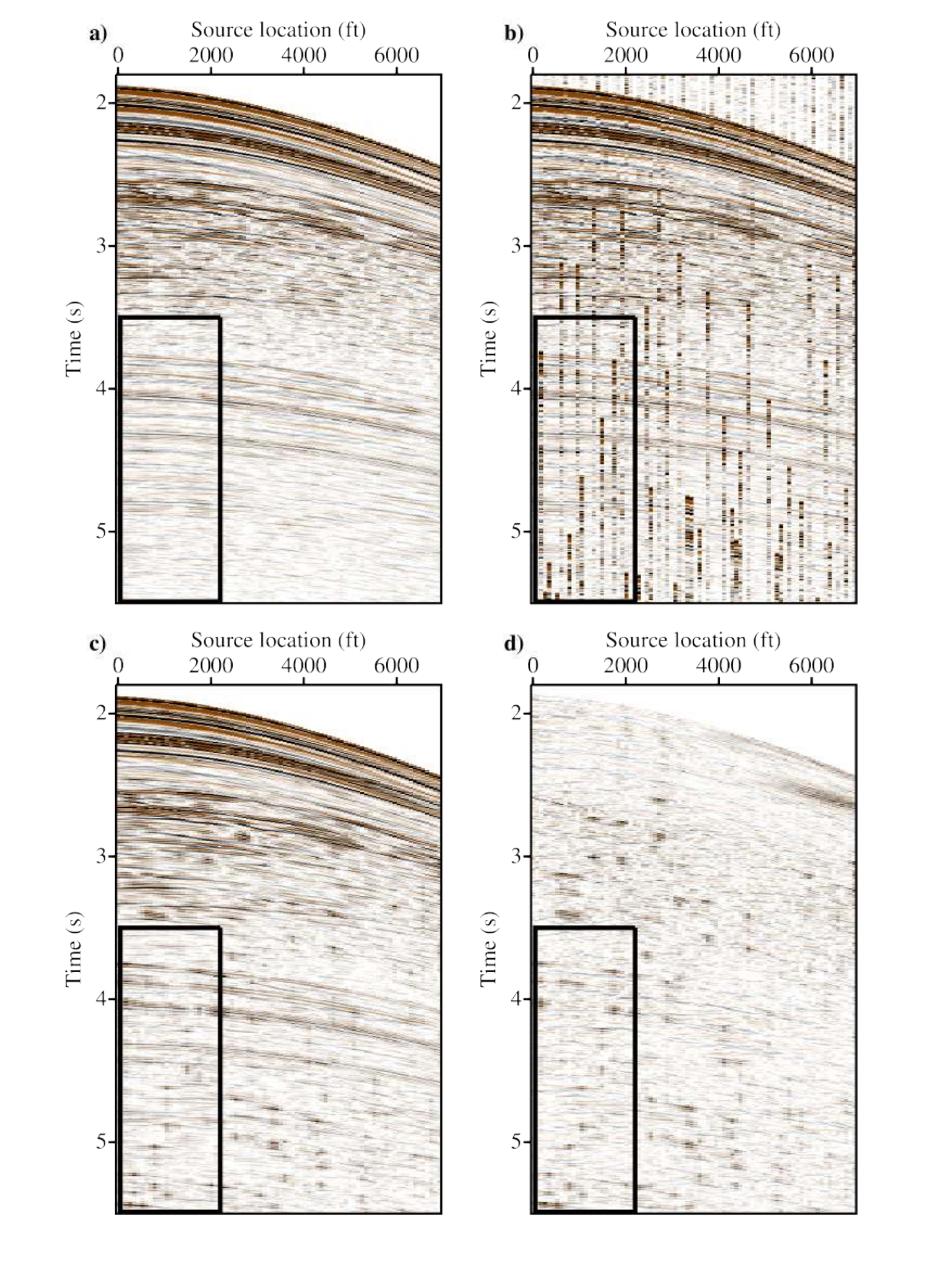}
	\reduceVspace
	\caption{Common receiver gather from the Gulf of Mexico data recovered from the Stolt-based ASHRT model estimated using $p=2$ and $q=1$ inversion.
	(a) Original gather.   (b) Pseudo-deblended gather.  (c) Recovered gather.   (d) Recovered gather error.}
	\label{ch5_GOM_CRG_L2L1}
\end{figure}

\begin{figure}[htp] 
	\centering
	\includegraphics{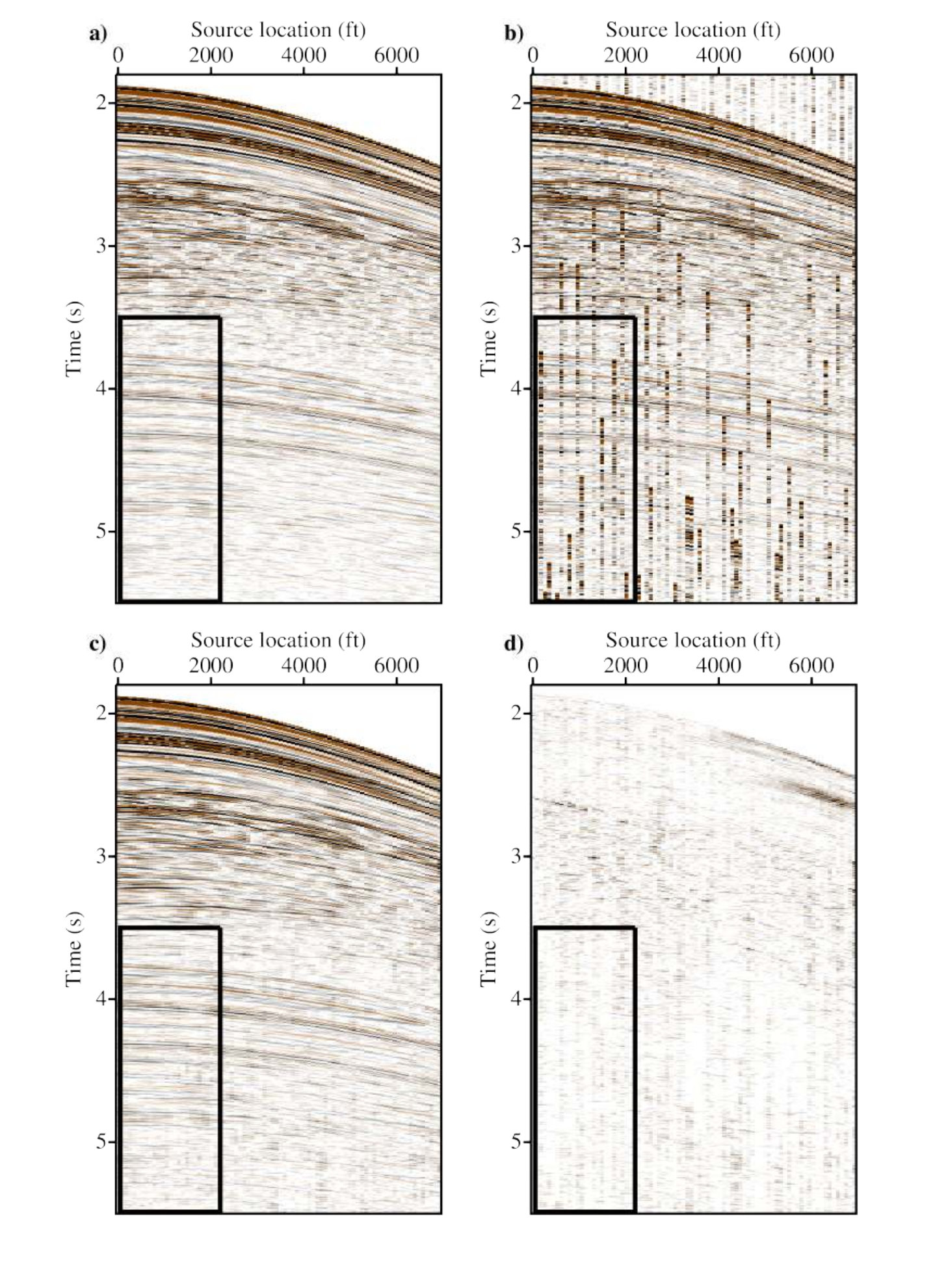}
	\reduceVspace
	\caption{Common receiver gather from the Gulf of Mexico data recovered from the Stolt-based ASHRT model estimated using $p=1$ and $q=2$ inversion.
	(a) Original gather.   (b) Pseudo-deblended gather.  (c) Recovered gather.   (d) Recovered gather error.}
	\label{ch5_GOM_CRG_L1L2}
\end{figure}

\begin{figure}[htp] 
	\centering
	\includegraphics{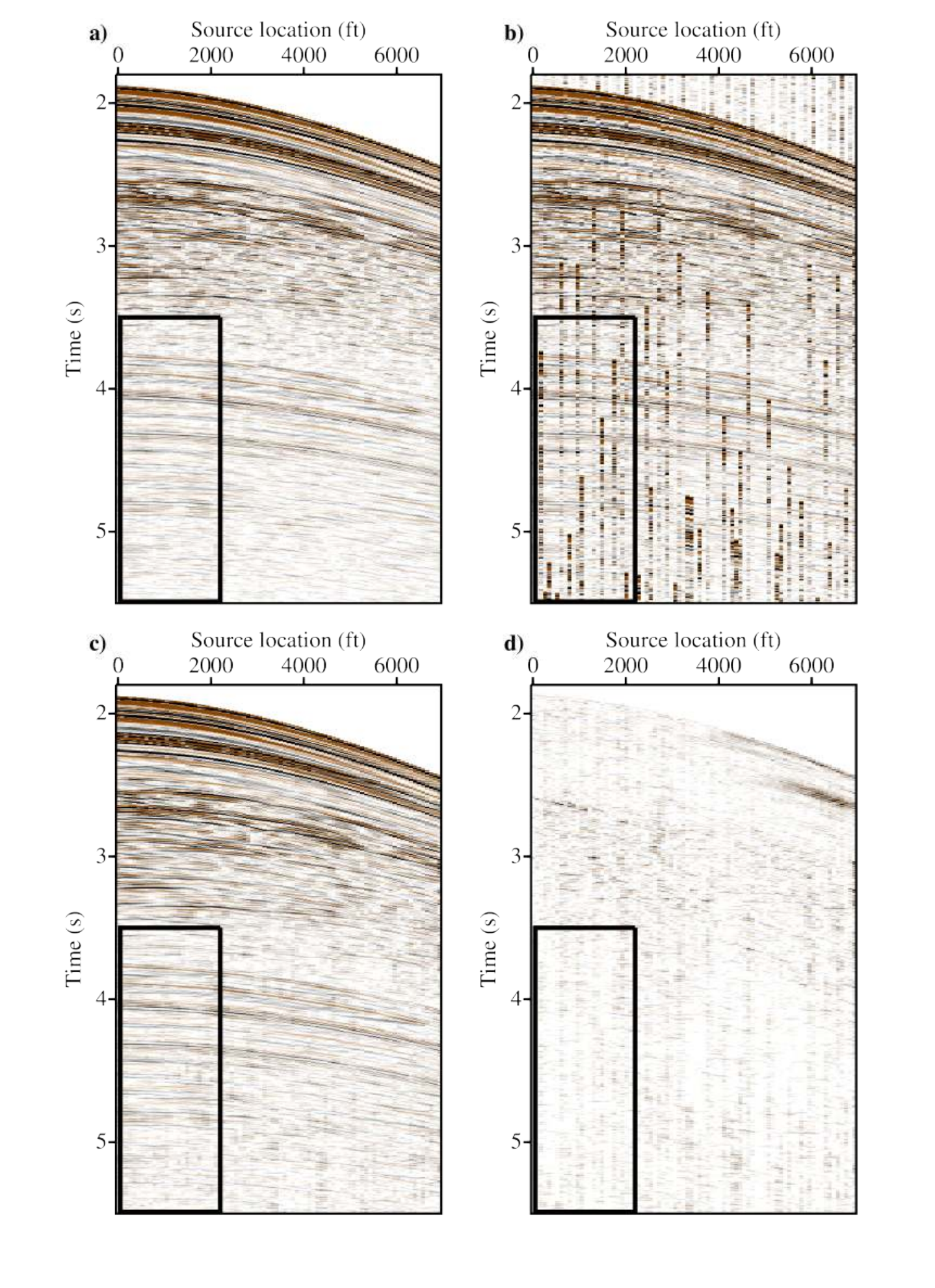}
	\reduceVspace
	\caption{Common receiver gather from the Gulf of Mexico data recovered from the Stolt-based ASHRT model estimated using $p=1$ and $q=1$ inversion.
	(a) Original gather.   (b) Pseudo-deblended gather.  (c) Recovered gather.   (d) Recovered gather error.}
	\label{ch5_GOM_CRG_L1L1}
\end{figure}

\begin{figure}[htp] 
	\centering
	\includegraphics{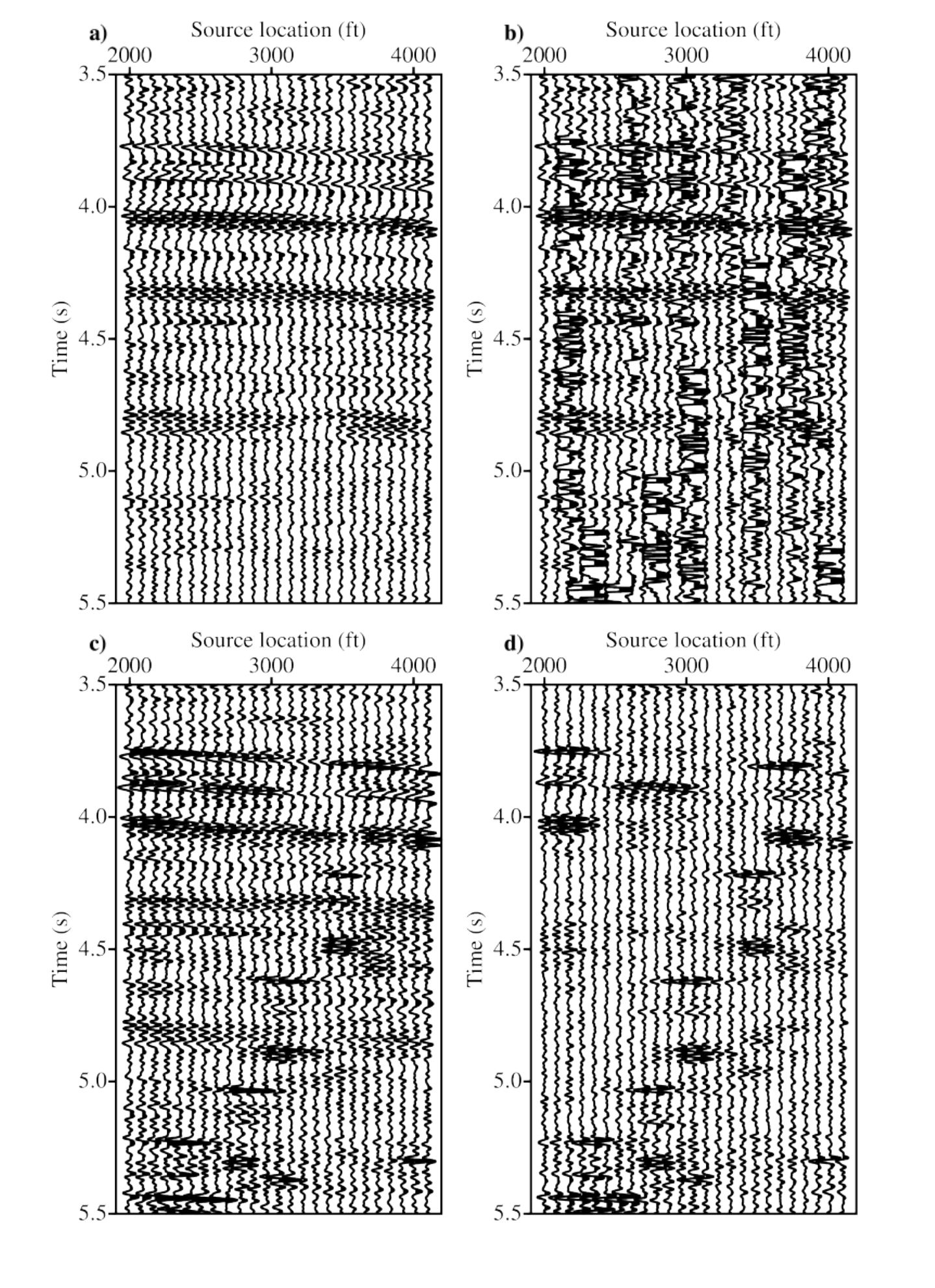}
	\reduceVspace
	\caption{Close-up of the common receiver gather from the Gulf of Mexico data recovered from the Stolt-based ASHRT model estimated using $p=2$ and $q=2$ inversion.
	(a) Original gather.   (b) Pseudo-deblended gather.  (c) Recovered gather.   (d) Recovered gather error.}
	\label{ch5_GOM_CRG_L2L2_closeup}
\end{figure}

\begin{figure}[htp] 
	\centering
	\includegraphics{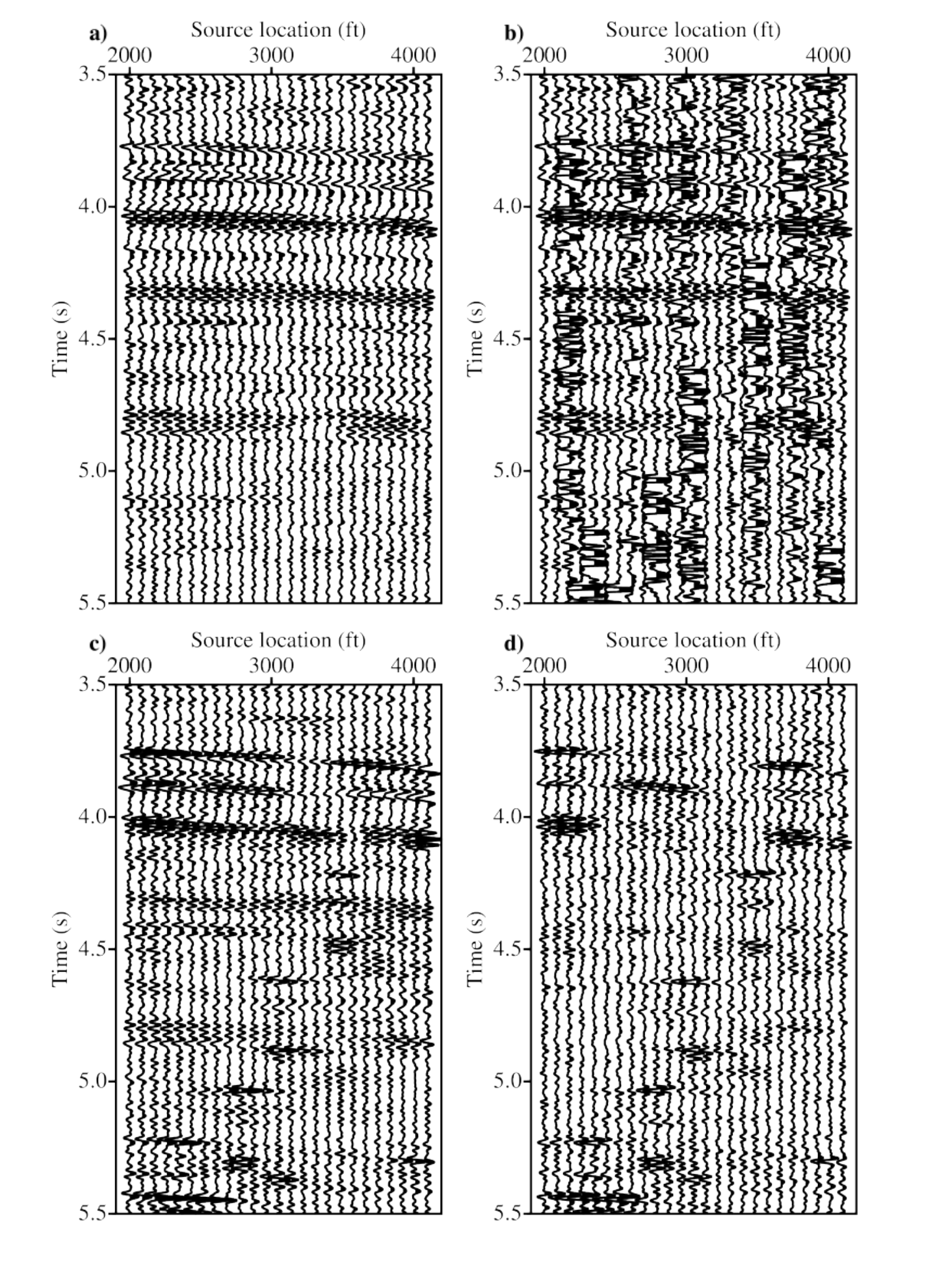}
	\reduceVspace
	\caption{Close-up of the common receiver gather from the Gulf of Mexico data recovered from the Stolt-based ASHRT model estimated using $p=2$ and $q=1$ inversion.
	(a) Original gather.   (b) Pseudo-deblended gather.  (c) Recovered gather.   (d) Recovered gather error.}
	\label{ch5_GOM_CRG_L2L1_closeup}
\end{figure}

\begin{figure}[htp] 
	\centering
	\includegraphics{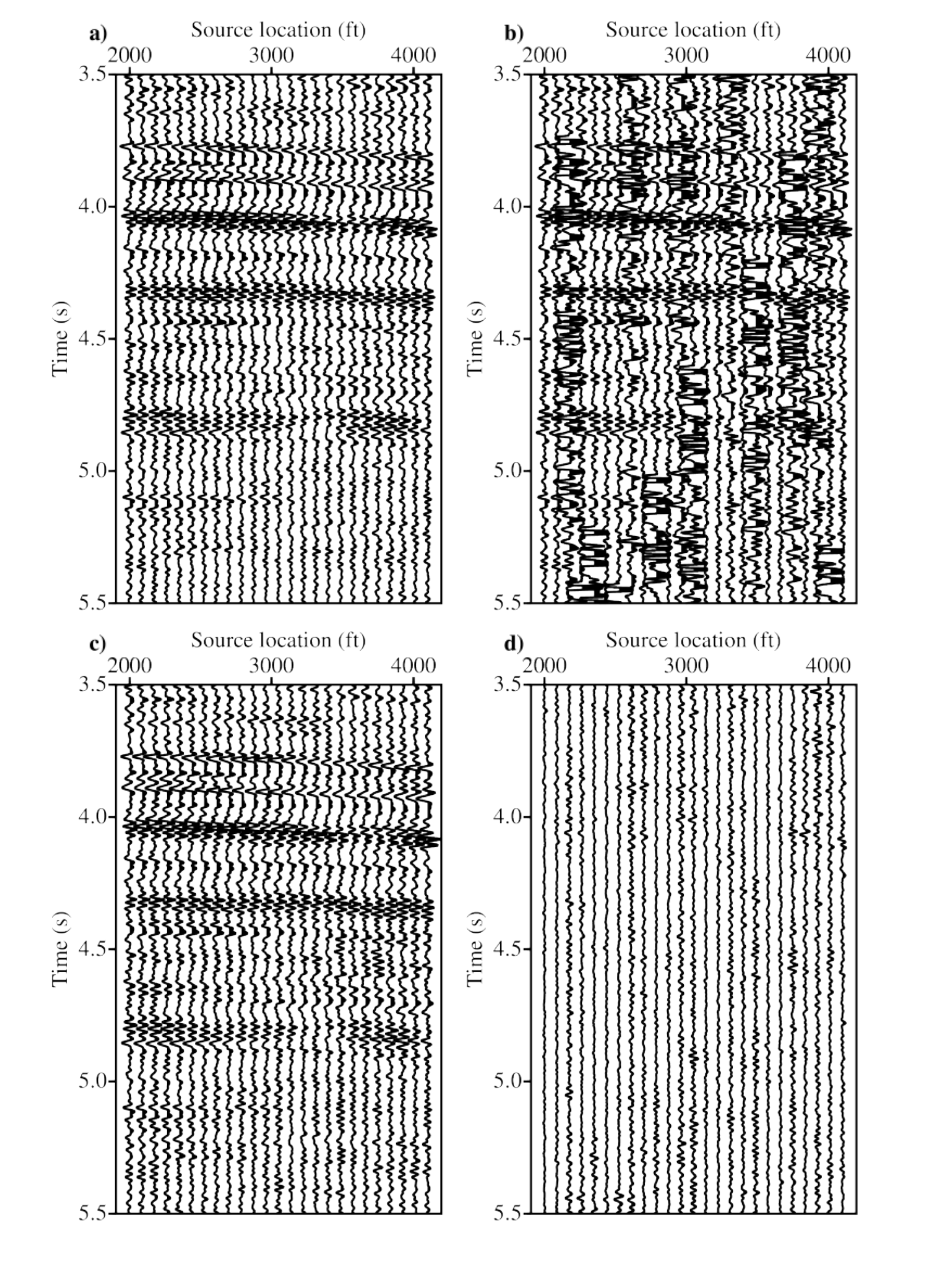}
	\reduceVspace
	\caption{Close-up of the common receiver gather from the Gulf of Mexico data recovered from the Stolt-based ASHRT model estimated using $p=1$ and $q=2$ inversion.
	(a) Original gather.   (b) Pseudo-deblended gather.  (c) Recovered gather.   (d) Recovered gather error.}
	\label{ch5_GOM_CRG_L1L2_closeup}
\end{figure}

\begin{figure}[htp] 
	\centering
	\includegraphics{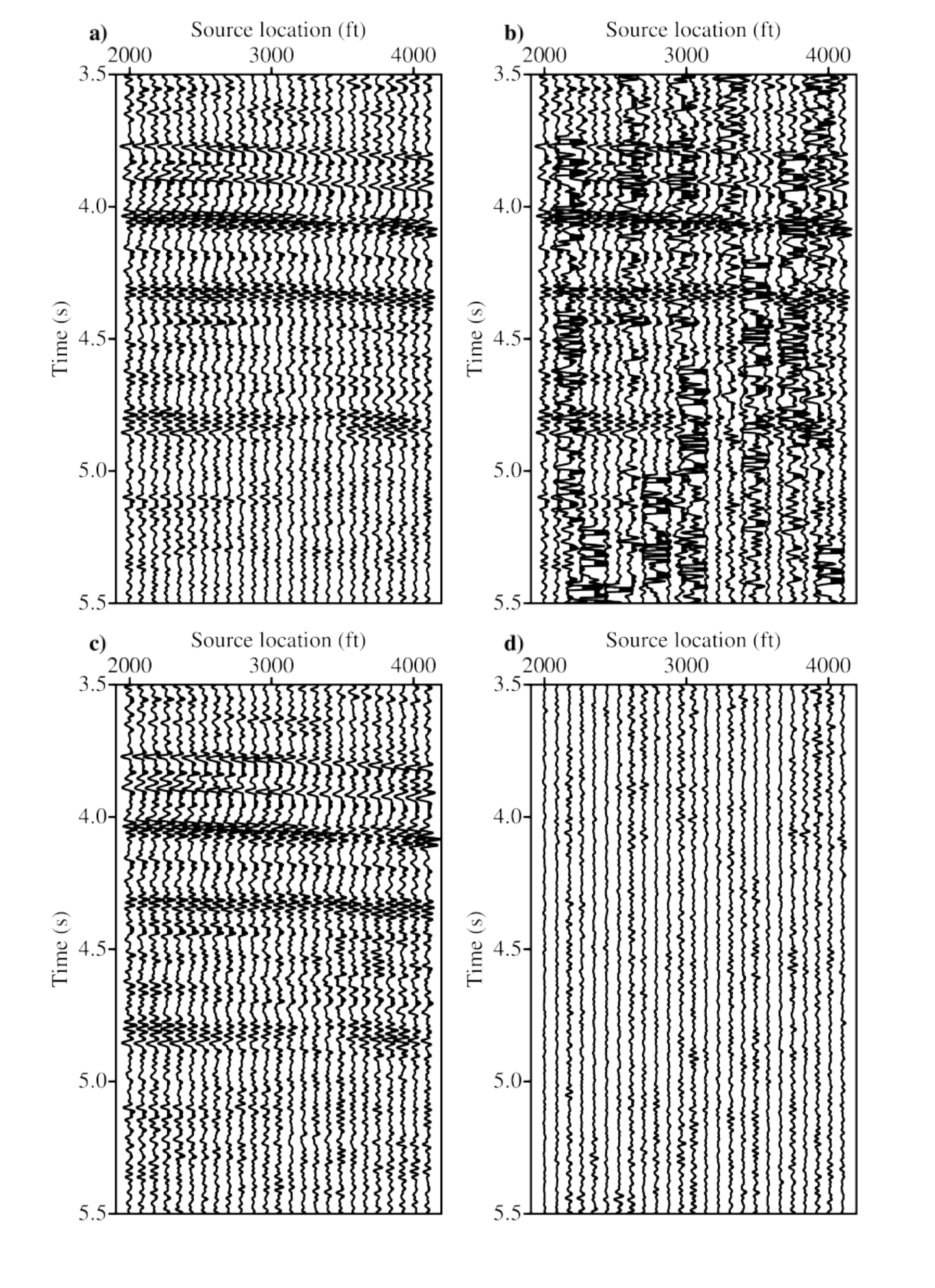}
	\reduceVspace
	\caption{Close-up of the common receiver gather from the Gulf of Mexico data recovered from the Stolt-based ASHRT model estimated using $p=1$ and $q=1$ inversion.
	(a) Original gather.   (b) Pseudo-deblended gather.  (c) Recovered gather.   (d) Recovered gather error.}
	\label{ch5_GOM_CRG_L1L1_closeup}
\end{figure}

The estimated Stolt-based ASHRT models are used to recover the common receiver gathers.  
Figures \ref{ch5_GOM_CRG_L2L2}, \ref{ch5_GOM_CRG_L2L1}, \ref{ch5_GOM_CRG_L1L2} and \ref{ch5_GOM_CRG_L1L1} show data recovered from 
Radon models estimated using ($p=2,q=2$), ($p=2,q=1$), ($p=1,q=2$) and ($p=1,q=1$) inversion, respectively. 
Similar to the results of the time domain ASHRT, these figures clearly show that the robust Radon transforms were able to attenuate interference while preserving the weak signals better than the non-robust transforms. 
The ability of robust Radon transforms to preserve weak reflections is quite clear in the close-up 
Figures \ref{ch5_GOM_CRG_L1L2_closeup} and \ref{ch5_GOM_CRG_L1L1_closeup}, which show data recovered from Radon models estimated using ($p=1,q=2$) and ($p=1,q=1$) robust inversion, respectively. 
The error of recovered data  in Figures \ref{ch5_GOM_CRG_L1L2_closeup}d and \ref{ch5_GOM_CRG_L1L1_closeup}d  confirm that the robust inversion represents an effective method for removing source interferences. 
On the other hand, Figures \ref{ch5_GOM_CRG_L2L2_closeup} and \ref{ch5_GOM_CRG_L2L1_closeup} show a close-up of data recovered from Radon models estimated using ($p=2,q=2$) and ($p=2,q=1$) non-robust inversion, respectively. 
The recovered data error in Figures \ref{ch5_GOM_CRG_L2L2_closeup}d and \ref{ch5_GOM_CRG_L2L1_closeup}d clearly show that non-robust inversions can remove source interferences effectively. 
For the field data example, imposing a strict sparsity constraint is not a simple task. 
The reflection hyperbola generated by the Stolt operator do not exactly match the reflection hyperbolas in the data.
This mismatch results from the approximated and coarse velocities that are used in the transform and by the presence of amplitude versus offset (AVO) changes which are  not included in the Stolt-based ASHRT operator.   
Therefore, the Stolt-based ASHRT transform does not focus reflections to sharp points and the estimated ASHRT model is not as sparse as the one obtained when using simple synthetic examples \citep{Ibrahim2014GEO}.
Additionally, field data could suffer from missing near offset, irregular sampling and feathering, which will reduce the operator ability to focus seismic  reflections.

\begin{figure}[htp] 
	\centering
	\includegraphics{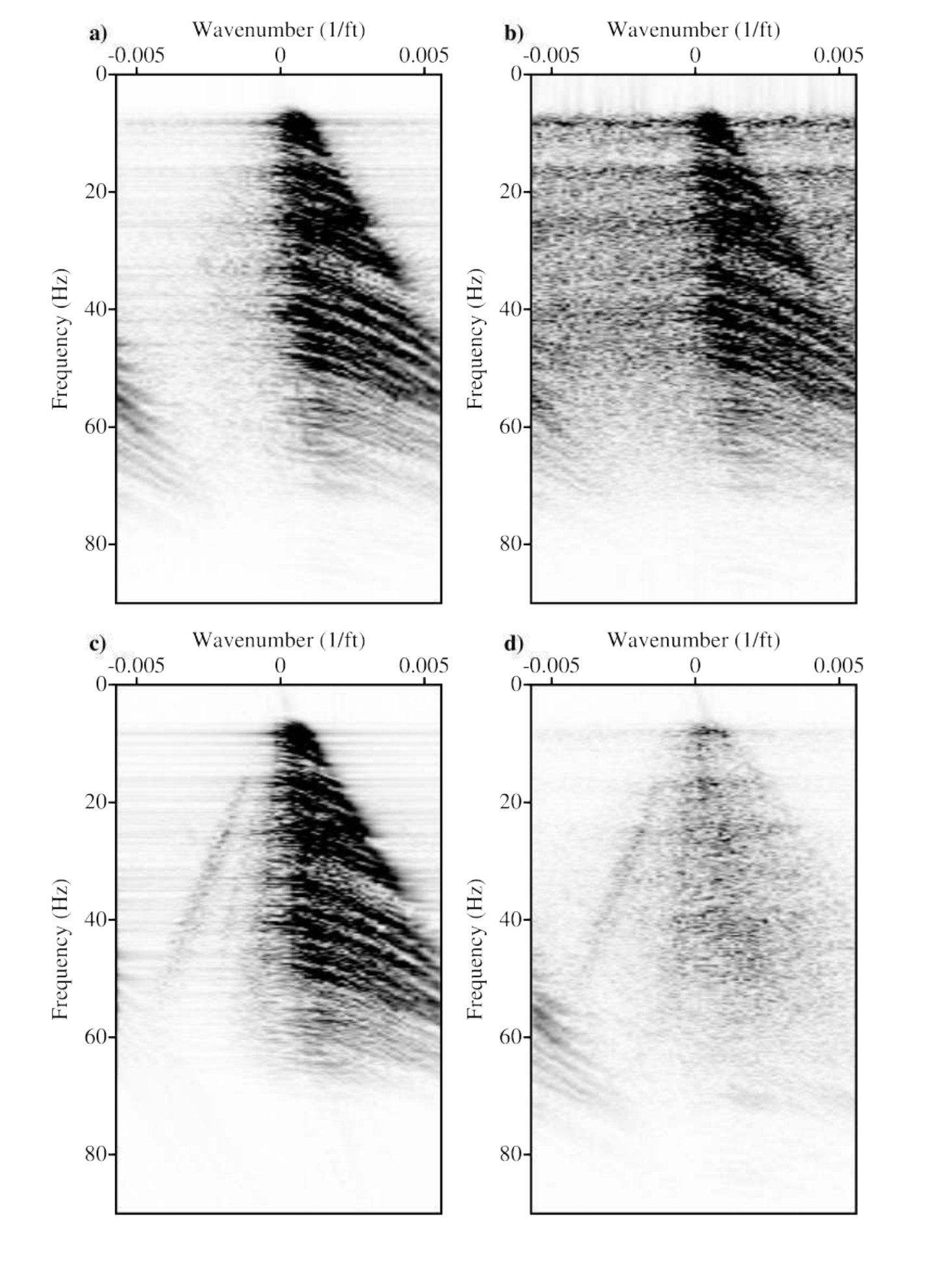}
	\reduceVspace
	\caption{The $f-k$ spectra of the common receiver gather from the Gulf of Mexico data  recovered from Stolt-based ASHRT model estimated using $p=2$ and $q=2$ inversion.
	(a) Original gather.   (b) Pseudo-deblended gather.  (c) Recovered gather.   (d) Recovered gather error.}
	\label{ch5_GOM_CRG_L2L2_fk}
\end{figure}

\begin{figure}[htp] 
	\centering
	\includegraphics{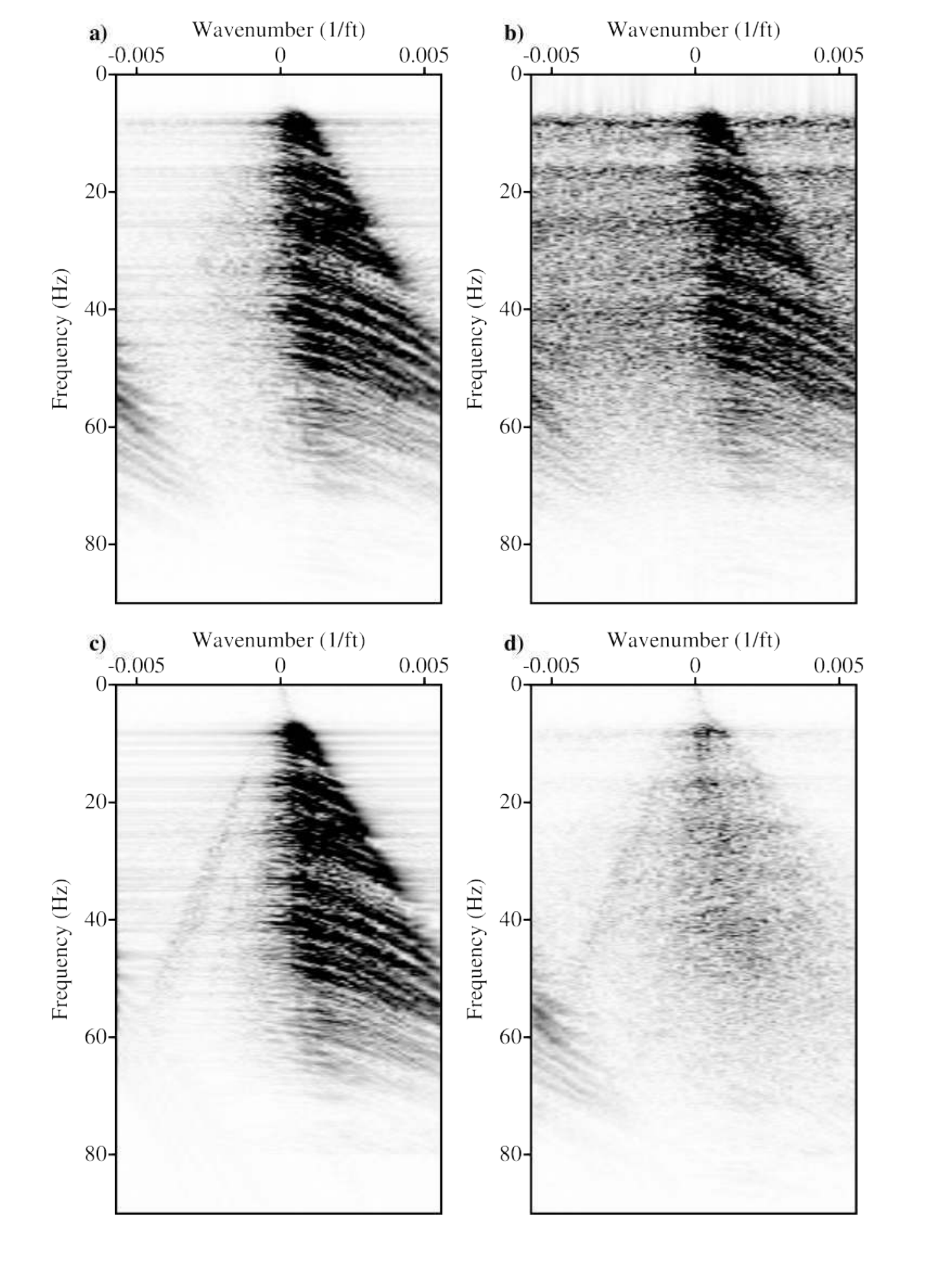}
	\reduceVspace
	\caption{The $f-k$ spectra of the common receiver gather from the Gulf of Mexico data  recovered from Stolt-based ASHRT model estimated using $p=2$ and $q=1$ inversion.
	(a) Original gather.   (b) Pseudo-deblended gather.  (c) Recovered gather.   (d) Recovered gather error.}
	\label{ch5_GOM_CRG_L2L1_fk}
\end{figure}

\begin{figure}[htp] 
	\centering
	\includegraphics{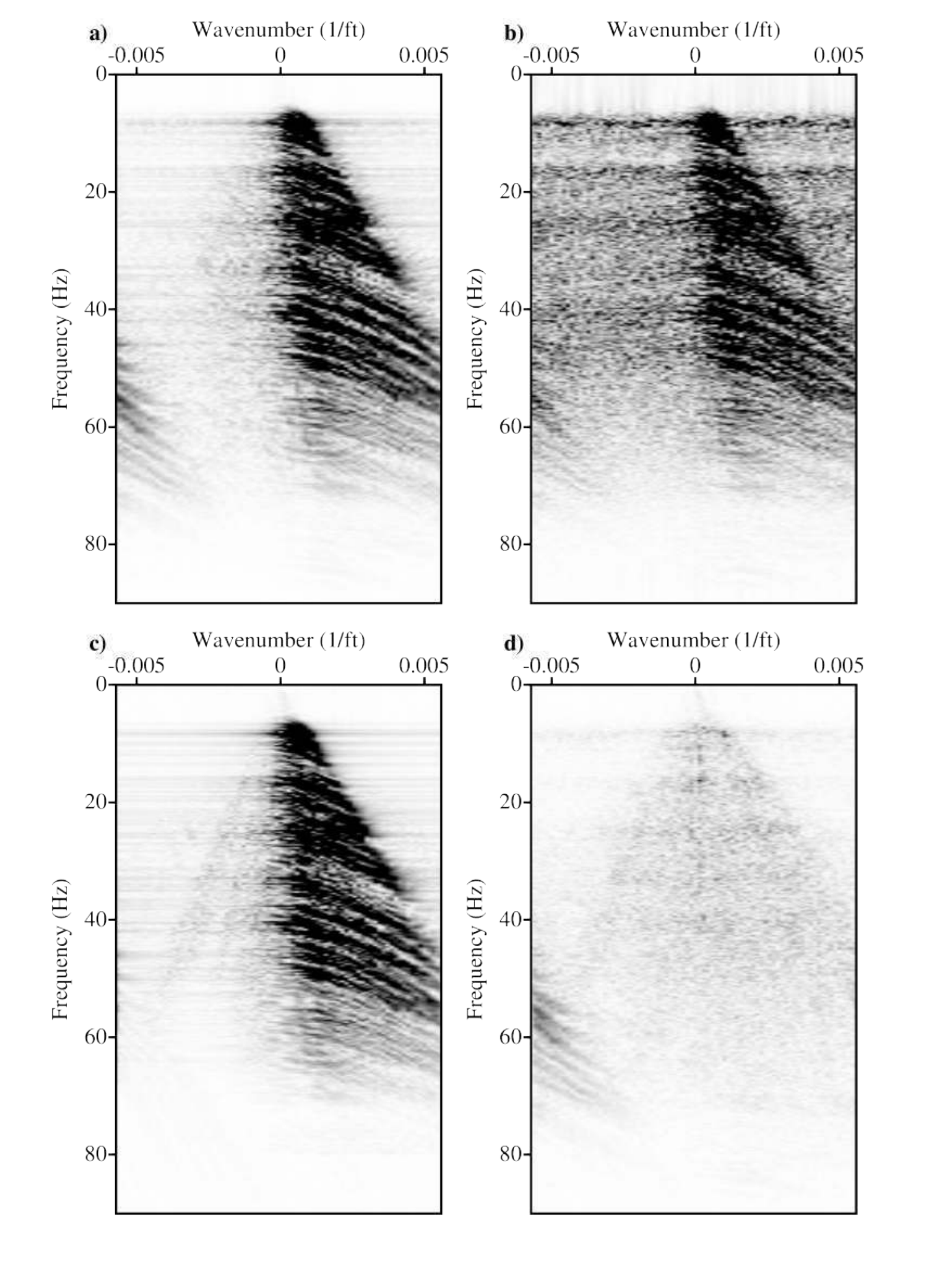}
	\reduceVspace
	\caption{The $f-k$ spectra of the common receiver gather from the Gulf of Mexico data  recovered from Stolt-based ASHRT model estimated using $p=1$ and $q=2$ inversion.
	(a) Original gather.   (b) Pseudo-deblended gather.  (c) Recovered gather.   (d) Recovered gather error.}
	\label{ch5_GOM_CRG_L1L2_fk}
\end{figure}

\begin{figure}[htp] 
	\centering
	\includegraphics{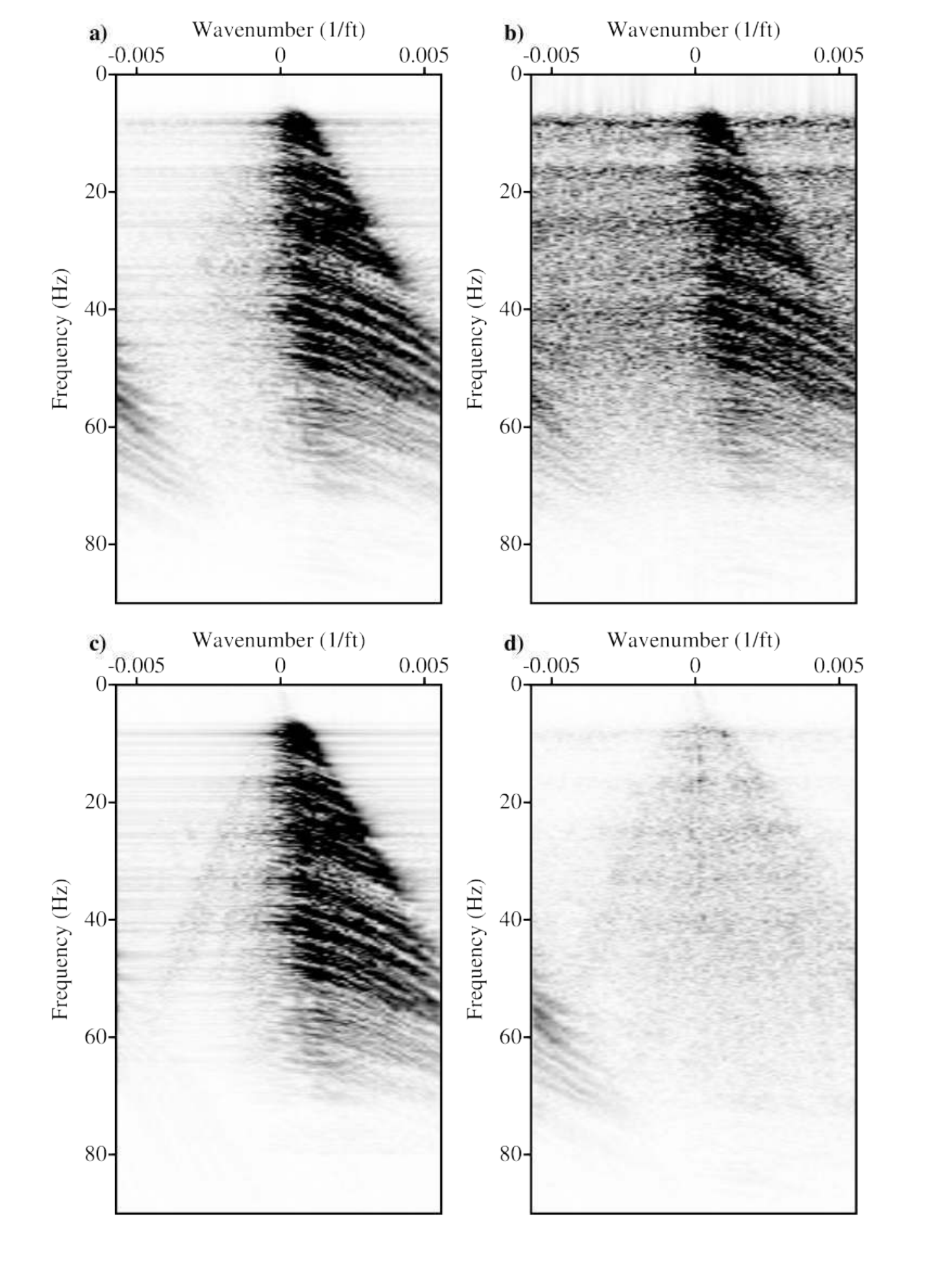}
	\reduceVspace
	\caption{The $f-k$ spectra of the common receiver gather from the Gulf of Mexico data  recovered from Stolt-based ASHRT model estimated using $p=1$ and $q=1$ inversion.
	(a) Original gather.   (b) Pseudo-deblended gather.  (c) Recovered gather.   (d) Recovered gather error.}
	\label{ch5_GOM_CRG_L1L1_fk}
\end{figure}

Figures \ref{ch5_GOM_CRG_L2L2_fk}, \ref{ch5_GOM_CRG_L2L1_fk}, \ref{ch5_GOM_CRG_L1L2_fk} and \ref{ch5_GOM_CRG_L1L1_fk} show the $f-k$ spectra of common receiver gathers recovered from Radon models estimated using ($p=2,q=2$), ($p=2,q=1$), ($p=1,q=2$) and ($p=1,q=1$) inversion, respectively. The $f-k$ spectra of the recovered common receiver gathers confirm that the robust inversion results in more accurate data recovery than the non-robust inversion. Comparing the results for the $f-k$ spectra of CRG recovered using the time domain ASHRT  in chapter 4 and the Stolt-based ASHRT in chapter 5 shows that the Stolt-based operator has removed the aliased energy in the Gulf of Mexico gathers. This anti-aliasing property is due to limiting the $f-k$ Stolt mapping to the signal cone in the $f-k$ domain.  Additionally, comparing Figure 4.40 and Figure \ref{ch5_GOM_CRG_L2L1_fk} shows that the Stolt-based ASHRT operator better preserves the $f-k$ spectral characteristics of  the common receiver gather.  


\begin{figure}[htp] 
	\centering
	\includegraphics{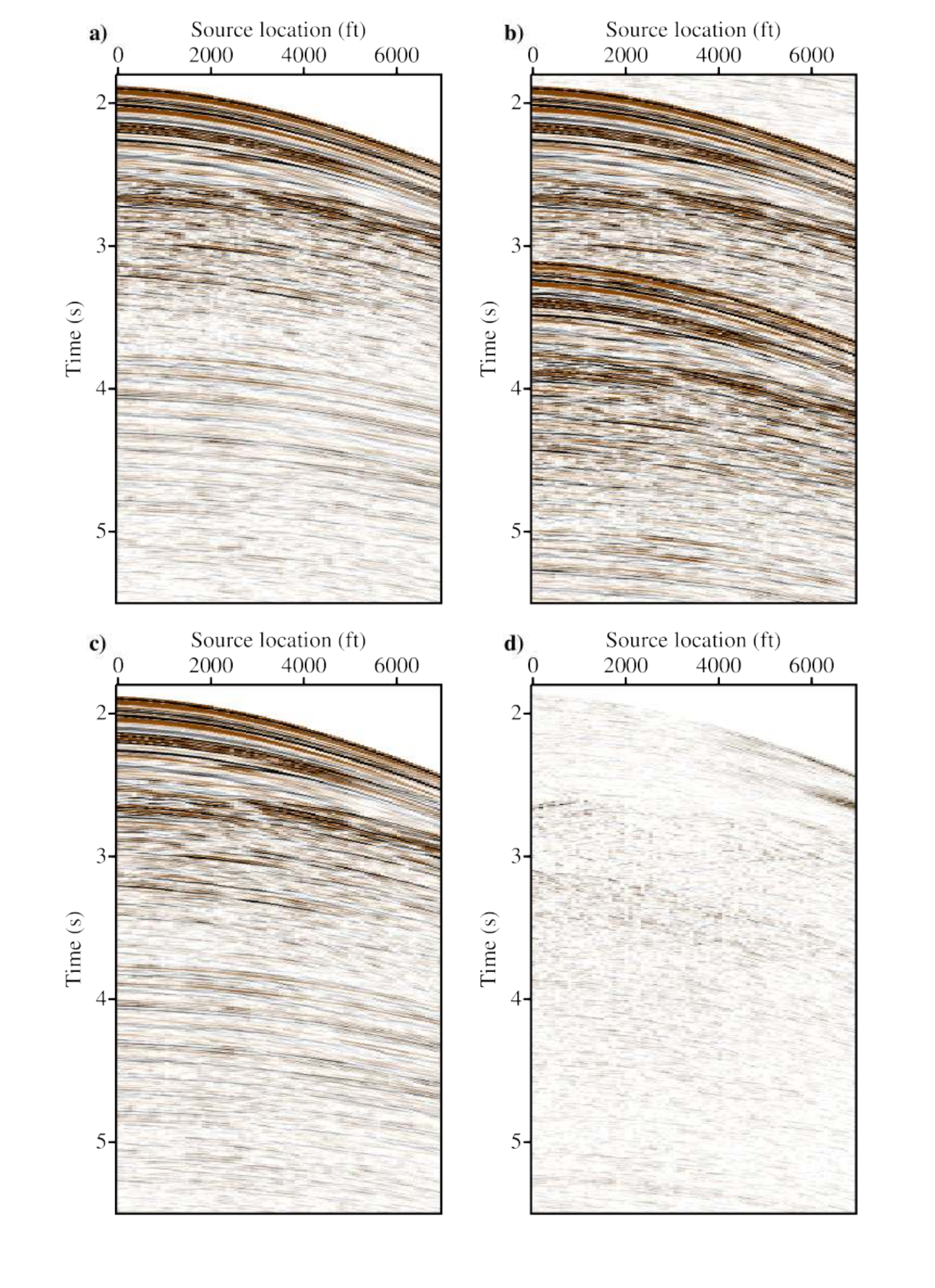}
	\reduceVspace
	\caption{Common source gather from the Gulf of Mexico data separated using  Stolt-based ASHRT models estimated by $p=1$ and $q=2$ inversion.
	(a) Original gather.   (b) Pseudo-deblended gather.  (c) Recovered gather.   (d) Recovered gather error.}
	\label{ch5_GOM_CSG_L1L2}
\end{figure}

\begin{sidewaysfigure}[htp] 
 		\includegraphics{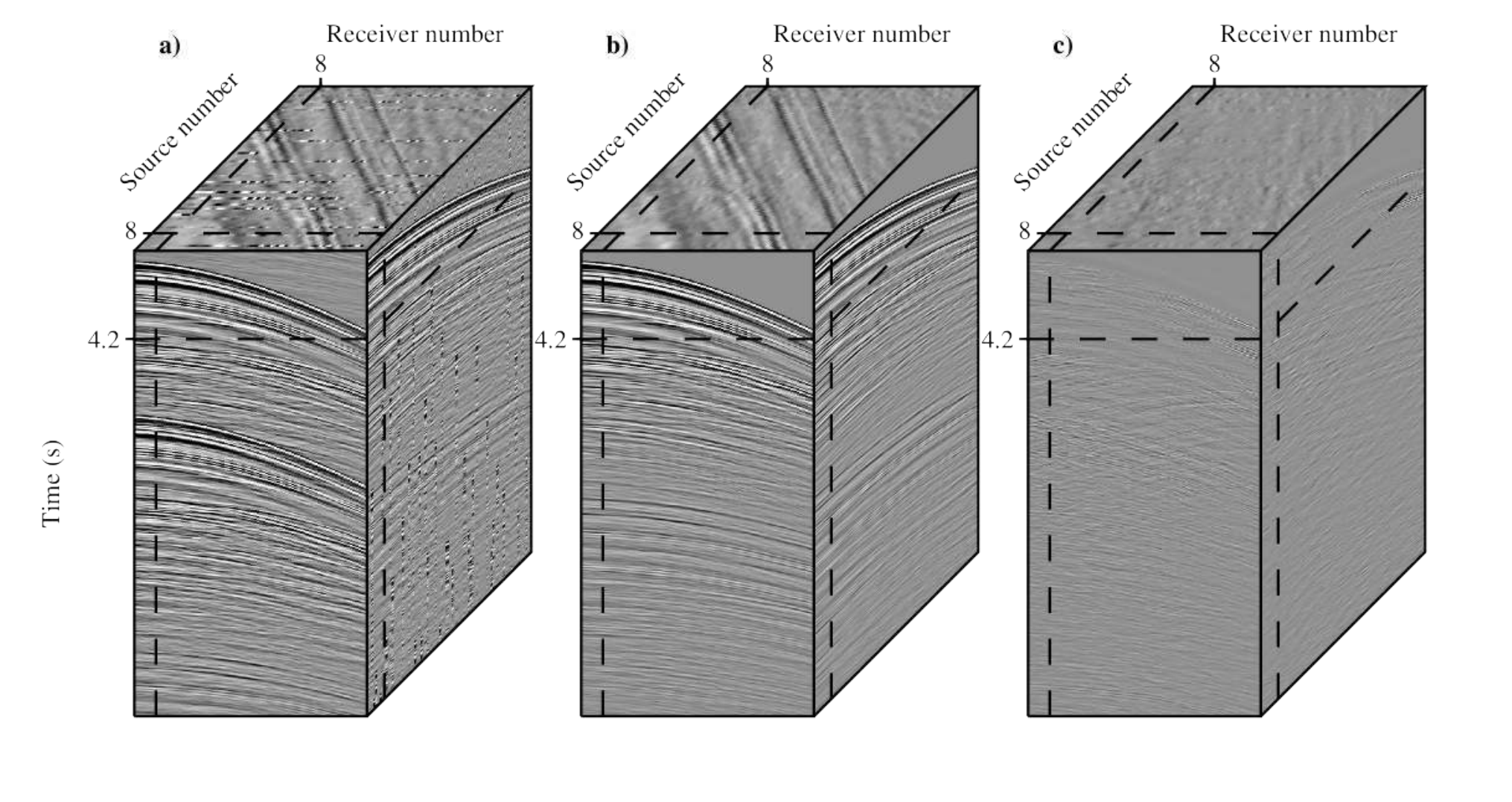}
 		\reduceVspace
 		\caption{Gulf of Mexico data cubes. 
   		(a) Pseudo-deblended.
   		(b) Data recovered by forward modeling from the Stolt-based ASHRT model estimated via $p=1$ and $q=2$ inversion.
   		(c) Difference between recovered and original data cubes. }
		\label{ch5_GOM_cubes_L1L2}
\end{sidewaysfigure}

The $Q$ values for the recovered Gulf of Mexico common receiver gathers are listed in Table \ref{table:ch5_quality}. 
Our tests show that inversion using both robustness and quadratic regularization $p=1$, $q=2$ produces the best results ($Q=11.95$ dB). 
For the Gulf of Mexico data, we also measure the quality for recovering the weak events window. 
The best quality weak events window was achieved using $p=1$, $q=2$ model inversion with $Q=14.07$ dB
This illustrates quantitatively that the robust Radon transform can remove source interferences effectively while preserving weak signals.
Figure \ref{ch5_GOM_CSG_L1L2} shows the common source gather recovered after denoising all common receiver gathers using $p=1,q=2$ inversion.  This figure shows that coherent interferences in the common source gather were removed effectively after denoising all common receiver gathers. Figure \ref{ch5_GOM_cubes_L1L2} shows the data cubes recovered using Radon models estimated using $p=1,q=2$ inversion.   

\clearpage
\clearpage

\setlength{\tabcolsep}{3.5pt}
\renewcommand{\arraystretch}{3.0}
\begin{table}[htp] 
\small{
\begin{center}
\caption{Quality of denoising CRG using Stolt-based ASHRT transform.}
\label{table:ch5_quality}
\begin{tabular}{| c | c | c | c | c |}
\hline 
{Model inversion} 	    							&$p=2$, $q=2$	     &$p=2$, $q=1$         &$p=1$, $q=2$	&$p=1$, $q=1$ \\ \hline  
{Synthetic data }   								& $10.0829$      	    & $13.3181$               &	$17.3538$	&	$22.3236$\\  \hline                    
{Field data  }										& $7.6898$			   &  $7.8917$      			  &   $11.9540$     &	$10.4278$	\\  \hline 
{Field data (weak events window)}		& $7.1227$          	   &  $7.2927$     			  &	$14.0716$		  &$11.4863$	\\ \hline    
\end{tabular}
\end{center}}
\end{table} 

\section{Conclusions}

We have implemented a fast Stolt-based transform that is similar to the Apex shifted Hyperbolic Radon Transform (ASHRT). Moreover, we used the Stolt-based ASHRT to remove source interferences in common receiver gathers of simultaneous source data. We showed that the Stolt-based transform can remove source interferences in common receiver gathers at a computational cost that is substantially below the computational cost of the classical ASHRT. The quality of source separation using the Stolt-based ASHRT is comparable to that of the classical ASHRT.  Since the Stolt operator is implemented in the $f-k$ domain, it can be used in combination with the non-uniform Fourier transform to interpolate missing traces. Future work entails generalizing the transform to a 3D shot distribution. 

\chapter[Asymptote and Apex shifted Radon transform]
{
Asymptote and Apex shifted Radon transform
\footnote{A version of this chapter has been published in Ibrahim , A, Trenghi, P. and Sacchi, M. D. 2015, Wavefield reconstruction using a Stolt-based asymptote and apex shifted hyperbolic Radon transform, 85th Annual International Meeting, SEG, Expanded Abstracts and an 
applications of the AASHRT transform is included in Hegge, R., Terenghi, P. Kaledtke, A. and Ibrahim, A., 2015, Prediction and subtraction of multiple diffraction, Provisional Patent application no. 62/195,459.}
}

\section{Motivation}

Improving the transform ability to focus seismic signals is vital for many processing applications that rely on sparsity such as interpolation. 
Seismic data usually contain low amplitude seismic diffractions that have hyperbolic travel times similar to seismic reflections. 
However,  hyperbolic travel times of seismic diffractions have its asymptote shifted in time. 
This shift is due to the asymmetric travel path of seismic diffractions in the subsurface. 
Seismic diffractions is an important part of seismic data since it can be used to map faults and increase imaging resolution.
In order to account for seismic diffractions, the Stolt-based ASHRT basis functions are extended to account for the time shift of diffraction hyperbolas. 
We named this new transform Asymptote and Apex Shifted Hyperbolic Radon Transform (AASHRT). 
This transform is used to interpolate data that contain significant amount of seismic diffractions. 
The results of the interpolation tests show that the AASHRT transform can be a powerful tool interpolating seismic diffractions \citep{Ibrahim2015SEG,Ibrahim2015patent}.

\section{Introduction}

In towed-streamer marine seismic acquisition, operational costs and entanglement issues pose substantial limits to sampling along the spatial cross-line axis.
This can lead to aliasing problems and lower the imaging resolution of the subsurface. 
Therefore, seismic data usually require interpolation to increase the spatial sampling prior to time-domain processing and imaging. 
Most interpolation algorithms rely on transforms that can focus seismic data in the transform domain.
These focusing abilities arise from the similarity between seismic data and the chosen transform dictionary. 
This justifies the ongoing research into new dictionaries that closely match seismic data for interpolation applications \citep{Paolo2014}.

Since the travel-times of a variety of seismic events can be approximated by hyperbolas, Radon transforms that use a hyperbolic dictionary represent a powerful tool for interpolation \citep{Sacchi1995,Trad2003b}.
The most common of these transforms is the Hyperbolic Radon Transform (HRT), which is used for processing common midpoint gathers, where the apexes of seismic reflection hyperbolas are usually located at zero offset \citep{Thorson1985}.
However, \citet{Trad2003b} proposed interpolating seismic data in the common shot gather domain using an Apex Shifted Hyperbolic Radon Transform (ASHRT), which extends the conventional HRT by scanning for the horizontal location of apexes.
The ASHRT has been applied in order to interpolate and/or denoise seismic data in the shot gather domain \citep{Trad2003,Ibrahim2014GEO} or as part of 3D multiple prediction algorithms \citep{vanDedem2000,vanDedem2005}.

In this paper, we present an additional extension to the ASHRT by scanning for both the apex and asymptote shifts to match both reflection and diffraction hyperbolas more closely. 
Seismic diffractions can provide important information about subsurface discontinuities such as faults, pinch outs and small size scattering objects that can be used in interpretation or imaging \citep{Khaidukov2004,Bansal2005,Klokov2012}.  
The travel time curve of diffracted waves can be represented by the double square root equation \citep{Landa1987,Kanasewich1988}
\begin{equation}
t=\sqrt{t_d^2+{\frac{(x_s-x_d)^2}{v^2}}}+\sqrt{t_d^2+\frac{(x_d-x_r)^2}{v^2}},
\end{equation}   
where $t_d$ is the one way travel time for the diffraction, $x_d$ is the diffraction location along the horizontal axis, $v$ is the velocity at the diffraction, $x_s$ is the source location and $x_r$ is the receiver location.  
If we are considering a common shot gather, the value of the first square root is constant for each diffraction hyperbola. 
Therefore, we introduce the new parameter, 
$${\tau}_0=\sqrt{t_d^2+\frac{(x_s-x_d)^2}{v^2}},$$
into the previous equation, 
\begin{equation}
t=\tau_0+\sqrt{t_d^2+\frac{(x_d-x_r)^2}{v^2}}.
\end{equation}
We use this equation to define the new Asymptote and Apex Shifted Hyperbolic Radon Transform (AASHRT) which scans for the asymptote origin time shift $\tau_0$.
This equation simplifies to the ASHRT definition when the asymptote origin time shift $\tau_0$ is set to zero and the diffraction time $t_d$ and location $x_d$ are replaced by the apex time $\tau$ and location $x_a$, respectively.

An efficient implementation of the AASHRT can be obtained using Stolt migration/demigration operators \citep{Stolt1978,Trad2003b,Ibrahim2014EAGE,Ibrahim2014SEG,Ibrahim2015GEO}.  The asymptote shift introduced by the AASHRT can be incorporated easily into the Stolt kernel using the Fourier time shift property. In case of uniform sampling, computational efficiency can be further increased using fast Fourier transforms and by scanning only the non-zeros elements of the data in the $\omega-k$ domain. 

\section{Asymptote and Apex Shifted \\ Hyperbolic Radon Transform}
The AASHRT models seismic data using a superposition of asymptote and apex shifted hyperbolas as follows
\begin{equation}\label{ch6_AASHRT_forward}
{d}(t,x_r)= \sum\limits_{\tau_0} \sum\limits_{x_a} \sum\limits_{v} m(\tau = \sqrt{t^2-\frac{(x_r-x_a)^2}{v^2}}-\tau_0,v,x_a,\tau_0) ,
\end{equation}
where $d(t,x_r)$ is the modelled seismic data and $m(\tau,v,x_a,\tau_0) $ is the AASHRT model. 
A low resolution AASHRT model can be estimated using the adjoint operation as follows  
\begin{equation}\label{ch6_AASHRT_adjoint}
\widetilde{m}(\tau,v,x_a,\tau_0) = \sum\limits_{x_r} {d}(t = \tau_0+\sqrt{\tau^2+\frac{(x_r-x_a)^2}{v^2}},x_r) ,
\end{equation} 
where $\widetilde{m}(\tau,v,x_a,\tau_0)$ is the estimated AASHRT model. 
These transforms can be rewritten in the operator format as 
\begin{align}
{\bf d} &= {\bf L} {\bf m},  \\
{\widetilde{\bf m}} &={\bf L}^T {\bf d}, 
\end{align}
where ${\bf d}$, ${\bf m}$ and ${\widetilde{\bf m}}$ represent the data, model and estimated model in vector form, respectively. 
The forward and adjoint AASHRT operators are represented by ${\bf L}$ and ${\bf L}^T$, respectively. 

\section{Stolt-based AASHRT} 
The time domain implementation of the AASHRT operator is computationally intensive. 
Fortunately, the AASHRT kernel can be computed efficiently in the $\omega-k$ domain using fast Stolt migration/demigration operators. 
These operators perform migration by mapping data in the $\omega-k_x$ domain into $\omega_{\tau}-k_x$ for a constant subsurface velocity using the dispersion relation \citep{Yilmaz2001,Ibrahim2015GEO}
\begin{equation}
\omega_{\tau} = \sqrt{\omega^2-(vk_x)^2}, 
\end{equation}
where $\omega_{\tau}$ is the Fourier dual of the apex time $\tau$, $k_x$ is the horizontal wavenumber and $v$ is the subsurface velocity.
Using the exploding reflector principle \citep{Claerbout1992} and the constant subsurface velocity assumption, the Stolt migration operator can be used to estimate the subsurface model.
Similarly, the Stolt migration operator can be used to estimate the AASHRT model ${\widetilde{m}}(\tau,v,x_a,\tau_0)$ as follows 
\begin{align}\label{ch6_AASHRT_adj}
\widetilde{m}(\tau,v,x_a,\tau_0)=&\int\int C\,{\exp{[i\omega_{\tau} \tau_0]}d}{(\omega=\sqrt{\omega_{\tau}^2+(vk_x)^2},k_x)}& \nonumber \\ 
& \times \exp{[-ik_xx-i\omega_{\tau}(v)\tau]}~d\omega_{\tau}~dk_x, 
\end{align} 
where $C=v~(\omega_{\tau}/\omega)$ is a scaling factor resulting from the change of variables.
There are two points worth emphasizing in deriving the Stolt-based AASHRT operator. First, every horizontal axis location $x$ is treated as a possible receiver and apex location (so $x_a \equiv x_r \equiv x$). Second, the periodicity of $\widetilde{m}$ along $x$ is implied.

The Stolt-based forward modelling can be written as 
\begin{align}\label{ch6_AASHRT_for}
{d}(t,x)=&\int\int\int\int~{m}(\omega_{\tau}=\sqrt{\omega^2-(vk_x)^2},v,k_x) \nonumber\\
& \times \exp{[-i\omega_{\tau} \tau_0]}\exp[{ik_xx+i\omega t}]~ d\omega~dk_x~dv~d\tau_0. 
\end{align}

The adjoint transforms in equation (\ref{ch6_AASHRT_adj}) can be rewritten in operator form as 
\begin{equation} \label{ch6_adj_operator}
{\bf L}^T={\bf F}_{\omega_{\tau},k_x}^{-1}~{\bf M}^{T}_{\omega,v,k_x}~{\bf T}^{T}~{\bf F}_{t,x}~{\bf A}^{T}{\bf S}^{T},
\end{equation} 
and similarly the forward (modelling) operator (equation \ref{ch6_AASHRT_for}) can be written as
\begin{equation} \label{ch6_forward_operator}
{\bf L}={\bf S}~{\bf A}~{\bf F}_{\omega,k_x}^{-1}~{\bf T}~{\bf M}_{\omega_{\tau},v,k_x}~{\bf F}_{\tau,x} ,
\end{equation} 
where, {\bf F} is the Fast Fourier Transform, ${\bf M}_{\omega_{\tau},v,k_x}$ is the Stolt mapping operator, ${\bf A}$ is a summation operator and its adjoint is a spraying operator \citep{Claerbout1992}. The operator ${\bf T}$ represents the time shift operator in the frequency domain while ${\bf S}$ represents the sampling operator used for interpolation \citep{Liu2004}. In case of non-uniform spatial sampling, the {\bf F} operator may be replaced by a Discrete Fourier Transform operator or a non-uniform Fast Fourier Transform operator.

\section{Sparse inversion}
The estimated model ${\widetilde{\bf m}}$ and the original model ${\bf m}$ are clearly not identical because Radon transforms are not an orthogonal transformations (${\bf L}{\bf L}^T \neq 1$).
Furthermore, seismic reflection data may be affected by a number of disturbing factors, including significant noise, limited spatial aperture, coarse and irregular spatial sampling, missing traces, etc.
The estimation of the Radon model must then be posed as an inversion problem \citep{Thorson1985} conditioned by a regularization (penalty) term.
The general form of the cost function to be minimized in order to obtain the Radon coefficients is given by  \citep{Ibrahim2014GEO}
\begin{align}\label{cost_fn}
J=\|{ {\bf d} -{\bf L}\, {\bf m}}\|_p^p+\mu \| {\bf m} \|_q^q 
\end{align}
where $\mu$ is the trade-off parameter that controls the relative weight between the model regularization term $\| {\bf m} \|_q^q$ and the misfit term $\|{ {\bf d} -{\bf L}\, {\bf m}}\|_p^p$. 
Furthermore, parameters $p$ and $q$ indicate the order of the norms used for the misfit and regularization terms.
Since we choose a Radon dictionary that closely matches the seismic data, the Radon coefficients should be sparse.
We therefore choose an $\ell_1$ norm ($q=1$) for the model regularization term and an $\ell_2$ norm ($p=2$) for data misfit term.
This cost function can be minimized using the Fast Iterative Shrinkage/Threshold Algorithm (FISTA) \citep{Beck2009}. 
The FISTA algorithm is used since it can estimate accurate sparse models with less sensitivity to the trade-off parameters than the IRLS used in previous examples. The FISTA algorithm requires an approximation for the largest eigenvalue of the ${\bf L^TL}$ operator which is calculated using Rayleigh's power method \citep{Larson2009}. For details about the FISTA algorithm and its application in geophysics refer to \cite{Beck2009}, \citet{Ismael2012} and \citet{Perez2013}. 

\section{Examples}

\begin{figure}[htp] 
	\centering
	\includegraphics{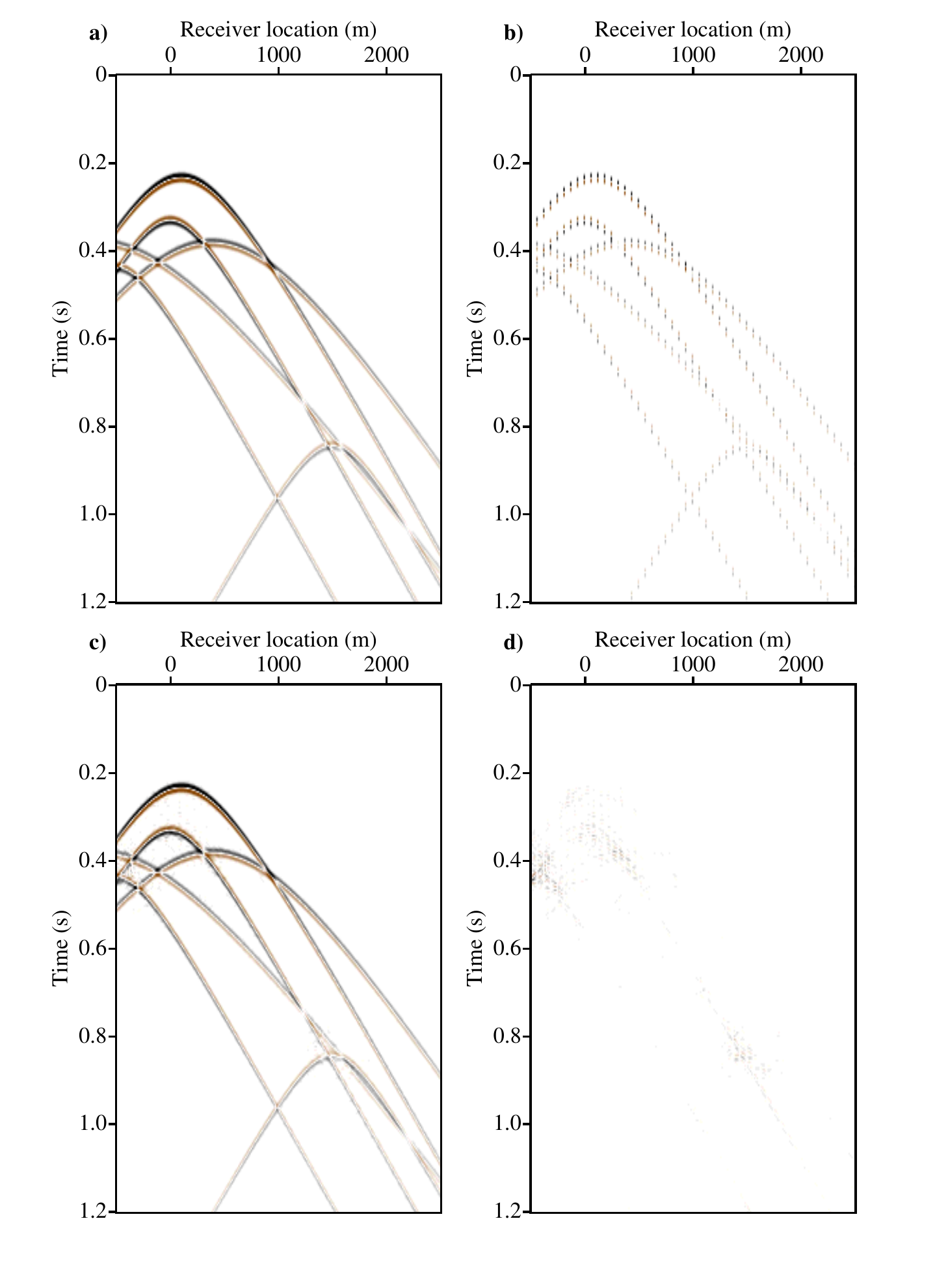}
	\reduceVspace
	\caption{Simplified synthetic common source gather example.
	(a) Original gather.   (b) Decimated gather.  (c) Interpolated gather.   (d) Interpolated gather error.}
	\label{ch6_synth2D}
\end{figure}

\begin{sidewaysfigure}[htp] 
	\centering
	\includegraphics[angle=90]{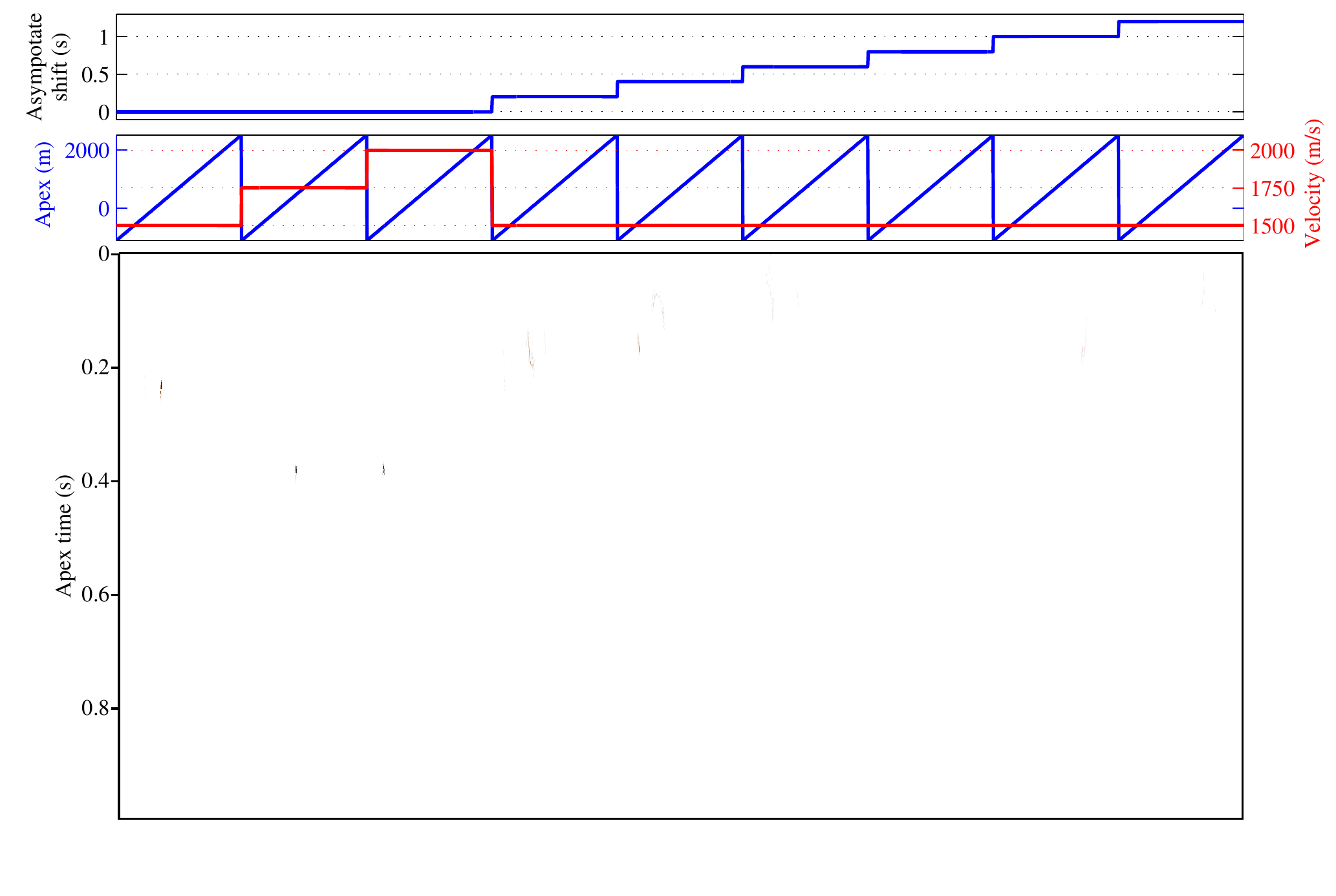}
	\caption{The Stolt-based AASHRT model of the decimated synthetic data estimated using sparse inversion ($p=2\&q=1$). }
	\label{ch6_synth2D_model}
\end{sidewaysfigure}

\begin{figure}[htp] 
	\centering
	\includegraphics{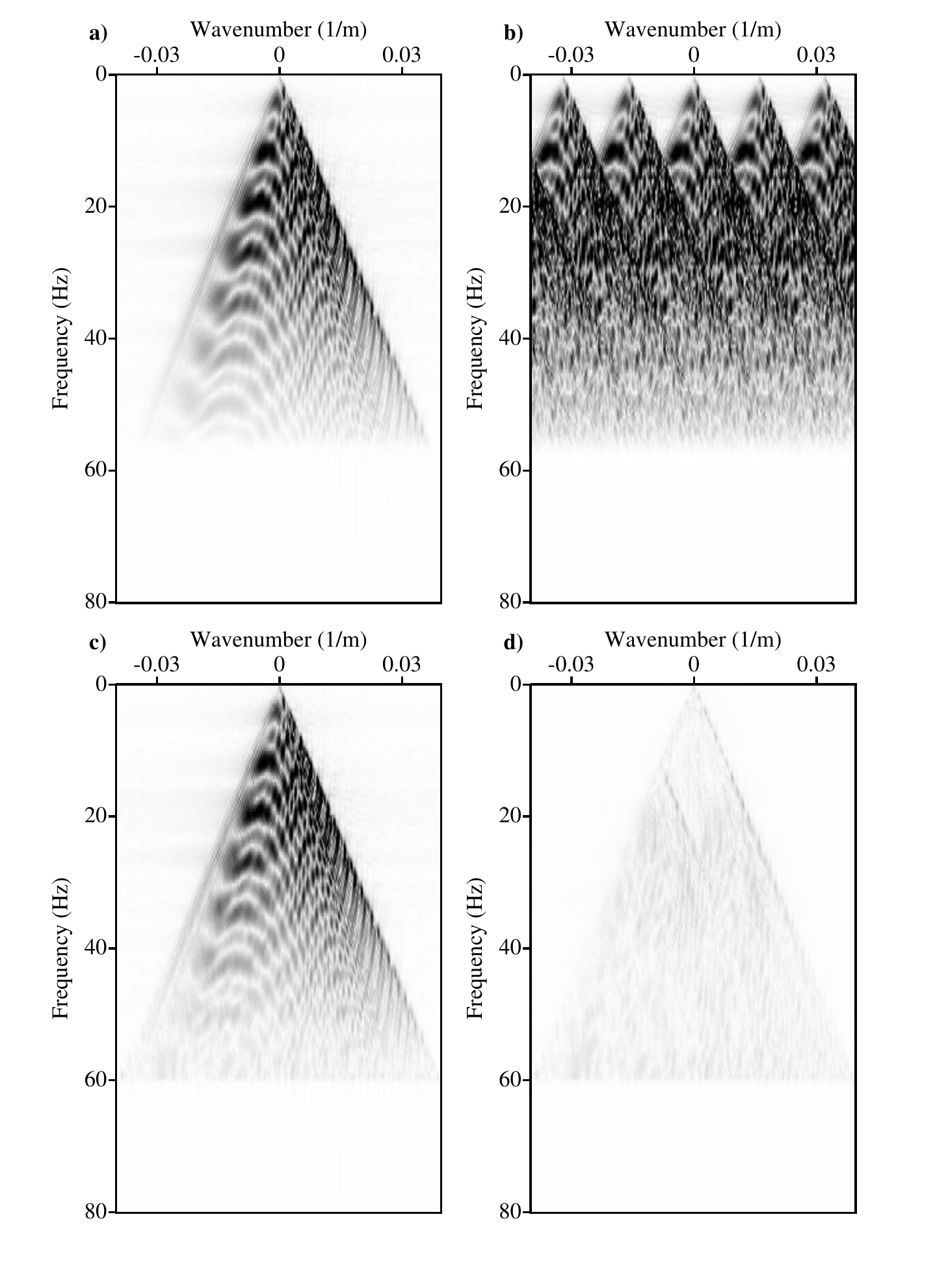}
	\reduceVspace
	\caption{The $f-k$ spectra for the simplified synthetic common source gather example.
	(a) Original gather.   (b) Decimated gather.  (c) Interpolated gather.   (d) Interpolated gather error.}
	\label{ch6_synth2D_fk}
\end{figure}

We evaluate the proposed AASHRT interpolation algorithm on a set of purpose-built synthetics composed of well distanced hyperbolic events, including 3 reflected and 3 diffracted events. Each reflection represents the acoustic response of a planar interface overlaid by a constant velocity medium.
The interfaces, characterized by dips of 10, 20 and -30 degrees, are located at depths of $300$, $600$ and $900$m below the source, respectively. 
The overlaying media are characterized by propagation velocities of $1500$, $1750$ and $2000$ m/s. The 3 diffractions originate from features located at a common depth of 400m and horizontal offsets of -500, 0, and 1500m with respect to the source. In its un-decimated version the test dataset is uniformly sampled with traces 12.5m apart (Figure \ref{ch6_synth2D}a), to ensure absence of spatial aliasing up to 60Hz (figure \ref{ch6_synth2D_fk}a). The subset used as input to the interpolation algorithm is obtained by uniformly under-sampling the initial data by a factor of 5. 
One in every five traces is kept as shown in Figure \ref{ch6_synth2D}b. The algorithm, parametrized to recover data at the initial spatial sampling rate of 12.5m produces the results shown in Figure \ref{ch6_synth2D}c. The AASHRT model scan three velocities ($1500, 1750$ and $2000m/s$) with zero asymptote shift and scans one velocity ($1500m/s$) with 6 asymptote shifts (from $0.2$ to $1.2$ s). The four dimensional AASHRT model of the synthetic data example is unfolded and shown in Figure \ref{ch6_synth2D_model}.  


\begin{figure}[htp] 
	\centering
	\includegraphics{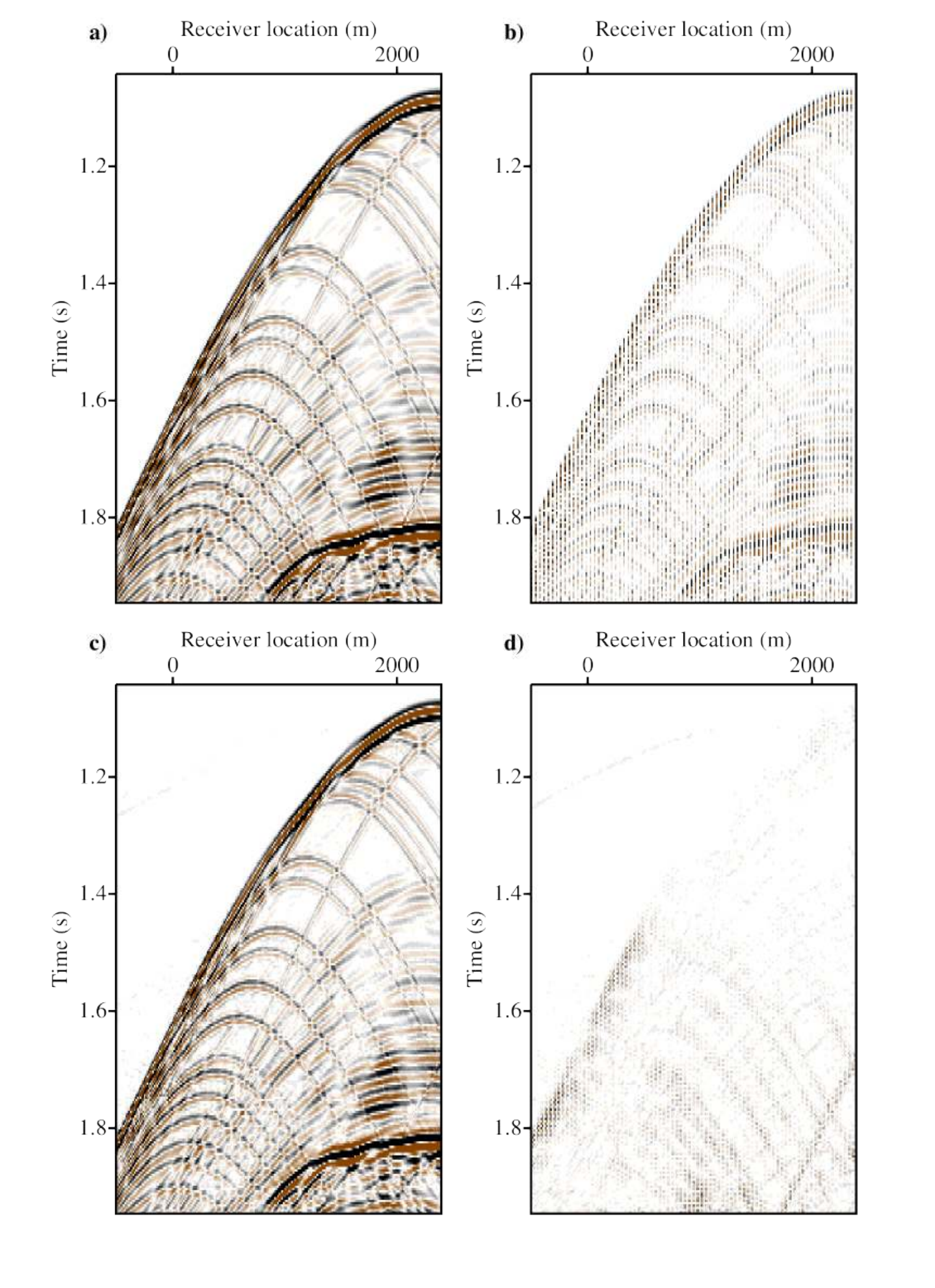}
	\reduceVspace
	\caption{Common source gather example of the BP/SEG velocity model data.
	(a) Original gather.   (b) Decimated gather.  (c) Interpolated gather.   (d) Interpolated gather error.}
	\label{ch6_BP2D}
\end{figure}

\begin{sidewaysfigure}[htp] 
	\centering
	\includegraphics{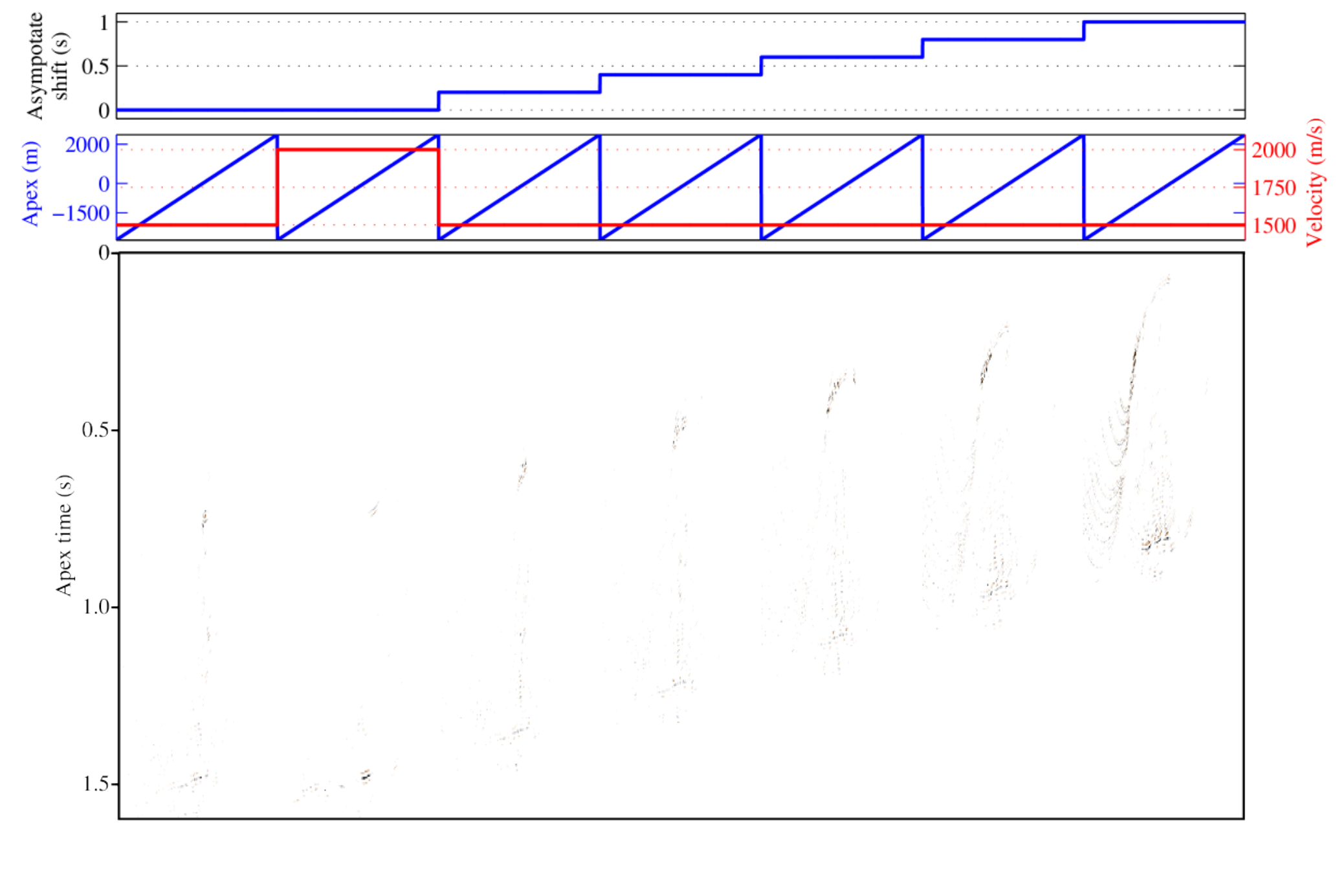}
	\caption{The Stolt-based AASHRT model of the decimated BP data estimated using sparse inversion ($p=2\&q=1$). }
	\label{ch6_BP2D_model}
\end{sidewaysfigure}

\begin{figure}[htp] 
	\centering
	\includegraphics{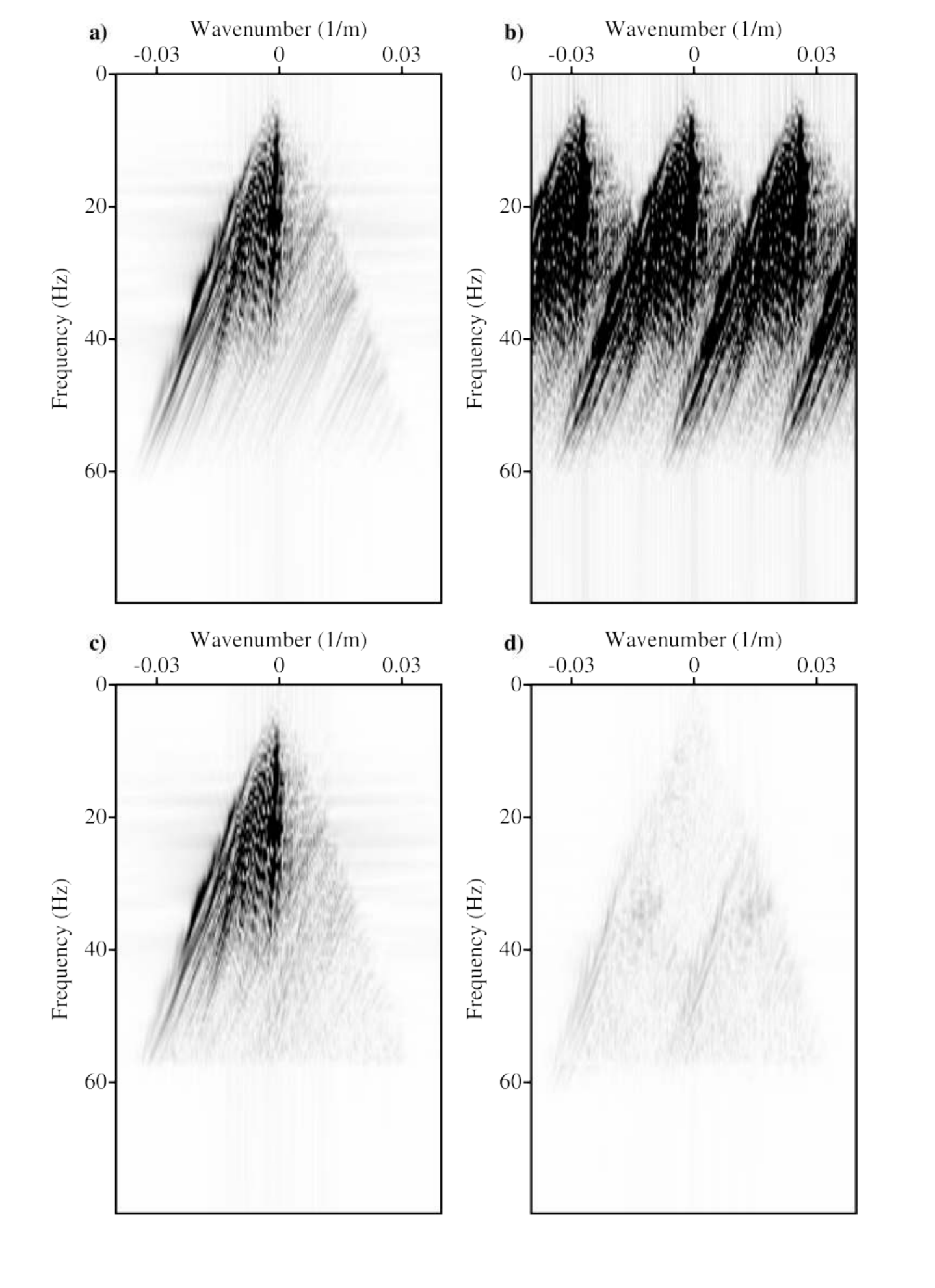}
	\reduceVspace
	\caption{The $f-k$ spectra for common source gather example of the BP/SEG velocity model.
	(a) Original gather.   (b) Decimated gather.  (c) Interpolated gather.   (d) Interpolated gather error.}
	\label{ch6_BP2D_fk}
\end{figure}

We also tested the new transform on synthetics modeled using the 2004 BP velocity benchmark \citep{BP2004}. 
Similar to the previous example, the data in its un-decimated version is uniformly sampled with traces 12.5m apart. 
The input to the interpolation algorithm is obtained by uniformly under-sampling the initial data by a factor of 3 (one in every three traces is kept).
We recover the data at sampling rate of 12.5 m producing the results shown in Figure \ref{ch6_BP2D}. In this example, the AASHRT transform scans two velocities ($1500$ and $2000$ m/s) with zero asymptote shift and scan one velocity ($1500$ m/s) with 5  asymptote shifts (from $0.2$ to $1.0$ s). The four dimensional AASHRT model of the BP data example is unfolded and shown in Figure \ref{ch6_BP2D_model}.  
The $f-k$ spectra of the original and interpolated gather are shown in Figure Figure \ref{ch6_BP2D_fk}.
The quality of the interpolation is calculated using the following formula
\begin{equation}
Q =10~log_{10}\left(\frac{\| {\bf d}_{original} \|_2^2}{\| {\bf d}_{original}-{\bf d}_{recovered} \|_2^2} \right).
\end{equation} 
The $Q$ value for the recovered simple synthetic shot gather is $16.25$ dB and for the BP/SEG model shot gather is $15.25$ dB.

\section{Conclusions}
We have implemented an asymptote and apex shifted hyperbolic Radon  transform with a Stolt migration/demigration operator as its kernel to speed up computations. The new transform dictionary is designed to closely match both reflections and diffractions. Our tests show that the new transform is a suitable tool for interpolation.  Since the new transform is implemented in $f-k$ domain, it can be used in combination with the non-uniform Fourier transform \citep{Dutt1993} to process data with non-uniform spatial sampling. Future work entails generalizing the problem to the 3D shot distribution  and application to simultaneous seismic sources separation.
\chapter[Conclusions]
{
Conclusions
}

\section{The thesis contributions}

In this thesis, a robust  misfit is incorporated into the denoising-based source separation problem. This is achieved by designing a robust Radon transform that use the robust $\ell_1$ norm misfit instead of the conventional $\ell_2$ norm misfit. Robust inversion is well suited for denoising source interferences in simultaneous seismic sources data. The time dithering of the source firing times produce incoherent source interferences in common receiver gathers. These interferences have non-Gaussian distribution and large amplitudes at the weak late arrivals of the primary source. Therefore, source interferences will appear as large fitting outliers that impact the accuracy of Radon denoising using the conventional least squares ($\ell_2$) misfit. Therefore, incorporating the robust $\ell_1$ norm misfit into the Radon transform will make more efficient denoising in common receiver gathers. We tested this hypothesis using numerically blended simple synthetics and field data from the Gulf of Mexico. Our tests showed that robust inversion is more effective in removing source interferences while preserving weak signals than the non-robust inversions. Robust inversion proved to be more significant in field data examples than the model sparsity constraint. This is due to the mismatch between the transform basis and the seismic reflections that limited the model sparsity. This mismatch arises from the coarse sampling of the Radon parameters, such as velocity and apex location. Radon transform parameters were coarsely sampled to reduce the computational cost and limit the model space.

In our tests, we used the apex-shifted hyperbolic Radon transform (ASHRT) as a denoising tool. The ASHRT transform basis functions resemble reflection hyperbolas in common receiver gathers. Tailoring the transform basis functions to match the data improves the transform resolution.  However, the ASHRT transform have high computational cost due to the extension of the model dimension.  Additionally, the ASHRT operator is a time variant Radon operator and cannot be computed efficiently in the frequency domain. Therefore, we designed a new ASHRT operator that uses the Stolt migration/demigration operators as its kernel. The computational efficiency of the Stolt operator results from employing Fast Fourier Transform (FFT) operators. Moreover, the band-limited structure of seismic data in the $f-k$ domain speeds up the Stolt $f-k$ mapping. We extended the single velocity Stolt operator to scan for multiple velocities and construct an ASHRT model. The Stolt-based ASHRT achieved better denoising accuracy. This could be attributed to the better matching between the operator basis and seismic reflections. The Stolt-based ASHRT operator includes a scaling factor that changes the amplitude with the incidence angle (obliquity factor). Again, we tested the Stolt-based ASHRT operator using numerically blended simple synthetics and field data from the Gulf of Mexico. Our tests showed that the Stolt-based ASHRT operator used in robust inversion can achieve both accurate and fast source separation. 

The Stolt-based ASHRT computational efficiency allows us to extend the model dimension in order to account for diffractions. Seismic diffractions have hyperbolic travel times that are shifted in time. The new AASHRT basis functions account for the time shift of the diffraction hyperbola. This new transform is used to interpolate data that data contain significant amount of seismic diffractions. The results of the interpolation tests show that the AASHRT transform can be a powerful tool interpolating seismic diffractions.

\section{Future developments}
  
The Stolt-based ASHRT transform can also be use in the inversion-based source separation. In this case, the misfit between modelled and observed simultaneous seismic sources data will not contain source interferences. This will allow us to only use the sparse model constraint without the need for a robust misfit function. This can be useful since imposing sparsity and robustness simultaneously is rather difficult, especially for complicated field data.   Also, the Stolt-based ASHRT can be used in compressive sensing algorithms  \citep{Wason2011,Chengbo2013b,Kumar2015} to separate and interpolate sources simultaneously. The Stolt-based ASHRT basis functions can be tailored to match seismic data closely which can increase the model sparsity that is needed in compressive sensing applications. The Stolt-based ASHRT and AASHRT transforms can also be used in combination with the non-uniform Fourier transform \citep{Dutt1993}. This can be used to interpolate/denoising seismic data with irregular sampling.

\section{Final remarks}

This thesis investigates the different inversion strategies for Radon transforms and its application in denoising source interferences and interpolating missing traces. 
Despite being popular in seismic processing, the majority of Radon algorithms use only time invariant Radon operators such as linear and parabolic.
This is mainly due to the slow implementation of the time variant Radon operators such as the hyperbolic Radon transform and apex shifted Radon transforms.  
However, these operators have basis functions that can be tailored to match seismic data more closely. 
We presented two Stolt-based transforms that are designed to match the data closely and can be computed efficiently. 
I believe these new transforms can be a very useful tool for many seismic data processing applications.

\makebib{references}
\end{document}